\pdfoutput=1
\documentclass[preprint2]{aastex61}
\received{December 7, 2016}
\revised{May 25, 2017}
\accepted{May 26, 2017}

\submitjournal{ApJ}

\shorttitle{Dust in the Winds of Proto-planetary Nebulae}
\shortauthors{Arneson et al.}

\begin{document}

\title{A SOFIA FORCAST Grism Study of the Mineralogy of Dust in the Winds of Proto-planetary Nebulae: RV Tauri Stars and SR\MakeLowercase{d} Variables}

\correspondingauthor{Ryan Arneson}
\email{arneson@astro.umn.edu}

\author{R. A. Arneson}
\affil{Minnesota Institute for Astrophysics, School of Physics and Astronomy, University of Minnesota, 106 Pleasant Street S.E., Minneapolis, MN 55455, USA}

\author{R. D. Gehrz}
\affil{Minnesota Institute for Astrophysics, School of Physics and Astronomy, University of Minnesota, 106 Pleasant Street S.E., Minneapolis, MN 55455, USA}

\author{C. E. Woodward}
\affil{Minnesota Institute for Astrophysics, School of Physics and Astronomy, University of Minnesota, 106 Pleasant Street S.E., Minneapolis, MN 55455, USA}

\author{L. A. Helton}
\affil{USRA-SOFIA Science Center, NASA Ames Research Center, Moffett Field, CA 94035, USA}

\author{D. Shenoy}
\affil{Minnesota Institute for Astrophysics, School of Physics and Astronomy, University of Minnesota, 106 Pleasant Street S.E., Minneapolis, MN 55455, USA}

\author{A. Evans}
\affil{Astrophysics Group, Lennard Jones Laboratory, Keele University, Keele, Staffordshire ST5, 5BG, UK}

\author{L. D. Keller}
\affil{Department of Physics and Astronomy, 264 Center for Natural Sciences, Ithaca College, Ithaca, NY 14850, USA}

\author{K. H. Hinkle}
\affil{National Optical Astronomy Observatory, P.O. Box 26732, Tucson, AZ 85726, USA}

\author{M. Jura}
\altaffiliation{\footnotesize Michael Jura died on 30 January, 2016 while this manuscript was being drafted.  He participated in writing the proposals to gather the data, and was aware of the importance of the results at the time of his death.}
\affiliation{Department of Physics and Astronomy, University of California, Los Angeles, CA 90095, USA}

\author{T. Lebzelter}
\affil{Institute for Astrophysics (IfA), University of Vienna, T{\"u}rkenschanzstrasse 17, 1180 Vienna, Austria}

\author{C. M. Lisse}
\affil{Solar System Exploration Branch, Space Department, Johns Hopkins University Applied Physics Laboratory, Laurel, MD 20723, USA}

\author{M. T. Rushton}
\affil{Astronomical Institute of the Romanian Academy, Str. Cutitul de Argint 5, Bucharest, Romania, 040557}

\author{J. Mizrachi}
\affil{Biomedical Engineering Department, Stony Brook University, Stony Brook, NY 11794, USA}

\begin{abstract}
We present a SOFIA FORCAST grism spectroscopic survey to examine the mineralogy of the circumstellar dust in a sample of post-asymptotic giant branch yellow supergiants that are believed to be the precursors of planetary nebulae.  Our mineralogical model of each star indicates the presence of both carbon rich and oxygen rich dust species--contrary to simple dredge-up models--with a majority of the dust in the form of amorphous carbon and graphite.  The oxygen rich dust is primarily in the form of amorphous silicates.  The spectra do not exhibit any prominent crystalline silicate emission features.  For most of the systems, our analysis suggests that the grains are relatively large and have undergone significant processing, supporting the hypothesis that the dust is confined to a Keplerian disk and that we are viewing the heavily processed, central regions of the disk from a nearly face-on orientation.  These results help to determine the physical properties of the post-AGB circumstellar environment and to constrain models of post-AGB mass loss and planetary nebula formation.
\end{abstract}


\section{Introduction}
\label{Intro}
RV Tauri and yellow semi-regular (SRd) variables are two classes of post-asymptotic giant branch (post-AGB) stars that lie along the horizontal track on the Hertzsprung-Russell (H-R) diagram linking AGB stars to planetary nebulae (PNe).  They are thought to be the immediate precursors of PNe and have been termed ``proto-planetary nebulae" (PPNe)\footnote{\footnotesize The phrase ``proto-planetary" is also widely used by the exoplanetary and planet formation communities to refer to dusty disks around young stars.  Also note that in the literature the terms preplanetary or proto-planetary nebulae have been used interchangeably with the term post-AGB objects.}.  RV Tauri stars are characterized by semi-regular, bimodal variability (possibly resulting from interaction with a binary companion; \citealp{Waelkens93, Percy93, Fokin94}), a high mass-loss rate, and often a prominent infrared (IR) excess.   SRd variables are similar to RV Tauri stars in many respects but are probably single star systems, as indicated by the absence of regular pulsations \citep{Percy06}.

RV Tauri stars are a loosely defined subclass of Population II Cepheid variables named after the prototype RV Tau.  They are defined as luminous (I-II) mid-F to K supergiants with a typical mass of $\sim 0.7\ \rm{M_{\sun}}$ \citep{Tuchman93, Fokin94} that show alternating deep and shallow minima in their light curves \citep{Preston63}.  They have formal periods (defined as the time between successive deep minimia) between 30 and 150 days, but cycle-to-cycle variability is common, and the amplitudes may reach up to 4 magnitudes in $V$ \citep{Percy93}.  RV Tauri stars are divided into two photometric classes (\lq{a}\rq\ and \lq{b}\rq) based on their light curves \citep{Kukarkin58}.  The RVa class contains constant mean magnitude stars, and the RVb  class contains stars that have a varying mean magnitude with a period of 600 to 1500 days.  There are several possible explanations for these light variations.  One explanation for the alternating minima is that there is a resonance between the fundamental period and the first overtone \citep{Takeuti83, Shenton92, Tuchman93, Fokin94}.  Another possibility is that the light variations are due to a geometrical projection effect where the pulsating star is periodically obscured by  a circumbinary disk \citep{vanWinckel99, Maas02}.

\citet{Preston63} classified the RV Tauri stars into three spectroscopic classes (\lq{A}\rq, \lq{B}\rq, \lq{C}\rq).  RVA stars are spectral type G--K, and show strong absorption lines and normal CN or CH bands while TiO bands sometimes appear at photometric minima.  RVB stars are generally hotter spectral types, weaker lined, and show enhanced CN and CH bands.  RVC stars are also weak lined but show normal CN and CH molecular bands.  There is no correlation between the photometric and spectroscopic classes.

It has long been known \citep{Gehrz70,Gehrz72a,Gehrz72b} that some RV Tauri stars (e.g. AC Her, U Mon, R Sct, R Sge) show very strong thermal IR emission from circumstellar dust.  Observations by the Infrared Astronomical Satellite (IRAS) confirmed these previous detections and increased the sample size.  IRAS detected considerable cool, circumstellar dust around many of the RV Tauri stars, which has been interpreted as being due to strong, dusty mass-loss during AGB evolution \citep{Jura86}.  From CO observations, \cite{Alcolea91} suggest a mass loss rate of $4 \times 10^{-7}$ to $7 \times 10^{-6}\ \rm{M_{\sun} yr}^{-1}$ within the last $10^2$ to $10^3$ years for most of the RV Tauri stars.  Because of their position on the H-R diagram, variability, high mass-loss rate and rarity (about 110 are known), it is generally believed that RV Tauri stars represent a relatively short-term, unstable transitional phase between the AGB and PNe phases of solar-mass stars.  If RV Tauri stars are assumed to be evolving on the post-AGB track, models predict that they spend about 200 years as variables and take about 1000 years to go from AGB to PNe \citep{Schoenberner83, Percy91, Fokin94}, however some (e.g. R Sct) are ``lazy", spend more time in the RV Tauri stage as evidenced by their mass loss history, have higher CO emission, and show a relatively low IR excess in the 5-20 \micron\ range but a larger IR excess for $\lambda \gtrsim 50$ \micron\ \citep{Bujarrabal88, Alcolea91}.  The lack of a near-IR excess and the presence of a large mid- and far-IR excess is evidence for a thick and extended dust envelope that is relatively cool.  The detection of SiO around R Sct could be an indication of on-going, weak mass loss \citep{Bujarrabal89}.  \citet{Alcolea91} estimate that $\sim$ 1/5 RV Tauri stars exhibit this ``lazy" evolutionary behavior.

Interestingly, RV Tauri stars do not have the same high C and s-process overabundances that are characteristic of other post-AGB stars.  Instead, their photospheres are ``depleted" of refractory elements \citep{Gonzalez97a, Gonzalez97b, Giridhar98, Giridhar00, Giridhar05, vanWinckel98, Molster02a, Molster02b, Molster02c, Maas02, Maas05, Deroo05, Gielen07, Gielen09, Rao14}.  This phenomenon is not fully understood, but is apparently due to a chemical rather than a nucleosynthetic process \citep{vanWinckel03}.  The refractory elements, which have a high dust condensation temperature, are separated from the volatiles as the condensed grains are driven away by radiation pressure in the circumstellar environment.  The depleted gas is then re-accreted onto the stellar photosphere leaving it with a peculiar composition similar to that of the depleted gas in the interstellar medium (ISM) \citep{Hinkle07}.  \citet{Waters92} suggested that these abundance anomalies are more likely to occur when the dust is trapped in a circumstellar disk.  These same depletion patterns are also observed in binary post-AGB stars with circumbinary disks \citep{vanWinckel95}.  This has led to the suggestion that the depleted RV Tauri stars are also binaries with a circumstellar disk \citep{vanWinckel99}.  In their study, \citet{deRuyter06} included many RV Tauri stars in the class of post-AGB binaries with disks and suggested that the thermal IR spectrum originates in the Keplerian circumstellar disk.  The narrow velocity width lines of the $\rm{^{12}CO\ and\ ^{13}CO\ J = 2\rightarrow1}$ and $\rm{J = 1\rightarrow0}$ transitions of AC Her, V441 Her, and R Sct and the $\rm{^{12}CO\ J = 3\rightarrow2}$ and $\rm{J = 2\rightarrow1}$ transitions of AR Pup are indicative of a Keplerian disk \citep{Bujarrabal13, Bujarrabal15}.  The trapping of grains in a disk could provide the environment needed to enable grains to grow to sizes significantly larger than those in the ISM or in the stellar outflows of single stars \citep{Shenton95, deRuyter05}.  The similar class of SRd stars, however, does not show evidence of disks and may evolve from single stars.

Depending on the oxygen and carbon abundances in the circumstellar gas, either oxygen- or carbon-rich dust is formed.  It has been found that many of the RV Tauri stars have very weak CO $\rm{J = 2\rightarrow1}$ lines, unlike the stronger CO $\rm{J = 2\rightarrow1}$ lines characteristic of other, more massive, post-AGB stars \citep{Alcolea91, He14, Bujarrabal15}.  It is possible that the unique circumstellar environment of RV Tauri stars either surpresses the formation of CO molecules (possibly through UV photodissociation) or underexcites them \citep{Alcolea91, McDonald15}.  However, the weakness may be due to the compactness of the disk, as has been suggested for the case of AC Her \citep{Bujarrabal15}.  In oxygen-rich gas, dust such as olivine ($\rm{Mg_{2(1-x)}Fe_{2x}SiO_4}$) and pyroxene ($\rm{Mg_{1-x}Fe_xSiO_3}$) is formed.  In carbon-rich gas, carbon rich dust particles such as SiC, amorphous carbon, and possible polycyclic aromatic hydrocarbons (PAHs) are formed.  

The amount of crystalline grain material compared to amorphous grain material is generally low, $\sim$10--15\%, and is dominant only in rare cases.  Crystalline material is also generally only detected in stars that have experienced high mass loss rates (above $10^{-5} \rm{M_{\sun} yr}^{-1}$) \citep{Cami98,Sylvester99, Sogawa99, Suh02}.  However, both crystalline and amorphous grains have been detected simultaneously in stellar outflows, both in the present study as well as in others.  \citet{Gielen08,Gielen09} showed that dust processing in circumstellar disk environments is conducive to creating large, crystalline grains.  As the disk is subjected to the hard radiation and stellar wind from the central source, the dust crystallization fraction increases \citep{Gielen11} and the disk dissipates \citep{Kastner04,Gezer15, Kastner16, Lisse17}. Thus, the IR excess associated with the dusty disk diminishes as the system ages and transitions to a PN.

In this work we present 5--40 \micron\ IR spectra on a diverse sample of RV Tauri and SRd variables from a grism spectroscopic study of suspected proto-planetary nebula precursors with the Faint Object infraRed CAmera for the SOFIA Telescope (FORCAST; \citealp{Herter12}) instrument on board the NASA Stratospheric Observatory for Infrared Astronomy (SOFIA; \citealp{Becklin07, Gehrz09, Young12}). With this rich data set, we produce spectroscopic sampling of these objects in the mid-IR.  By modeling the emission we can determine the source of the IR-excess, identify the dust species present and quantify fundamental dust properties, such as the grain size distribution and dust temperature. These parameters help to determine the physical properties of the post-AGB circumstellar environment and to constrain models of post-AGB mass loss and planetary nebula formation.

In Section \ref{program} we summarize the stars observed by our program.  An overview of the observations and data reduction strategies is given in Section \ref{obs}.  Section \ref{sed} contains the construction of the spectral energy distributions and the spectral decomposition model we used to measure the mineralogy of the program stars.  The results of our model, a discussion of our results and our conclusions are presented in Sections \ref{results}--\ref{conclusion}.

\section{Program Stars}
\label{program}
We have selected a sample of RV Tauri and related SRd stars based upon: 1) their availability for SOFIA flights from Palmdale, CA and Christchurch, NZ, 2) diversity of their IR spectral energy distributions (SEDs), 3) our ability to obtain a signal-to-noise ratio compatible with our science objectives in a reasonable integration time.  

The properties of the 18 RV Tauri and SRd variables presented in this work are summarized in Table \ref{tab:program}.  TX Per is sometimes categorized as an RV Tauri star and sometimes as an SRd variable.  We concur with \citet{Percy05}, which refers to TX Per as being a ``mild" RV Tauri as the consecutive minima are very similar in depth, and categorize TX Per as an RV Tauri variable.  

The mineralogy of many of these systems has been studied previously \citep{Molster02c, Deroo06, Gielen07, Gielen11, Blommaert14, Hillen15}.  These studies have mostly focused on crystalline silicates.  Most of the studies found evidence for large, crystalline dust grains indicative of highly processed material.  Some of the studies suggest that the crystalline and amorphous silicates are at different temperatures suggesting that the two species are spatially separated and have different formation histories. 

\begin{deluxetable*}{cccccccccccc}
\tablecaption{\label{tab:program}Properties of the RV Tauri and SRd Variables in this Survey}
\tablewidth{\textwidth}
\tabletypesize{\scriptsize}
\tablecolumns{12}
\tablehead{\colhead{Name} & \colhead{Type} & \colhead{Spectral Type} & \colhead{Period (d)\tablenotemark{a}} & \colhead{$\rm{[Fe/H]_0}\tablenotemark{b}$} & \colhead{PC\tablenotemark{c}} & \colhead{SC\tablenotemark{c}} & \colhead{SED\tablenotemark{d}} & \colhead{$\rm{T_{eff} (K)}$} & \colhead{Binarity\tablenotemark{e}} & \colhead{Chemical Type\tablenotemark{f}} & \colhead{Ref.}}

\startdata
TW Cam &  RV &  F8IbG8Ib &  87 &  -0.40 &  a &  A &  Disk &  4800 &   &   &  1\\
UY CMa &  RV &  G0 &  114 &  -0.50 &  a &  B & &  5500 &   &   &  2\\
$\rm{o^1\ Cen}$ &  SRd &  G3Ia0 &  200 &   &   &   &   &   &   &   &  3\\
RU Cen &  RV &  A7IbG2pe &  65 &  -1.10 &  a &  B &  Disk &  6000 &  Y &   &  4, 5\\
SX Cen &  RV &  F5G3/5Vp &  33 &  -0.30 &  b &  B &  Disk &  6250 &  Y &   &  4, 5\\
SU Gem &  RV &  F5M3 &  50 &  0.00 &  b &  A &  Disk &  5250 &   &   &  6\\ 
AC Her &  RV &  F2pIbK4e &  75 &  -0.90 &  a &  B &  Disk &  5900 &  Y &  O &  7\\
V441 Her &  SRd &  F2Ibe &  70 &   & & &  Disk &   &  Y &  O &  8, 9\\
U Mon &  RV &  F8IbeK0pIb &  91 &  -0.50 &  b &  A &  Disk &  5000 &  Y &  O &  1, 10\\
CT Ori &  RV &  F9 &  136 &  -0.60 &  a &  B &  Disk &  5500 &   &   &  10, 11\\
TV Per &  SRd &  K0 &  358 &   & & & &   &   &   &  12\\
TX Per &  RV &  Gp(M2)K0e(M2) &  78 &  -0.60 &  a &  A & &  4250 &   &   &  6\\
AR Pup &  RV &  F0IF8I &  76 &  0.40 &  b &  B &  Disk &  6000 &   &  O &  10, 13\\
R Sge &  RV &  G0IbG8Ib &  71 &  0.10 &  b &  A &  Disk &  5100 &   &   &  13\\
AI Sco &  RV &  G0K2 &  71 &  -0.30 &  b &  A &  Disk &  5300 &   &  C? &  2, 10\\
R Sct &  RV &  G0IaeK2p(M3)Ibe &  147 &  -0.20 &  a &  A &  Uncertain &  4500 &   &   &  1\\
RV Tau &  RV &  G2IaeM2Ia &  79 &  -0.40 &  b &  A &  Disk &  4500 &   &  C &  1\\
V Vul &  RV &  G4eK3(M2) &  76 &  0.10 &  a &  A &  Disk &  4500 &   &   &  2, 6 \\
\enddata
\tablenotetext{a}{Pulsation period in days}
\tablenotetext{b}{The estimated initial metallicity obtained via the Zn or S abundance \citep{Gezer15}}
\tablenotetext{c}{Photometric class (PC) and spectroscopic class (SC)}
\tablenotetext{d}{Spectral energy distribution classification from \citet{Gezer15}}
\tablenotetext{e}{Y indicates confirmed binarity based on radial velocity measurements.  Confirming binarity using this method is difficult because the photospheres of these variables have large amplitude radial pulsations.}
\tablenotetext{f}{Stellar chemical type from \citet{He14} and references therein}
\tablerefs{(1) \citet{Giridhar00}; (2) \citet{Giridhar05}; (3) \citet{O'Connell61}; (4) \citet{Maas02}; (5) \citet{Maas05}; (6) \citet{Rao14}: (7) \citet{Giridhar98}; (8) \citet{Waters93}; (9) \citet{deRuyter06}; (10) \citet{Kiss07}; (11) \citet{Gonzalez97b}; (12) \citet{Payne-Gaposchkin52}; (13) \citet{Gonzalez97a}}

\end{deluxetable*}

\section{Observations and Data Reduction}
\label{obs}
The targets were observed with SOFIA during Guest Investigator (GI) Cycles 2, 3, and 4.  Descriptions of the SOFIA Observatory and its science instrument (SI) suite have been given by \citet{Becklin07}, \citet{Gehrz09}, and \citet{Young12}.  All of the targets in our survey were observed using FORCAST.

FORCAST is a dual-channel mid-IR camera and spectrograph operating between 5--40 \micron.  Each channel consists of a 256 $\times$ 256 pixel array that yields a 3.4\arcmin\ $\times\ $3.2\arcmin\ field-of-view with a square plate scale of 0.768\arcsec, after distortion correction. The Short Wave Camera (SWC) uses a Si:As blocked-impurity band (BIB) array optimized for $\lambda < 25$ \micron, while the Long Wave Camera's (LWC) Si:Sb BIB array is optmized for $\lambda > 25$ \micron.  Observations can be made through either of the two channels individually or, by use of a dichroic mirror, with both channels simultaneously across the entire range.  All of the observations presented in this work were taken in the single channel, long-slit mode.  We utilized FORCAST's suite of grisms, which provided low spectral resolution (R $\approx$ 200) over the 5--40 \micron\ range.  The following grisms were used for our observations: G1 covering 4.9--8.0 \micron, G3 covering 8.4--13.7 \micron, G5 covering 17.6--27.7 \micron, and G6 covering 28.7--37.1 \micron.  All of the observations were taken in the ``nod match chop" mode (C2N) which used a chop throw of 30\arcsec, a chop angle of either 0\degr\ or 30\degr, and no dithering.

The data were reduced by the SOFIA Science Center using FORCAST Redux v1.5.0 and v1.2.0 pipeline versions \citep{Clarke15} and released to the authors as level 3 results.  We stacked the spectra when there were multiple observations of a given target.  We did not use any of the data points between 9.19--10.0 \micron\ as these are strongly affected by telluric ozone absorption.  The spectra were smoothed with a 3 point un-weighted boxcar to emphasize spectral features.

\section{Spectral Energy Distributions of the Survey Objects}
\label{sed}
\begin{figure*}
\gridline{\leftfig{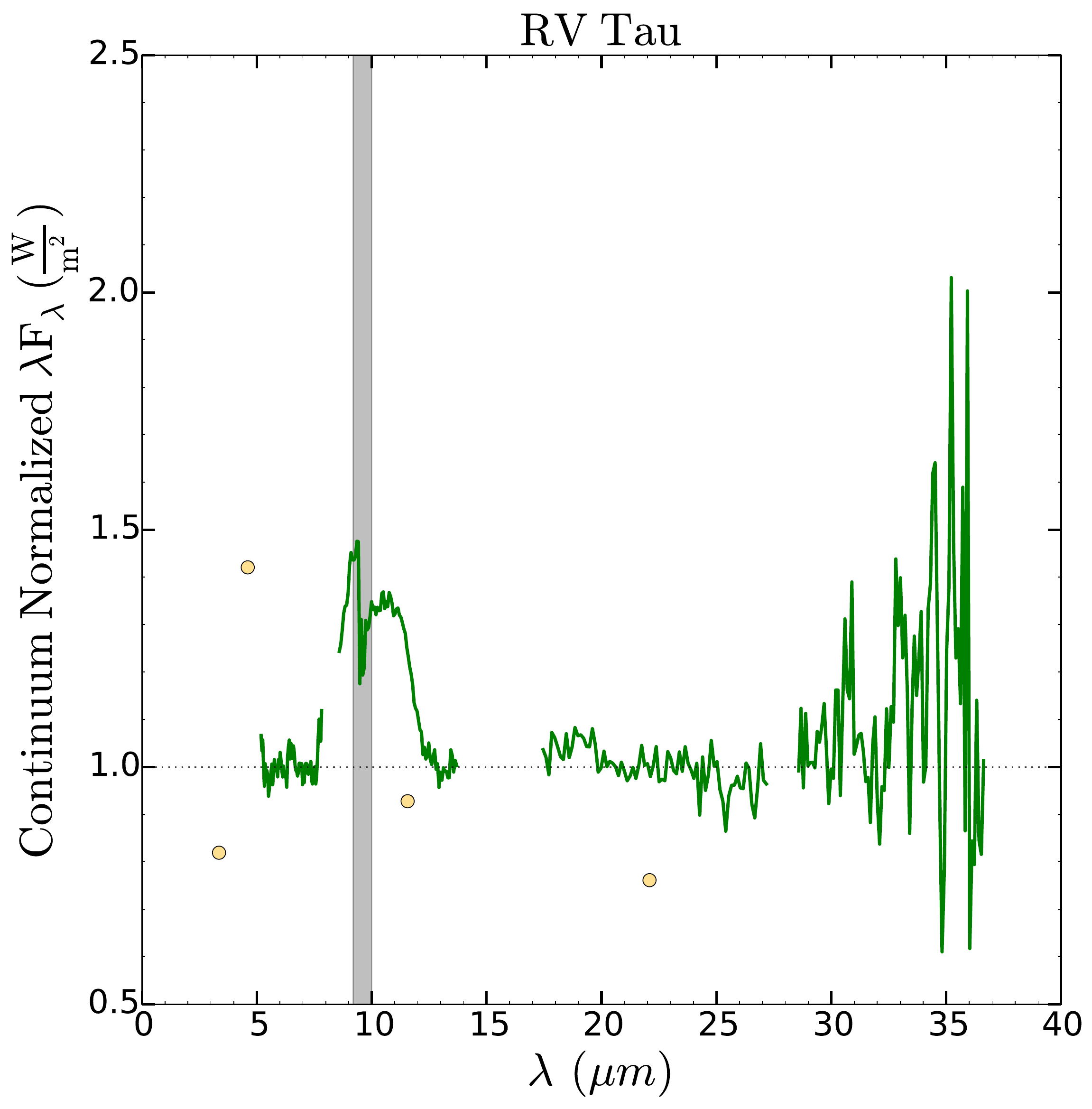}{0.5\textwidth}{(q) RV Tau}
	     \rightfig{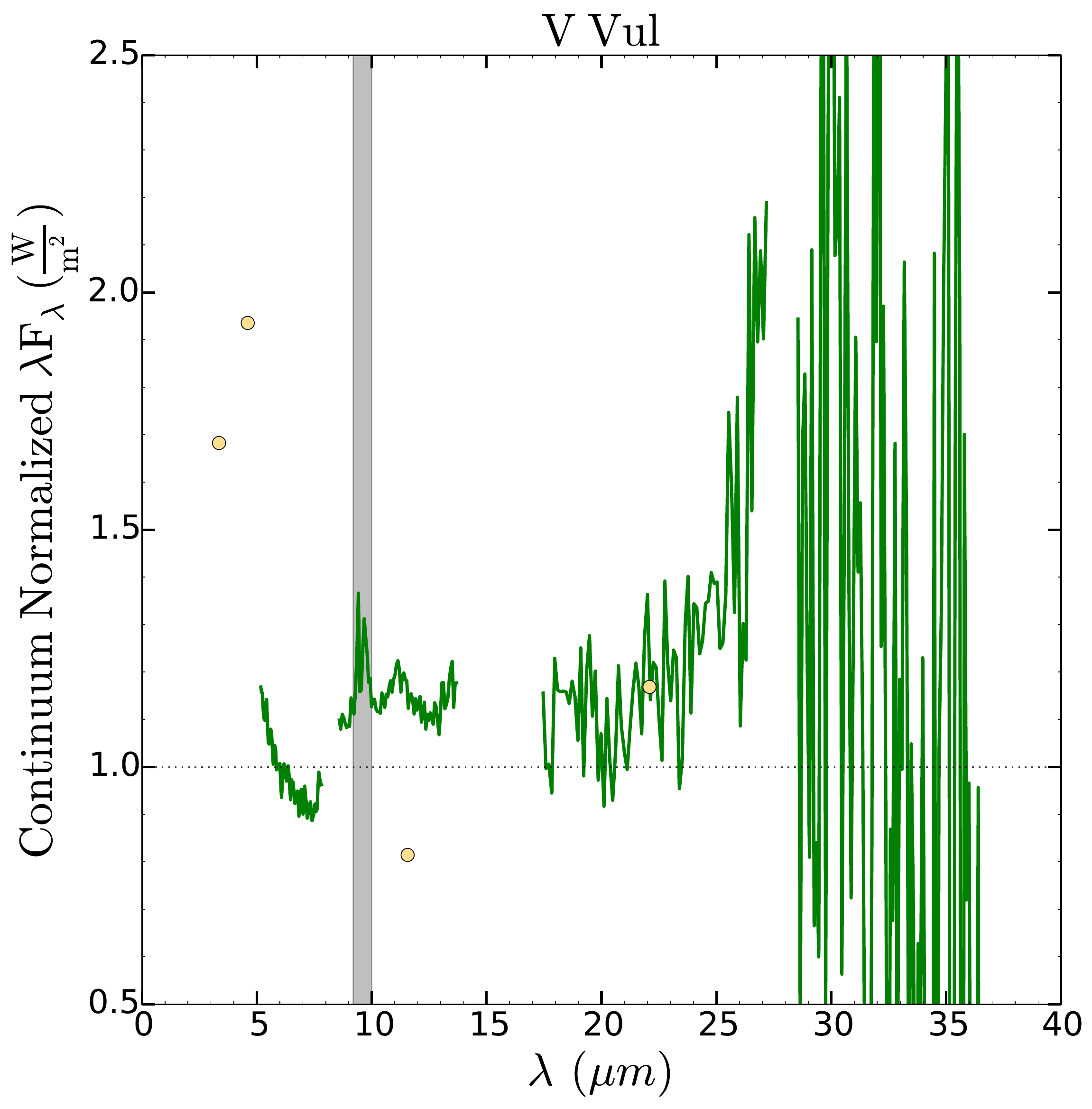}{0.5\textwidth}{(r) V Vul}}
\caption{Continuum normalized SOFIA FORCAST spectrum (green curve) after dividing by the best fitting continuum of our sample of stars showing the WISE archival photometry (yellow points) and telluric ozone region (gray band).}
\label{spectra}
\end{figure*}

We present the IR SEDs of the survey objects in Figure~\ref{data}.  As can be seen in Figure \ref{spectra}, all of the program stars except for TX Per were observed with the G1, G3, and G5 grisms, and only 7 of the 18 stars were also observed with the G6 grism.  TX Per was fainter than expected and we were only able to obtain the G1 grism spectrum (4.9--8.0 \micron).  We were unable to model the spectral energy distribution of TX Per with such a limited wavelength range.  The continuum normalized SOFIA grism spectra are shown in Figure~\ref{spectra}.  We have also gathered archival broadband IR photometry for comparison with the SOFIA spectra.  Photometry from the Two Micron All-Sky Survey (2MASS; \citealp{Skrutskie06}) at 1.25, 1.65, and 2.17 \micron, the AKARI satellite \citep{Murakami07} at 9 and 18 \micron, the Wide-field Infrared Survey Explorer (WISE; \citealp{Wright10}) at 3.4, 4.6, 12 and 22 \micron, and the Infrared Astronomical Satellite (IRAS; \citealp{Neugebauer84}) at 12, 25, 60, and 100 \micron\ were all taken from the  NASA/IPAC Infrared Science Archive (IRSA; \citealp{Berriman08}) database.  Photometry from the Herschel \citep{Pilbratt03} Photoconductor Array Camera and Spectrometer (PACS; \citealp{Poglitsch10}) at 70, 100, and 160 \micron\ and Spectral and Photometric Imaging Receiver (SPIRE; \citealp{Griffin10}) at 250, 350, and 500 \micron\ were obtained by using aperture photometry after sky background subtraction.  The archival photometry data points were \emph{not} used in any of the least squares fitting routines and are only plotted to visualize the SED of each star. 

\subsection{Dust Species}
Previous studies have shown that the most common dust species present in circumstellar environments are amorphous and crystalline silicates with olivine and pyroxene stoichiometries \citep{Molster02a, Molster02b, Molster02c, Gielen11}.   Amorphous olivine has very prominent broad features around 9.8 and 18 \micron.  These features arise from the Si-O stretching and O-Si-O bending modes.  Amorphous pyroxene shows a 10 \micron\ feature similar to that of amorphous olivine, but shifted towards shorter wavelengths.  Crystalline forsterite has prominent emission features at 11.2, 23.7, and 33.7 \micron.  Many of the stars in our sample are known to be oxygen rich (see Table \ref{tab:program}) however an increased
abundance of carbon dredged up as the star evolves is possible \citep{Iben81, Chan90}.  Models suggest that stars with main sequence masses less than 1.5 $\rm{M}_\sun$ do not experience third dredge-up.  Therefore, the surface composition of these stars is fixed by the first dredge-up and red giant branch extra-mixing, and they are oxygen rich AGB stars.  For main sequence stars with masses over 6 $\rm{M}_\sun$ the stars undergo shallow third dredge-up episodes.  Stars in the intermediate main sequence mass range undergo repeated third dredge-up episodes that bring carbon to the surface to become carbon stars (for a review see \citealt{Straniero06, Karakas14}).  The interpulse time for the third dredge-up is a few times $10^4$ years with carbon stars undergoing several pulses after the carbon abundance first exceeds the oxygen abundance \citep{Straniero97}.  The interpulse time exceeds the expansion time for circumstellar shells and as expected nearly all normal AGB stars have shells of the same composition as the star.  Carbon rich AGB stars are expected to have circumstellar shells dominated by amorphous carbon or graphite grains with some silicon carbide (SiC) possibly present \citep{Suh00, Speck05, Speck09}.  Amorphous carbon does not have prominent IR features but contributes to the dust continuum emission, however, graphitic carbon and silicon carbide have emission features at 11.53 \micron\ and in the 10--13 \micron\ region, respectively.  The strong depletion of iron in the photospheres of RV Tauri stars suggests that metallic iron may be present in the circumstellar environment.  Iron grains can form at temperatures 50--100 K lower than silicates and are stable in O-rich environments above 700 K \citep{Kemper02}.  Like amorphous carbon, metallic iron lacks prominent IR features and contributes to the overall dust continuum.  Therefore, the dust species we included in our model are crystalline forsterite, amorphous olivine, amorphous pyroxene, amorphous carbon, silicon carbide, graphite, and metallic iron.  We only considered the magnesium-rich crystalline species of olivine (forsterite) as it has been found that the iron content of circumstellar crystalline olivine around evolved stars is lower than $10\%$ \citep{Tielens98, Molster02c}.  In addition, we only included amorphous enstatite and pyroxene with an iron content of 0\% as these species were found to be better fits to the spectra than the same species with an iron content of 50\%.  We tried including Mg-righ crystalline olivine, Mg-rich crystalline enstatite, crystalline bronzite, crystalline fayalite, iron oxide, amorphous alumina, and amorphous silica in our model, however none of these dust species were significantly present.  We discuss the addition of some of these species more in section \ref{limits}.

Mass absorption coefficients for the different dust species are calculated from optical constants using a homogeneous sphere approximation \citep{Min05}.  Although the continuous distribution of ellipsoids approximation (CDE; \citealp{Bohren83}) is widely used, it is only valid in the Rayleigh limit for small grain sizes.  Because we are interested in the grain size distribution we did not use the CDE approximation.  The details of the different optical constants that we used can be found in Table \ref{tab:minerals}.  In cases where the refractive index was reported for the three crystallographic directions, we assumed even distributions of each orientation.  There are many laboratory measurements of IR optical constants available, corresponding to different material compositions, crystal structures, annealing temperatures, measurement environments, grain sizes, and grain orientations.  These different measurements produce spectra with similar global features but with unique differences.  While the minerals we have chosen to use in our model may result in different relative abundances compared to those derived in previous studies, we are less concerned with making comparisons with those works and more concerned with drawing comparisons between the program stars in the present study.

\begin{deluxetable*}{cccccc}
\tablecaption{Dust Species and Properties Used in this Work \label{tab:minerals}}
\tablecolumns{6}
\tablehead{\colhead{Dust Species} & \colhead{Composition} & \colhead{Structure} & \colhead{Density (g/cm$^3$)} & \colhead{Grain Size (\micron)} & \colhead{Reference}} 

\startdata
Forsterite &  $\rm{Mg_2SiO_4}$ &  C &  3.27 &  0.1 &  \citet{Koike03}\\
Olivine &  $\rm{Mg_2SiO_4}$ &  A &  3.71 &  0.1, 2.0 &  \citet{Dorschner95}\\
Pyroxene &  $\rm{MgSiO_3}$ &  A &  3.20 &  0.1, 2.0 &  \citet{Dorschner95}\\
Carbon &  $\rm{Pyrolized\ at\ 400^\circ\ C}$ &  A &  1.435 &  0.1, 2.0 &  \citet{Jaeger98b}\\
Silicon Carbide &  $\alpha$-SiC &  C &  3.26 &  0.1, 2.0 &  \citet{Pegourie88}\\
Graphite &   &  C &  2.24 &  0.1, 2.0 &  \citet{Draine84}\\
Metallic Iron & Fe & C & 7.87 & 0.1, 2.0 & \citet{Pollack94}\\
\enddata

\tablecomments{The mineral structure is denoted as either amorphous (A) or crystalline (C).}
\end{deluxetable*}

\subsection{Spectral Decomposition Model}
\label{model}
To identify the minerals present and to quantify the grain size distributions, we fit the observed SOFIA spectra with synthetic spectra of various mineral species.  The synthetic spectra were calculated from the optical constants of each mineral.  The conversion from laboratory measured optical constants of dust to mass absorption coefficients is not straightforward.  Several factors affect the observed emission features, including the chemical composition of the dust, the grain size, and the grain shape \citep{Min03,Min05}.  We constructed a basic model to fit the full FORCAST wavelength range.  Given that we see the silicate features in emission, we assume that the dust features are in an optically thin part of the disk and, therefore, we approximate the spectrum as a linear combination of dust absorption profiles.  The emission model is given by
\begin{equation}
\label{flux_model}
\lambda F_\lambda \propto \sum_i{c_i \mu_i(\lambda)} \times \sum_j{a_j \lambda B_{\lambda}(T_j)},
\end{equation}
where $\mu_i(\lambda) (\rm{cm^{-1}})$ is the absorption coefficient of dust component $i$ and $c_i$ gives the volume fraction of that dust component, $B_{\lambda}(T_j)$ {$(\rm{W\,sr^{-1}\,m^{-3}})$} denotes the Planck function at temperature $T_j$ and $a_j$ is the scaling factor for the $j^{\rm{th}}$ Planck function.  The absorption coefficient is related to the mass absorption coefficient (opacity) by $\mu = \kappa\rho$ where $\kappa\,(\rm{cm^{2}\,g^{-1}})$ is the mass absorption coefficient (opacity) and $\rho\,(\rm{g\, cm^{-3}})$ is the density of the dust component.  The formulation of our model assumes that the stellar contribution to the SED in this range is negligible.  We further assume that all of the dust in a population is in thermal equilibrium with all of the other dust species, regardless of particle sizes or the ability to absorb and re-emit starlight.  We first used a least squares minimization to fit the Planck functions to the FORCAST continuum.  We fit the functions to the entire FORCAST wavelength range available except for the $8-12.5$ \micron\ range, which is dominated by silicate emission.  Two Planck functions were used for all of the spectral models except for $\rm{o}^1$ Cen and V Vul where only a single Planck function was needed to fit the underlying continuum.  The best fitting Planck function parameters are shown in Figure \ref{data} and summarized in Table~\ref{tab:data}, however, the sum of the Planck scaling factors, $a_j$, have been normalized to unity in order to represent the fraction of dust at each temperature $T_j$.  The best fitting Planck functions are plotted with the FORCAST spectra in Figure~\ref{data}.  It is probably not realistic to model the dust as a single temperature component or even as a two-temperature component--a temperature gradient is probably more realistic--but in order to keep the number of parameters at a minimum we only use a maximum of two Planck functions in our model.

After finding the best fitting Planck functions, the best fitting dust fraction coefficients, $c_i$, were found by a non-negative least squares minimization of the model to the entire observed FORCAST wavelength range.  The reduced $\chi^2$ of the spectrum is given by
\begin{equation}
\label{chi}
\resizebox{.41 \textwidth}{!}
{$\chi^2_{\rm{red}} = \frac{1}{N-M} \sum\limits_{i=1}^N \left | \frac{F_{\rm{model}}(\lambda_i) - F_{\rm{obs}}(\lambda_i)}{\sigma_i}\right |^2,$}
\end{equation}
where $N$ is the number of wavelength points, $M$ the number of fit parameters, $F_{\rm{model}}(\lambda_i)$ is the model flux at a given wavelength, $F_{\rm{obs}}(\lambda_i)$ is the observed flux at a given wavelength, and $\sigma_i$ the absolute error of the observed flux at each wavelength $\lambda_i$.  The reduced $\chi^2$ values are summarized in Table~\ref{tab:data}.

\begin{figure*}
\gridline{\leftfig{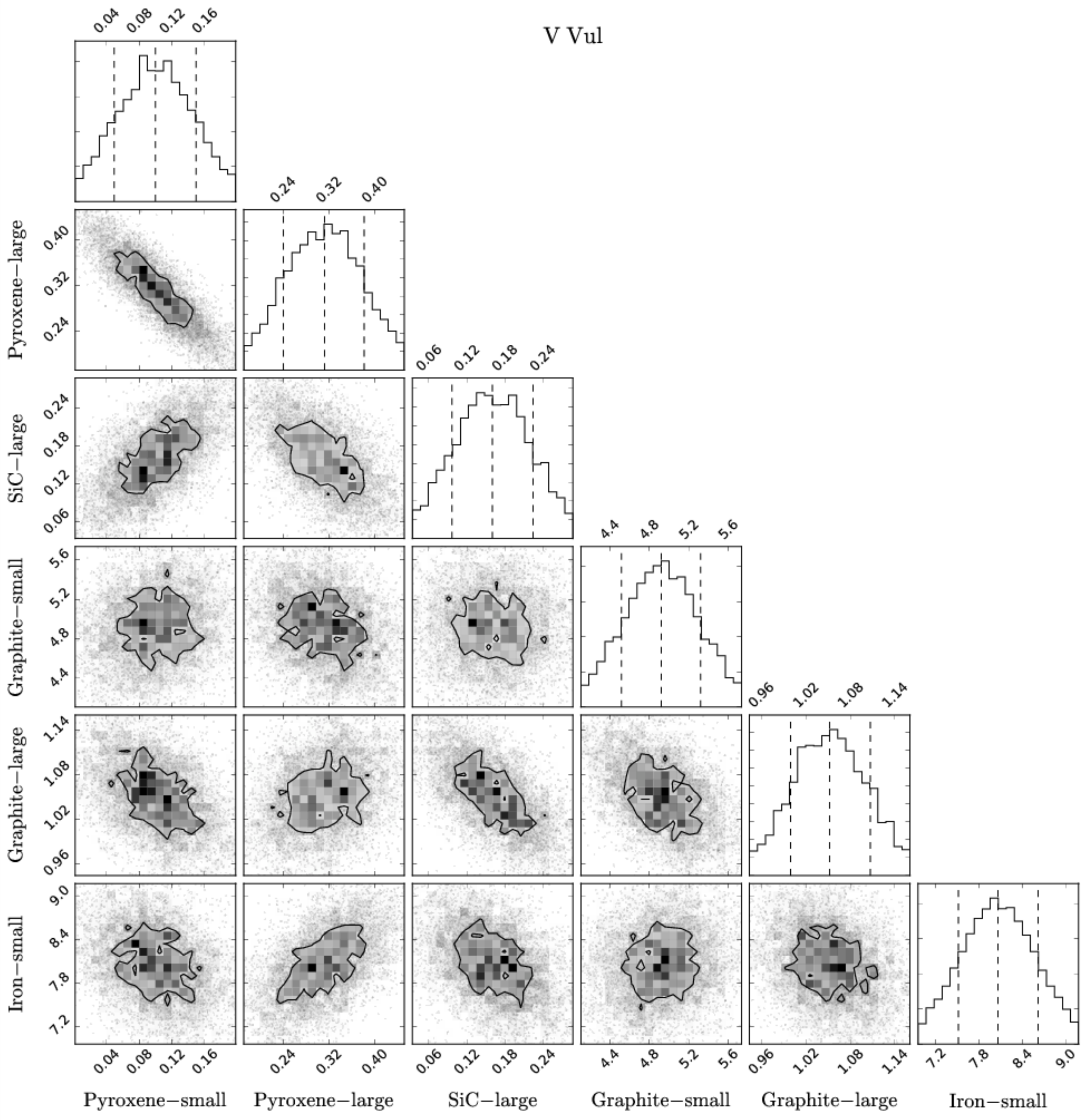}{0.5\textwidth}{(q) V Vul}}
\caption{Normalized probability distribution functions of the best fit coefficients, $c_i$, of each mineral species after 5000 realizations of a Monte Carlo simulation with Gaussian noise distributions and the covariance between the coefficients.  The dashed lines show the mean ($\overline{c_i}$) and the $1\sigma$ ($\sigma_{\overline{c_i}}$) confidence levels.  The contours show the $1\sigma$ confidence levels.}
\label{covariance}
\end{figure*}

Errors on the dust fraction coefficients were calculated from 5000 realizations of a Monte Carlo simulation with Gaussian noise distributions.  We omitted a mineral species from the fit if the error, $\sigma_{\overline{c_i}}$, on $\overline{c_i}$ is greater than the value of $\overline{c_i}$ itself.  Figure \ref{covariance} illustrates the distributions of the best fitting dust fraction coefficients, $c_i$, and the covariance between the coefficients  \citep{corner}.  Most of the coefficients show little to no correlation with the exception of the carbon, graphite, and metallic iron species which are strongly anti-correlated in most cases.  Even though this model is only an approximation, it gives a good fit overall to the observed spectra (see Figure \ref{spec_decomp}).  A full radiative transfer model that could account for a temperature gradient in the disk that may produce a better fitting result is beyond the scope of this work.
\\
\subsection{Grain Size Distribution}
To study the grain size distribution, we used two dust grain sizes in our model with radii of 0.1 and 2.0 \micron.  These sizes  were chosen based on the work of \citet{Bouwman01} and \citet{Honda04} which found that, in the 10 \micron\ spectral region, 0.1 \micron\ grains sufficiently describe grains with $a <$ 1.0 \micron\ while $1.5-2.0$ \micron\  grains sufficiently describe grains with $a >$ 1.5 \micron.  Larger sized grains were not considered as the emission features from larger grains become too weak to distinguish from the continuum emission.  Because the grains with radii of 2.0 \micron\ are in the Mie scattering regime, we used the python module \emph{pymiecoated} \citep{pymiecoated} to calculate the mass absorption coefficients for the 2.0 \micron\ grains.  \emph{pymiecoated} computes the scattering properties of single- and dual-layered spheres in the Mie regime using the results of \citet{Bohren83} and the optical constants of bulk materials.  The mass absorption coefficient for crystalline forsterite was taken directly from \citet{Koike03}.  Because we did not have the optical constants for this mineral we were unable to calculate the mass absorption coefficient for larger sized grains in the Mie scattering regime.

\section{Results}
\label{results}
\begin{figure*}
\gridline{\leftfig{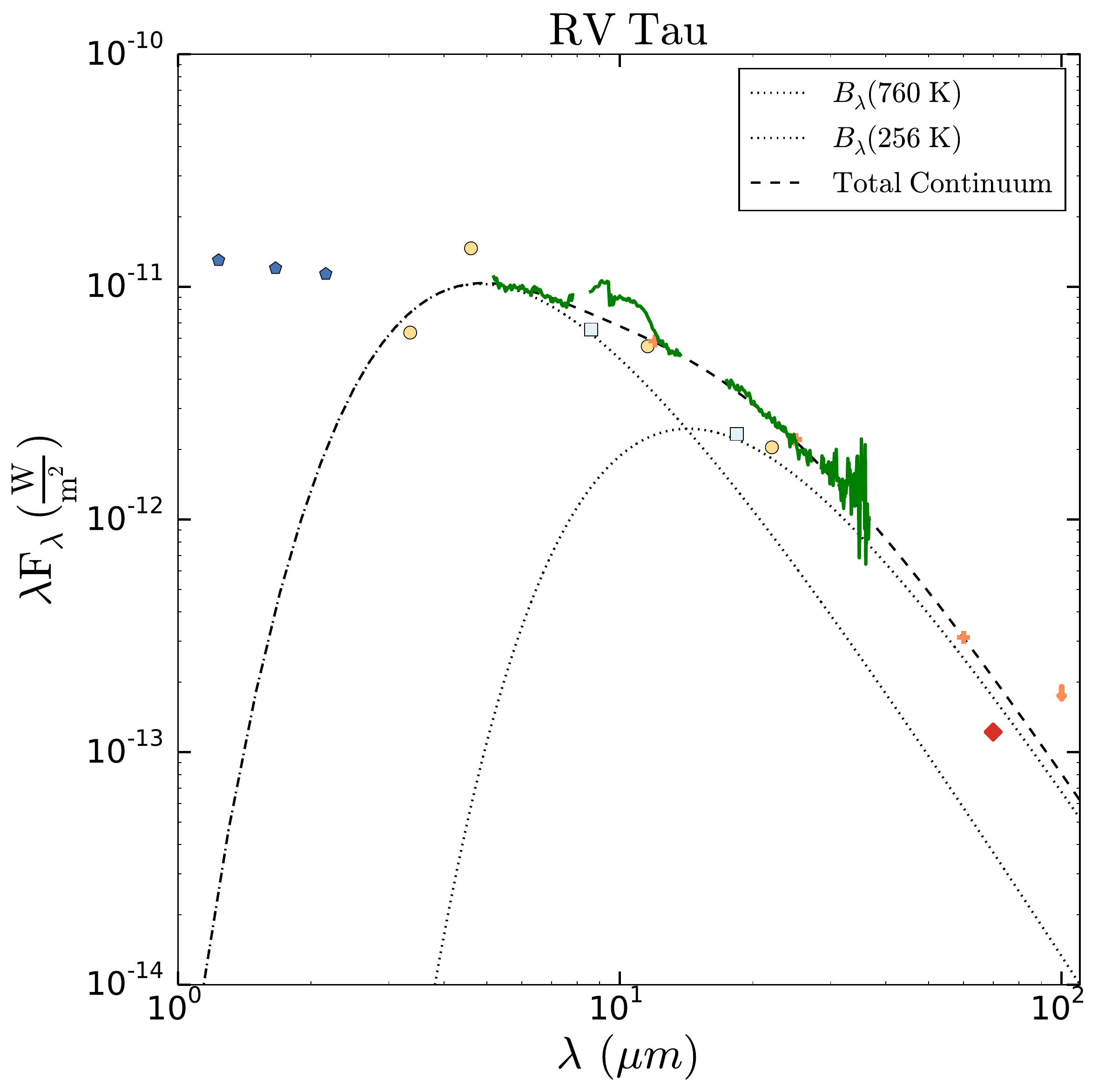}{0.5\textwidth}{(q) RV Tau}
	     \rightfig{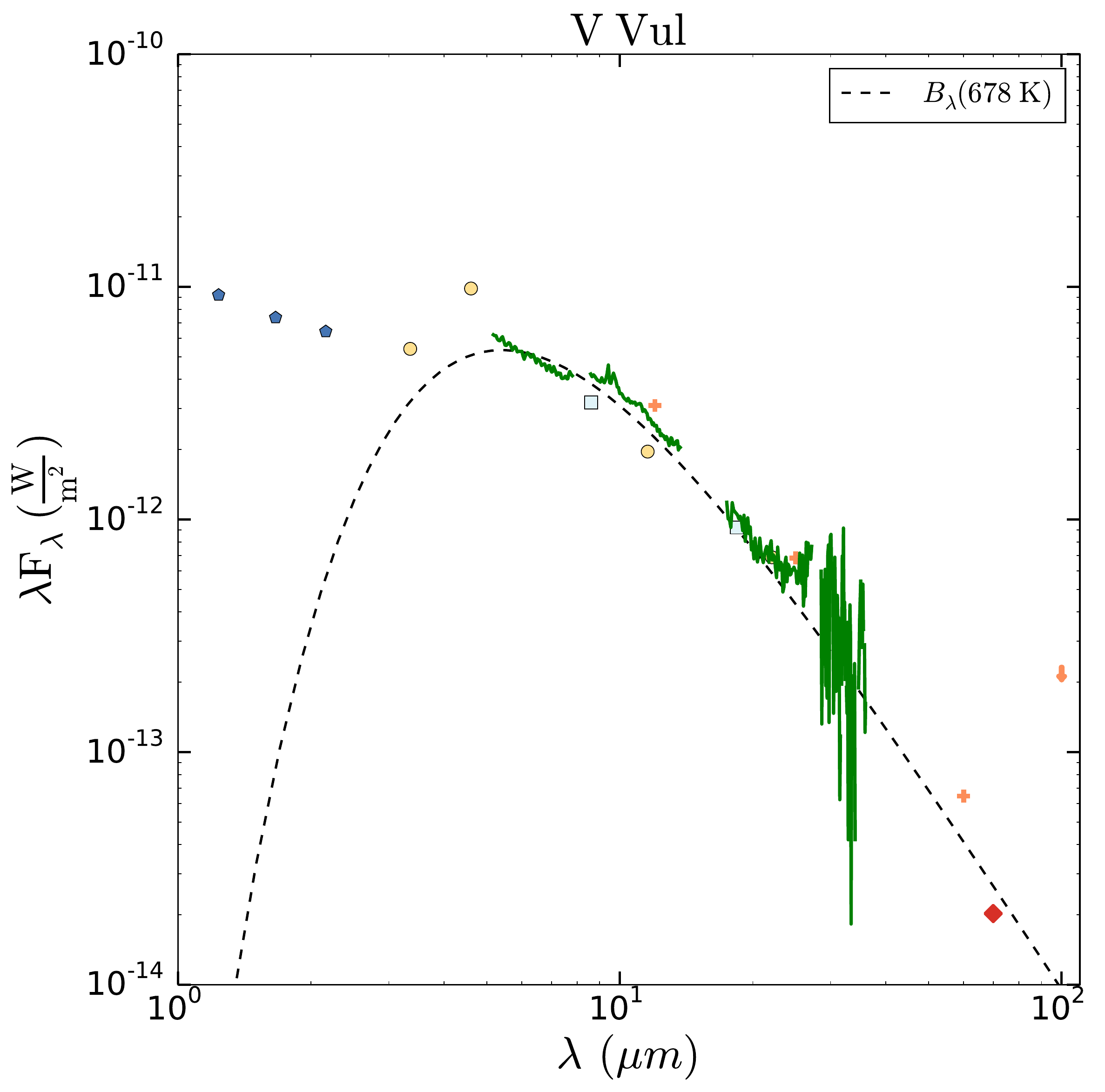}{0.5\textwidth}{(r) V Vul}}
\caption{The observed SOFIA FORCAST spectrum (green curve) of our sample of stars is plotted together with the archival photometry from 2MASS (blue pentagons), MSX (blue triangles), AKARI (light blue squares), WISE (yellow circles), IRAS (orange crosses), Herschel (red diamonds), and the best fitting Planck functions (black dotted and dashed curves).  WISE photometry upper limits are depicted as orange downward arrows.}
\label{data}
\end{figure*}

\begin{figure*}
\gridline{\leftfig{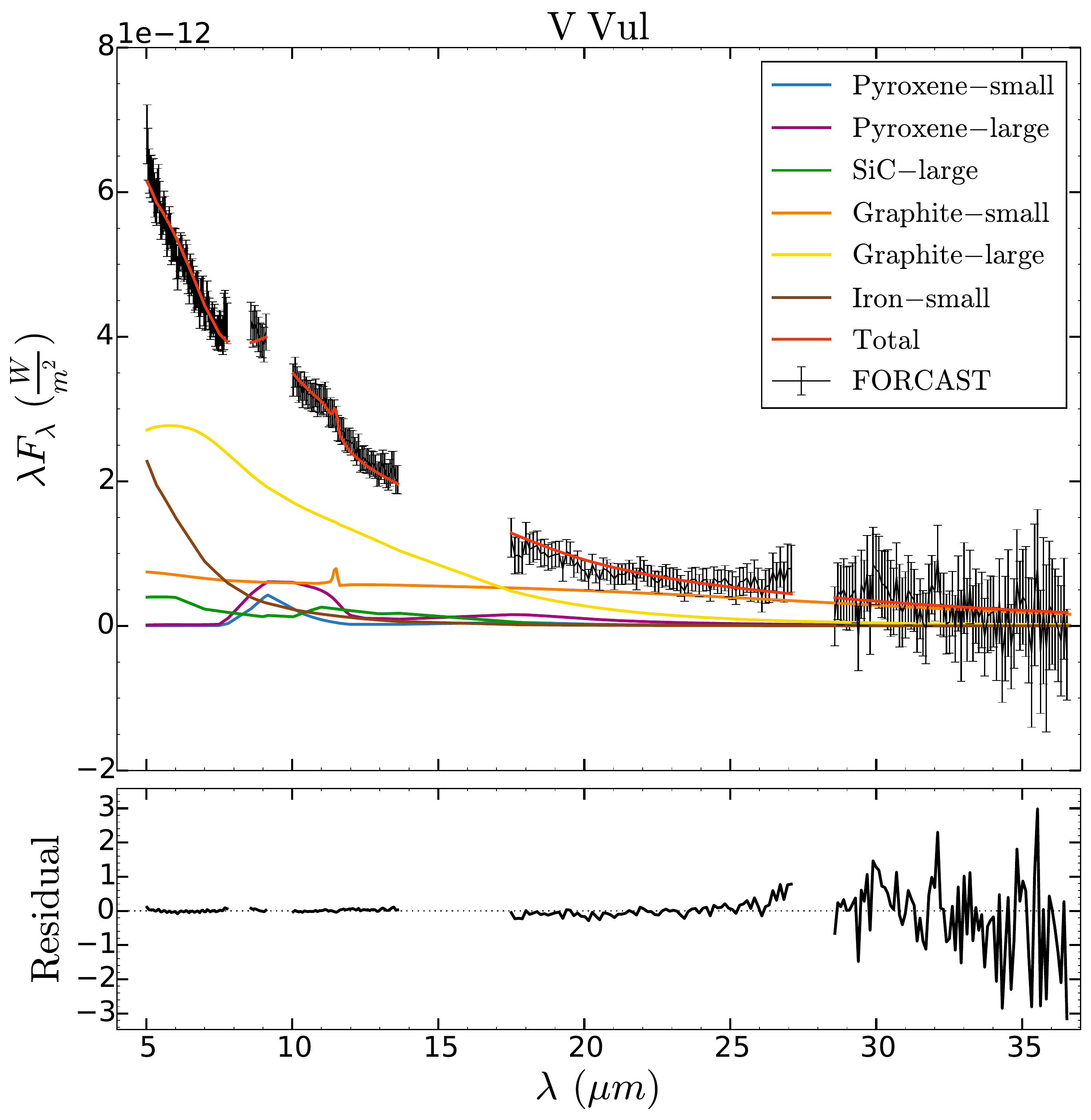}{0.5\textwidth}{(q) V Vul}}
\caption{Best model fits for our sample of stars, showing the contribution of the different mineral species.
\emph{Top}: the observed SOFIA FORCAST spectra and 1$\sigma$ errors (black points) are plotted together with the best model fit (red curve) and the mineral species (colored curves).  The data points between 9.19--10.0 \micron\ have been removed as these are strongly affected by telluric ozone absorption.
\emph{Bottom}: the normalized residual spectra after dividing by the best model of the observed spectra.}
\label{spec_decomp}
\end{figure*}

The results of the spectral decomposition modeling are shown in Figure \ref{spec_decomp}. Details of individual sources are provided in Table \ref{tab:data}.  The spectroscopic features present in these sources span a broad range of dust properties and characteristics. All of the RV Tauri stars in our sample, with the exceptions of UY CMa, TX Per, and R Sct have been reported as disk sources \citep{Gezer15}.  Two sources, $\rm{o^1\ Cen}$ and V Vul display a simple blackbody continuum. However, model fits demonstrate that they both also exhibit a weak IR excess (see Figure \ref{data} (c) and (r)), suggesting that these systems may be in the final stages of dissipation.  The remaining RV Tauri stars all exhibit emission from carbon-rich minerals with varying degrees of amorphous and crystalline silicates.

Two of our SRd sources, V441 Her and TV Per, show prominent silicate features. The presence of these features suggests that these sources, too, may have dusty disks akin to the disks of the RV Tauri stars.  Alternatively, these features might also arise from normal dusty outflows.  In the case of V441 Her, the 10 \micron\ feature is strong while the 20 \micron\ feature is weak, as is expected for relatively fresh and unprocessed amorphous silicates \citep{Nuth90}. In contrast, TV Per exhibits strong 10 and 20 \micron\ silicate emission, suggestive of prolonged exposure to hard radiation.  We discuss this more fully in Section \ref{discussion}.

\startlongtable
\begin{deluxetable*}{c|c|ccccccccc}
\tablecaption{\label{tab:data}RV Tauri and SRd Star Mineralogy}
\tabletypesize{\scriptsize}
\tablewidth{\textwidth}
\tablecolumns{11}
\tablehead{\colhead{$\phantom{bl}$ Star $\phantom{bl}$} & \colhead{$\chi_{\rm{red}}^2$} & \colhead{Mineral} & \colhead{$\overline{{c_i}}$} & \colhead{$\sigma_{\overline{c_i}}$} & \colhead{$\rm{V_f}$} & \colhead{$\sigma_{\rm{V_f}}$} & \colhead{$\rm{T_1 (K)}$} & \colhead{$\rm{F_{T_1}}$} & \colhead{$\phantom{a}\rm{T_2 (K)}$} & \colhead{$\rm{F_{T_2}}$}}
\startdata
\hline 
TW Cam & 0.26 & Graphite-small & 3.75 & 0.15 & 0.58 & 2.6E-2 & 1365 $\pm$ 105 & 0.019 $\pm$ 3.6E-3 & 332 $\pm$ 8 & 0.981 $\pm$ 3.6E-3 \\ 
 &  & Graphite-large & 1.14 & 0.06 & 0.18 & 1.1E-2 &  &  &  &  \\ 
 &  & Carbon-large & 1.08 & 0.10 & 0.17 & 1.7E-2 &  &  &  &  \\ 
 &  & Pyroxene-small & 0.26 & 0.03 & 0.04 & 4.1E-3 &  &  &  &  \\ 
 &  & SiC-small & 0.20 & 0.04 & 0.03 & 5.7E-3 &  &  &  &  \\ 
\hline 
UY CMa & 0.08 & Graphite-small & 1.76 & 0.49 & 0.36 & 1.0E-1 & 1251 $\pm$ 201 & 0.010 $\pm$ 4.1E-3 & 337 $\pm$ 7 & 0.990 $\pm$ 4.1E-3 \\ 
 &  & Graphite-large & 1.29 & 0.11 & 0.26 & 3.6E-2 &  &  &  &  \\ 
 &  & Carbon-large & 1.20 & 0.21 & 0.24 & 4.9E-2 &  &  &  &  \\ 
 &  & Pyroxene-small & 0.47 & 0.06 & 0.10 & 1.7E-2 &  &  &  &  \\ 
 &  & Forsterite-small & 0.21 & 0.04 & 0.04 & 8.6E-3 &  &  &  &  \\ 
\hline 
$\rm{o}^1$ Cen & 0.32 & Iron-small & 7.01 & 0.48 & 0.49 & 3.8E-2 & 3780 $\pm$ 193 & 1.0 & - & -\\ 
 &  & Graphite-small & 5.96 & 0.49 & 0.42 & 3.7E-2 &  &  &  & \\ 
 &  & Graphite-large & 1.17 & 0.06 & 0.08 & 5.5E-3 &  &  &  & \\ 
 &  & Pyroxene-small & 0.17 & 0.03 & 0.01 & 2.1E-3 &  &  &  & \\ 
\hline 
RU Cen & 0.23 & Graphite-small & 3.53 & 0.20 & 0.45 & 2.8E-2 & 535 $\pm$ 16 & 0.010 $\pm$ 1.4E-3 & 203 $\pm$ 2 & 0.990 $\pm$ 1.4E-3 \\ 
 &  & Carbon-large & 2.01 & 0.18 & 0.25 & 2.4E-2 &  &  &  &  \\ 
 &  & Pyroxene-small & 0.89 & 0.05 & 0.11 & 8.0E-3 &  &  &  &  \\ 
 &  & Graphite-large & 0.86 & 0.08 & 0.11 & 1.1E-2 &  &  &  &  \\ 
 &  & SiC-small & 0.50 & 0.06 & 0.06 & 8.3E-3 &  &  &  &  \\ 
 &  & Forsterite-small & 0.11 & 0.02 & 0.01 & 3.0E-3 &  &  &  &  \\ 
\hline 
SX Cen & 0.24 & Graphite-small & 2.65 & 0.26 & 0.41 & 4.5E-2 & 715 $\pm$ 14 & 0.042 $\pm$ 4.8E-3 & 244 $\pm$ 7 & 0.958 $\pm$ 4.8E-3 \\ 
 &  & Carbon-large & 1.77 & 0.14 & 0.27 & 2.7E-2 &  &  &  &  \\ 
 &  & Graphite-large & 0.95 & 0.08 & 0.15 & 1.5E-2 &  &  &  &  \\ 
 &  & Pyroxene-large & 0.72 & 0.19 & 0.11 & 3.0E-2 &  &  &  &  \\ 
 &  & Olivine-large & 0.32 & 0.18 & 0.05 & 2.8E-2 &  &  &  &  \\ 
 &  & Pyroxene-small & 0.08 & 0.06 & 0.01 & 9.8E-3 &  &  &  &  \\ 
\hline 
SU Gem & 0.40 & Graphite-small & 4.06 & 0.12 & 0.59 & 1.9E-2 & 1444 $\pm$ 83 & 0.023 $\pm$ 3.1E-3 & 349 $\pm$ 5 & 0.977 $\pm$ 3.1E-3 \\ 
 &  & Carbon-large & 1.16 & 0.07 & 0.17 & 1.1E-2 &  &  &  &  \\ 
 &  & Graphite-large & 1.07 & 0.05 & 0.15 & 7.5E-3 &  &  &  &  \\ 
 &  & Pyroxene-small & 0.38 & 0.02 & 0.06 & 2.9E-3 &  &  &  &  \\ 
 &  & SiC-small & 0.27 & 0.03 & 0.04 & 3.9E-3 &  &  &  &  \\ 
\hline 
AC Her & 1.43 & Graphite-small & 3.43 & 0.06 & 0.48 & 9.4E-3 & 489 $\pm$ 4 & 0.022 $\pm$ 9.4E-4 & 207 $\pm$ 1 & 0.978 $\pm$ 9.4E-4 \\ 
 &  & Carbon-large & 1.51 & 0.05 & 0.21 & 6.6E-3 &  &  &  &  \\ 
 &  & Graphite-large & 0.92 & 0.02 & 0.13 & 3.6E-3 &  &  &  &  \\ 
 &  & Pyroxene-small & 0.76 & 0.01 & 0.11 & 2.3E-3 &  &  &  &  \\ 
 &  & SiC-small & 0.50 & 0.02 & 0.07 & 2.6E-3 &  &  &  &  \\ 
 &  & Forsterite-small & 0.09 & 0.01 & 0.01 & 9.8E-4 &  &  &  &  \\ 
\hline 
V441 Her & 0.15 & Graphite-small & 3.14 & 0.14 & 0.40 & 2.2E-2 & 1607 $\pm$ 68 & 0.019 $\pm$ 1.8E-3 & 363 $\pm$ 3 & 0.981 $\pm$ 1.8E-3 \\ 
 &  & Carbon-small & 2.76 & 0.20 & 0.35 & 2.6E-2 &  &  &  &  \\ 
 &  & Graphite-large & 1.07 & 0.05 & 0.14 & 8.3E-3 &  &  &  &  \\ 
 &  & Olivine-small & 0.44 & 0.08 & 0.06 & 1.0E-2 &  &  &  &  \\ 
 &  & Pyroxene-small & 0.23 & 0.04 & 0.03 & 5.2E-3 &  &  &  &  \\ 
 &  & SiC-small & 0.21 & 0.04 & 0.03 & 5.3E-3 &  &  &  &  \\ 
\hline 
U Mon & 0.50 & Carbon-small & 3.55 & 0.50 & 0.30 & 6.3E-2 & 772 $\pm$ 10 & 0.020 $\pm$ 1.0E-3 & 254 $\pm$ 1 & 0.980 $\pm$ 1.0E-3 \\ 
 &  & Graphite-small & 3.17 & 0.13 & 0.27 & 4.5E-2 &  &  &  &  \\ 
 &  & Iron-large & 2.94 & 1.88 & 0.25 & 1.6E-1 &  &  &  &  \\ 
 &  & Pyroxene-small & 0.87 & 0.06 & 0.07 & 1.3E-2 &  &  &  &  \\ 
 &  & SiC-small & 0.56 & 0.05 & 0.05 & 9.0E-3 &  &  &  &  \\ 
 &  & Graphite-large & 0.42 & 0.15 & 0.04 & 1.4E-2 &  &  &  &  \\ 
 &  & Olivine-small & 0.38 & 0.10 & 0.03 & 9.8E-3 &  &  &  &  \\ 
\hline 
CT Ori & 0.25 & Graphite-small & 3.91 & 0.17 & 0.53 & 2.5E-2 & 684 $\pm$ 13 & 0.041 $\pm$ 4.0E-3 & 258 $\pm$ 5 & 0.959 $\pm$ 4.0E-3 \\ 
 &  & Carbon-large & 1.58 & 0.12 & 0.21 & 1.7E-2 &  &  &  &  \\ 
 &  & Graphite-large & 0.97 & 0.07 & 0.13 & 1.1E-2 &  &  &  &  \\ 
 &  & Pyroxene-small & 0.57 & 0.03 & 0.08 & 5.0E-3 &  &  &  &  \\ 
 &  & SiC-small & 0.38 & 0.04 & 0.05 & 5.7E-3 &  &  &  &  \\ 
\hline 
TV Per & 0.58 & Carbon-small & 3.75 & 0.58 & 0.29 & 6.5E-2 & 2340 $\pm$ 12 & 0.002 $\pm$ 4.0E-5 & 297 $\pm$ 1 & 0.998 $\pm$ 4.0E-5 \\ 
 &  & Iron-large & 3.24 & 2.20 & 0.25 & 1.7E-1 &  &  &  &  \\ 
 &  & Graphite-small & 2.47 & 0.16 & 0.19 & 3.5E-2 &  &  &  &  \\ 
 &  & Pyroxene-small & 1.47 & 0.07 & 0.11 & 2.0E-2 &  &  &  &  \\ 
 &  & Olivine-small & 1.04 & 0.14 & 0.08 & 1.7E-2 &  &  &  &  \\ 
 &  & SiC-small & 0.87 & 0.07 & 0.07 & 1.3E-2 &  &  &  &  \\ 
 &  & Graphite-large & 0.30 & 0.17 & 0.02 & 1.4E-2 &  &  &  &  \\ 
\hline 
AR Pup & 1.93 & Graphite-small & 4.07 & 0.11 & 0.48 & 1.4E-2 & 734 $\pm$ 7 & 0.062 $\pm$ 3.0E-3 & 280 $\pm$ 3 & 0.938 $\pm$ 3.0E-3 \\ 
 &  & Carbon-small & 2.88 & 0.06 & 0.34 & 8.3E-3 &  &  &  &  \\ 
 &  & Graphite-large & 1.02 & 0.02 & 0.12 & 2.8E-3 &  &  &  &  \\ 
 &  & Olivine-large & 0.23 & 0.01 & 0.03 & 1.5E-3 &  &  &  &  \\ 
 &  & Pyroxene-small & 0.19 & 0.01 & 0.02 & 6.9E-4 &  &  &  &  \\ 
\hline 
R Sge & 0.17 & Graphite-small & 3.66 & 0.54 & 0.43 & 6.8E-2 & 862 $\pm$ 29 & 0.026 $\pm$ 3.6E-3 & 270 $\pm$ 7 & 0.974 $\pm$ 3.6E-3 \\ 
 &  & Carbon-small & 3.02 & 0.42 & 0.36 & 5.5E-2 &  &  &  &  \\ 
 &  & Graphite-large & 0.97 & 0.13 & 0.11 & 1.8E-2 &  &  &  &  \\ 
 &  & Pyroxene-small & 0.51 & 0.03 & 0.06 & 6.2E-3 &  &  &  &  \\ 
 &  & SiC-small & 0.34 & 0.05 & 0.04 & 6.8E-3 &  &  &  &  \\ 
\hline 
AI Sco & 0.32 & Graphite-small & 3.94 & 0.17 & 0.59 & 3.0E-2 & 1049 $\pm$ 74 & 0.030 $\pm$ 6.1E-3 & 322 $\pm$ 7 & 0.970 $\pm$ 6.1E-3 \\ 
 &  & Carbon-large & 1.15 & 0.14 & 0.17 & 2.1E-2 &  &  &  &  \\ 
 &  & Graphite-large & 1.10 & 0.08 & 0.16 & 1.3E-2 &  &  &  &  \\ 
 &  & Pyroxene-small & 0.30 & 0.03 & 0.04 & 4.2E-3 &  &  &  &  \\ 
 &  & SiC-small & 0.19 & 0.04 & 0.03 & 6.3E-3 &  &  &  &  \\ 
\hline 
R Sct & 1.18 & Graphite-small & 5.42 & 0.23 & 0.70 & 3.1E-2 & 1857 $\pm$ 44 & 0.017 $\pm$ 7.0E-3 & 158 $\pm$ 18 & 0.983 $\pm$ 7.0E-3 \\ 
 &  & Graphite-large & 1.10 & 0.05 & 0.14 & 8.0E-3 &  &  &  &  \\ 
 &  & Carbon-large & 1.04 & 0.07 & 0.13 & 1.0E-2 &  &  &  &  \\ 
 &  & Pyroxene-small & 0.23 & 0.02 & 0.03 & 2.2E-3 &  &  &  &  \\ 
\hline 
RV Tau & 0.26 & Graphite-small & 3.70 & 0.13 & 0.43 & 1.9E-2 & 760 $\pm$ 10 & 0.052 $\pm$ 2.9E-3 & 256 $\pm$ 3 & 0.948 $\pm$ 2.9E-3 \\ 
 &  & Carbon-small & 3.21 & 0.21 & 0.37 & 2.5E-2 &  &  &  &  \\ 
 &  & Graphite-large & 0.92 & 0.06 & 0.11 & 7.1E-3 &  &  &  &  \\ 
 &  & Pyroxene-small & 0.47 & 0.03 & 0.05 & 3.3E-3 &  &  &  &  \\ 
 &  & SiC-small & 0.38 & 0.03 & 0.04 & 3.9E-3 &  &  &  &  \\ 
\hline 
V Vul & 0.36 & Iron-small & 8.06 & 0.55 & 0.55 & 4.1E-2 & 678 $\pm$ 4 & 1.0 & - & -\\ 
 &  & Graphite-small & 4.92 & 0.41 & 0.34 & 3.1E-2 &  &  &  & \\ 
 &  & Graphite-large & 1.05 & 0.05 & 0.07 & 5.0E-3 &  &  &  & \\ 
 &  & Pyroxene-large & 0.31 & 0.07 & 0.02 & 4.9E-3 &  &  &  & \\ 
 &  & SiC-large & 0.16 & 0.06 & 0.01 & 4.4E-3 &  &  &  & \\ 
 &  & Pyroxene-small & 0.10 & 0.05 & 0.01 & 3.3E-3 &  &  &  & \\ 
\enddata
\tablecomments{$\overline{c_i}$ is the average best fit coefficient of a given mineral species.  $\sigma_{\overline{c_i}}$ is the standard deviation of the average best fit coefficient, $\rm{V_f}$ is the volume fraction, $\sigma (\rm{V_f})$ is the error in the volume fraction, $\rm{T_{1,2}}$ are the blackbody dust temperatures, and $\rm{F_{T_{1,2}}}$ is the fraction of dust at those temperatures, respectively.  `small' refers to 0.1 \micron\ spherical grains, `large' designates 2.0 \micron\ spherical grains.}
\end{deluxetable*}

\begin{figure*}
\gridline{\leftfig{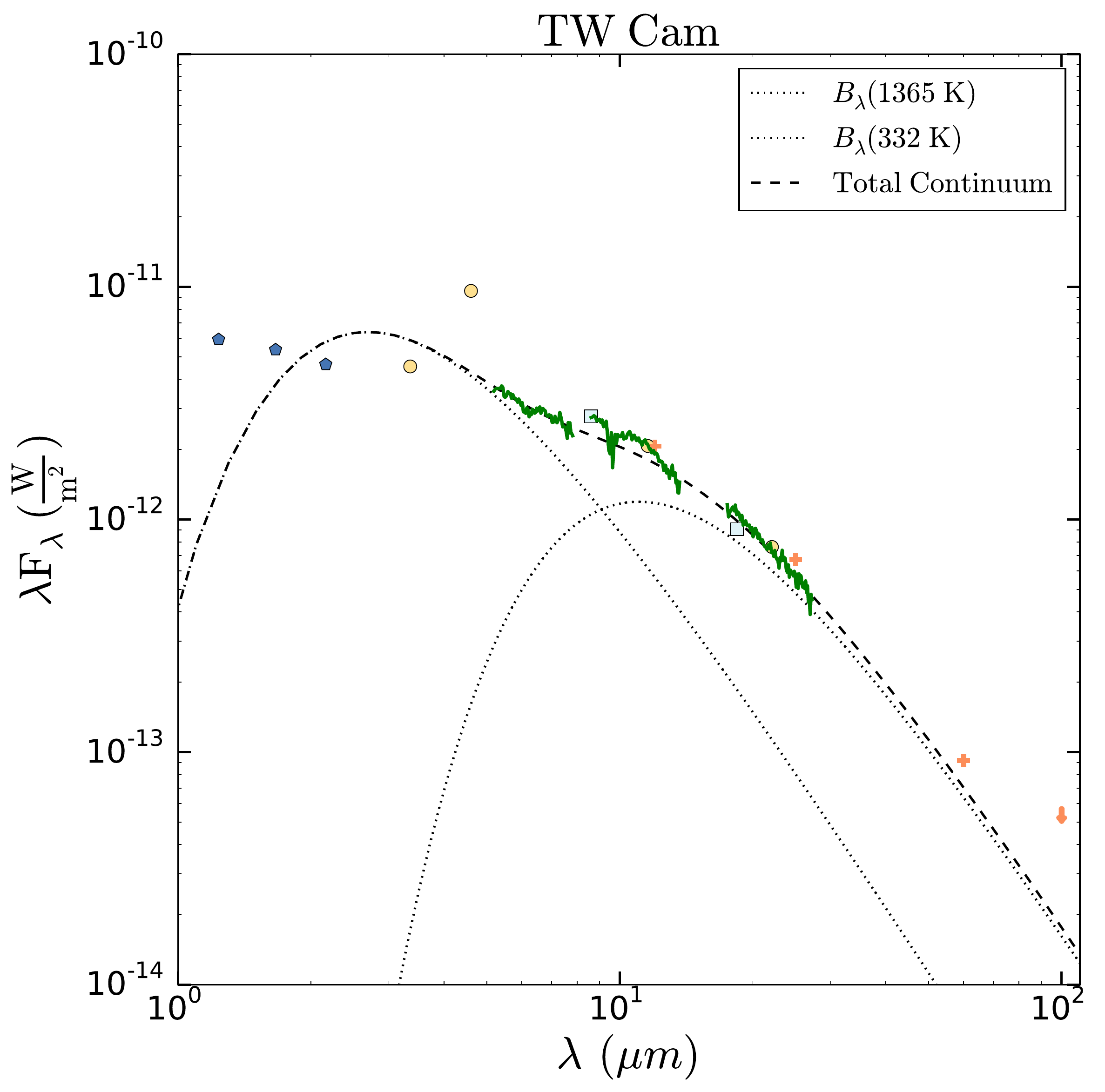}{0.5\textwidth}{(a) TW Cam}
          \rightfig{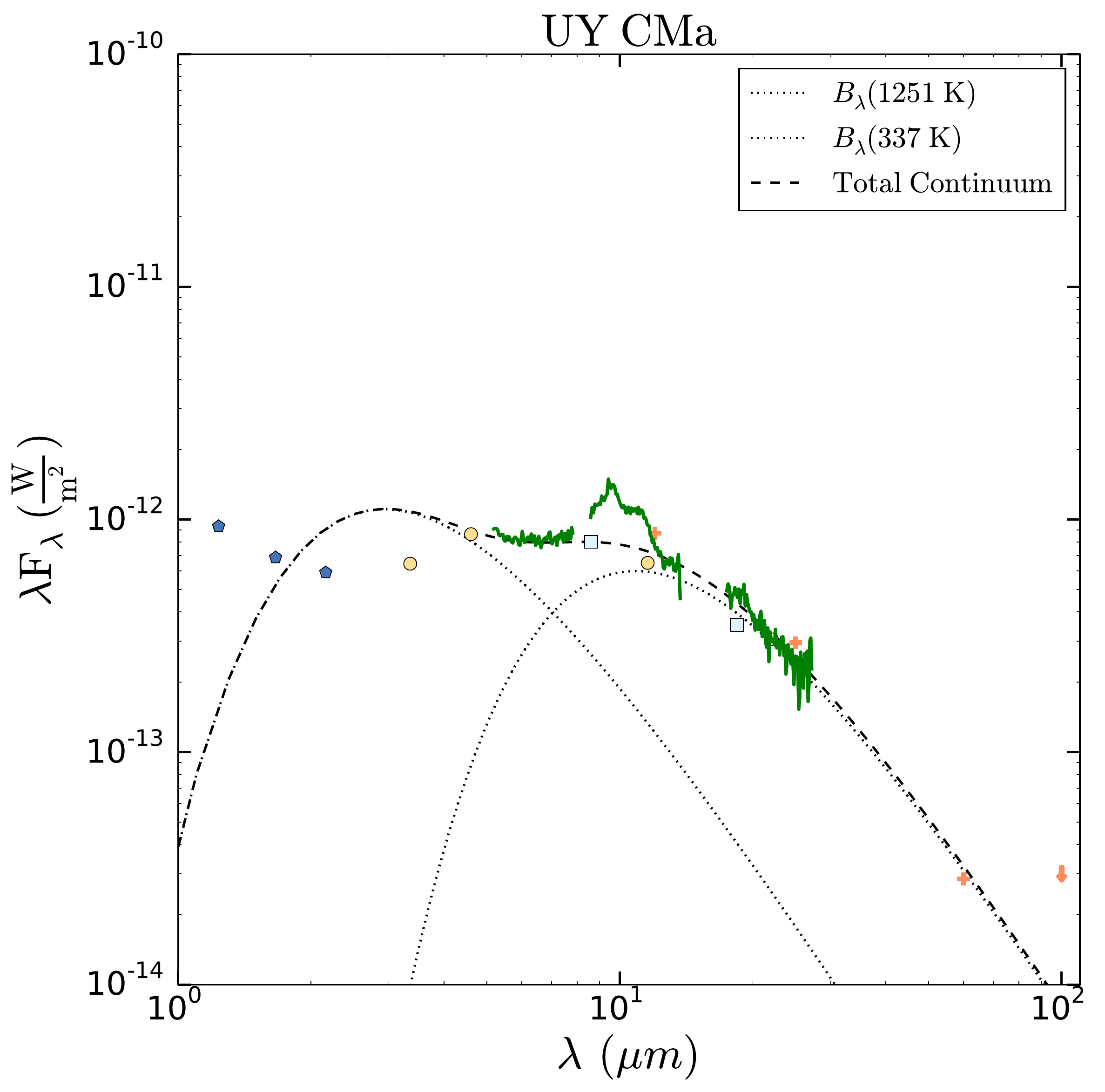}{0.5\textwidth}{(b) UY CMa}}
\gridline{\leftfig{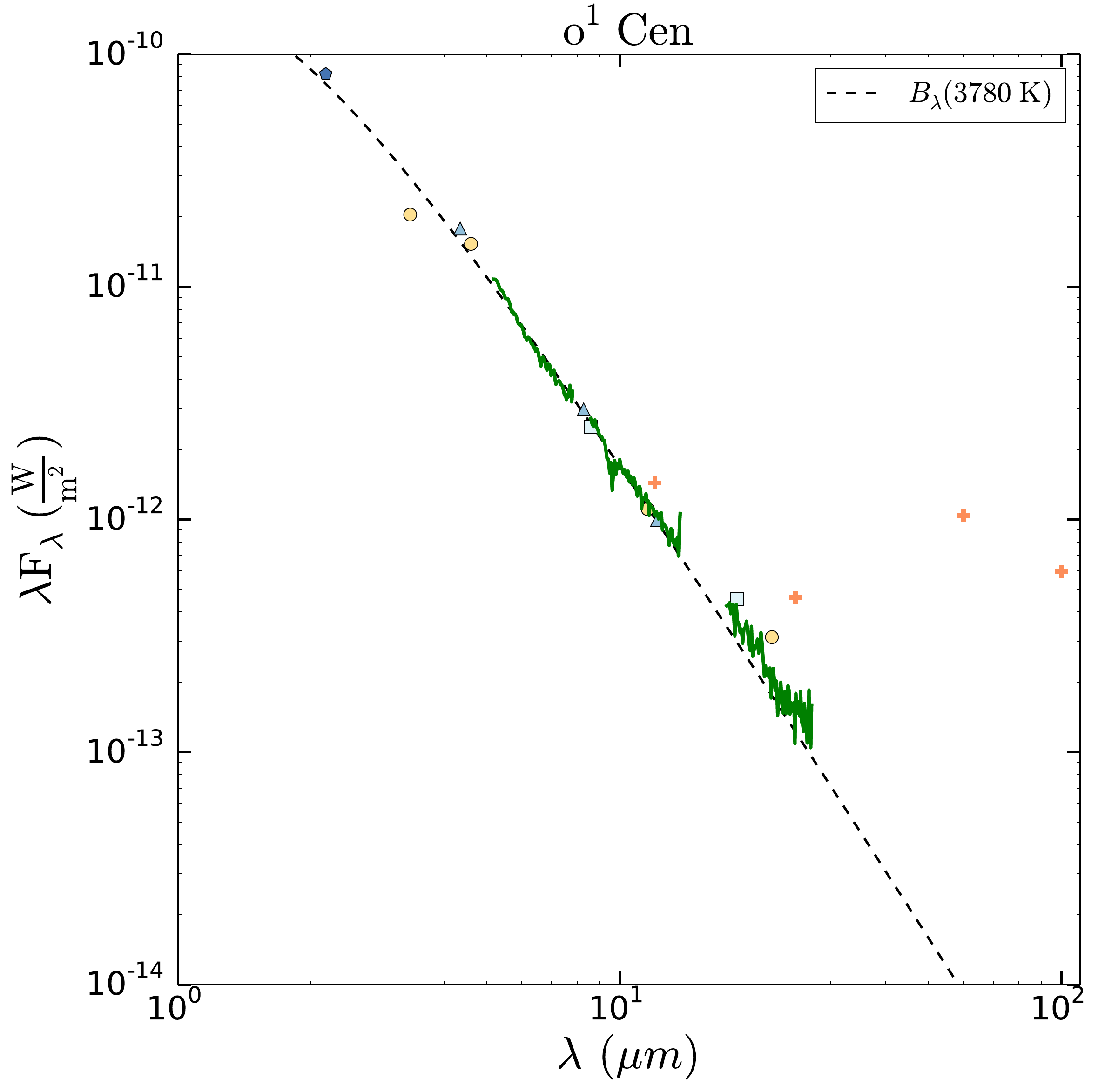}{0.5\textwidth}{(c) $\rm{o}^1$ Cen}
          \rightfig{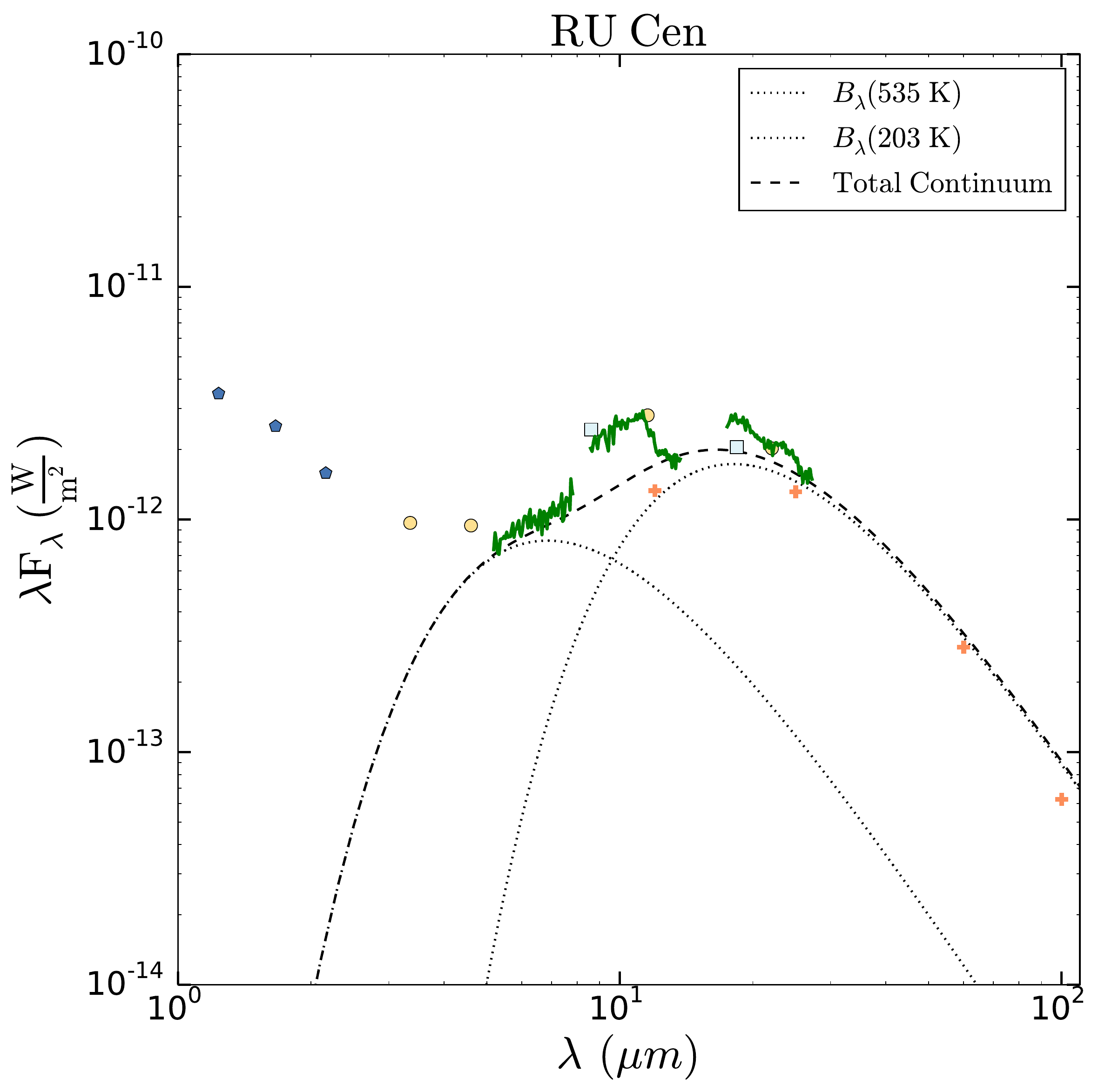}{0.5\textwidth}{(d) RU Cen}}
\end{figure*}
\begin{figure*}
\gridline{\leftfig{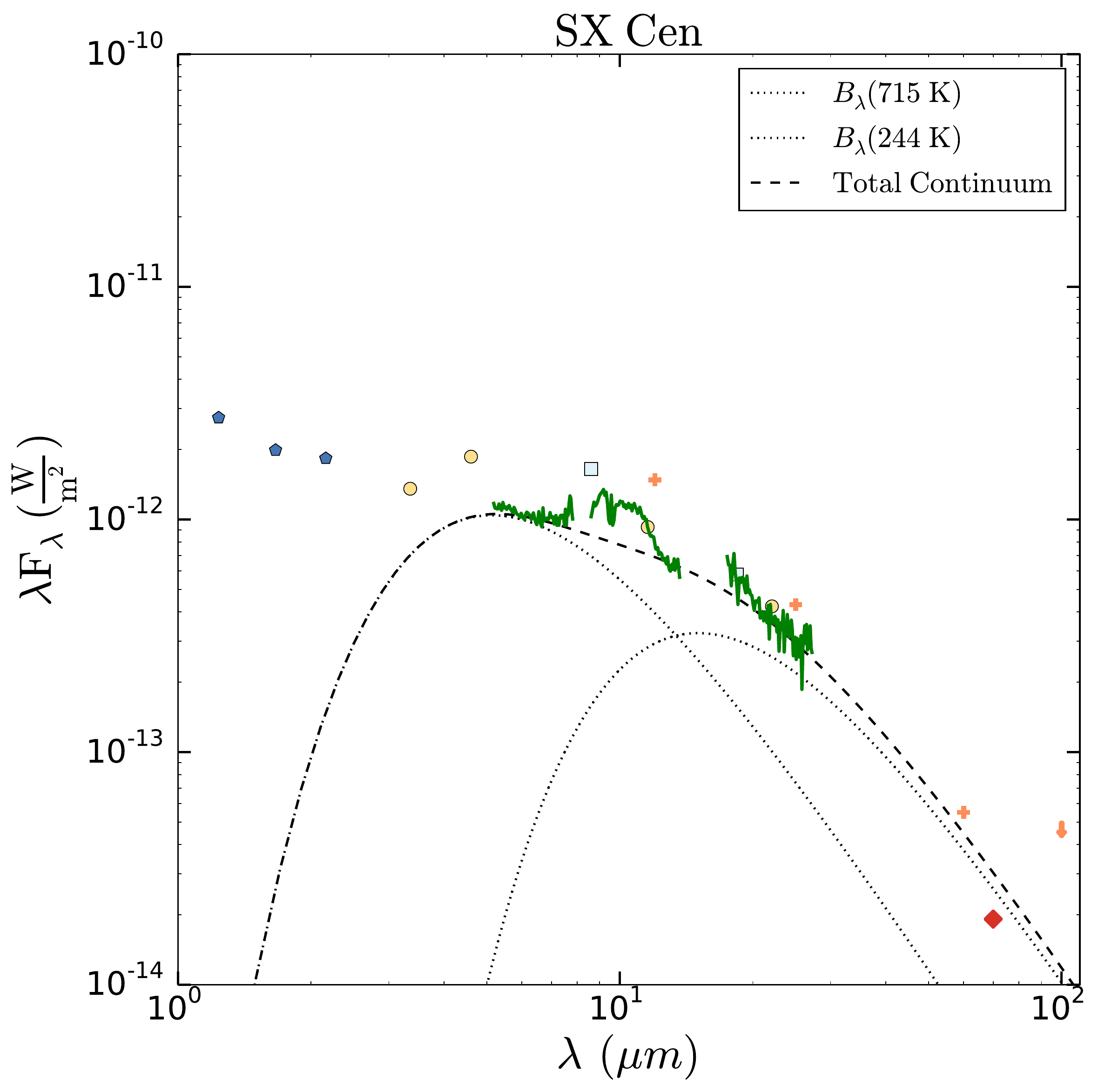}{0.5\textwidth}{(e) SX Cen}
          \rightfig{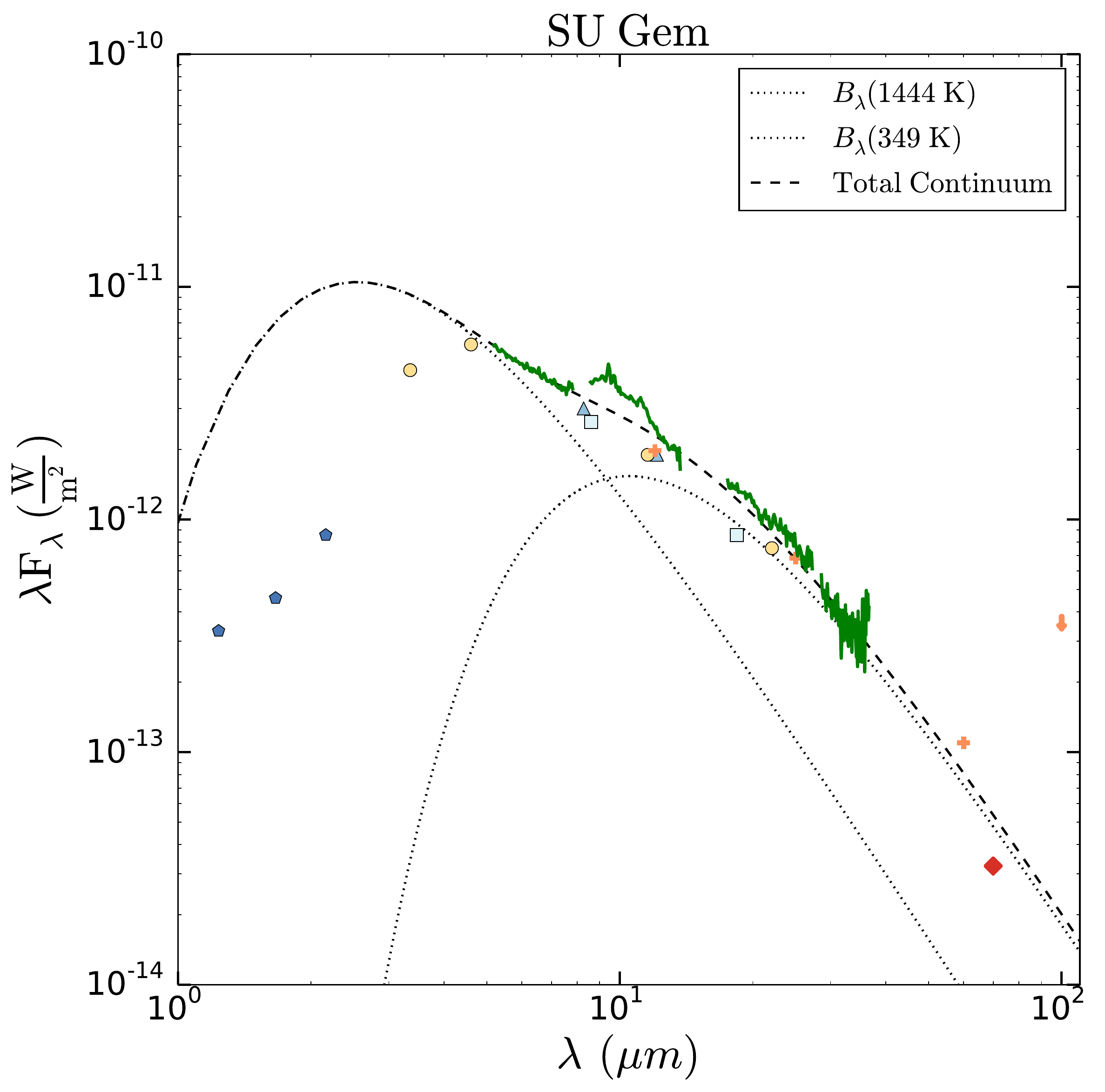}{0.5\textwidth}{(f) SU Gem}}
\gridline{\leftfig{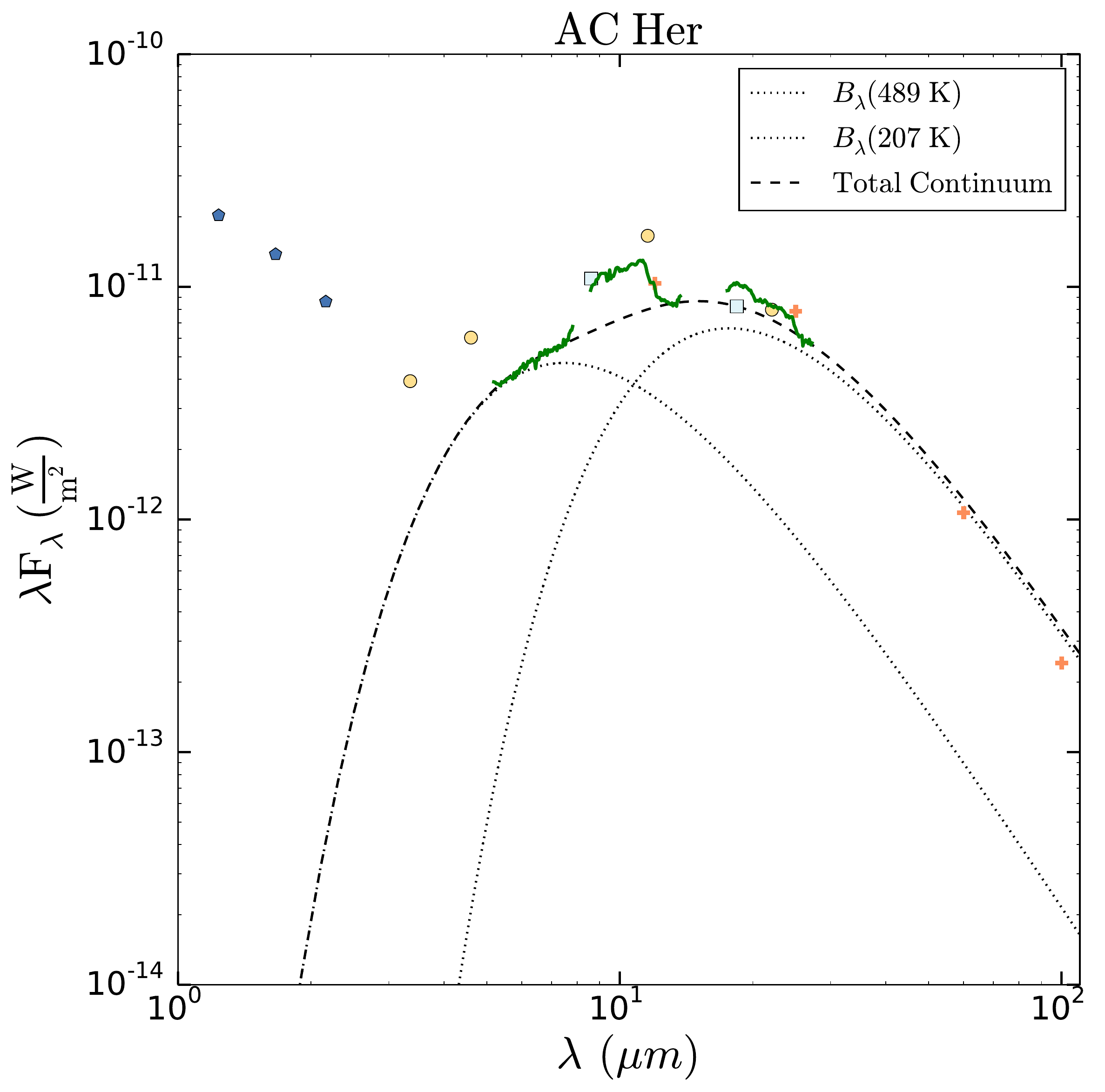}{0.5\textwidth}{(g) AC Her}
          \rightfig{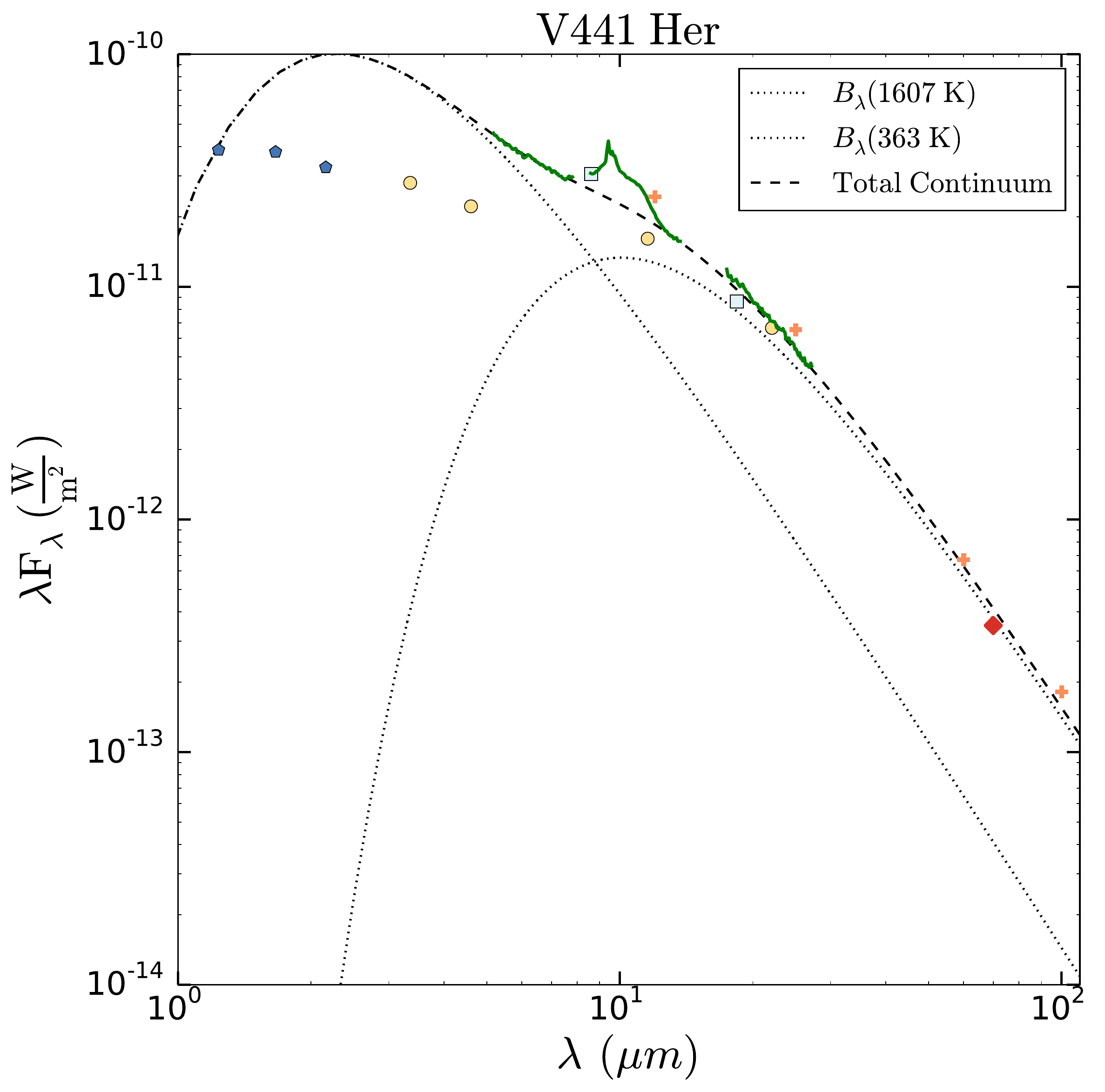}{0.5\textwidth}{(h) V441 Her}}
\end{figure*}
\begin{figure*}
\gridline{\leftfig{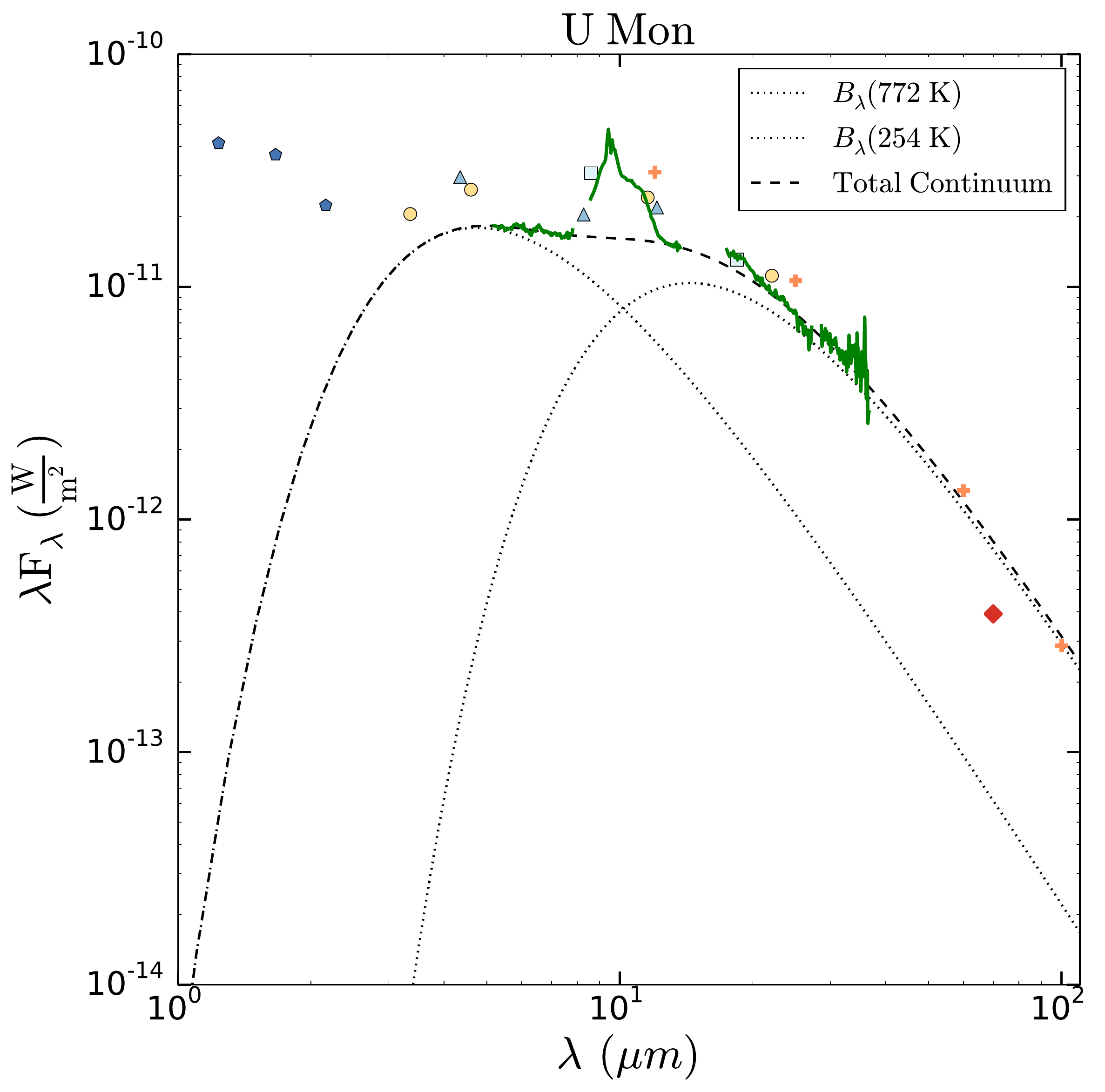}{0.5\textwidth}{(i) U Mon}
          \rightfig{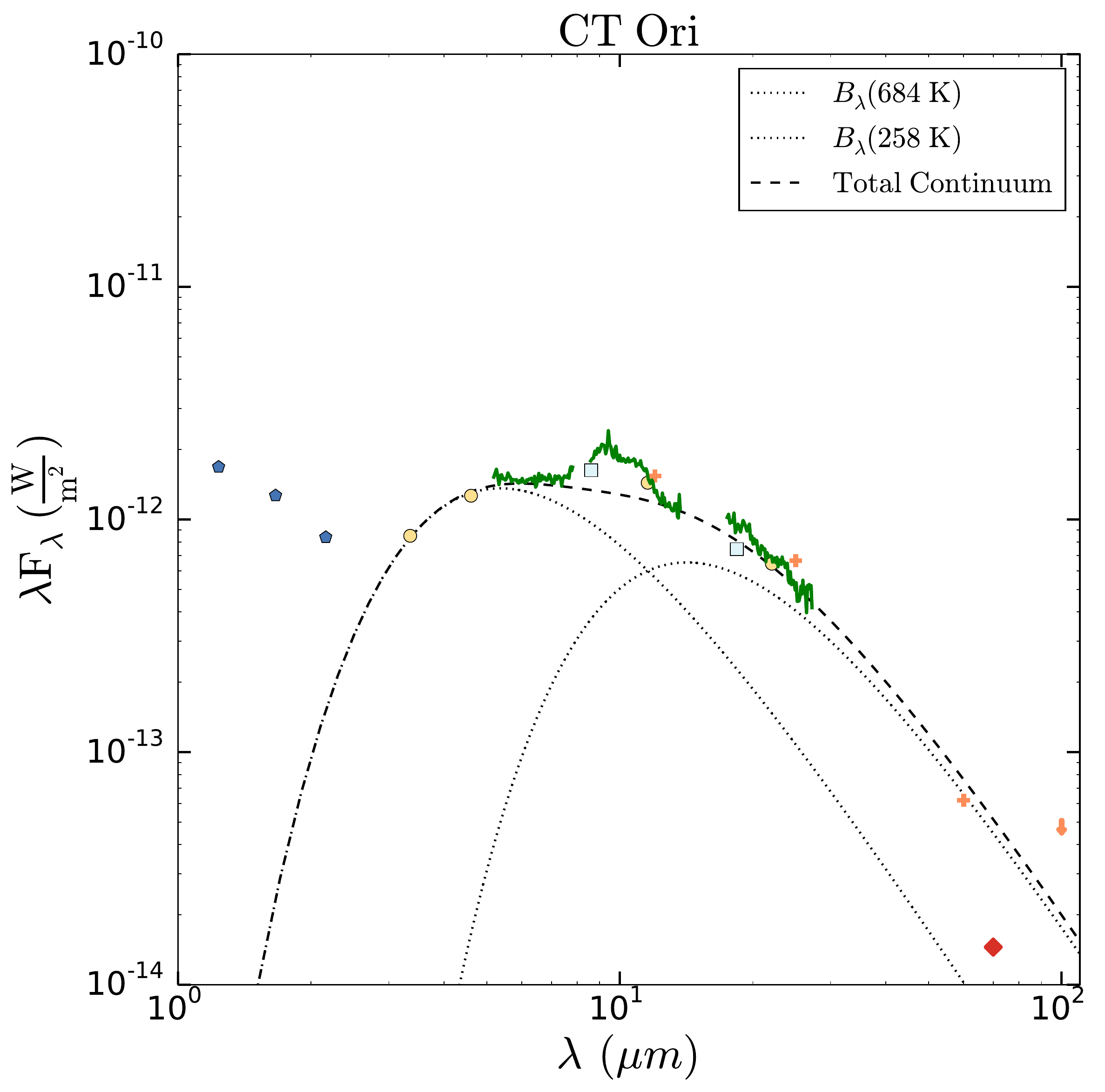}{0.5\textwidth}{(j) CT Ori}}
\gridline{\leftfig{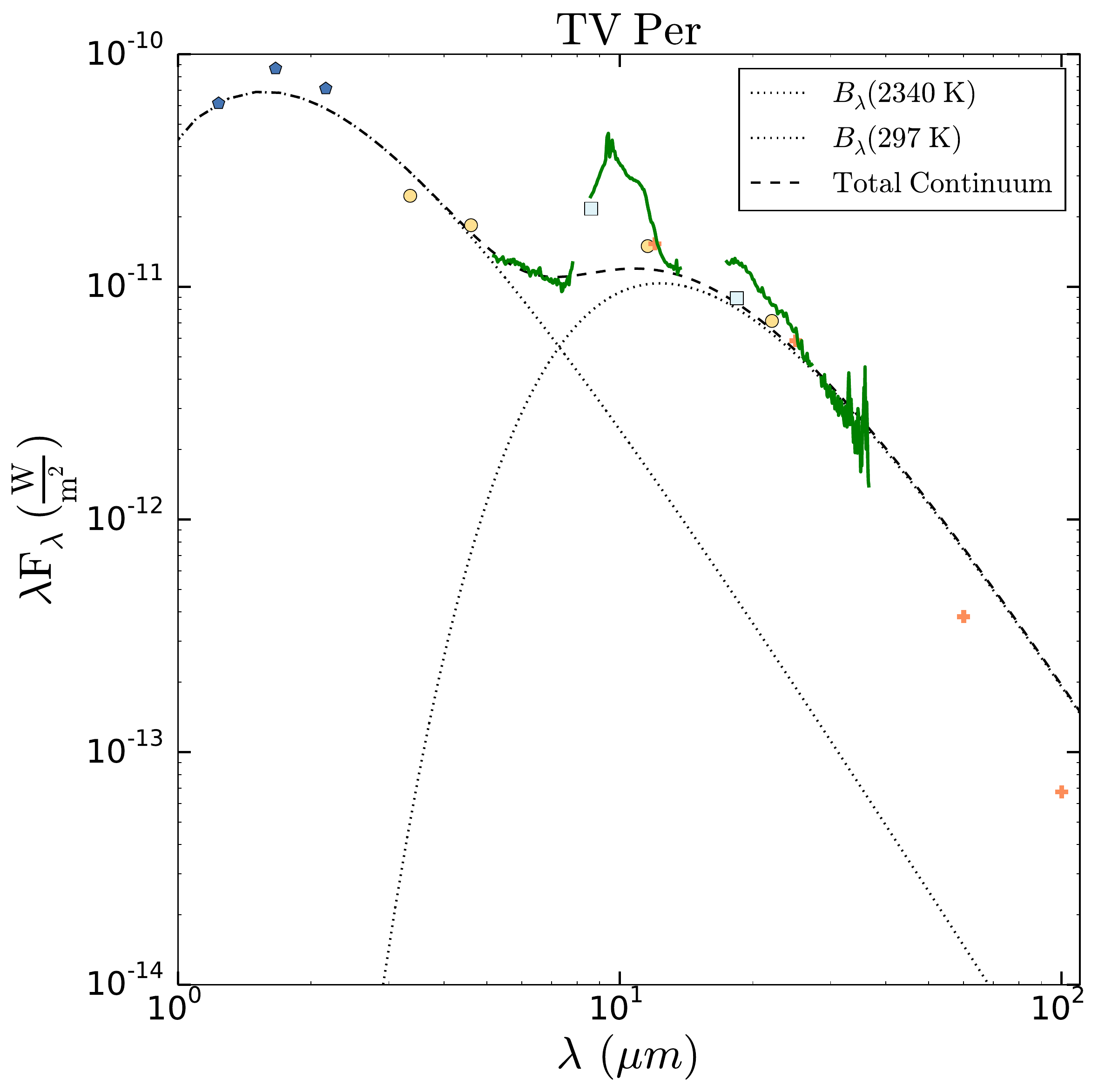}{0.5\textwidth}{(k) TV Per}
	  \rightfig{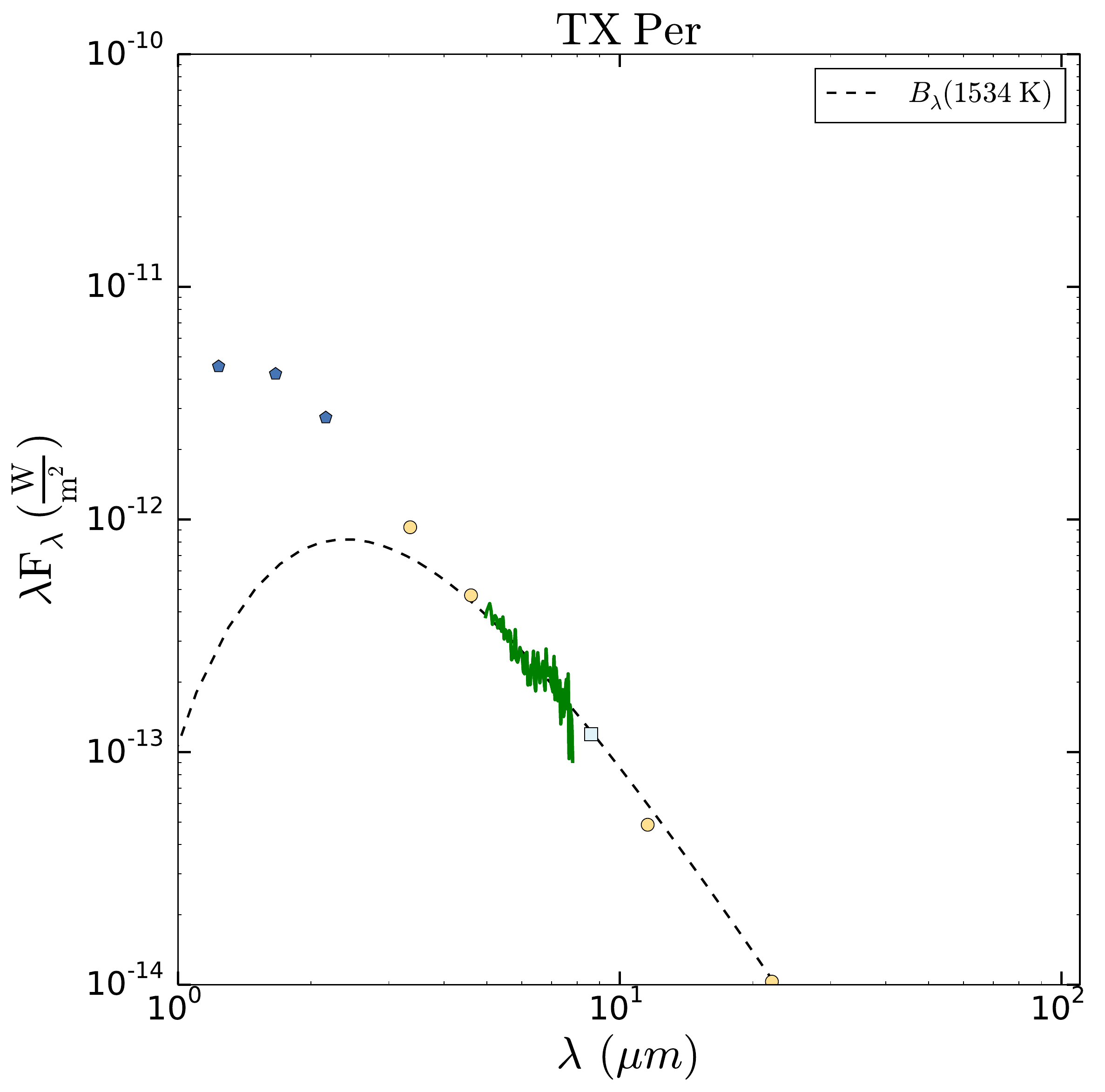}{0.5\textwidth}{(l) TX Per}}
\end{figure*}
\begin{figure*}
\gridline{\leftfig{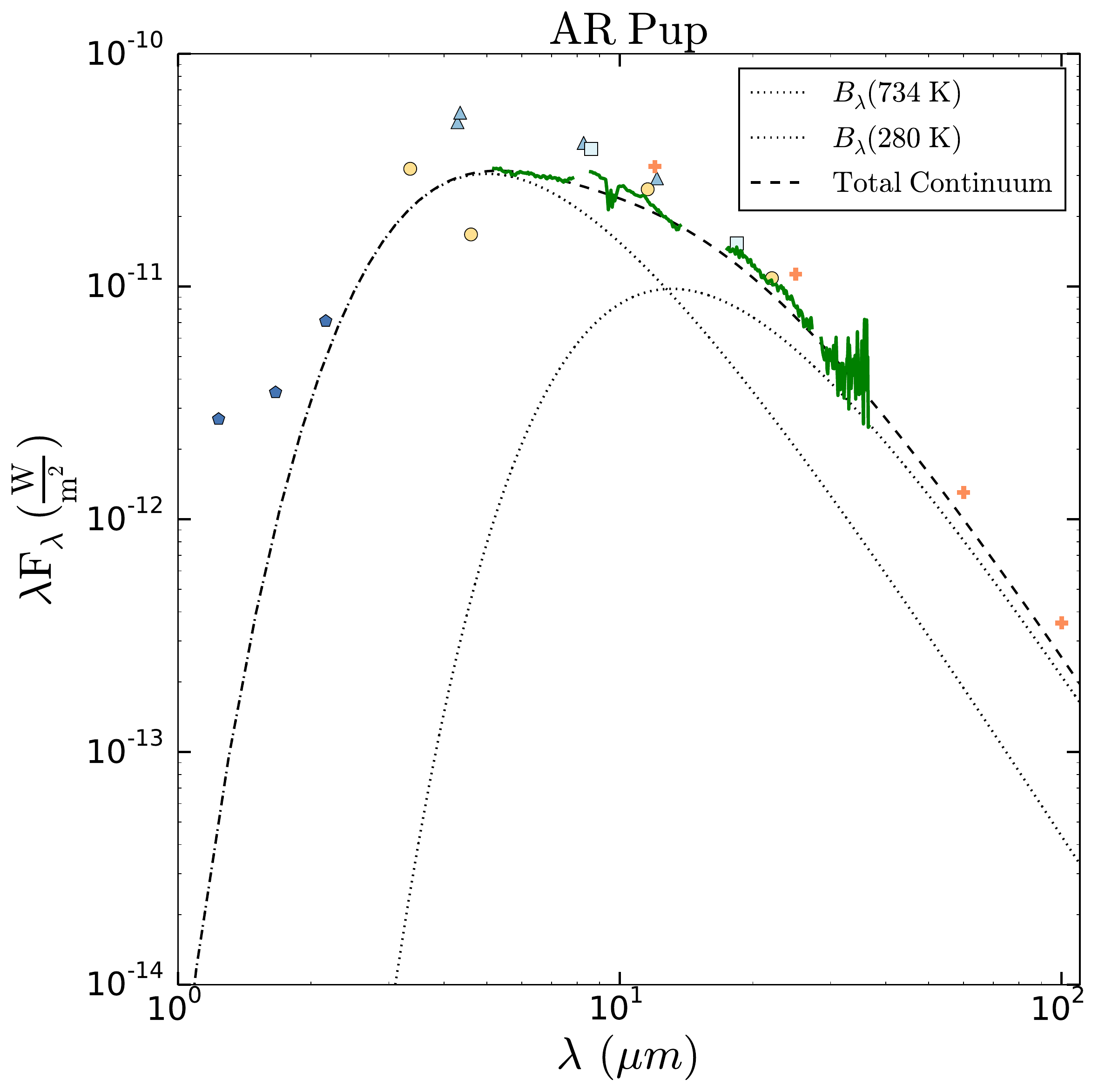}{0.5\textwidth}{(m) AR Pup}
	      \rightfig{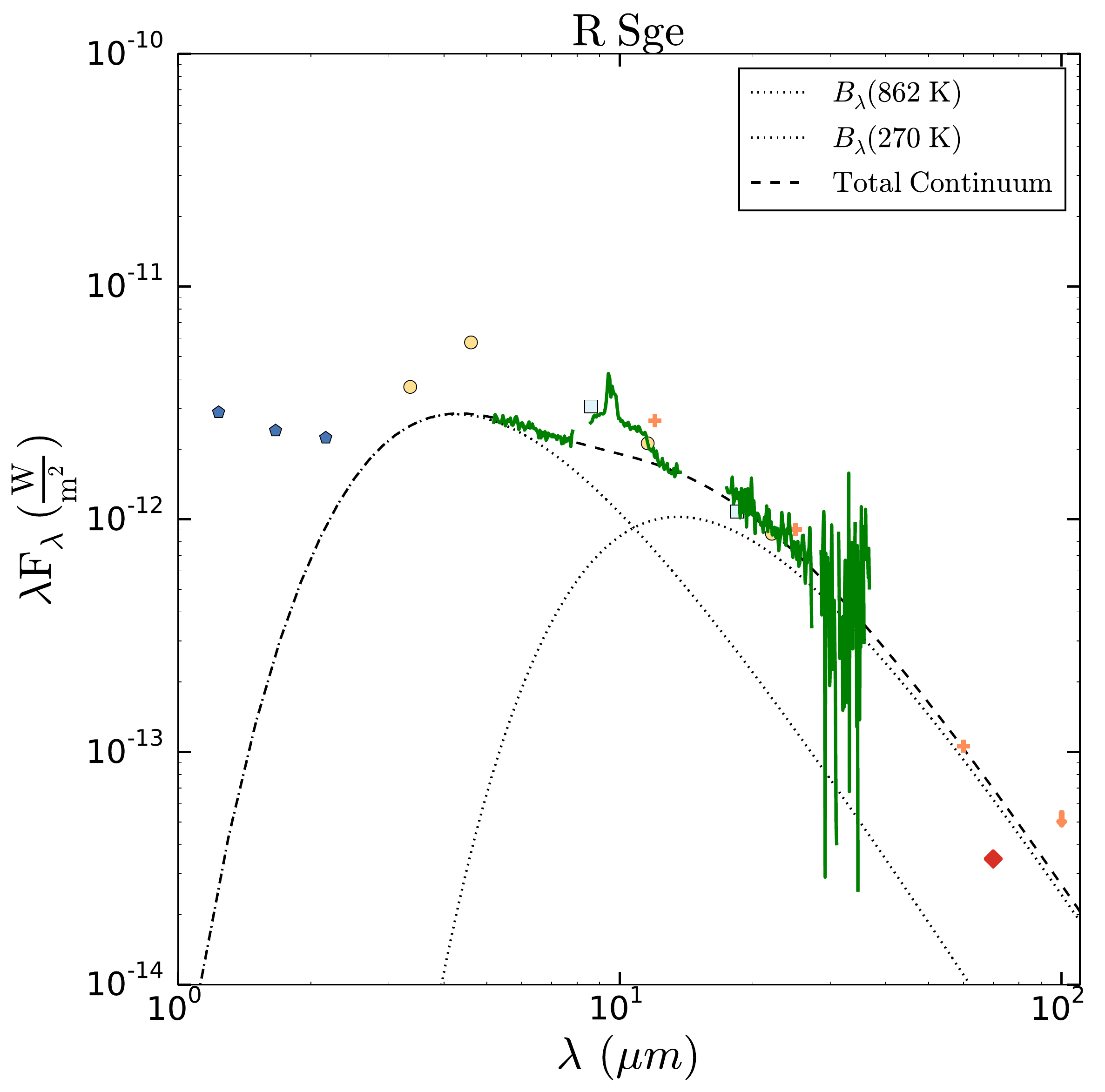}{0.5\textwidth}{(n) R Sge}}
\gridline{\leftfig{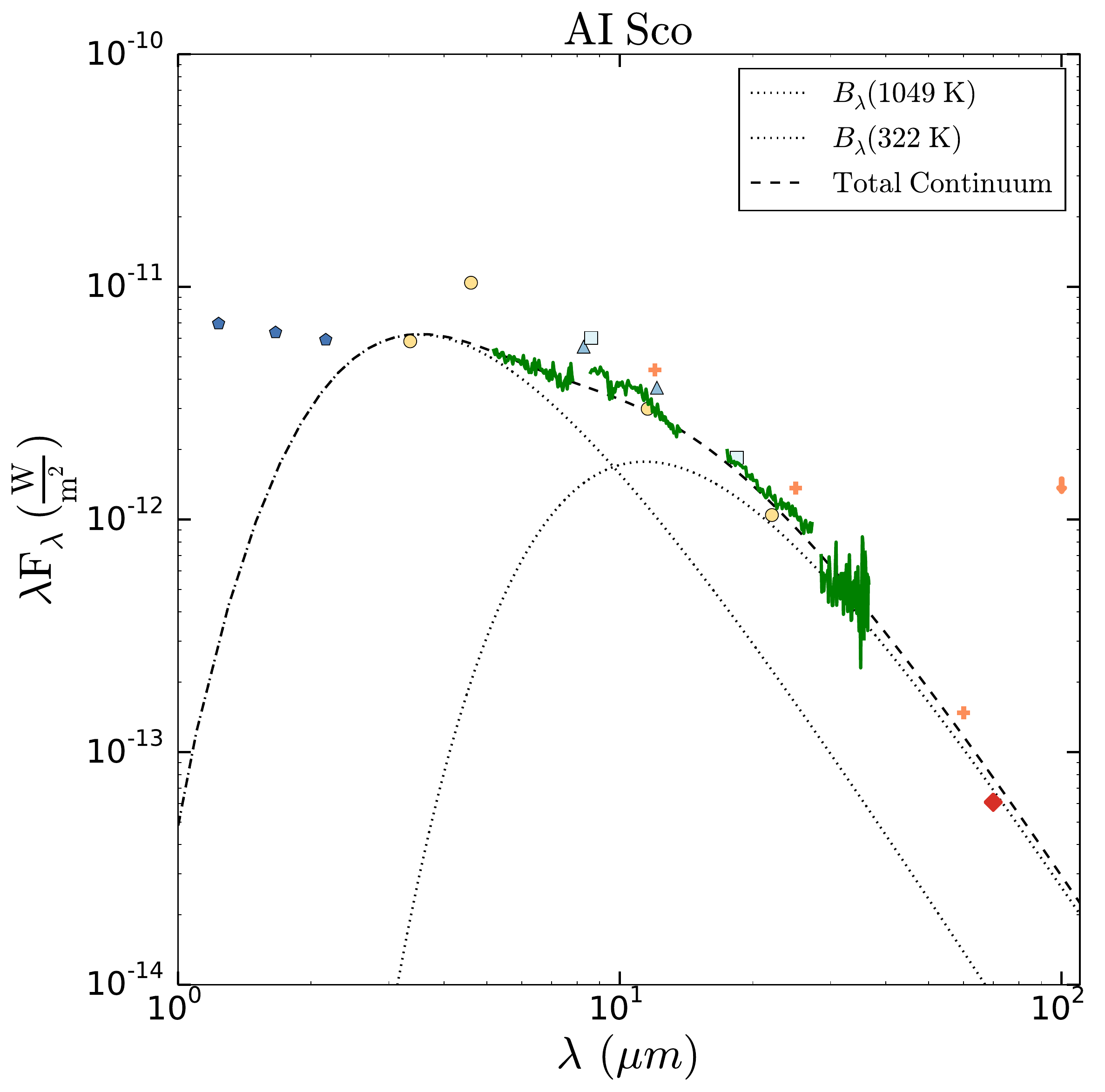}{0.5\textwidth}{(o) AI Sco}
	     \rightfig{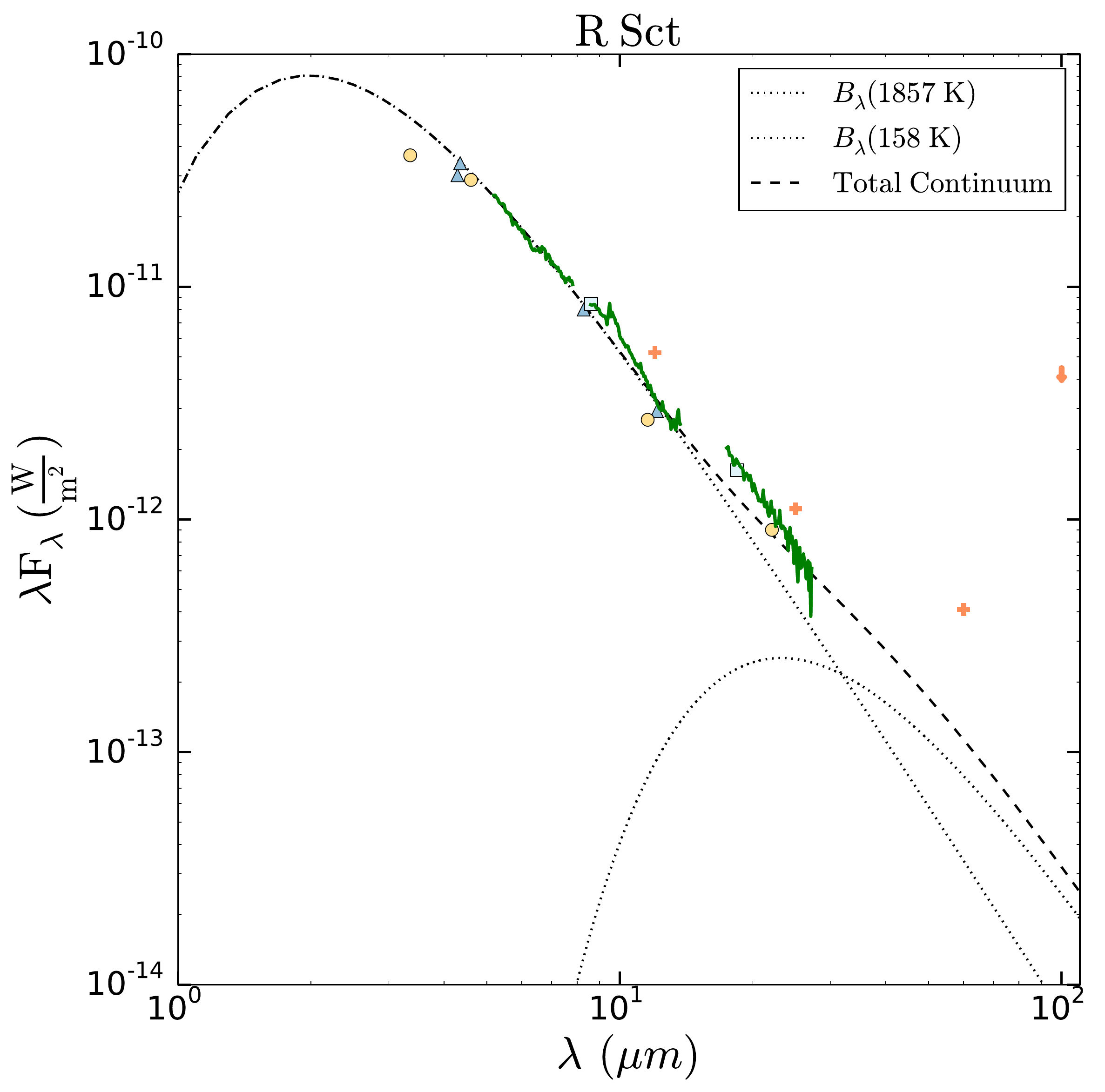}{0.5\textwidth}{(p) R Sct}}
\end{figure*}
\begin{figure*}
\gridline{\leftfig{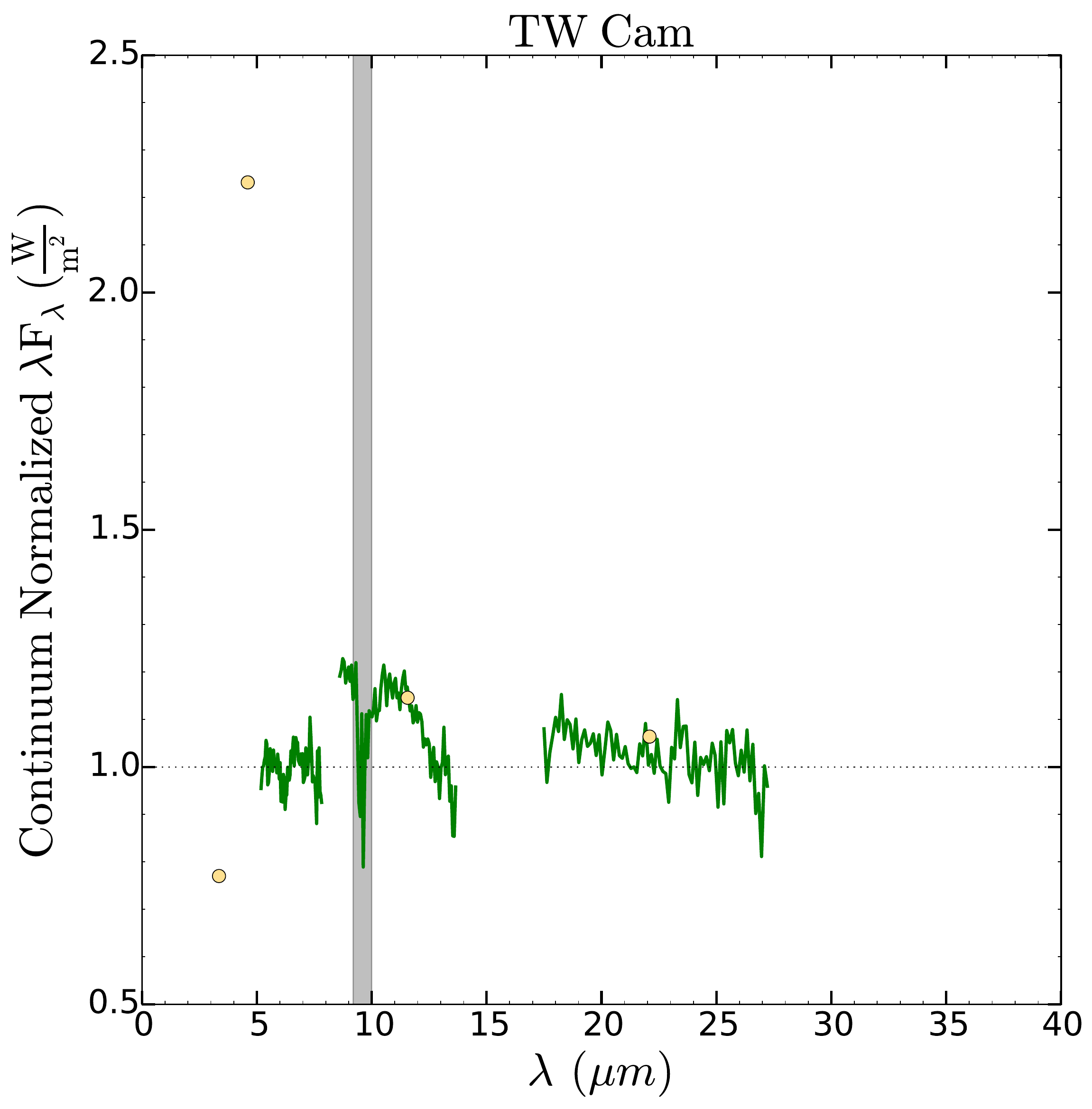}{0.5\textwidth}{(a) TW Cam}
          \rightfig{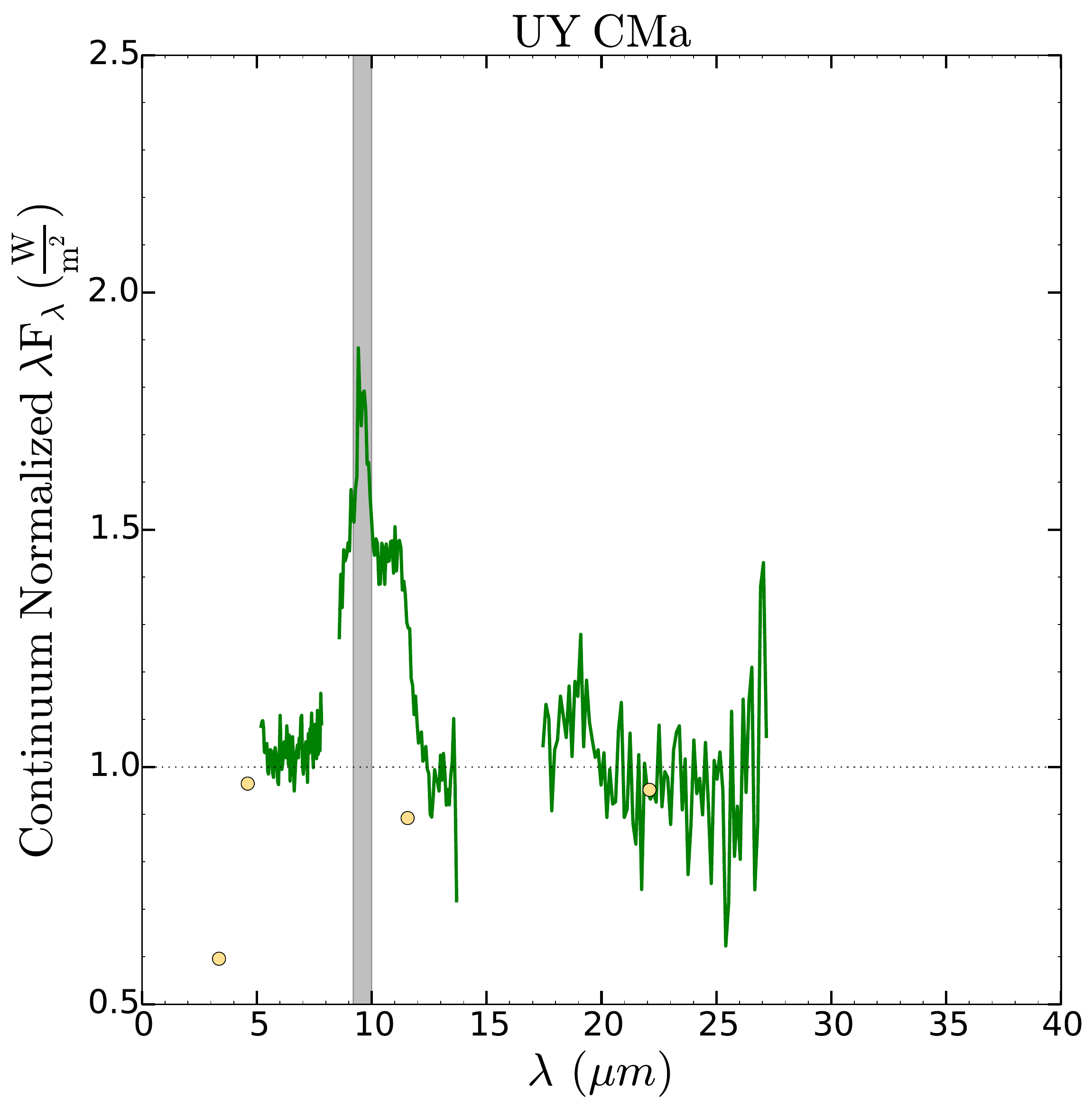}{0.5\textwidth}{(b) UY CMa}}
\gridline{\leftfig{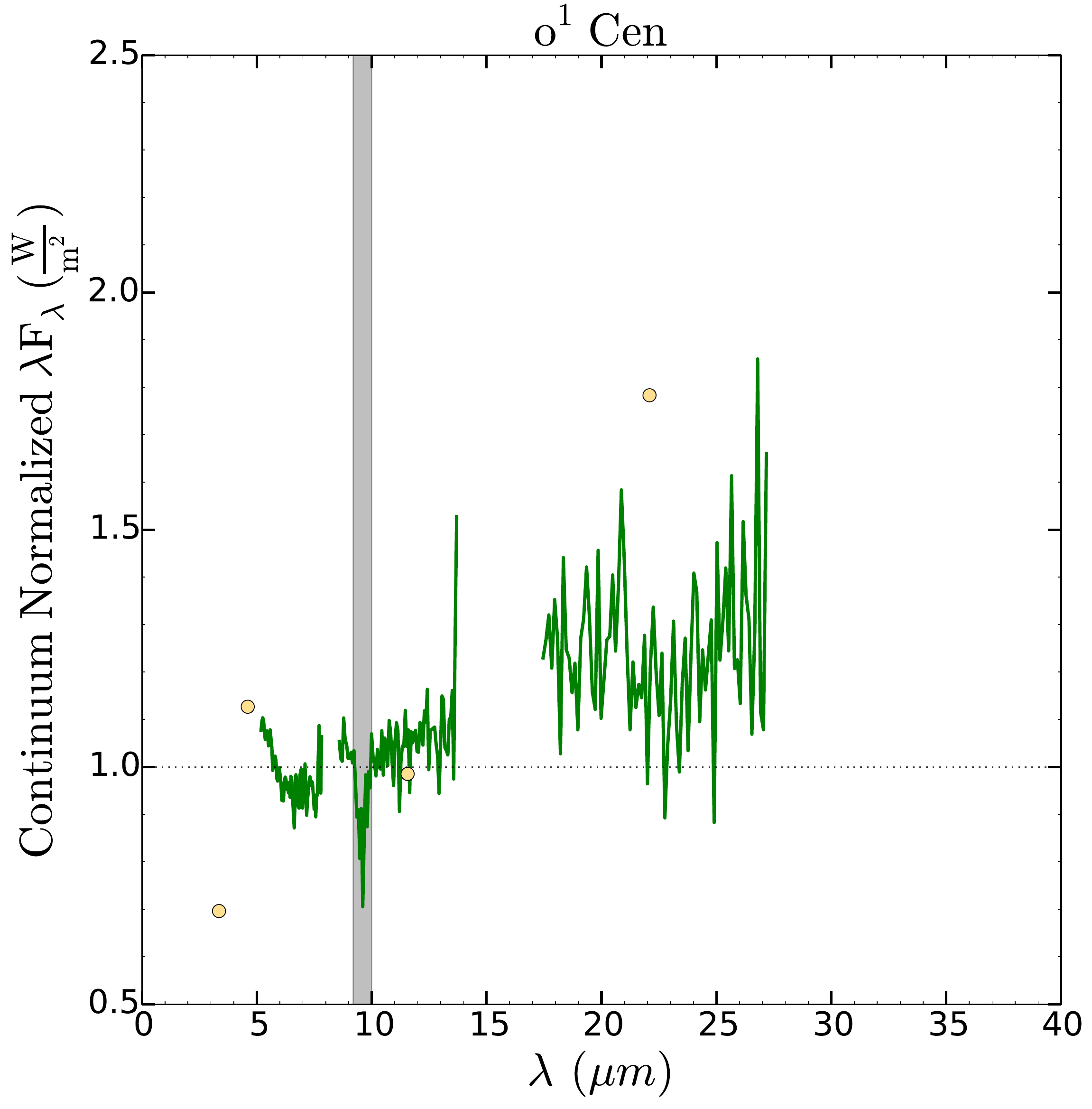}{0.5\textwidth}{(c) $\rm{o}^1$ Cen}
          \rightfig{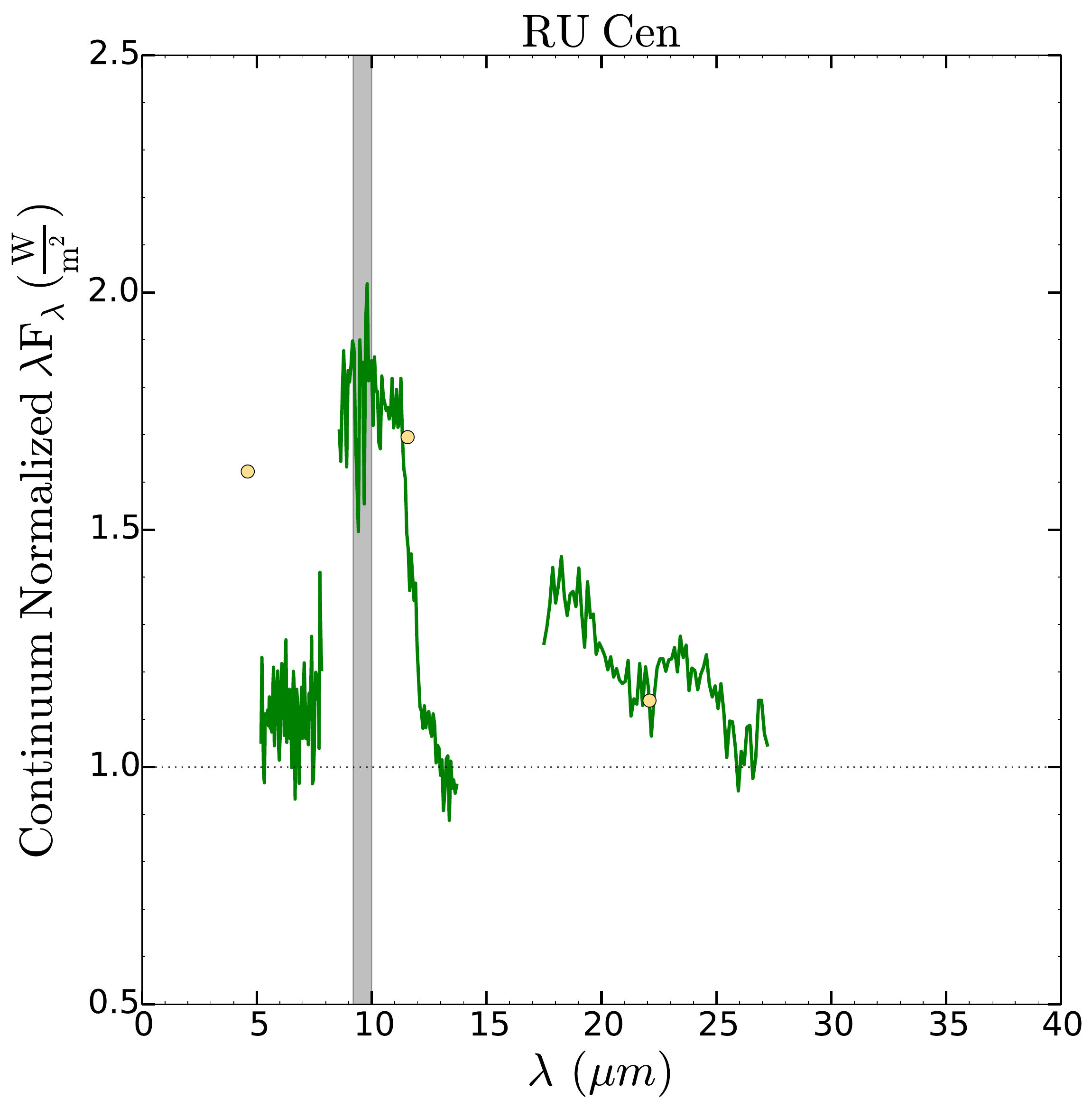}{0.5\textwidth}{(d) RU Cen}}
\end{figure*}
\begin{figure*}
\gridline{\leftfig{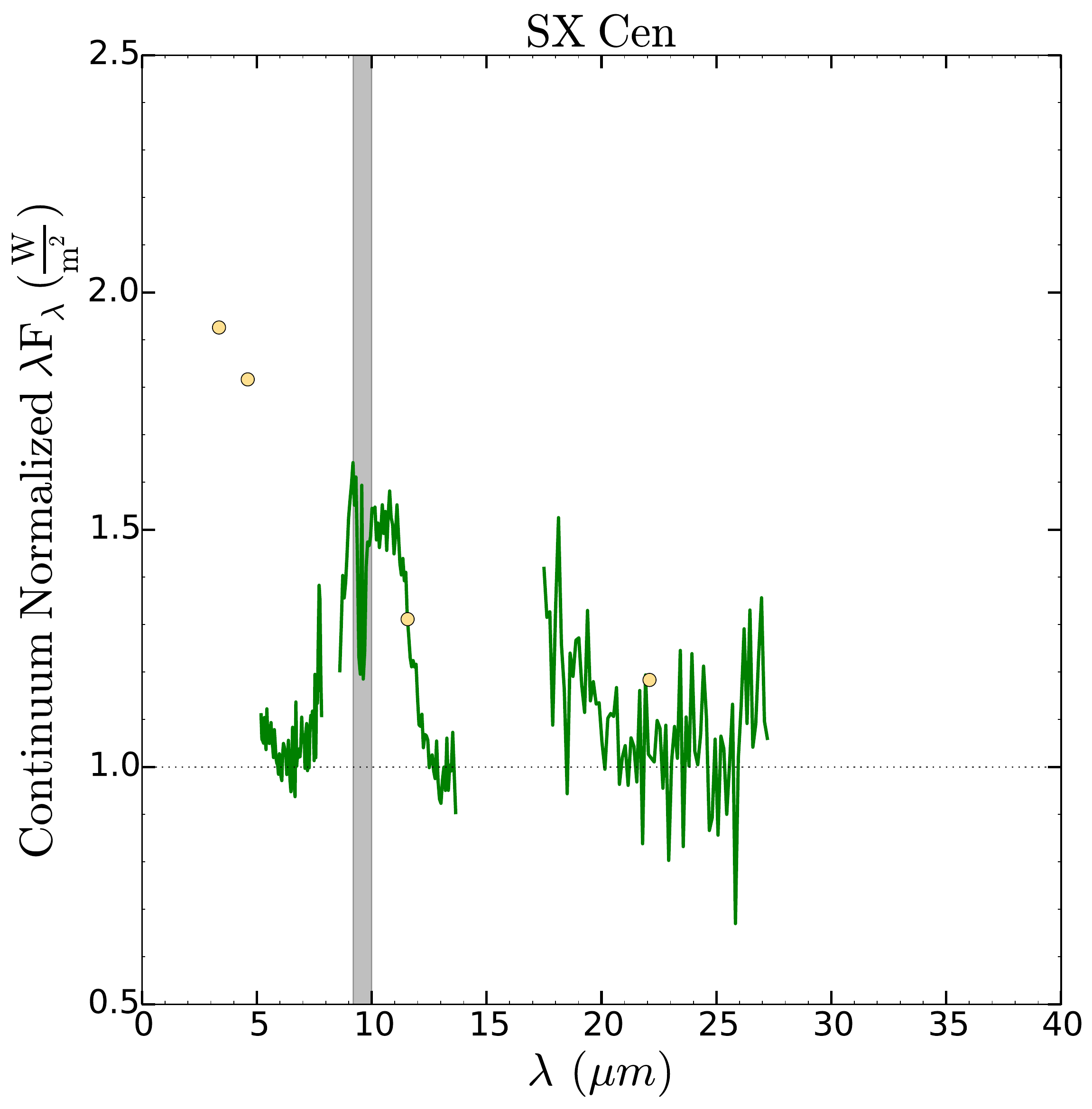}{0.5\textwidth}{(e) SX Cen}
          \rightfig{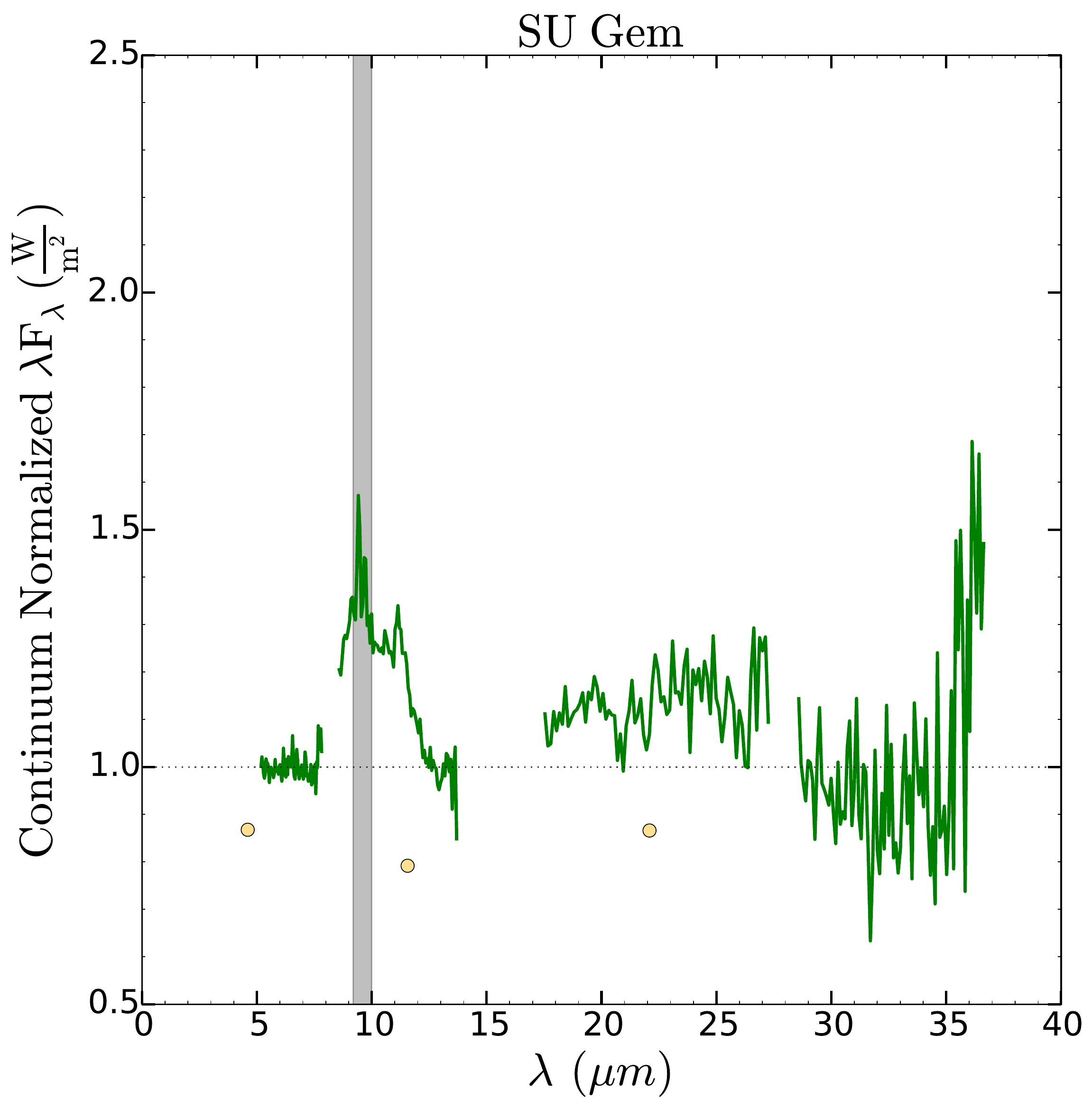}{0.5\textwidth}{(f) SU Gem}}
\gridline{\leftfig{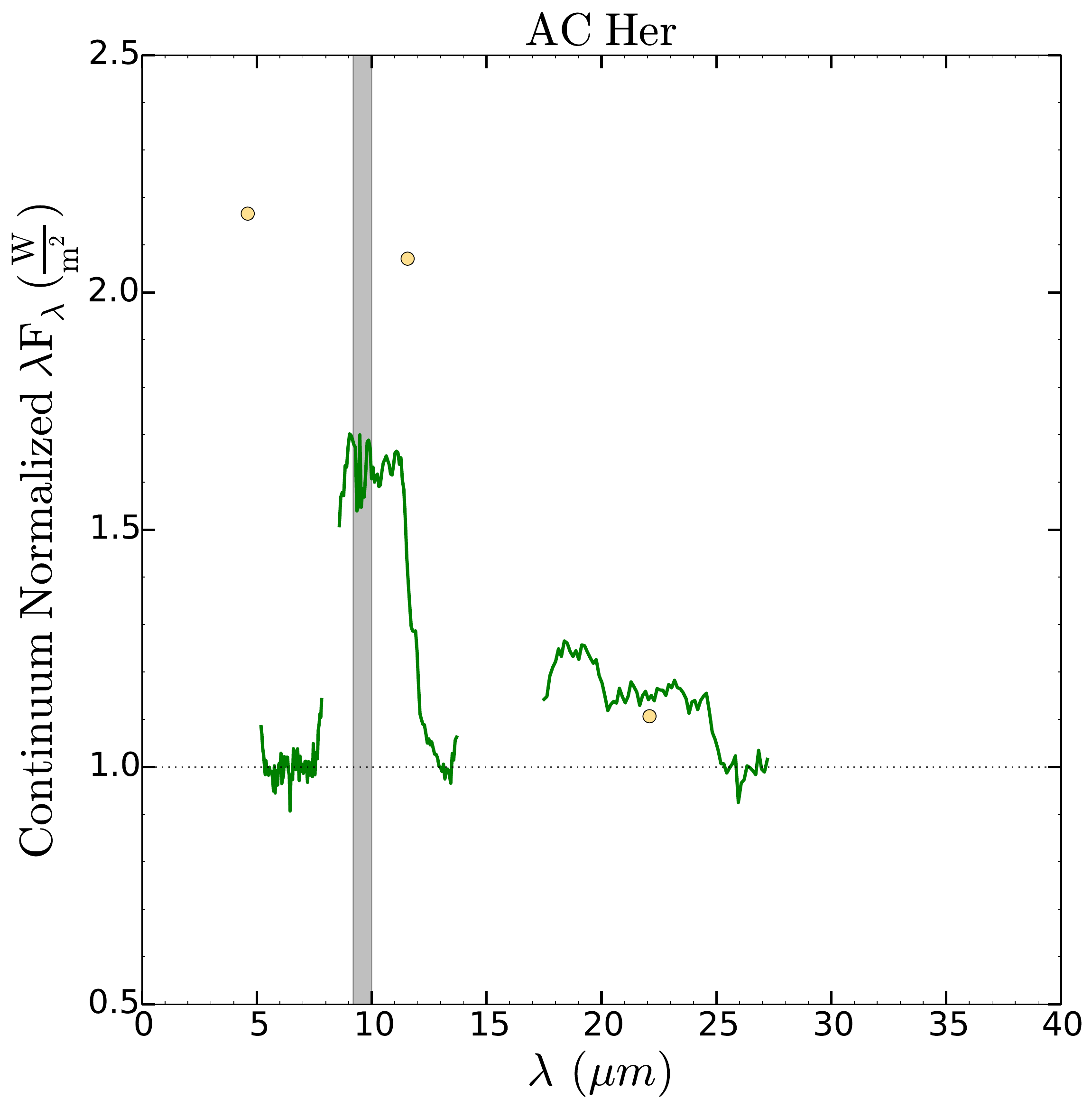}{0.5\textwidth}{(g) AC Her}
          \rightfig{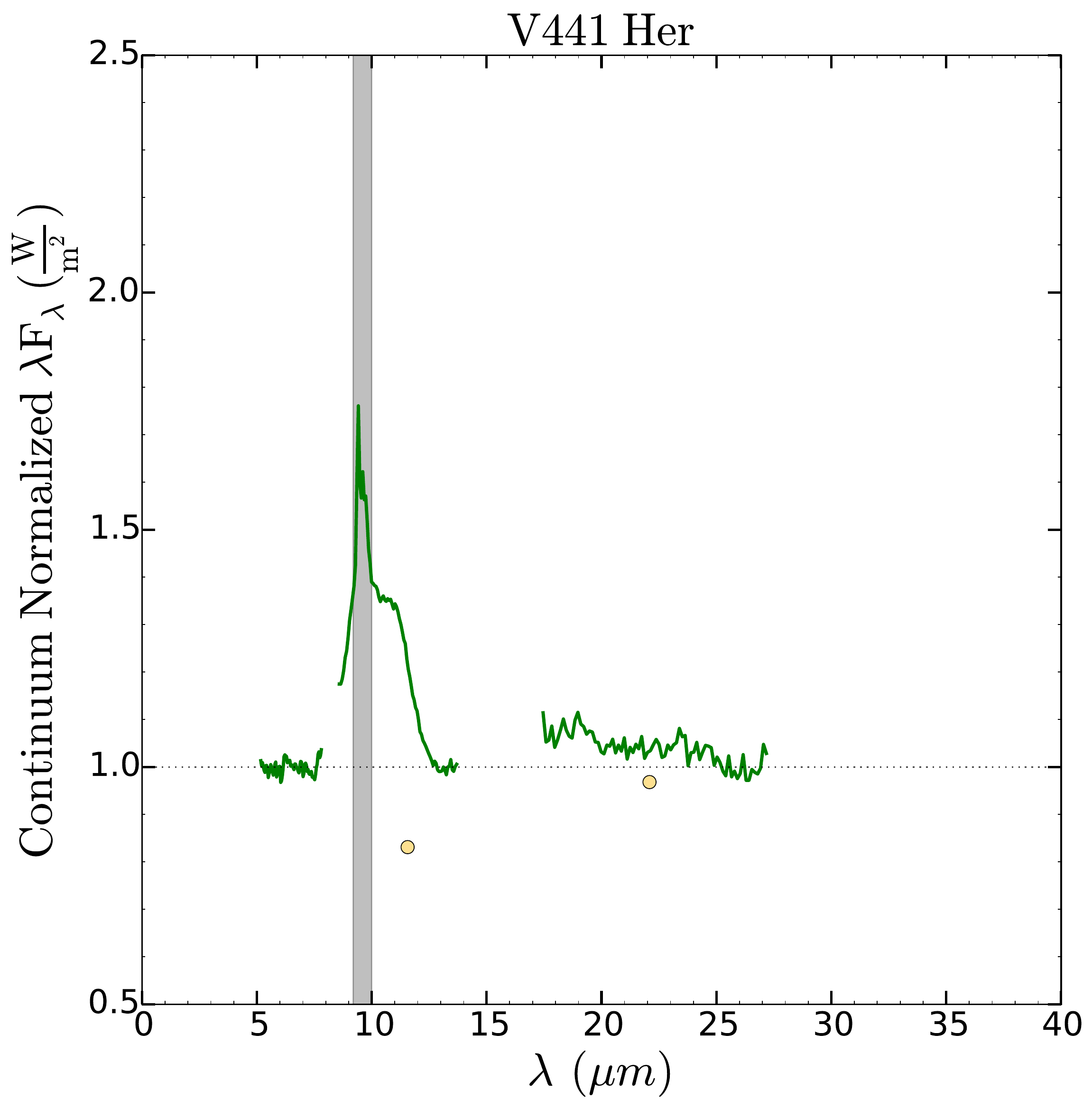}{0.5\textwidth}{(h) V441 Her}}
\end{figure*}
\begin{figure*}
\gridline{\leftfig{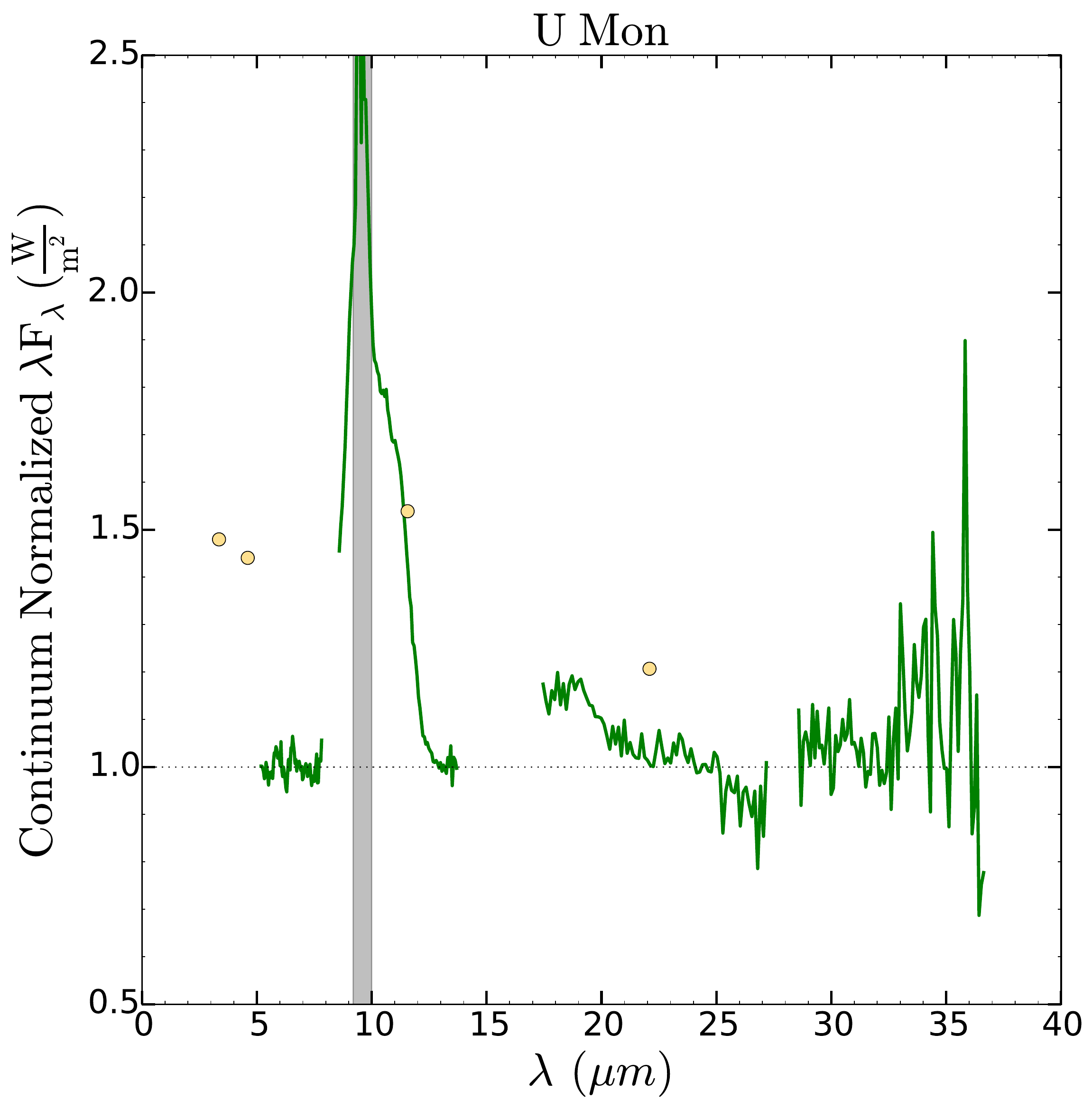}{0.5\textwidth}{(i) U Mon}
          \rightfig{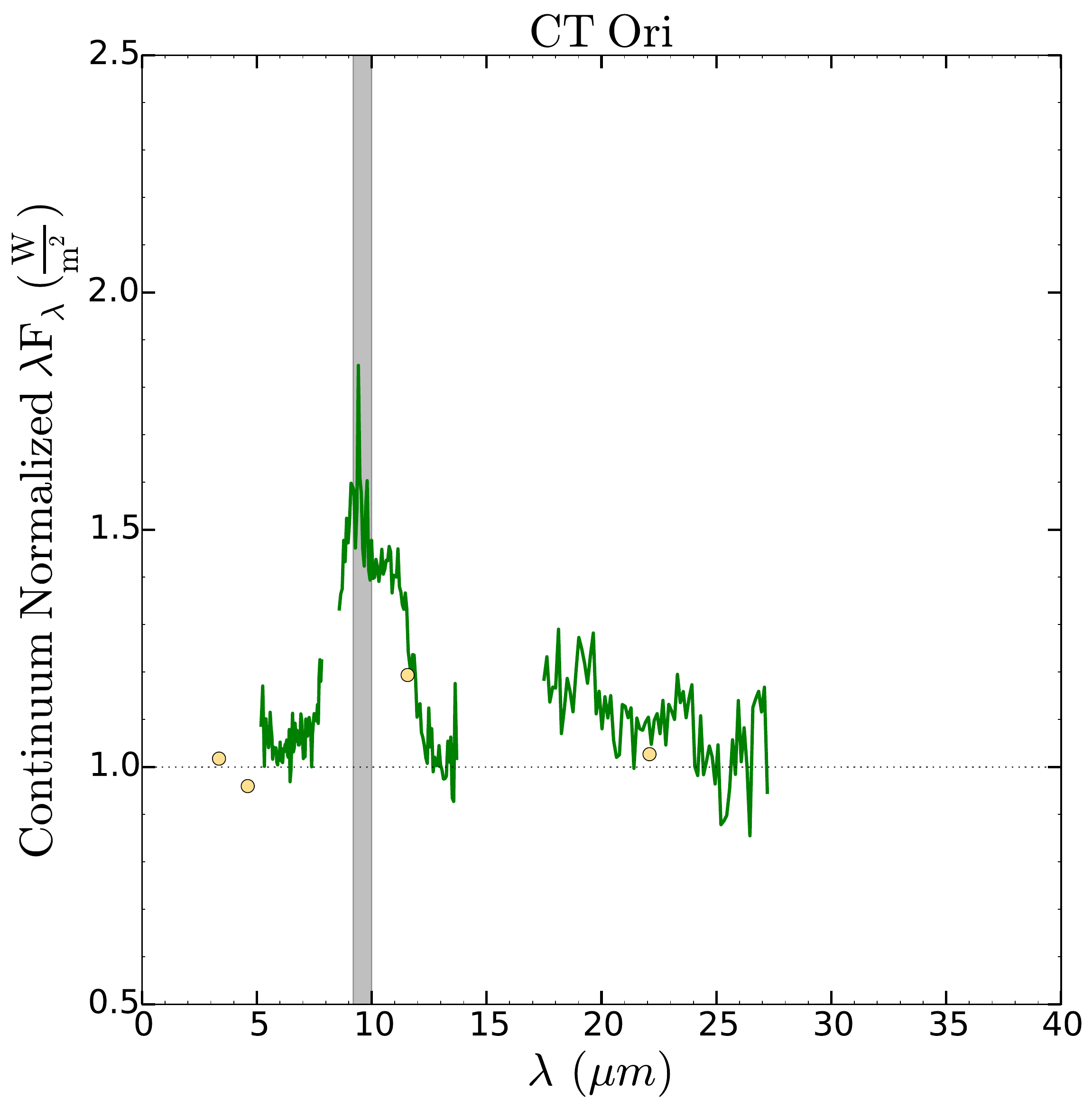}{0.5\textwidth}{(j) CT Ori}}
\gridline{\leftfig{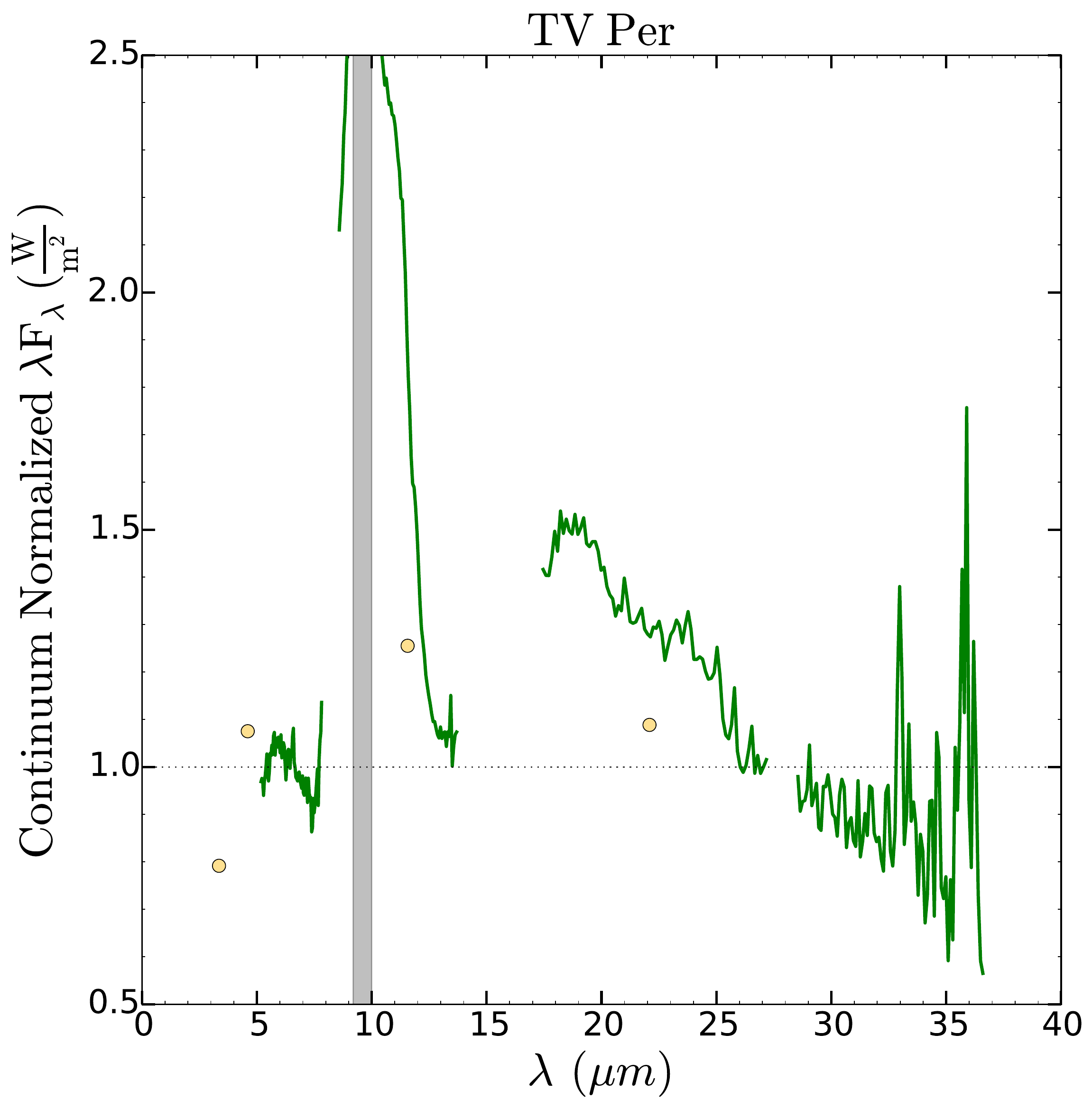}{0.5\textwidth}{(k) TV Per}
	  \rightfig{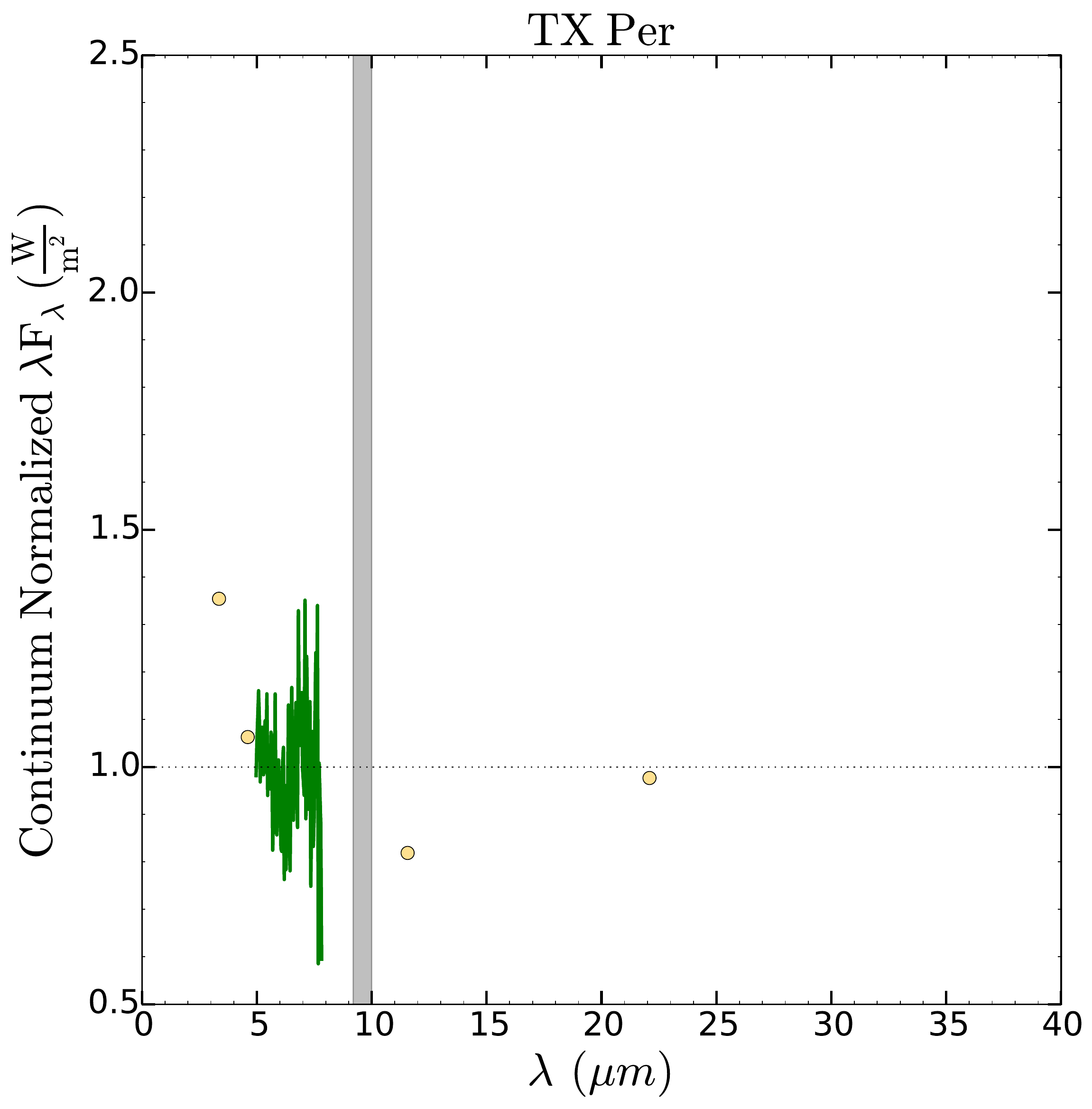}{0.5\textwidth}{(l) TX Per}}
\end{figure*}
\begin{figure*}
\gridline{\leftfig{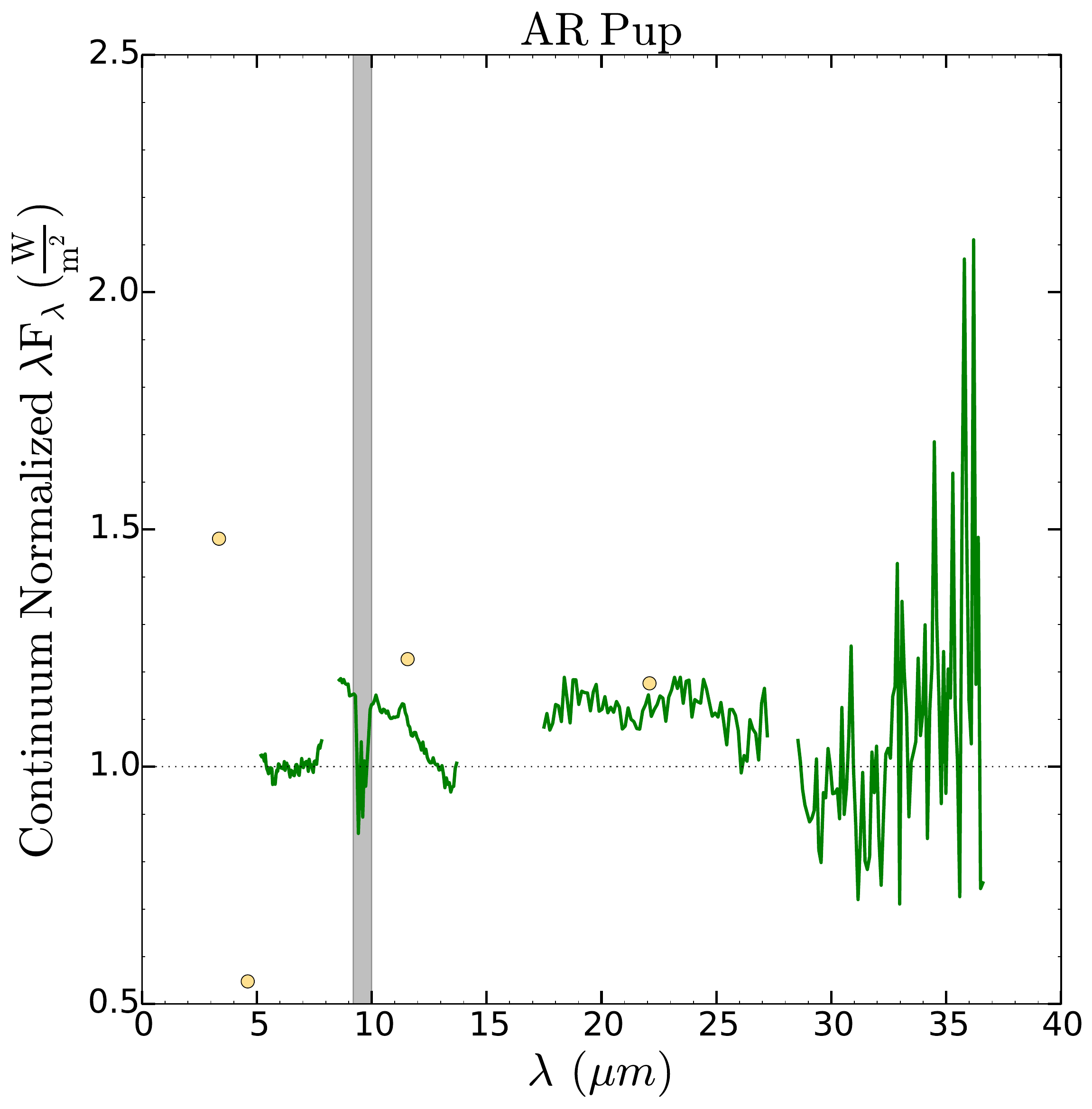}{0.5\textwidth}{(m) AR Pup}
	      \rightfig{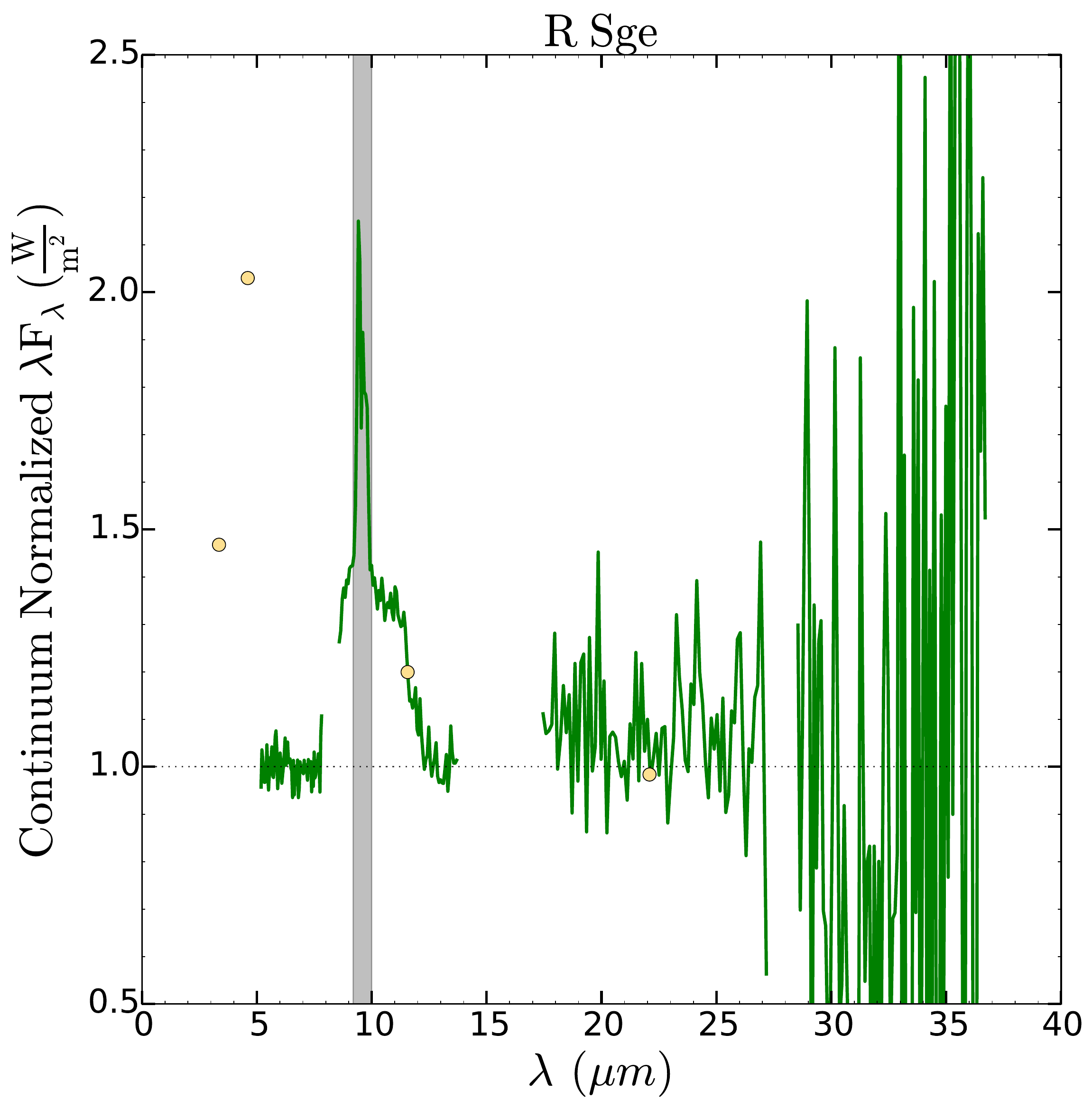}{0.5\textwidth}{(n) R Sge}}
\gridline{\leftfig{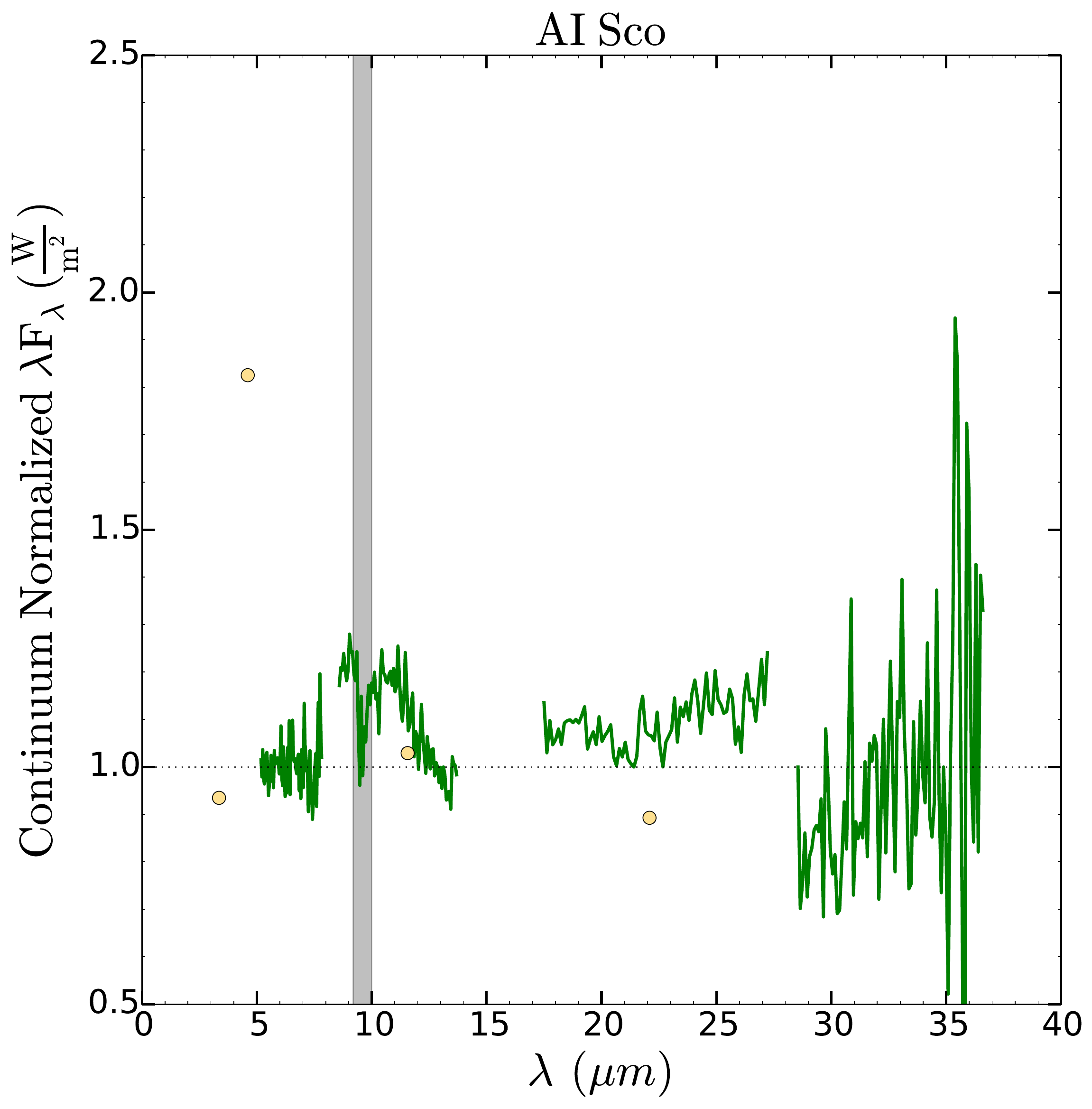}{0.5\textwidth}{(o) AI Sco}
	     \rightfig{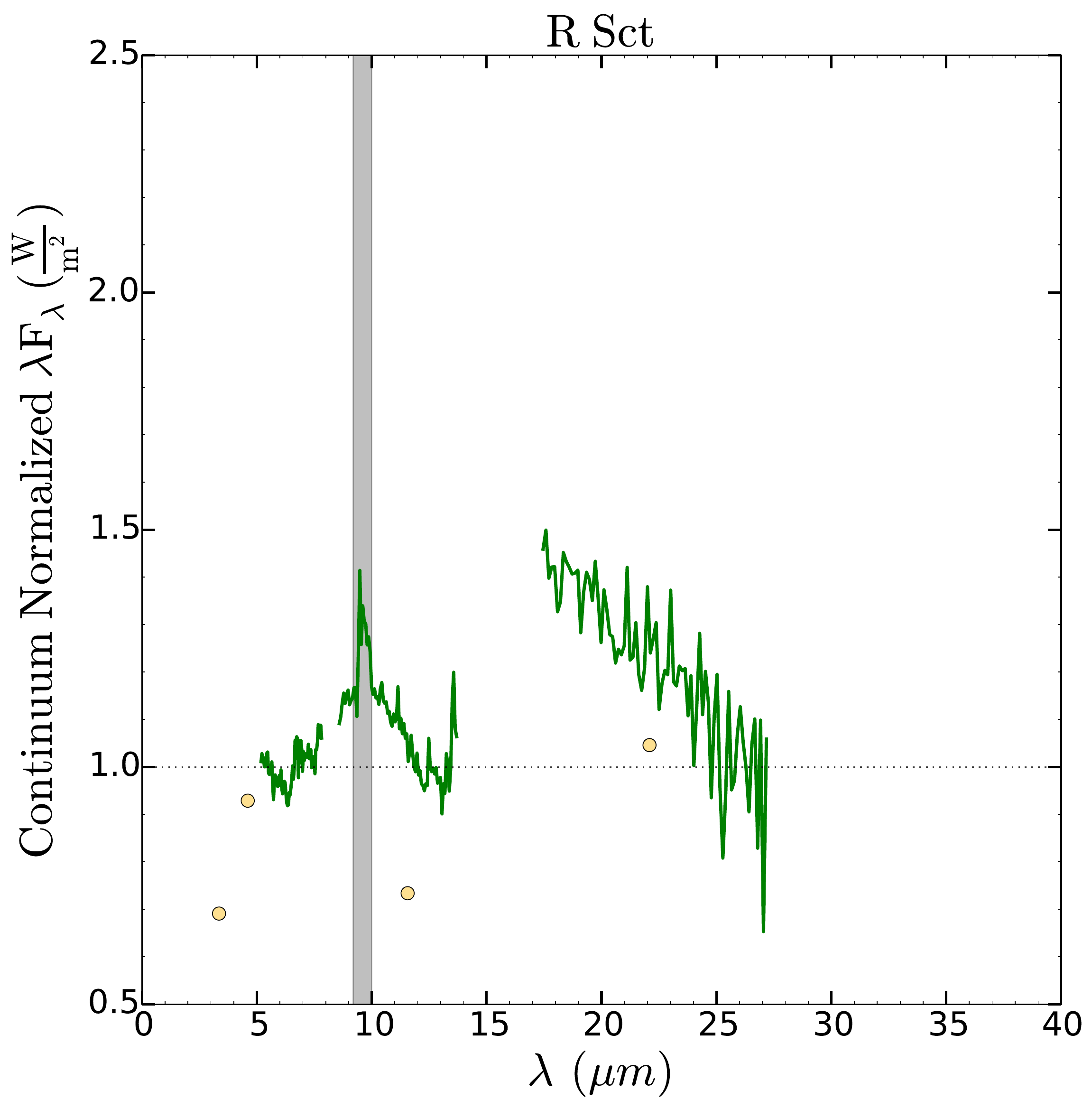}{0.5\textwidth}{(p) R Sct}}
\end{figure*}
\begin{figure*}
\gridline{\leftfig{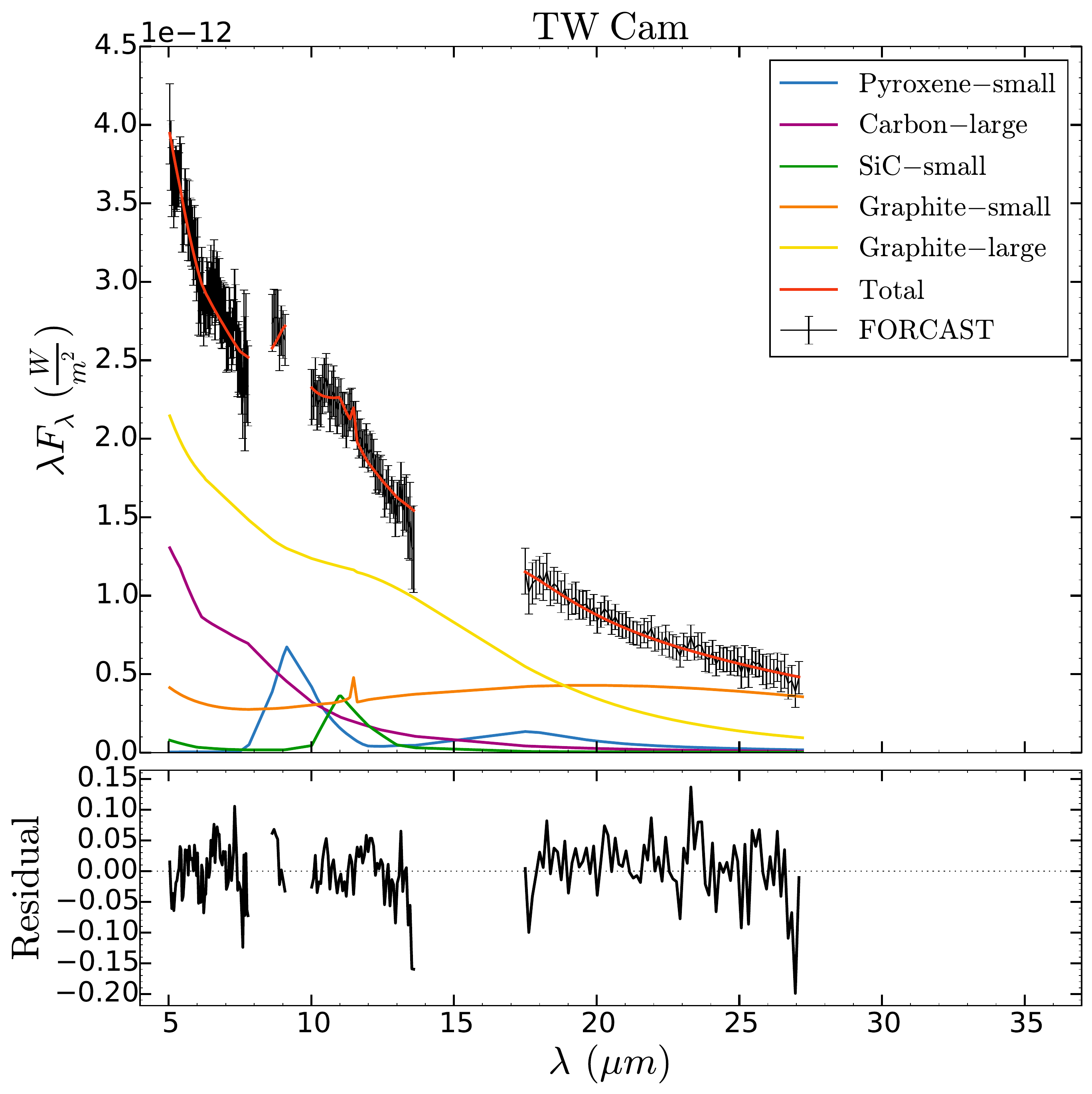}{0.5\textwidth}{(a) TW Cam}
          \rightfig{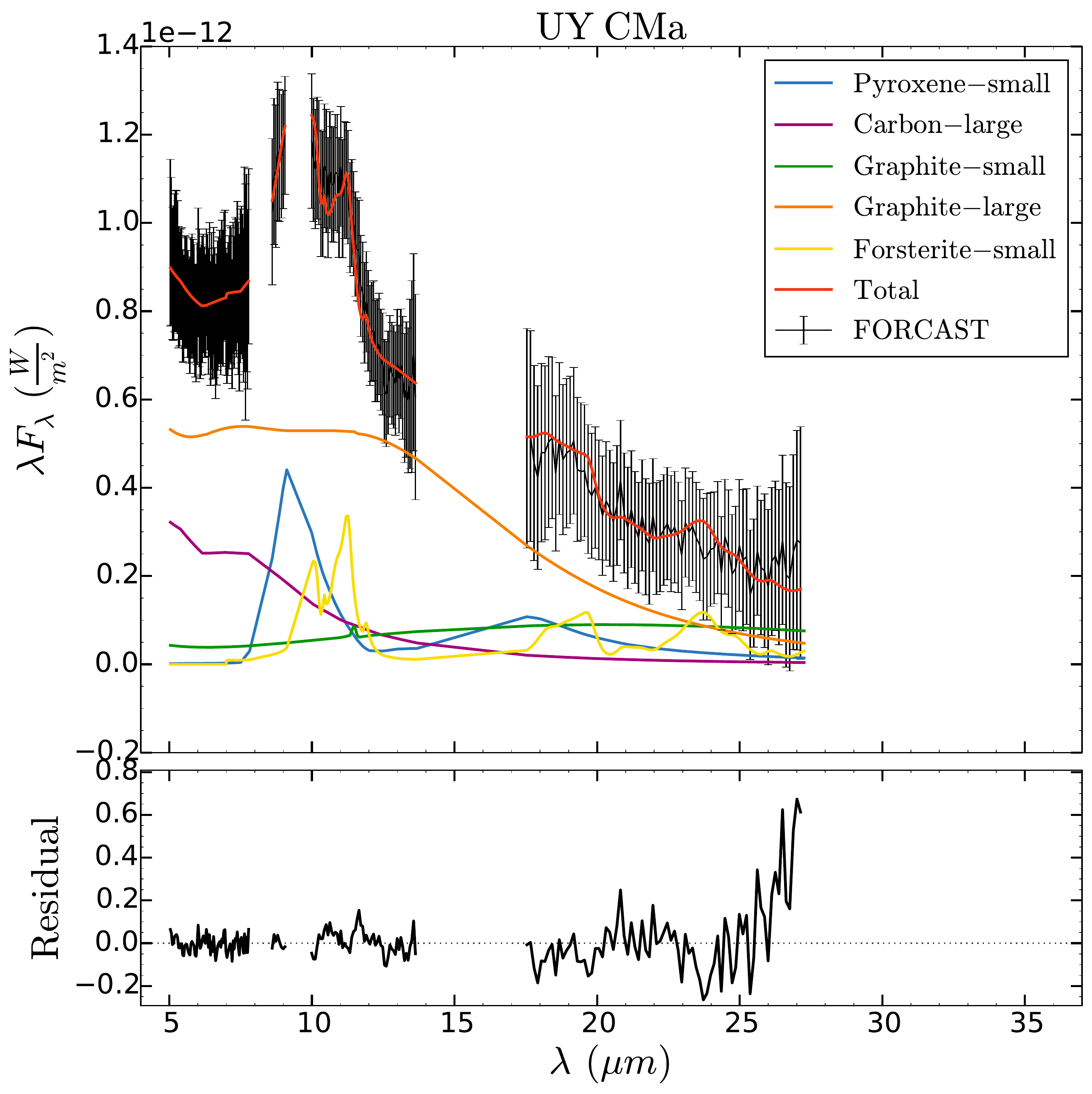}{0.5\textwidth}{(b) UY CMa}}
\gridline{\leftfig{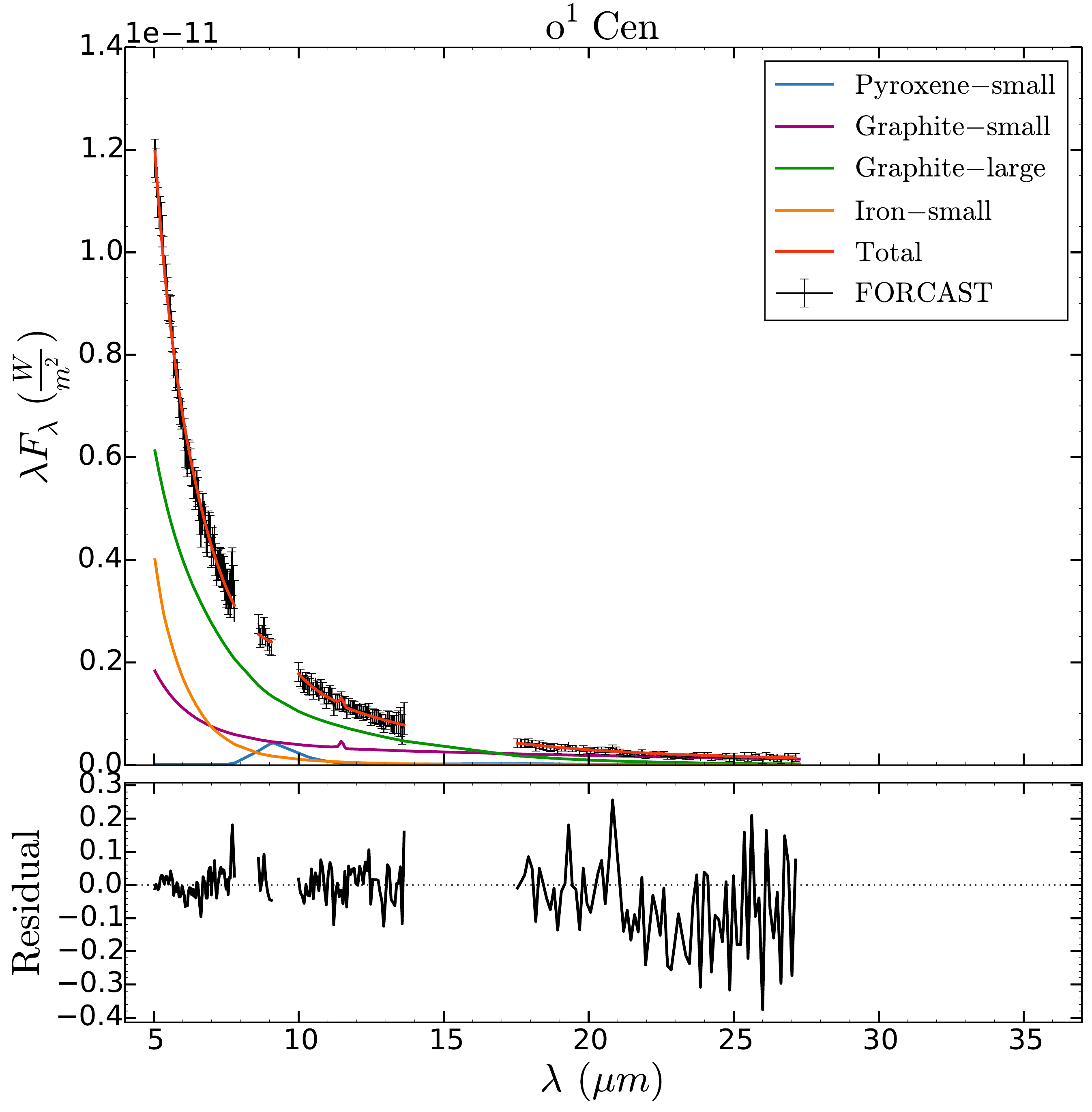}{0.5\textwidth}{(c) $\rm{o}^1$ Cen}
          \rightfig{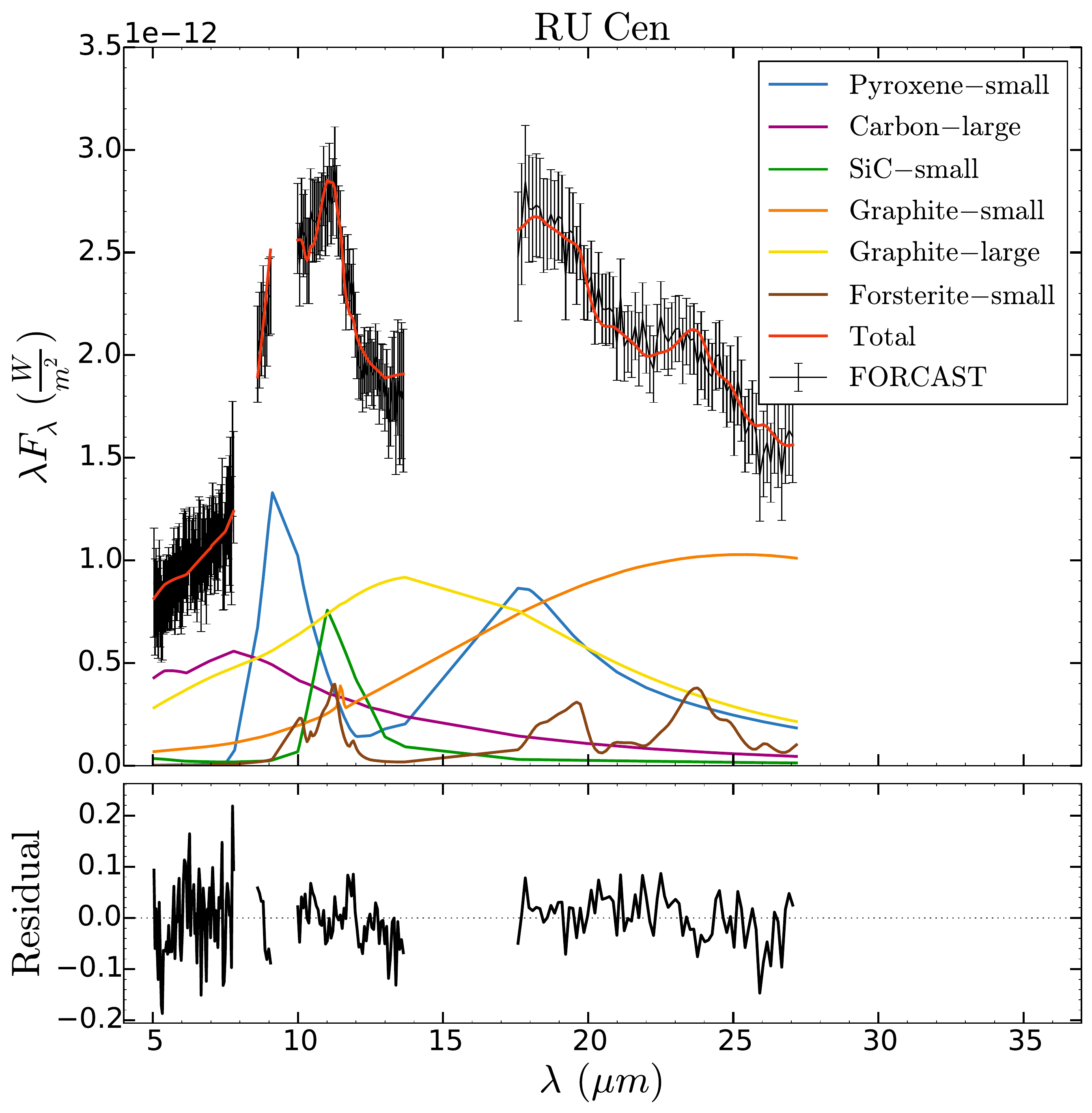}{0.5\textwidth}{(d) RU Cen}}
\end{figure*}
\begin{figure*}
\gridline{\leftfig{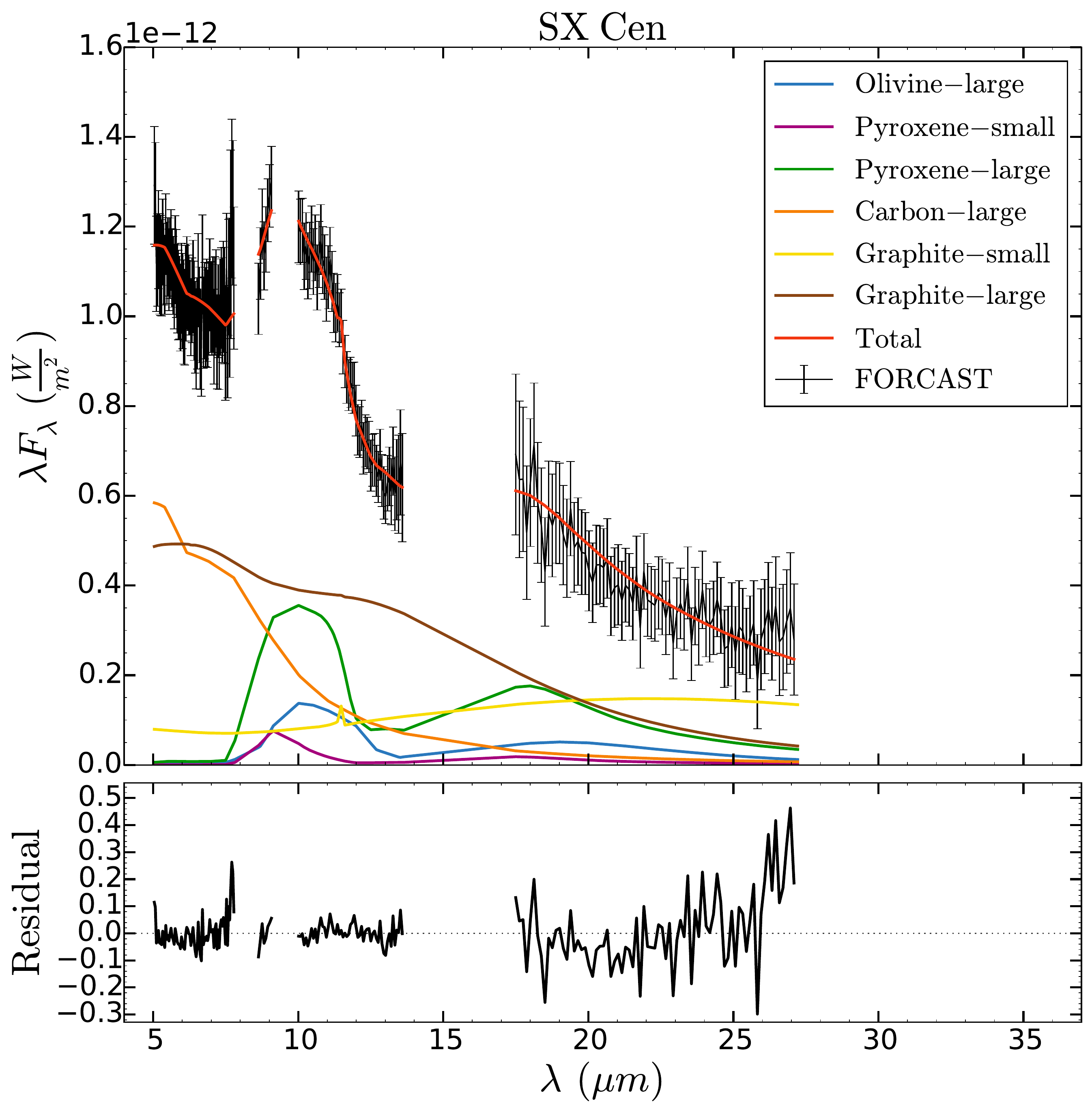}{0.5\textwidth}{(e) SX Cen}
          \rightfig{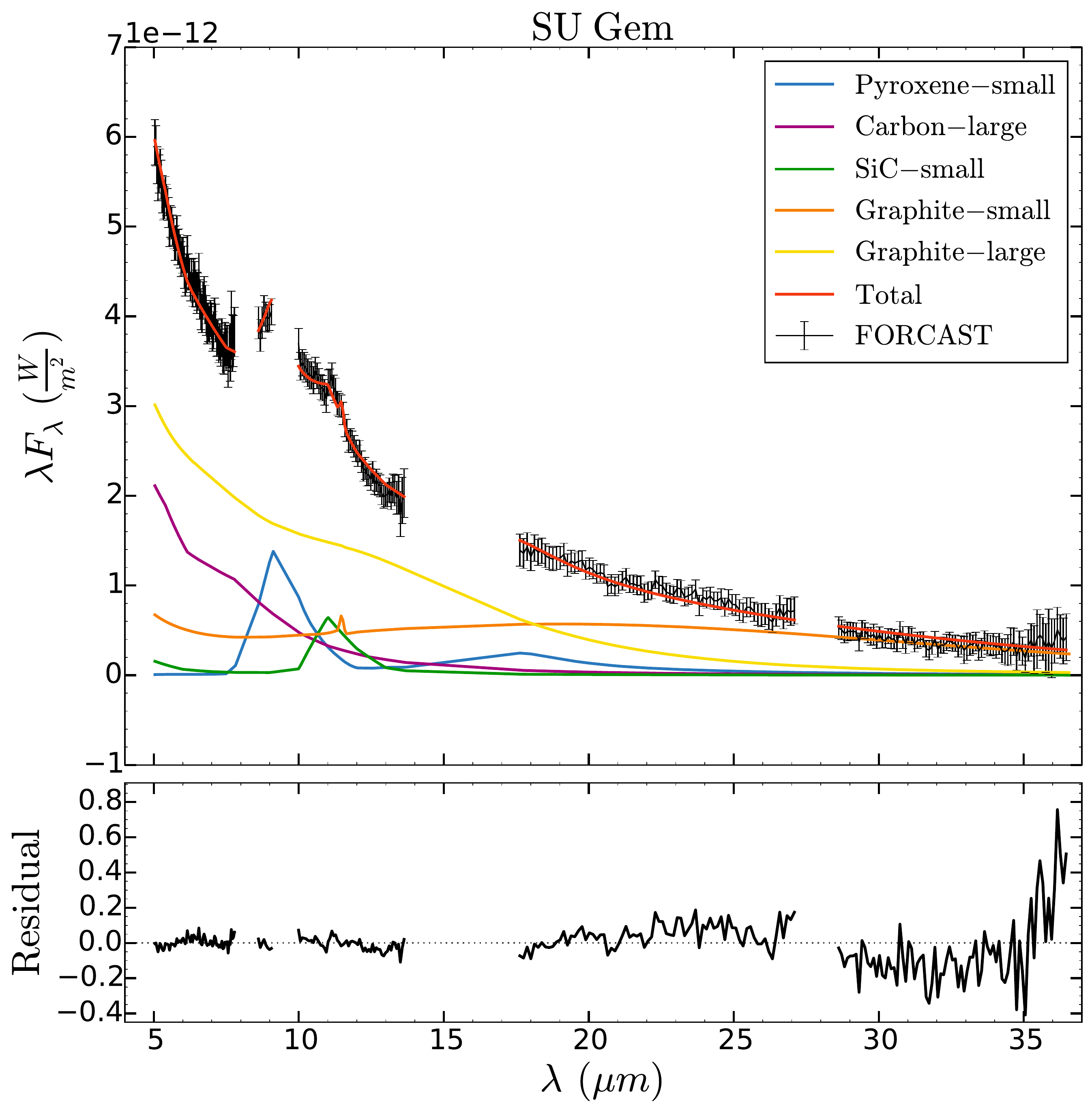}{0.5\textwidth}{(f) SU Gem}}
\gridline{\leftfig{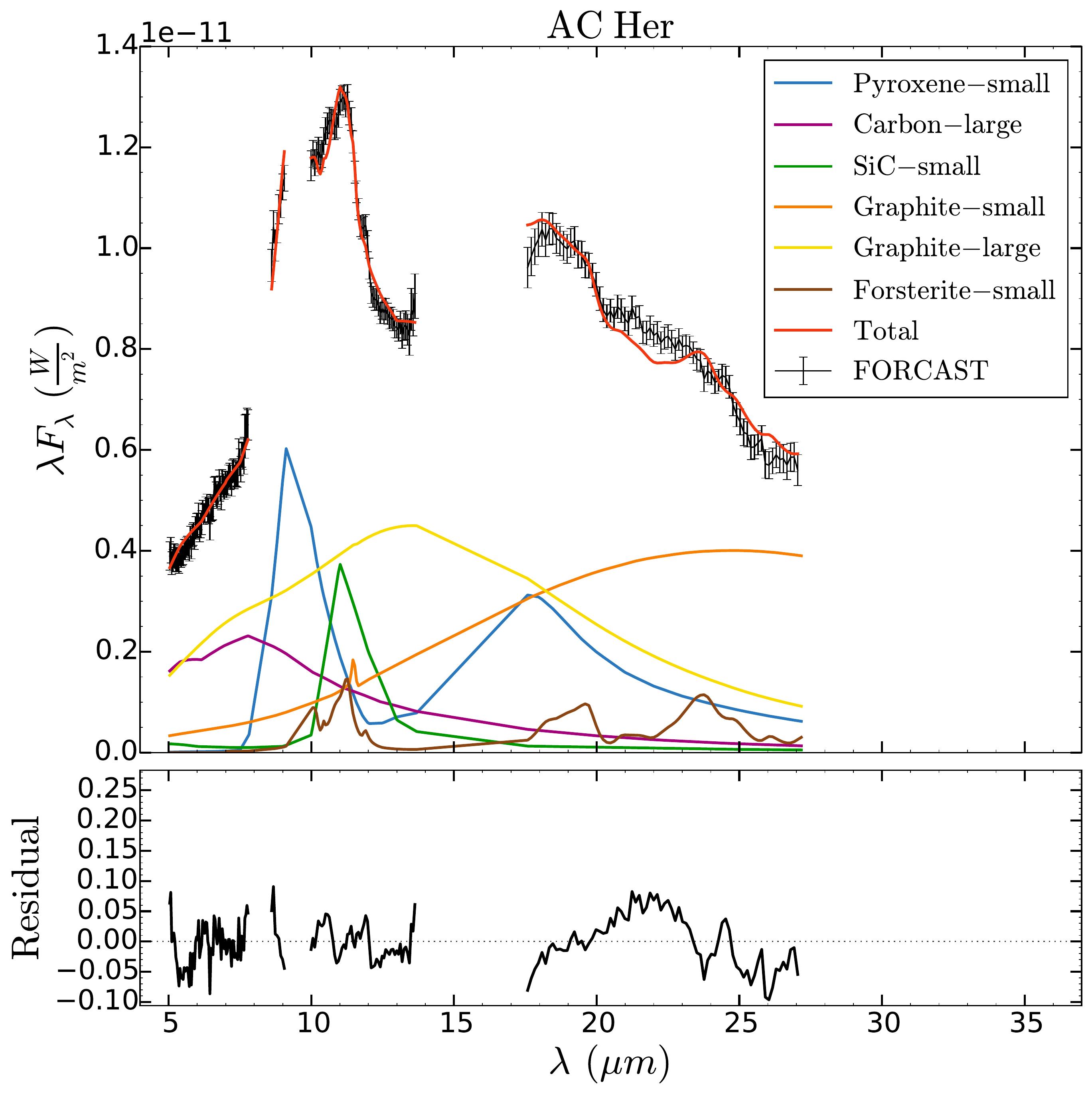}{0.5\textwidth}{(g) AC Her}
          \rightfig{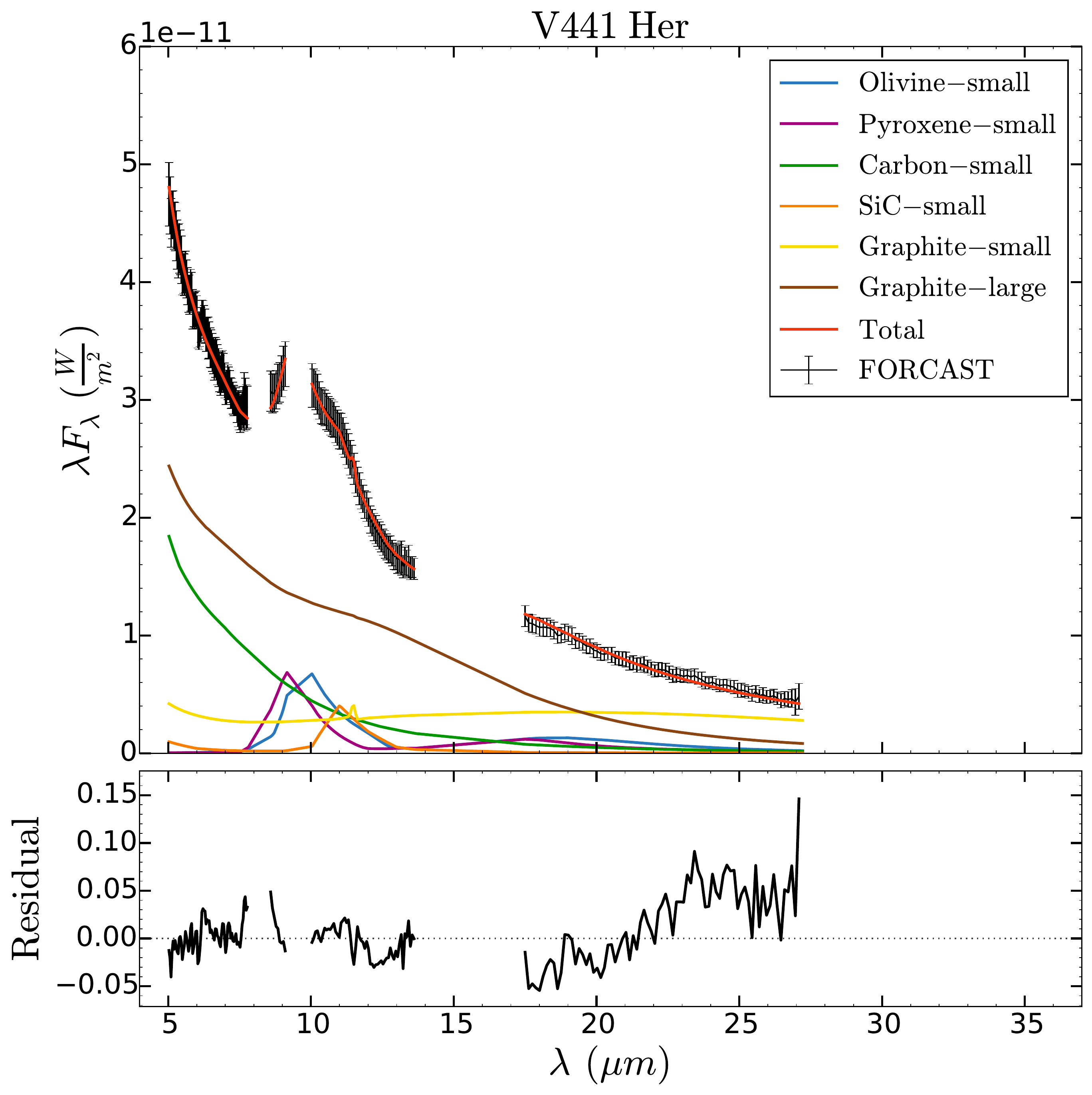}{0.5\textwidth}{(h) V441 Her}}
\end{figure*}
\begin{figure*}
\gridline{\leftfig{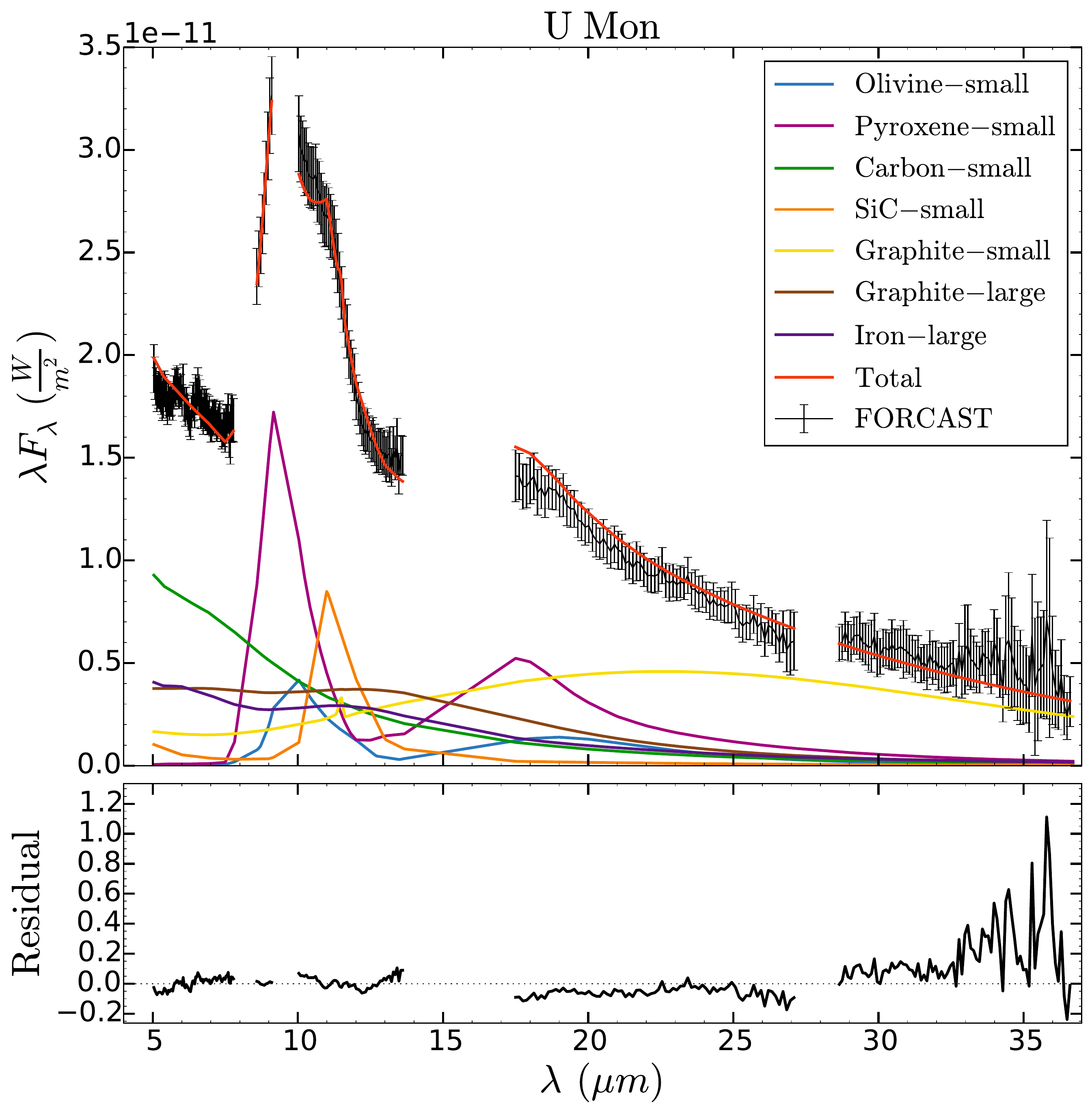}{0.5\textwidth}{(i) U Mon}
          \rightfig{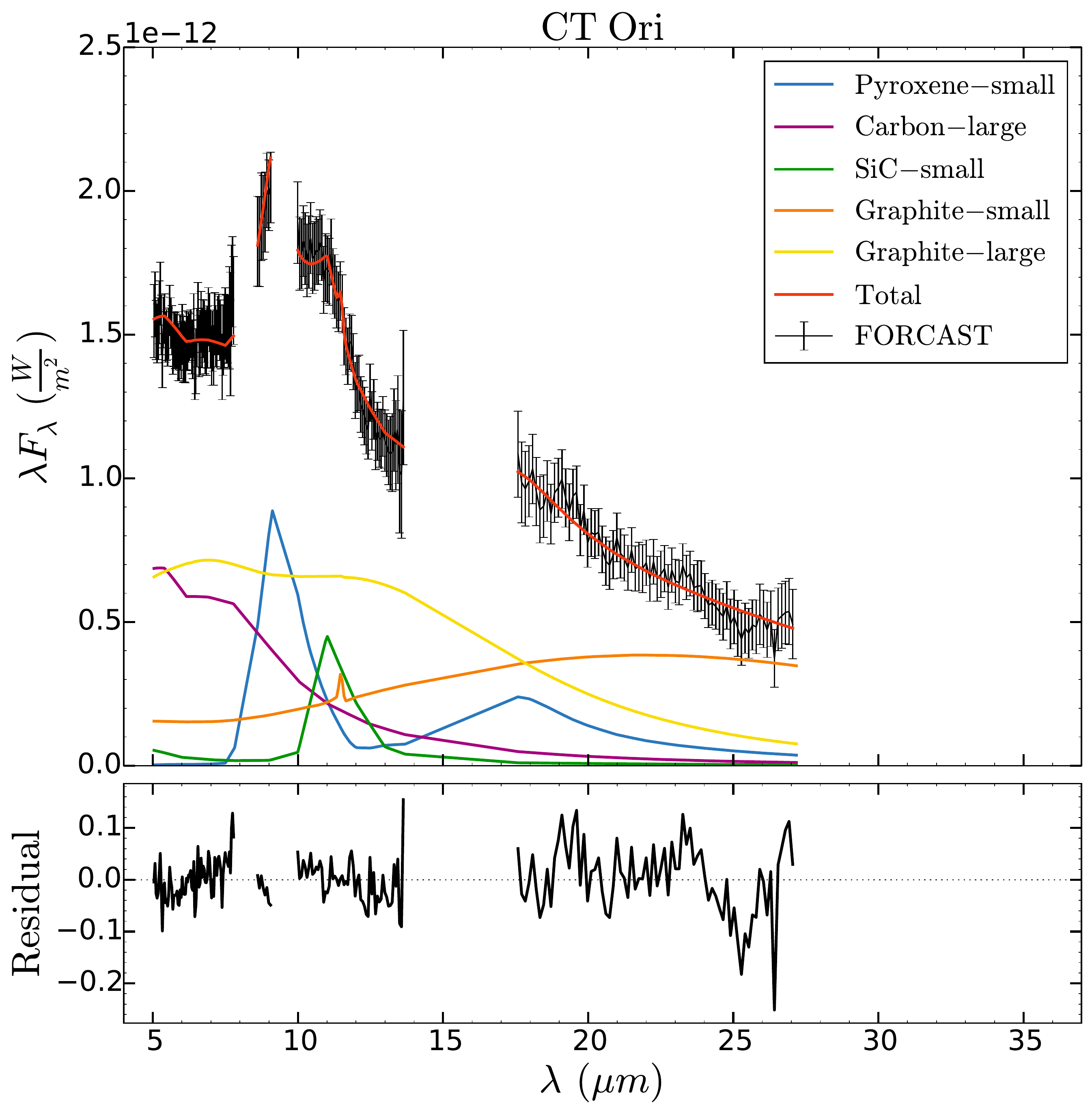}{0.5\textwidth}{(j) CT Ori}}
\gridline{\leftfig{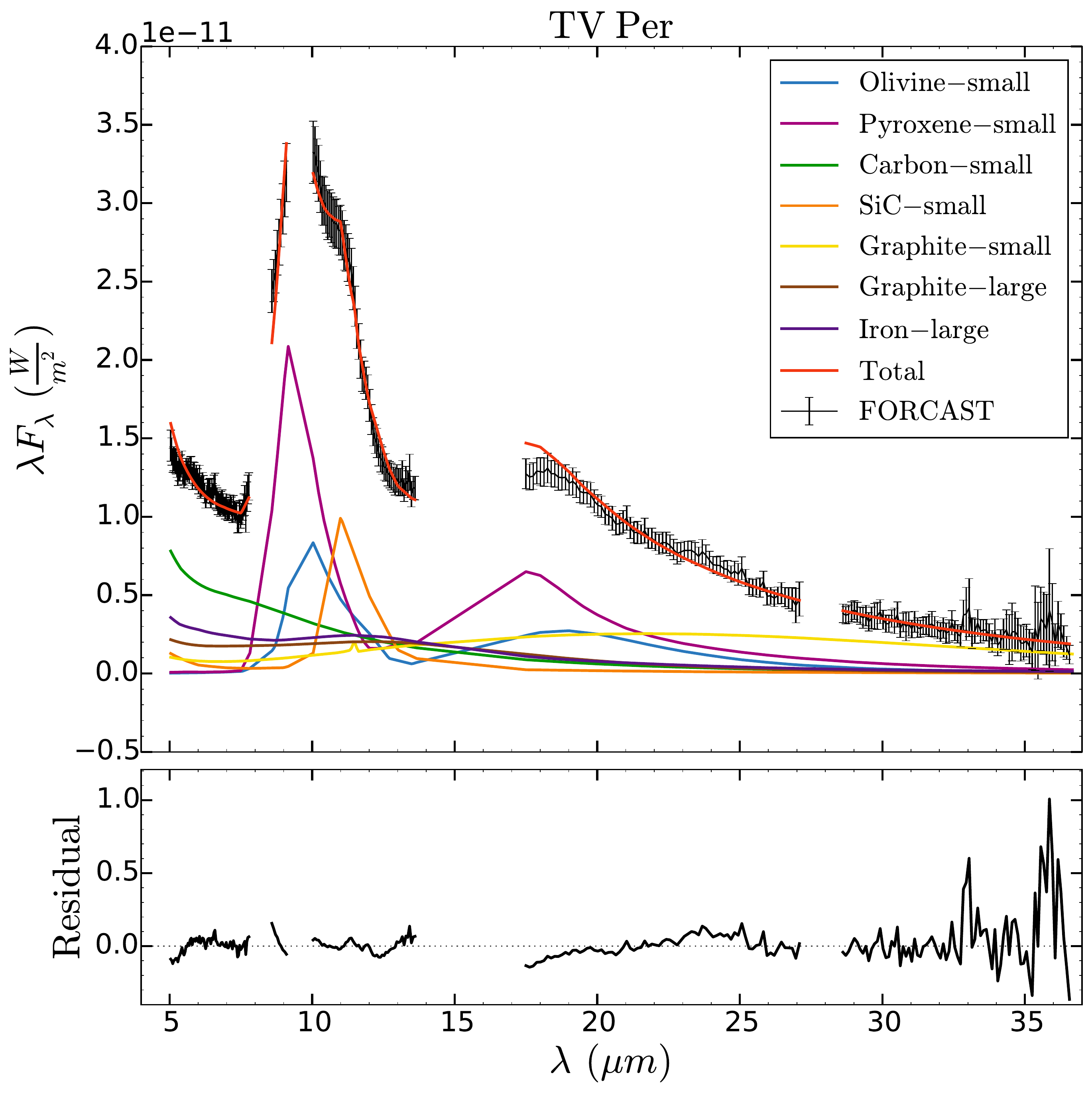}{0.5\textwidth}{(k) TV Per}
          \rightfig{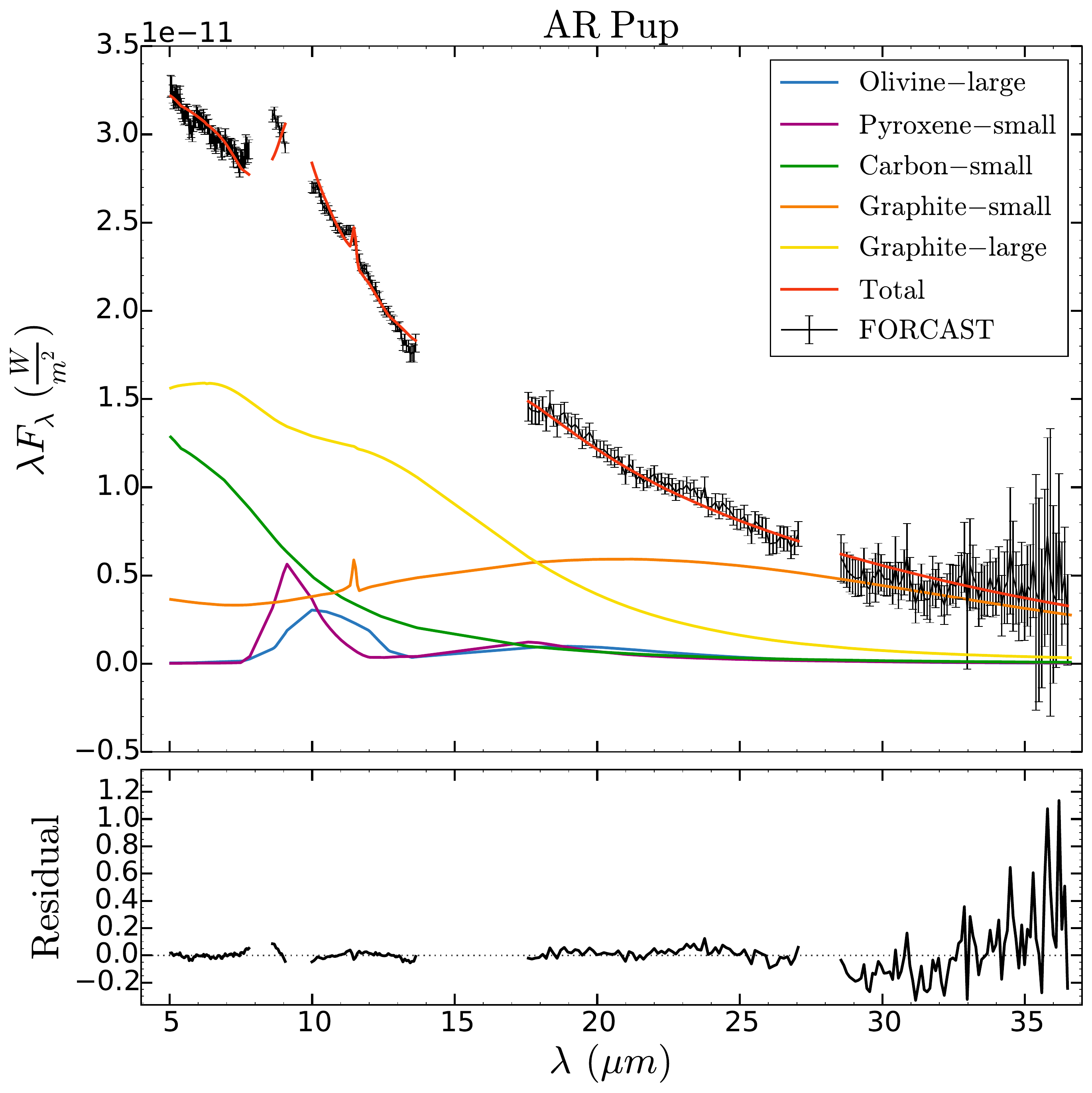}{0.5\textwidth}{(l) AR Pup}}
\end{figure*}
\begin{figure*}
\gridline{\leftfig{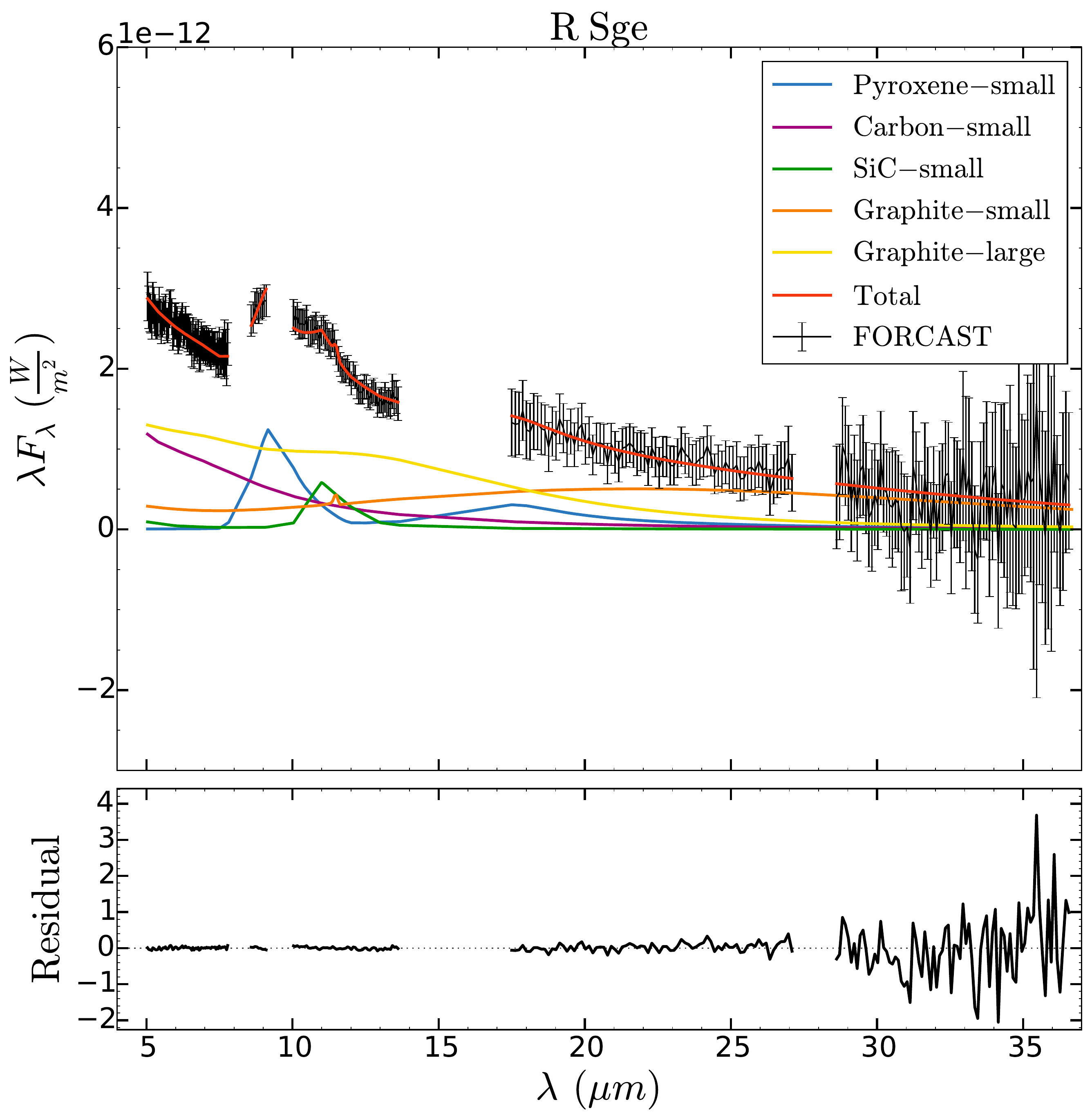}{0.5\textwidth}{(m) R Sge}
          \rightfig{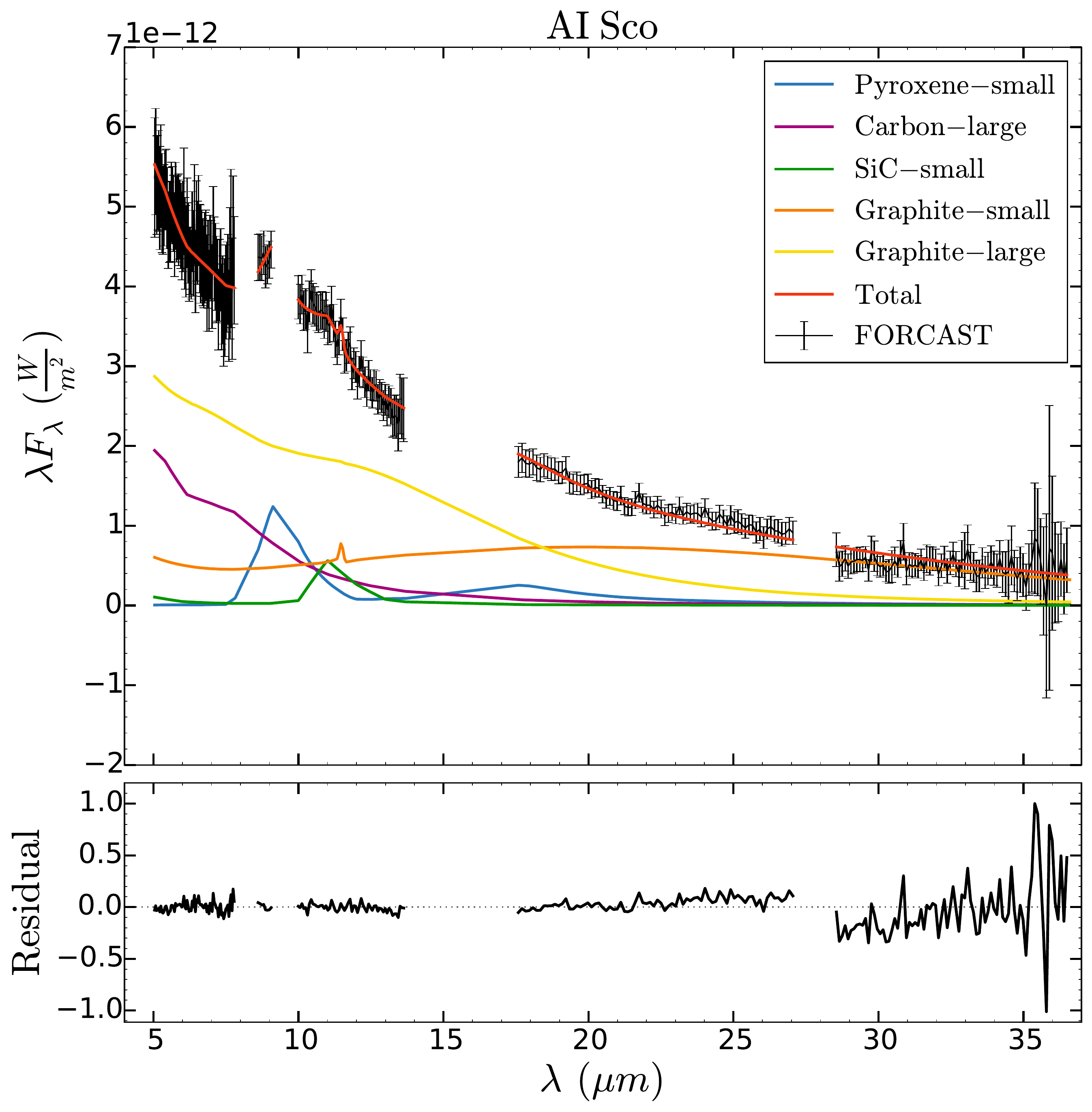}{0.5\textwidth}{(n) AI Sco}}
\gridline{\leftfig{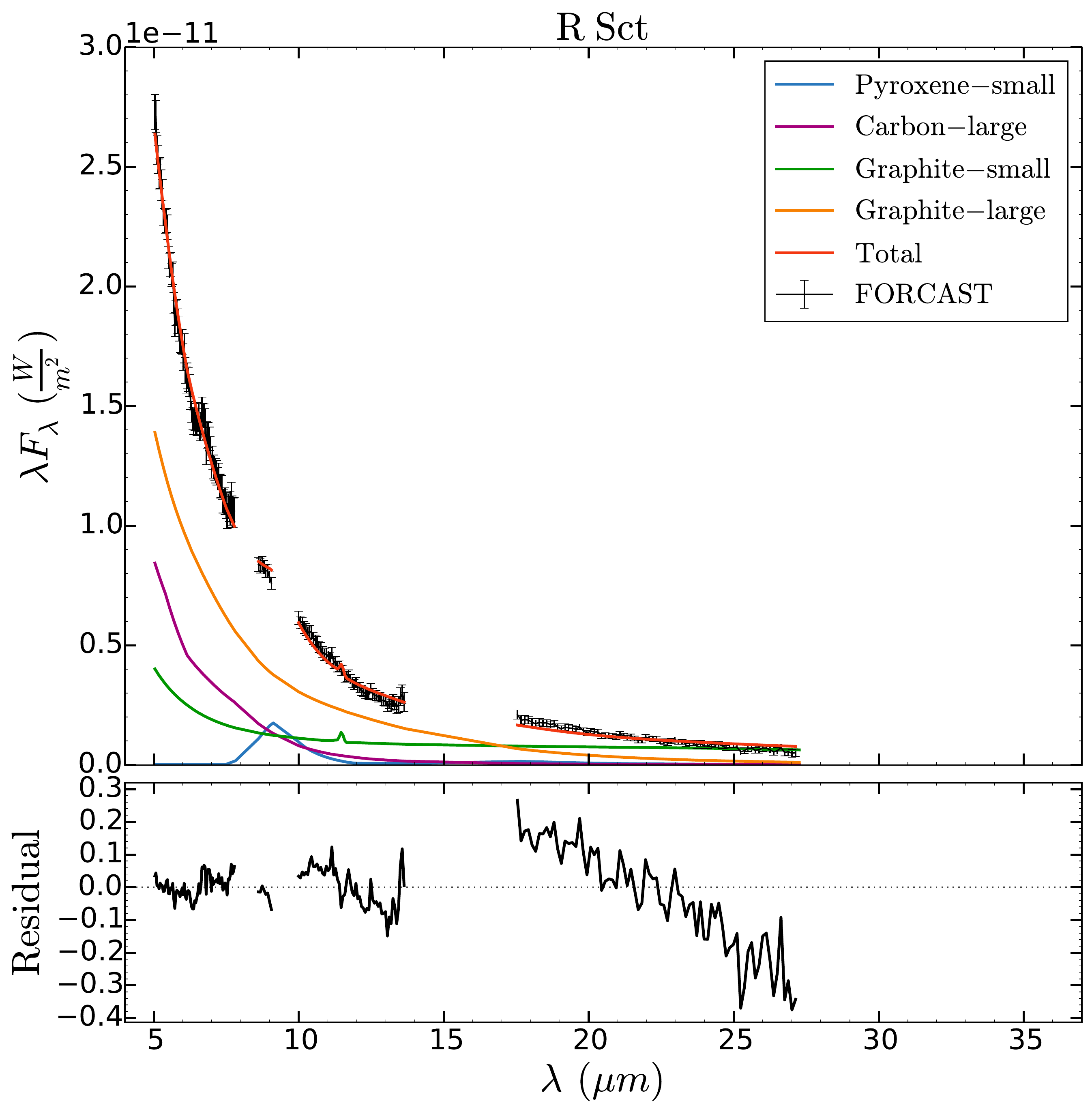}{0.5\textwidth}{(o) R Sct}
          \rightfig{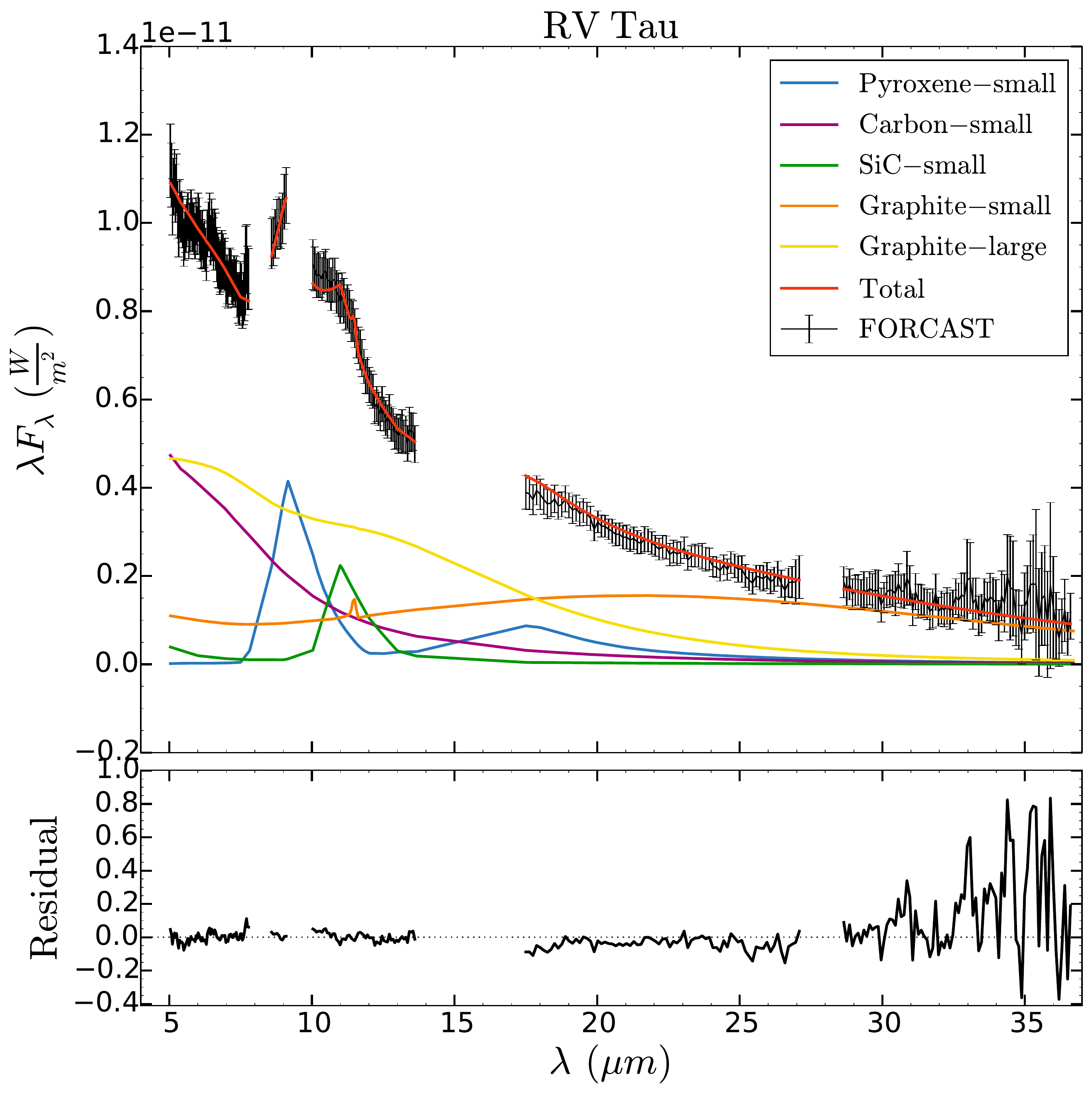}{0.5\textwidth}{(p) RV Tau}}
\end{figure*}
\clearpage
\begin{figure*}
\gridline{\leftfig{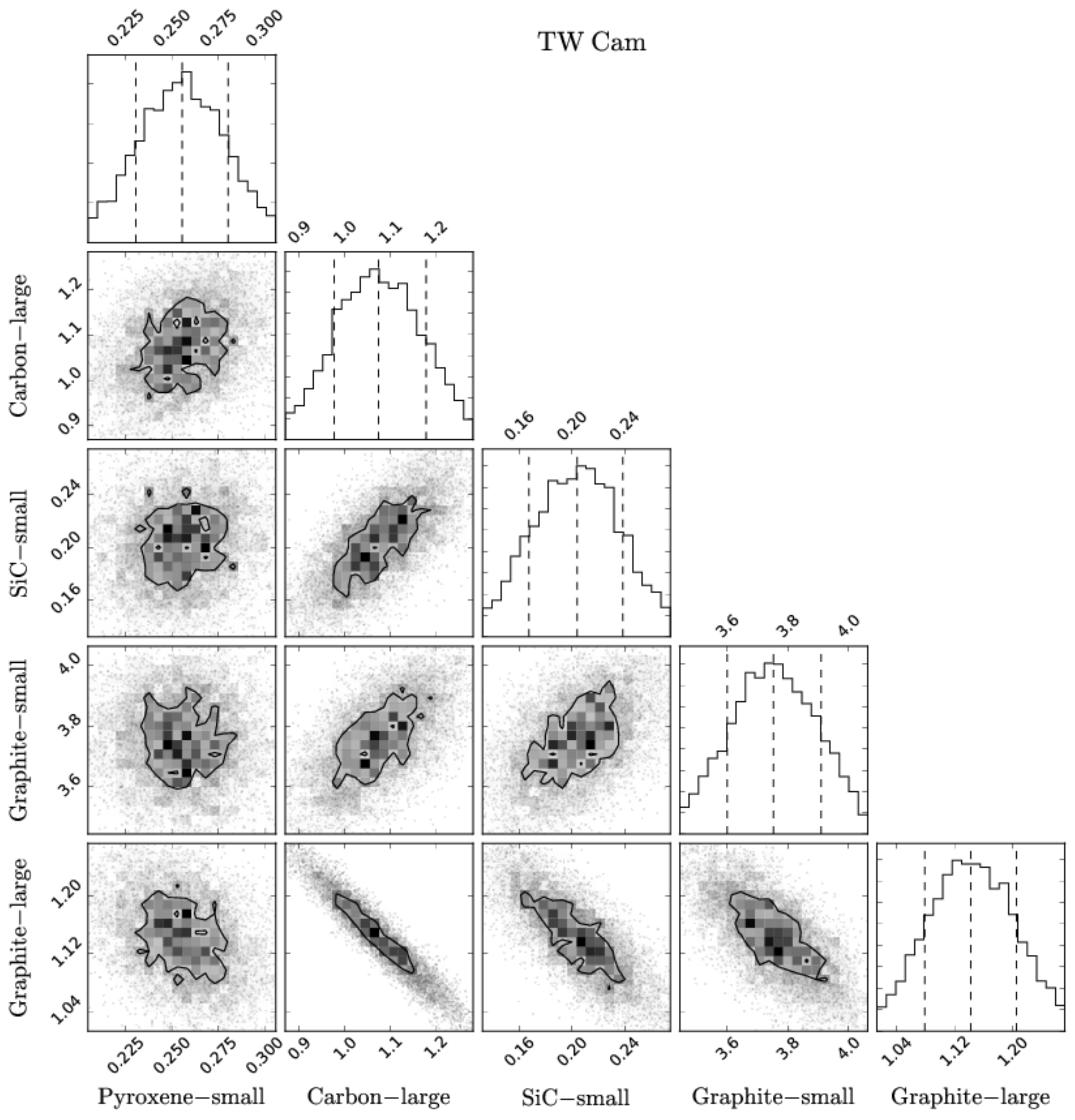}{0.5\textwidth}{(a) TW Cam}
          \rightfig{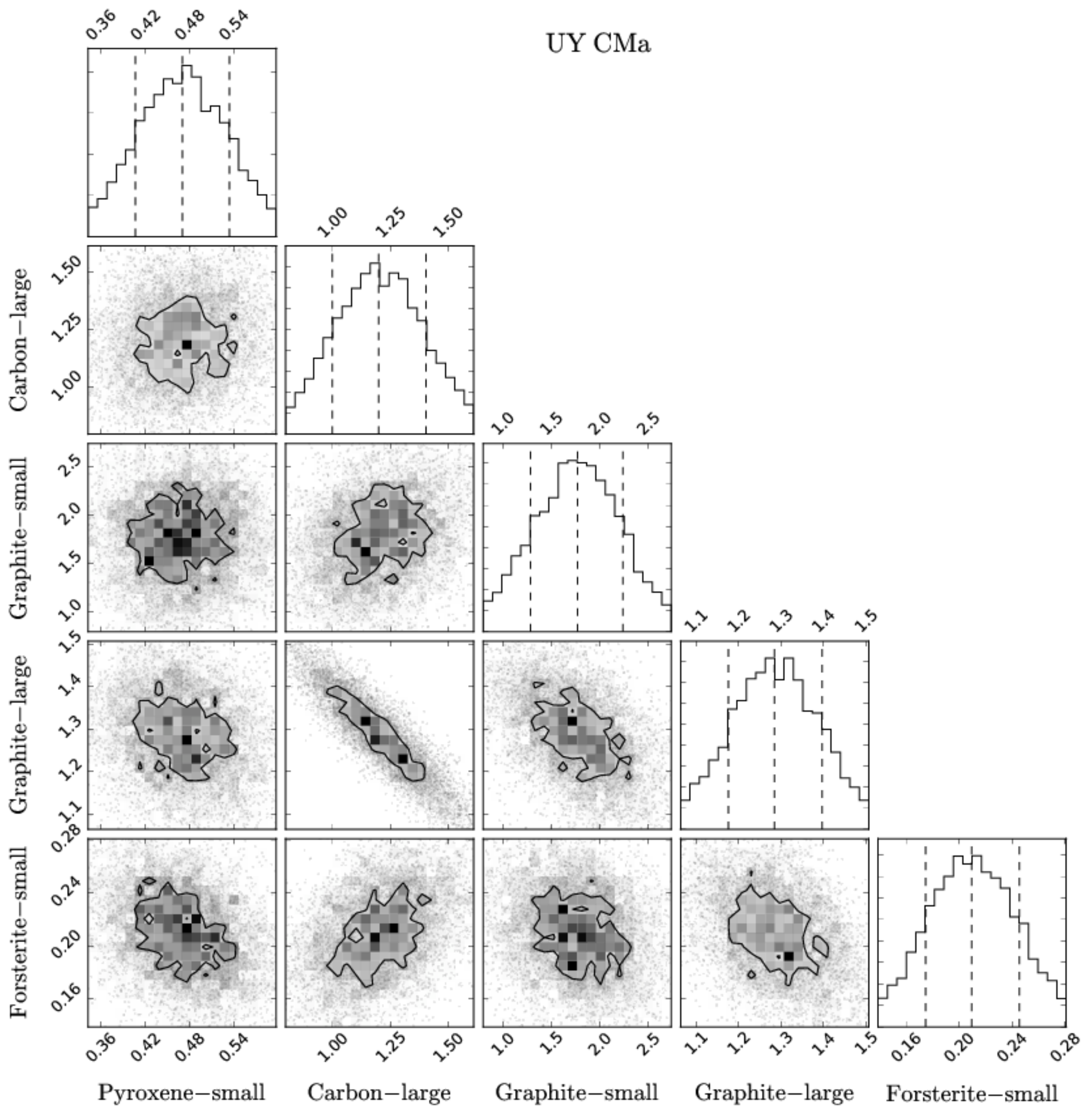}{0.5\textwidth}{(b) UY CMa}}
\gridline{\leftfig{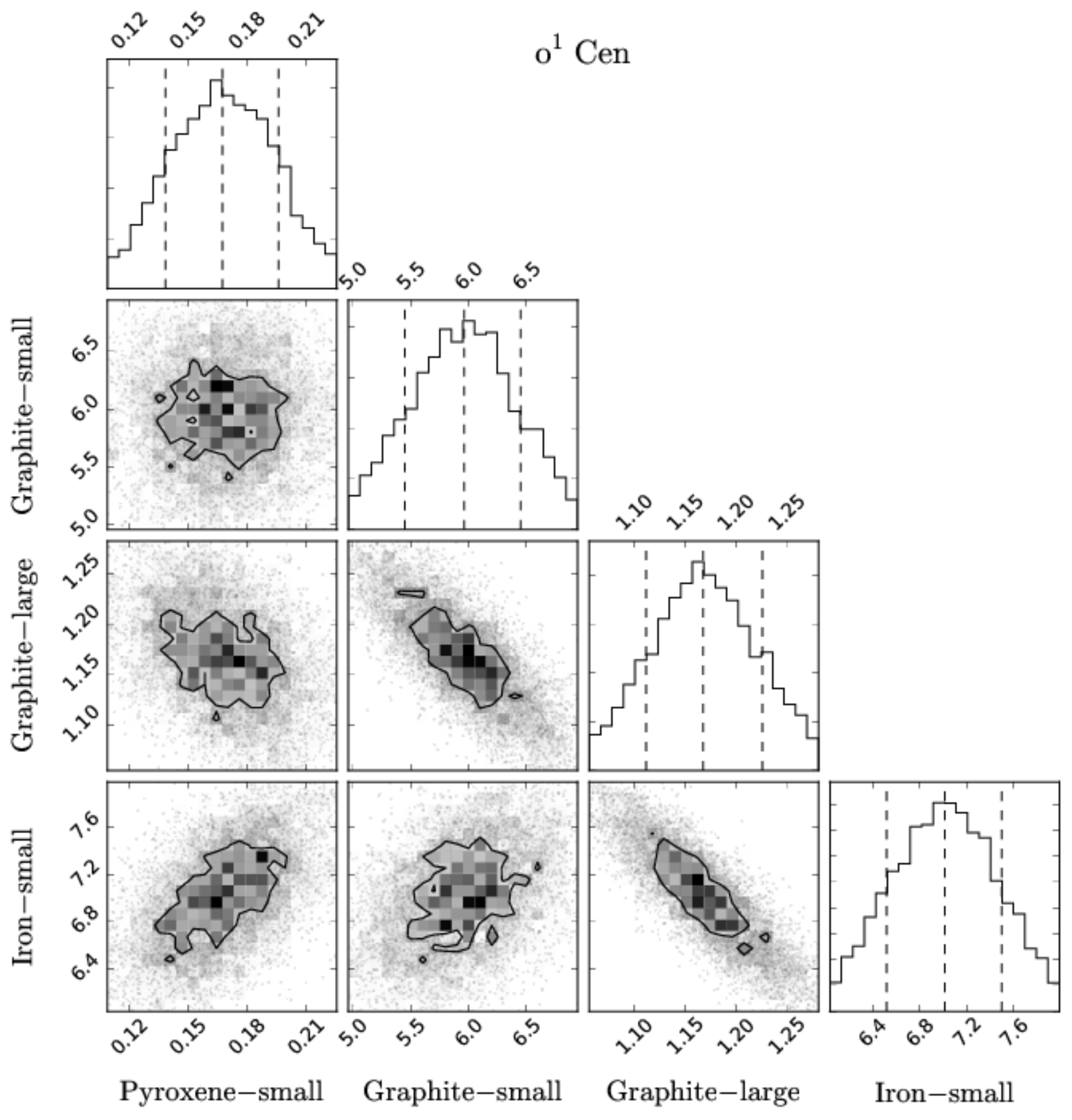}{0.5\textwidth}{(c) $\rm{o}^1$ Cen}
          \rightfig{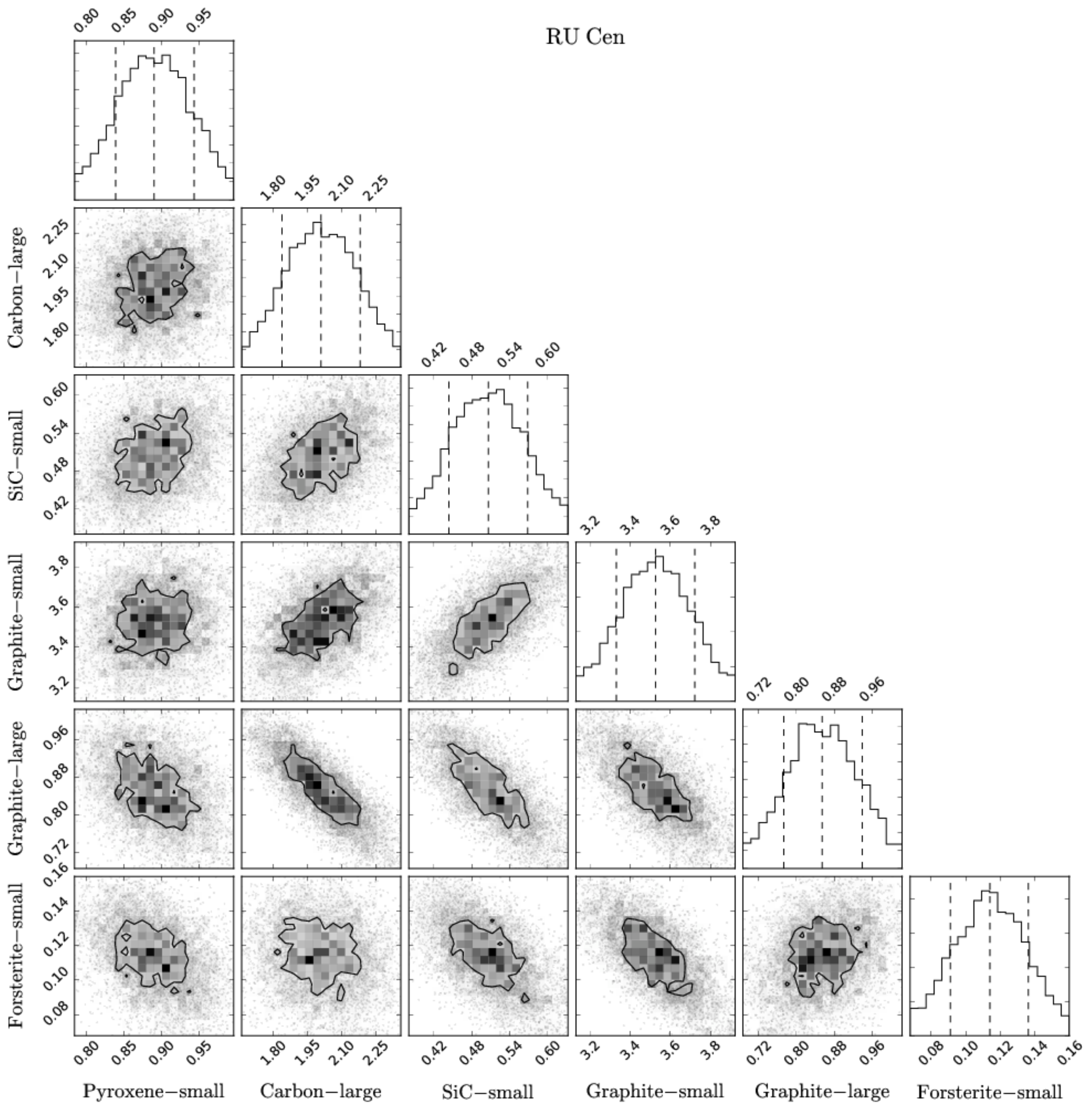}{0.5\textwidth}{(d) RU Cen}}
\end{figure*}
\begin{figure*}
\gridline{\leftfig{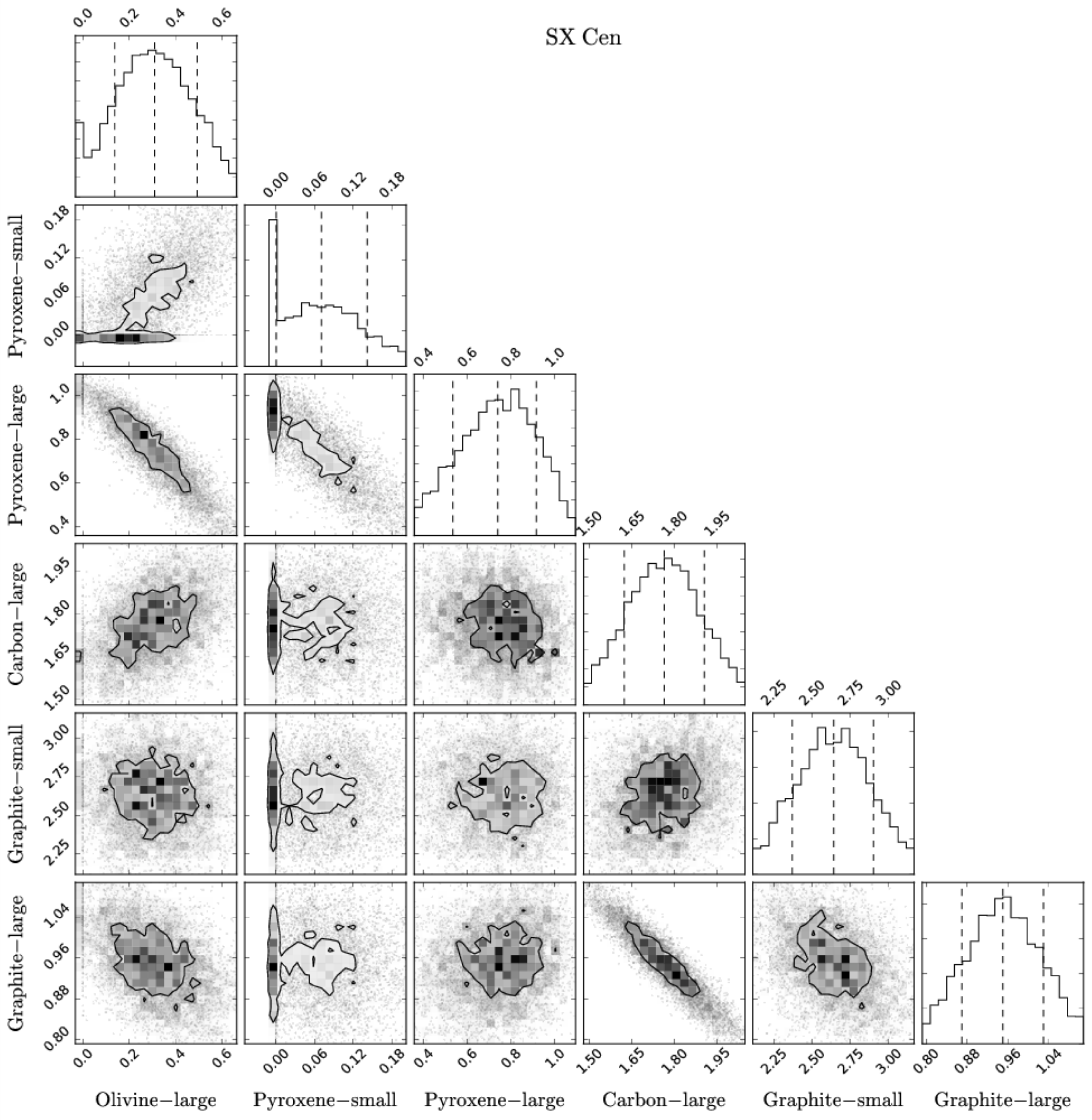}{0.5\textwidth}{(e) SX Cen}
          \rightfig{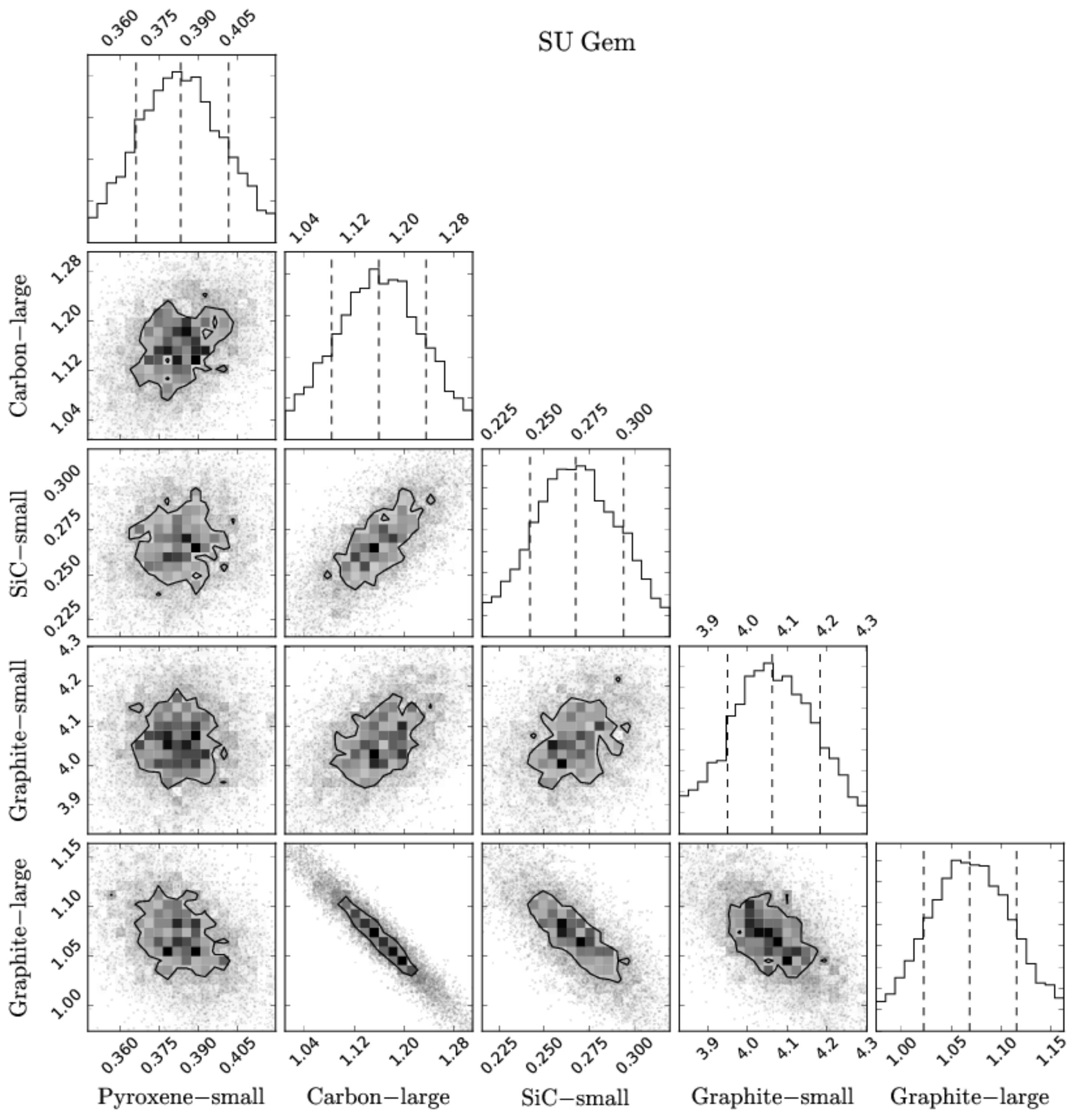}{0.5\textwidth}{(f) SU Gem}}
\gridline{\leftfig{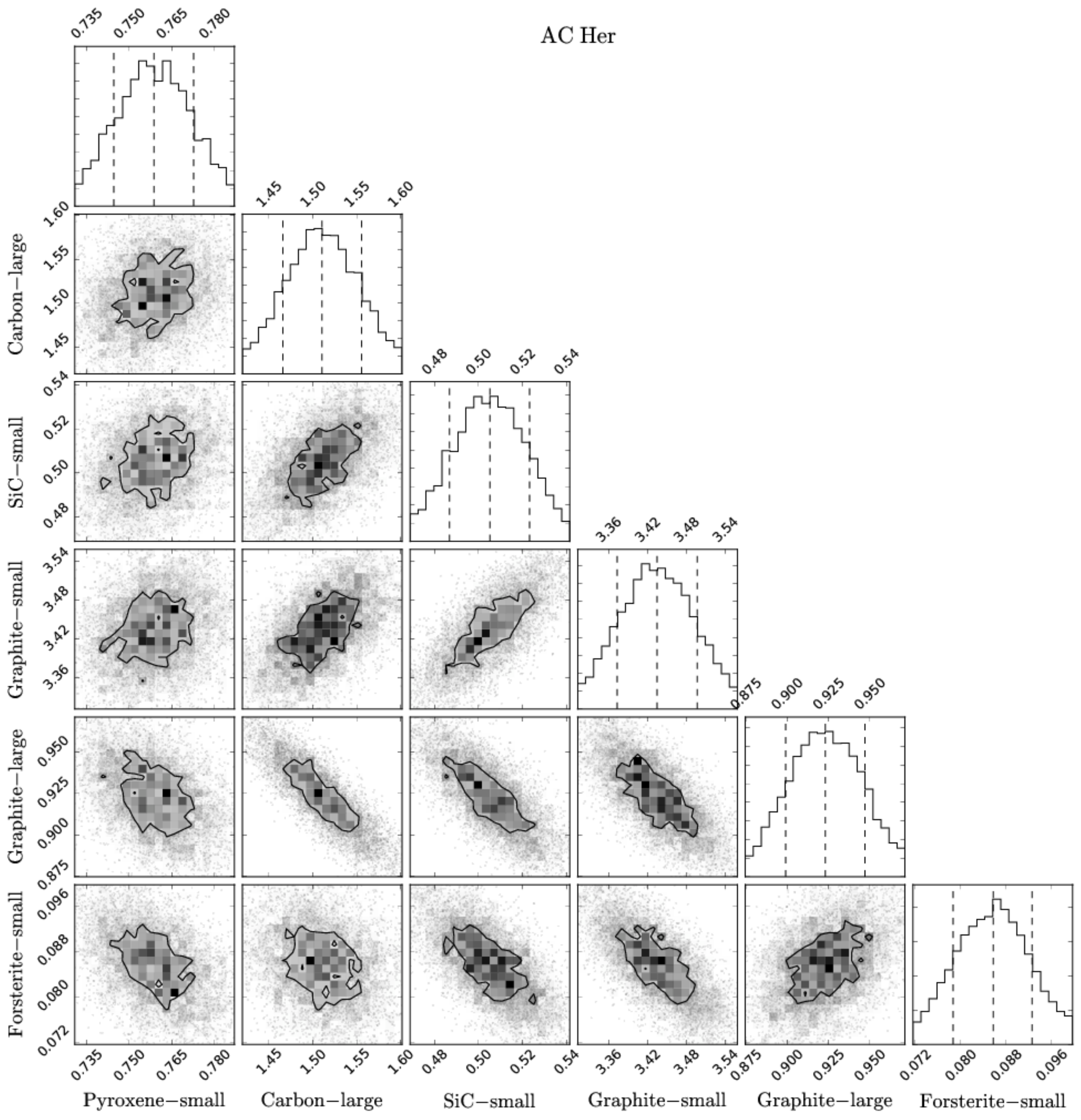}{0.5\textwidth}{(g) AC Her}
          \rightfig{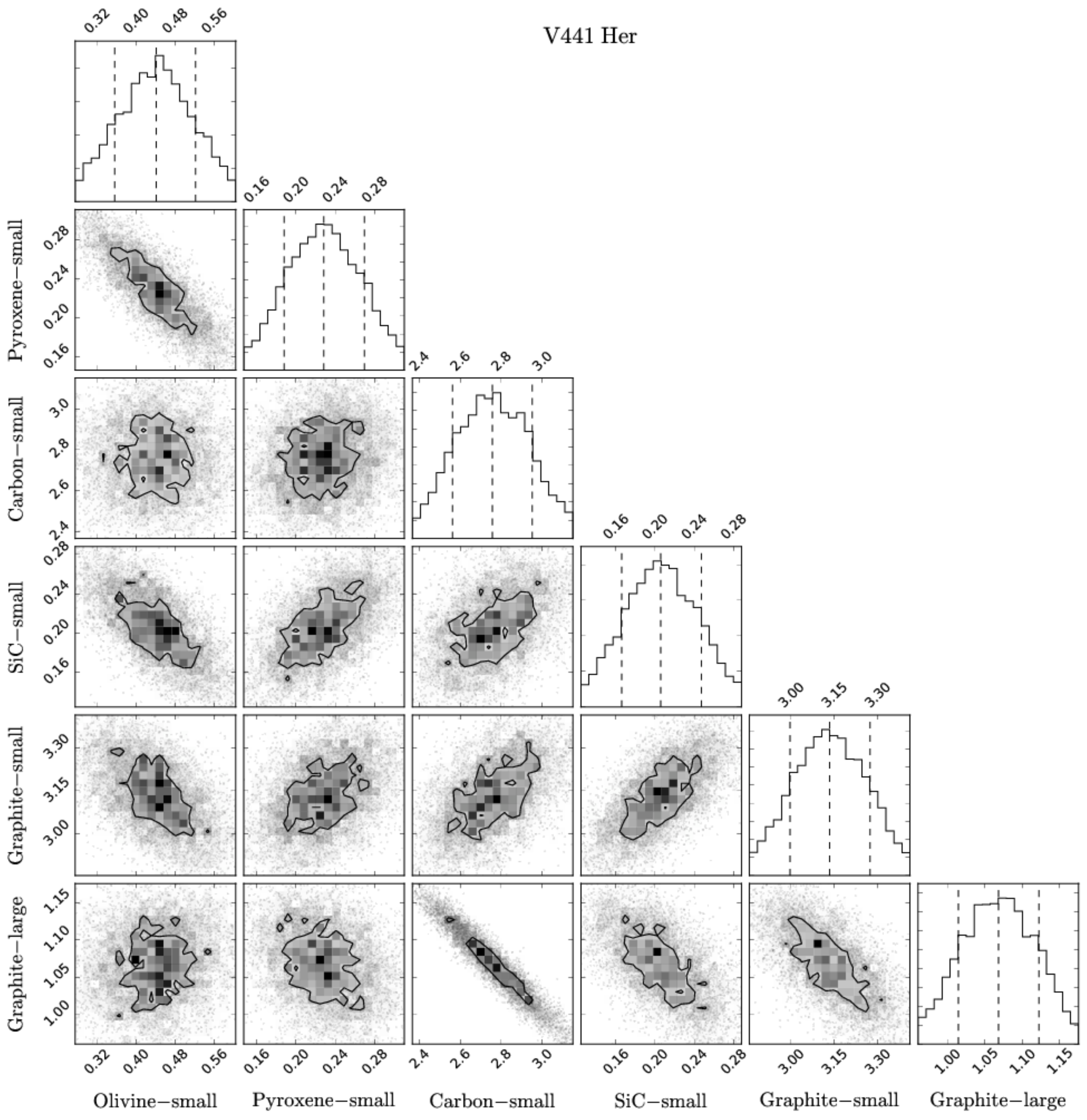}{0.5\textwidth}{(h) V441 Her}}
\end{figure*}
\begin{figure*}
\gridline{\leftfig{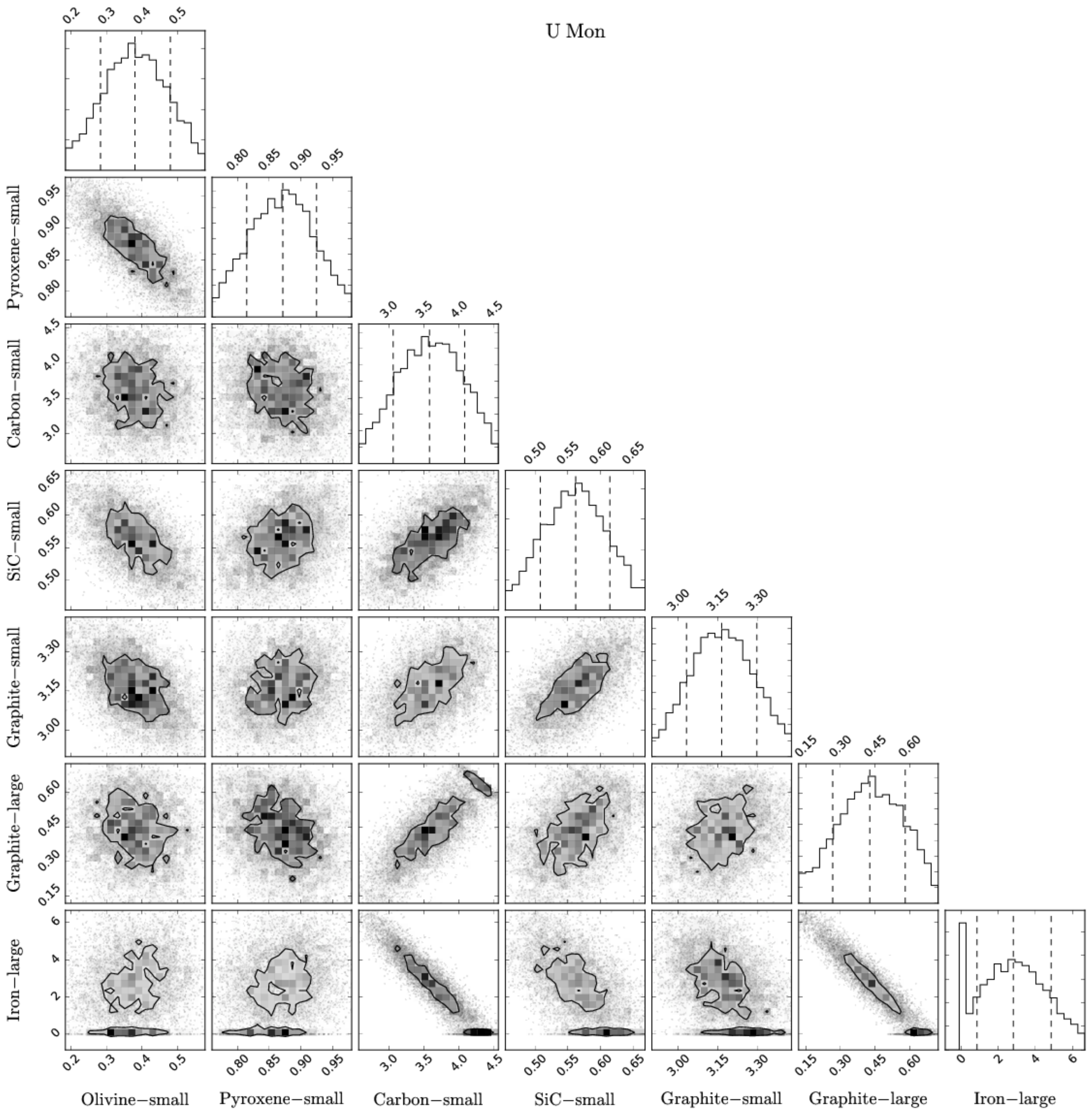}{0.5\textwidth}{(i) U Mon}
          \rightfig{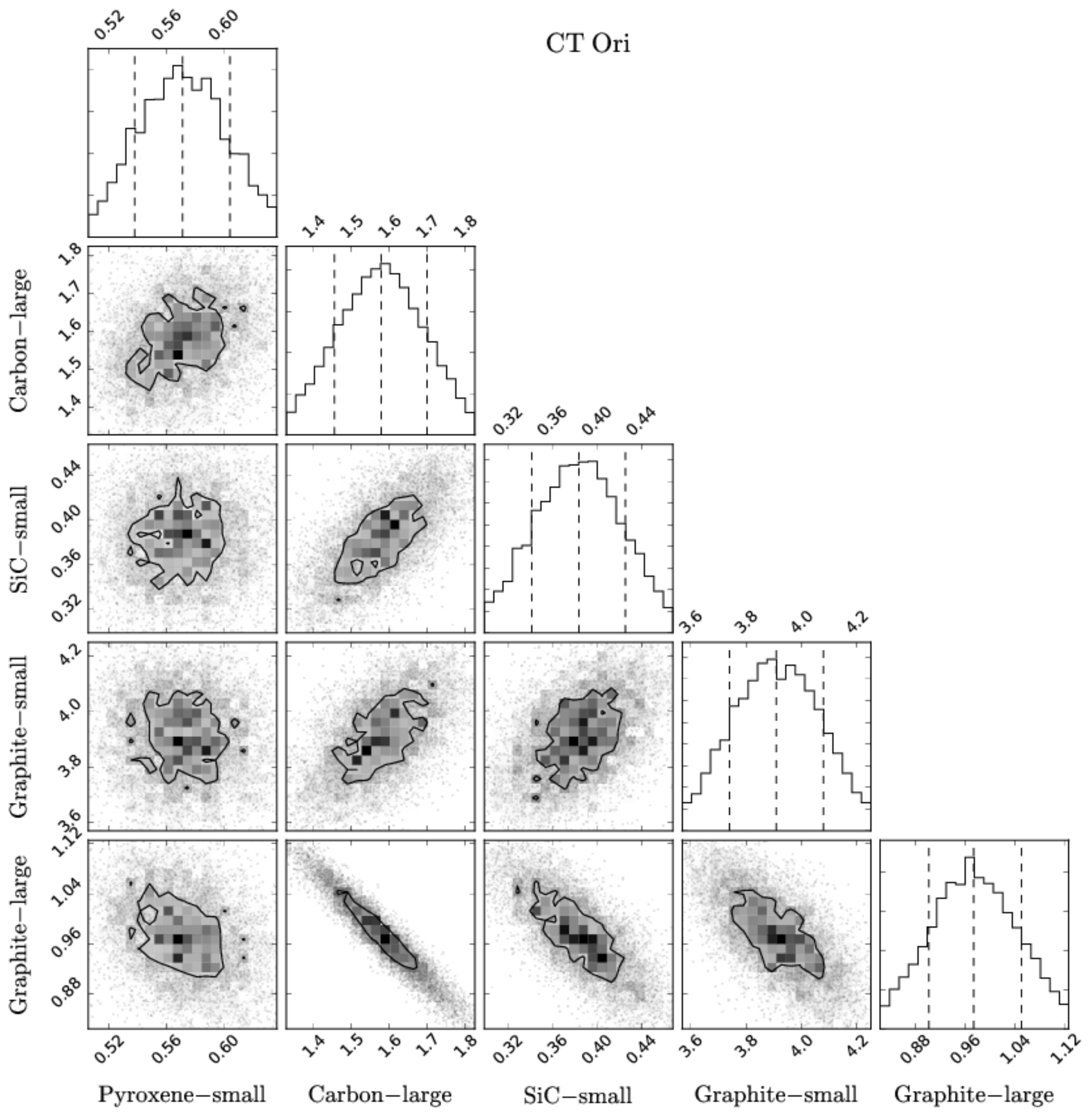}{0.5\textwidth}{(j) CT Ori}}
\gridline{\leftfig{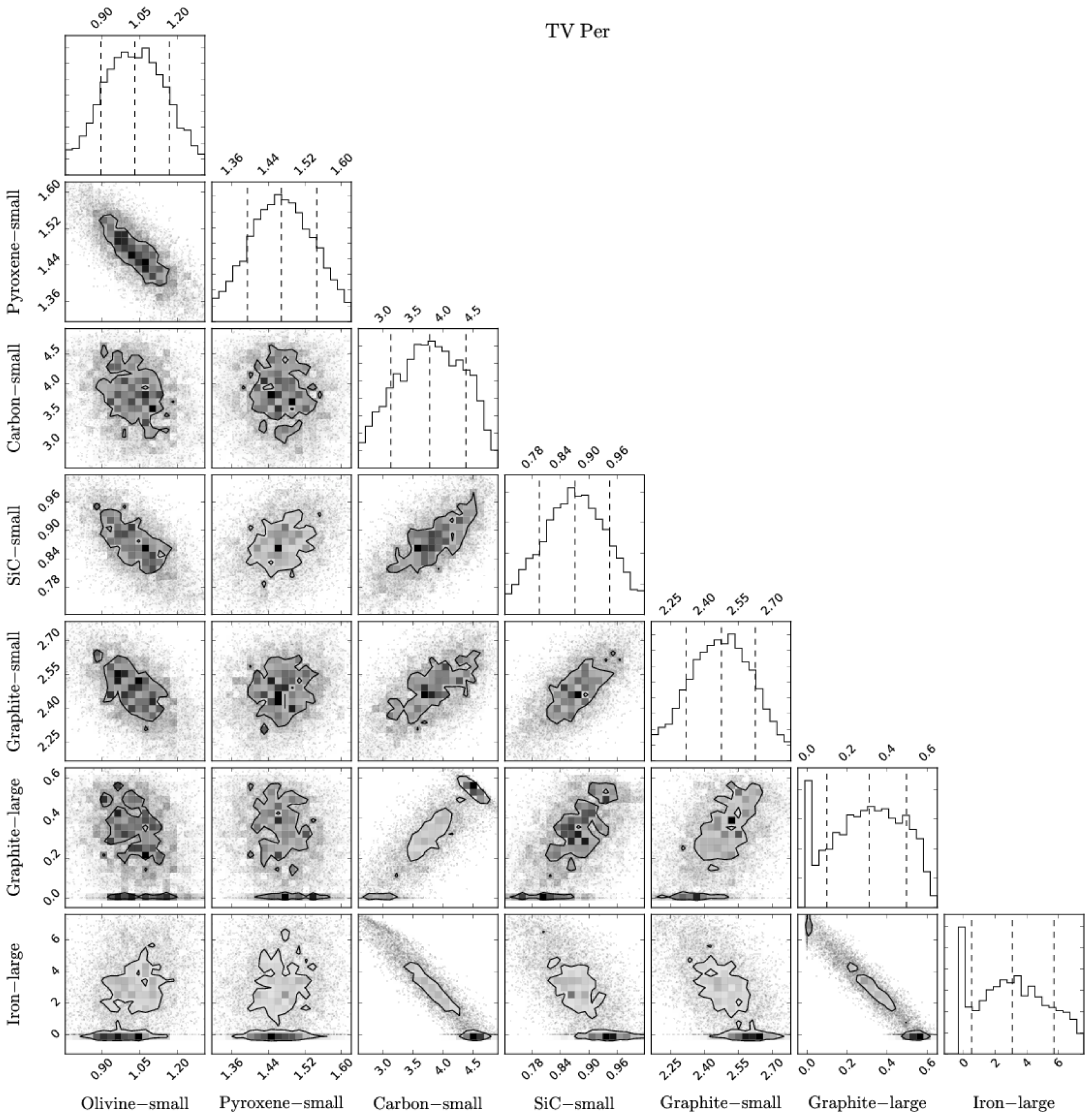}{0.5\textwidth}{(k) TV Per}
          \rightfig{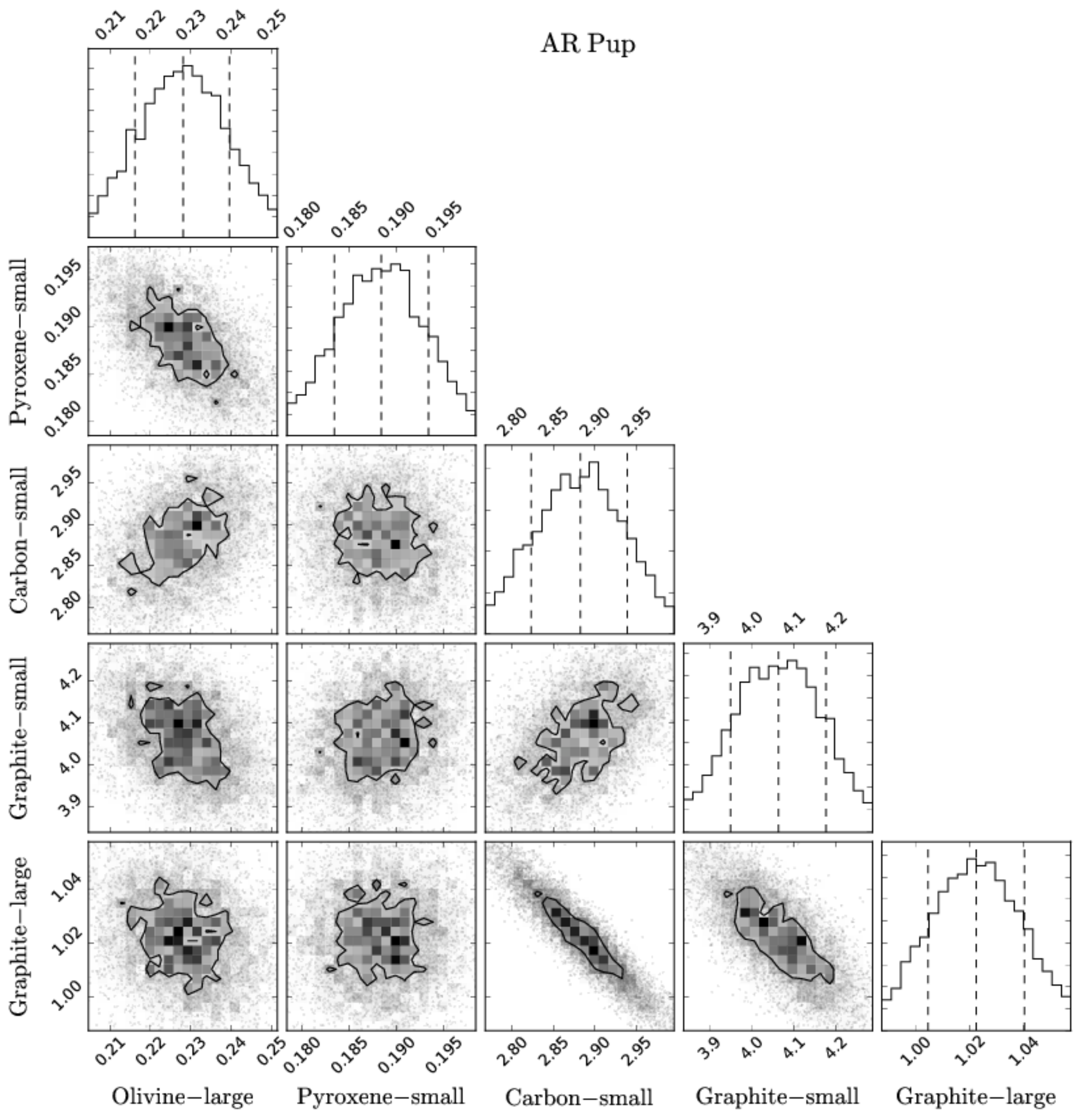}{0.5\textwidth}{(l) AR Pup}}
\end{figure*}
\begin{figure*}
\gridline{\leftfig{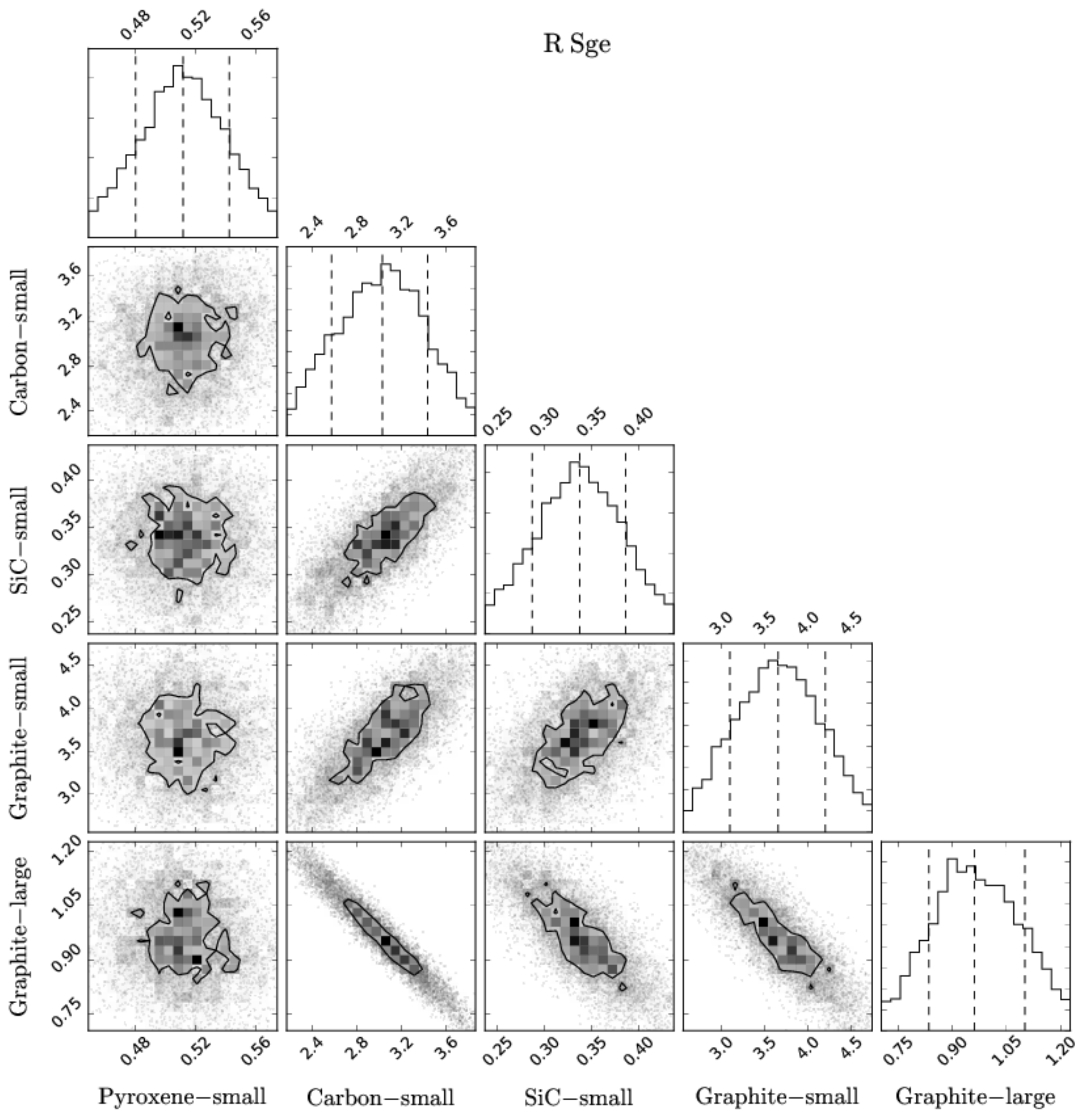}{0.5\textwidth}{(m) R Sge}
          \rightfig{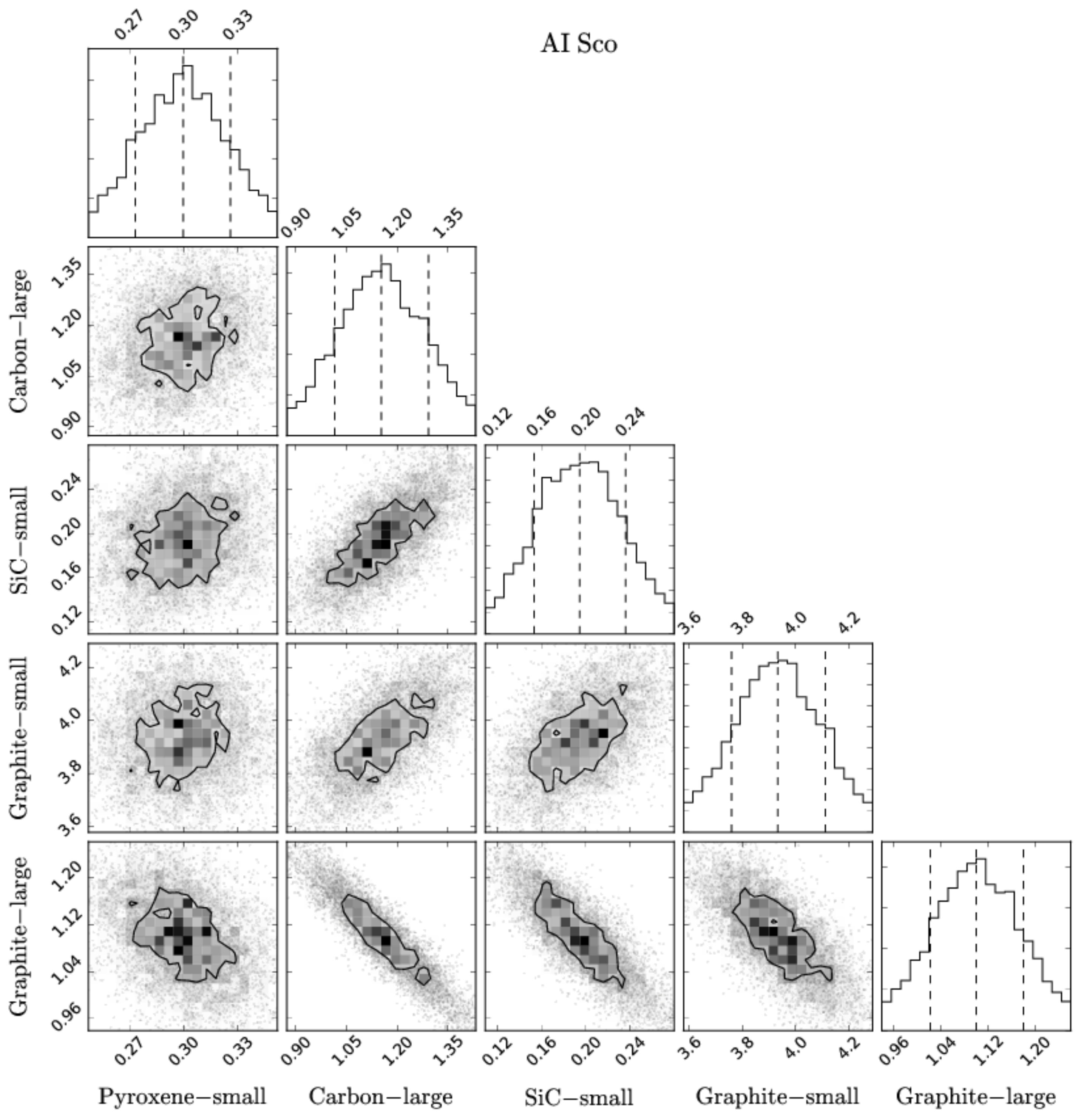}{0.5\textwidth}{(n) AI Sco}}
\gridline{\leftfig{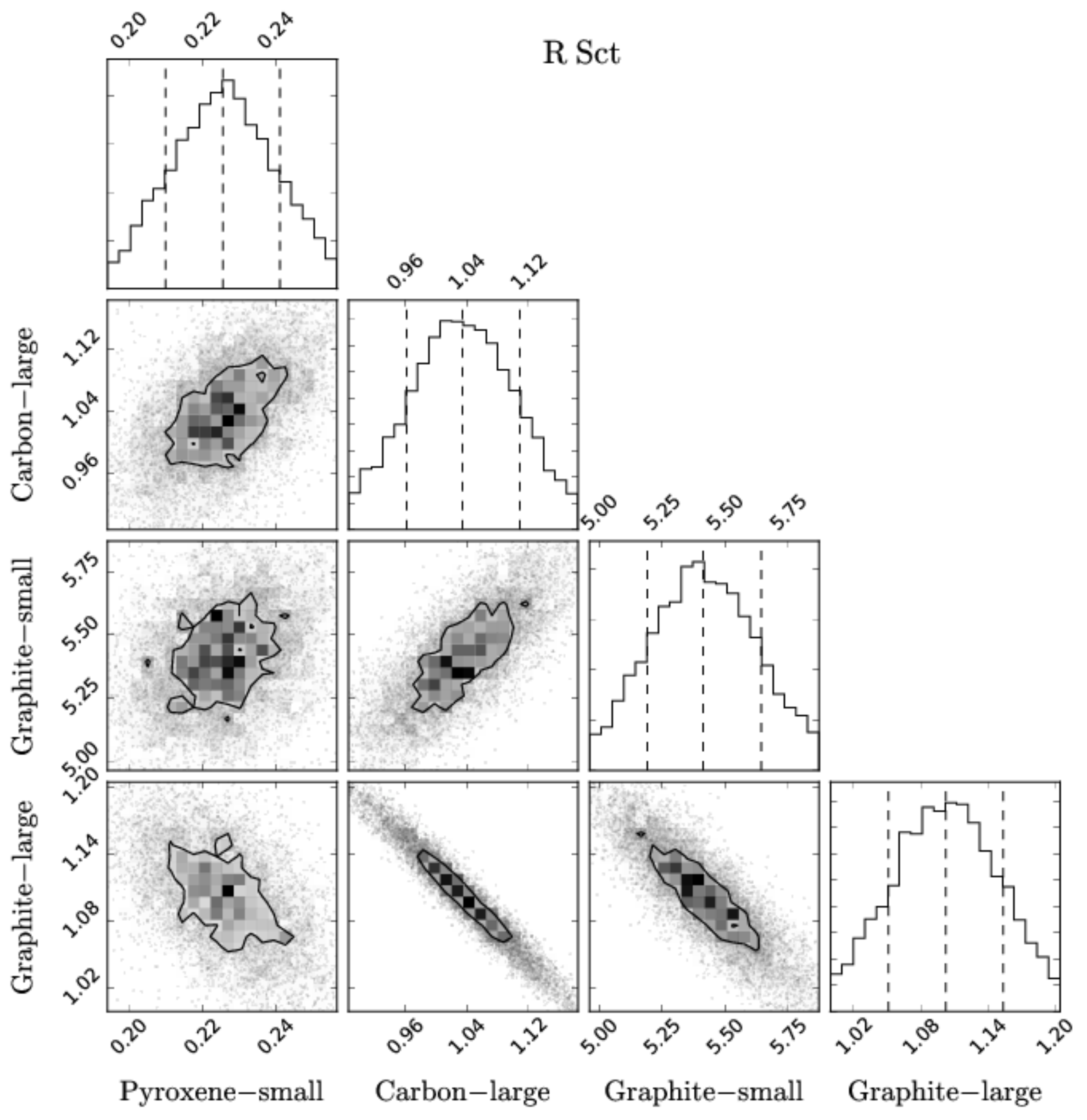}{0.5\textwidth}{(o) R Sct}
          \rightfig{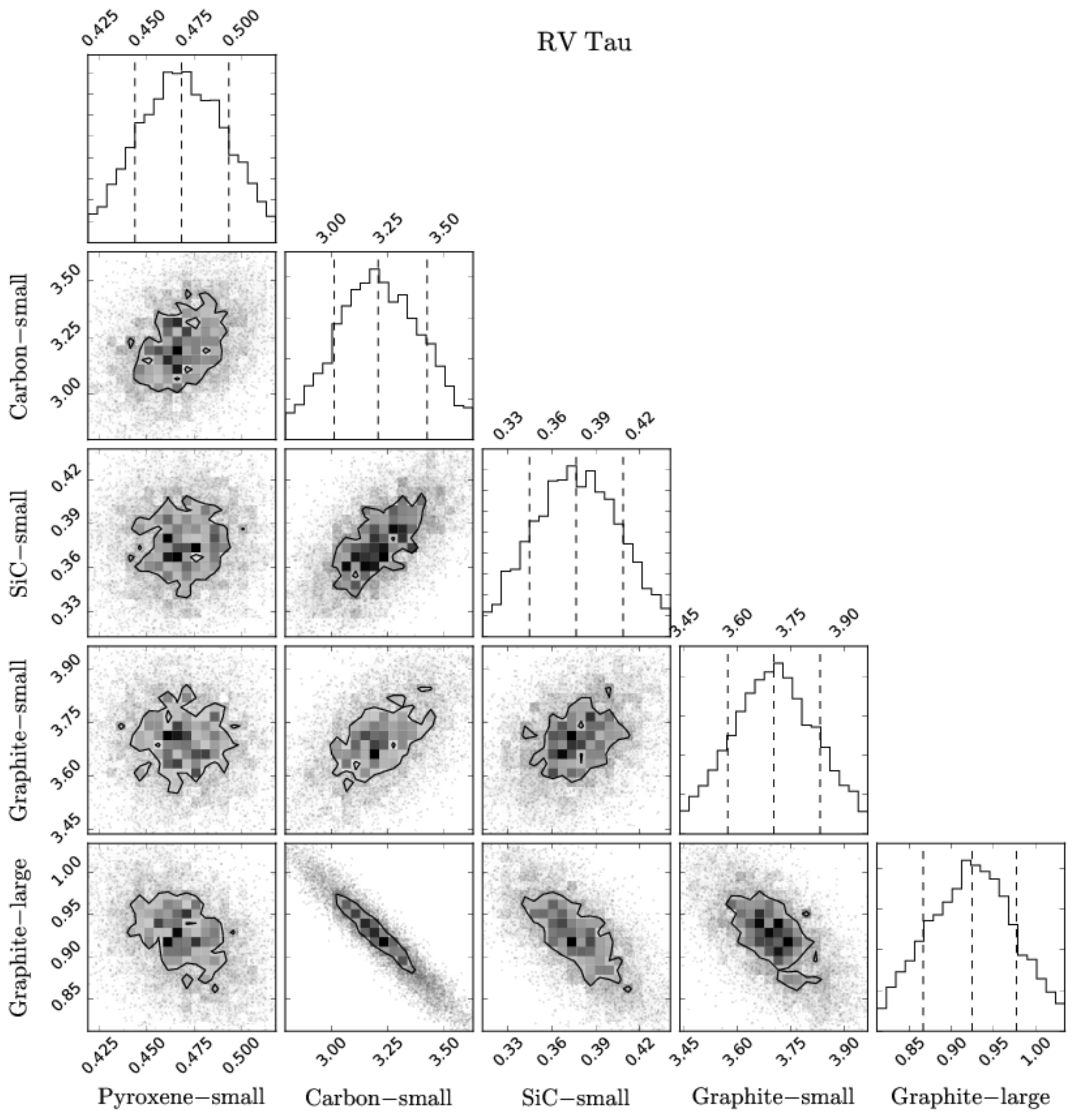}{0.5\textwidth}{(p) RV Tau}}
\end{figure*}

\section{Discussion}
\label{discussion}

\begin{figure*}
\plotone{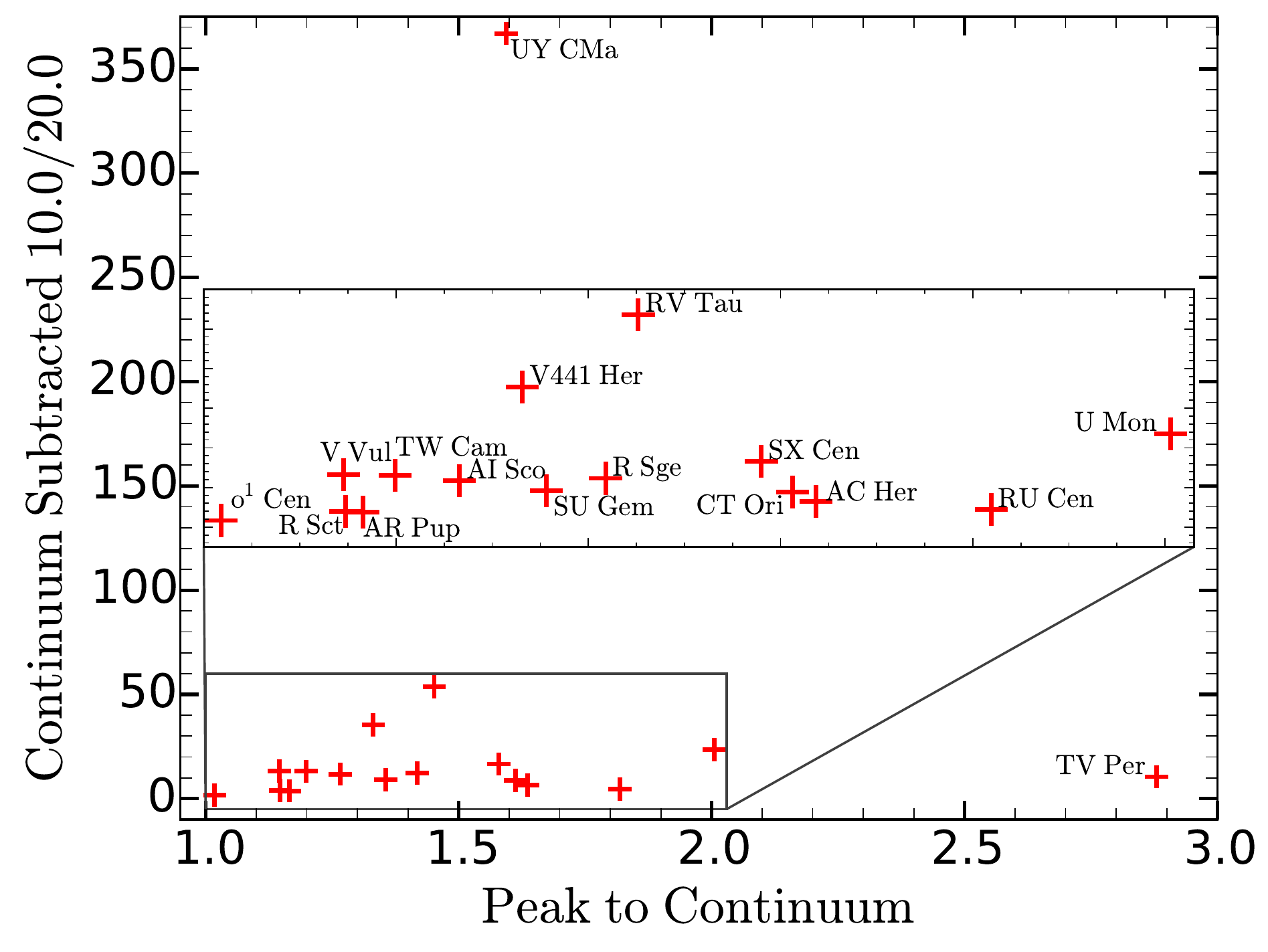}
\figcaption{\label{fluxratio_plot2}Ratio of the continuum subtracted flux at 10 and 20 \micron\ versus the peak to continuum ratio of the 10 \micron\ silicate feature.}
\end{figure*}

The 10 and 20 \micron\ emission features can be used to quantify the grain size and age of the circumstellar dust \citep{vanBoekel03, vanBoekel05, Juhasz10}.  The peak-to-continuum ratio of the 10 \micron\ feature can be used as a measure of the amount of grain growth because larger grains will produce a less prominent feature.  In addition, the continuum subtracted 10/20 \micron\ flux ratio has been shown to decrease monotonically with increased processing and therefore can be used to indicate the age of the circumstellar silicates \citep{Nuth90}.  Older, more processed grains will have a lower 10/20 \micron\ ratio.  A plot of these two ratios for all our program stars can be seen in Figure~\ref{fluxratio_plot2}.  Most of the sources show a low peak to continuum value (i.e. $<\,2.0$) and a low 10/20 \micron\ ratio (i.e. $<\,50$) indicating that the grains are relatively large and have undergone significant processing.  This supports the idea that the dust is constrained to a Keplerian disk.  There are two outliers in Figure~\ref{fluxratio_plot2}, TV Per and UY CMa.  TV Per has a high peak to continuum value and a small 10/20 \micron\ ratio indicating that the dust grains are small and old.  The small grain size is consistent with our model, which predicts a small spherical grain volume fraction of $\sim 74\%$ for TV Per.  This suggests that the circumstellar environment around TV Per is such that the grains are unable to grow to large sizes.  UY CMa has a low peak to continuum value and a large 10/20 \micron\ ratio indicating that the dust grains are both large and young.  This is also consistent with our model which predicts a small grain volume fraction of $\sim 50\%$ around UY CMa.  \citet{Robinson77} and \citet{Mitchell81} found that for low optical depths, some of the circumstellar silicate dust may be in absorption rather than emission if it is at a low temperature.  The viewing angle of the disks will also affect the peak-to-continuum ratio of the 10 \micron\ feature \citep{Crapsi08}  As mentioned in Section \ref{model}, factors like the grain size and shape will affect the observed emission features.  The optical depth, viewing angle, temperature, and particle size of the grains may mask the 10/20 \micron\ ratio and could contribute to the high volume fraction of small grains found by our model.

\subsection{Crystallinity}
Although \citet{Molster02a, Molster02b} showed that the silicate crystallinity fraction in disk sources was much higher than that observed in outflow sources, we do not see a high silicate crystallinity fraction in any of the objects in our sample listed as ``disk" SEDs in Table \ref{tab:program} and for all of the FORCAST spectra there are no obvious crystalline emission features present.   One explanation is that crystalline silicate material is not abundant in any of the stars observed.  However, crystalline olivines have been detected around AC Her and are thought to be present at the $\sim10-50\%$ level around AR Pup and U Mon \citep{Blommaert14, deRuyter05}.  The lack of strong crystalline silicate emission features does not necessarily indicate a lack of presence.  If a temperature difference exists between the amorphous and crystalline silicates, it is possible to include up to $40\%$ of crystalline silicates in the circumstellar dust without seeing crystalline features in the spectra \citep{Kemper01}.  Another possible explanation for this observation is that crystalline silicates are generally colder than amorphous silicates, which could mean that the grains are not co-spatial or that they have different optical properties.  In fact, when modeling the circumstellar material around AC Her, \citet{Hillen15} found the spatial distribution of the forsterite to be different from the amorphous dust.  The difference in optical properties could be due in part to the different iron content of each material, which increases the opacity in the near-IR significantly \citep{Molster02c, Dorschner95}.  Additionally, the spectral features will be less prominent if the crystalline grains are larger than the amorphous grains.  Therefore, if the crystalline silicates are only moderately abundant (i.e. $ \lesssim 40\%$), cooler, and larger than the amorphous silicates, the spectral features of the crystalline silicates could easily be masked by the amorphous silicates.  Interestingly, \citet{Blommaert14} did not detect crystalline olivines around AR Pup and U Mon in the mid-IR and interpreted this as an indication that the crystalline olivines are hot ($\sim 600$ K) around these two stars.  If they are indeed hot, than the crystalline olivine abundances must be relatively low or the grains must be large around these two stars for them to go undetected by FORCAST.

\subsection{Dual Chemistry}
Our model predicts that most of the dust is carbon rich with some oxygen rich silicates.  This dual formation of carbon and oxygen rich minerals has been observed in several classical novae, namely V1280 Sco \citep{Sakon16}, V705 Cas \citep{Evans05}, V842 Cen \citep{Smith94}, and QV Vul \citep{Gehrz92} as well as IRAS 09425-6040, a carbon AGB star which shows circumstellar silicate dust features \citep{Suh16}.  The formation of both carbon rich and oxygen rich dust could be due to a chemical gradient in the wind as the stars evolve from oxygen rich to carbon rich after undergoing C dredge-up processes due to a recent AGB thermal pulse.  \citet{Suh16} successfully modeled the dust envelope around IRAS 09425-6040 with an outer oxygen rich shell and an inner carbon rich shell, validating this hypothesis.  Similarly, the carbon-rich planetary nebula BD +30\degr\ 3639 exhibits spatially separated carbon-rich polycyclic aromatic hydrocarbons and oxygen-rich silicate dust \citep{Guzman-Ramirez15}.  The post-AGB binary HR 4049 is a peculiar example of a depleted oxygen rich star with a featureless mid-IR spectrum possibly resulting from amorphous carbon masking the spectral features from silicates \citep{Acke13}.  While it is possible we may be observing the stars in transition from oxygen rich to carbon rich, it would require that all of these stars result from a narrow range of masses that terminate AGB evolution just as the carbon exceeds the oxygen abundance.  A more plausible explanation for the dual chemistry mineralogy is that the dust formed in a common envelope environment of a binary system where the carbon and oxygen abundances can rapidly change.  This mechanism has been invoked as the possible origin of post-AGB disks \citep{Kashi11, Lu13, Hardy16}.

\subsection{Viewing Effects}
Most of the FORCAST continua are well described by two Planck functions, suggesting that we are viewing the systems from a nearly face-on orientation and see both the inner ($\sim1000$ K) and outer ($\sim250$ K) regions of the disks.  Our results are corroborated by \citet{Hillen15} who used a radiative transfer code to model the dust around AC Her as a highly evolved (i.e. mm sized grains), circumstellar disk with an inclination of $50 \pm 8\degr$ and found good agreement with observations.  \citet{Bujarrabal07} detected an extended bipolar outflow and an unresolved, compact (presumably disk) component around V441 Her with an inclination of $\sim 75\degr$.  The typical uniform disk diameter of the N band emission region of the RV Tauri stars interferometrically observed by \citet{Hillen17} is $\sim 40$ mas. making it difficult to verify the inclination angle of other systems.

\subsection{Limitations of the Fit}
\label{limits}
It is worth mentioning the various difficulties encountered in the modeling of the dust species present around our program stars and the limitations of our simplified model.  As mentioned in Section \ref{obs}, less than half of the program stars were observed with the G6 grism (28.7--37.1 \micron).  The addition of this data would have aided the modeling and identification of the minerals as crystalline silicates have prominent emission features in this region.  Similarly, the lack of coverage from 14--17 \micron\ between grisms G3 and G5 made it more difficult to constrain the abundance of amorphous silicates which have emission features around 17 \micron.  Amorphous carbon and metallic iron, on the other hand, lack prominent IR features and our model could be fitting these species to the overall dust continuum or temperature gradient thereby increasing the relative abundances.

\begin{figure*}
\gridline{\leftfig{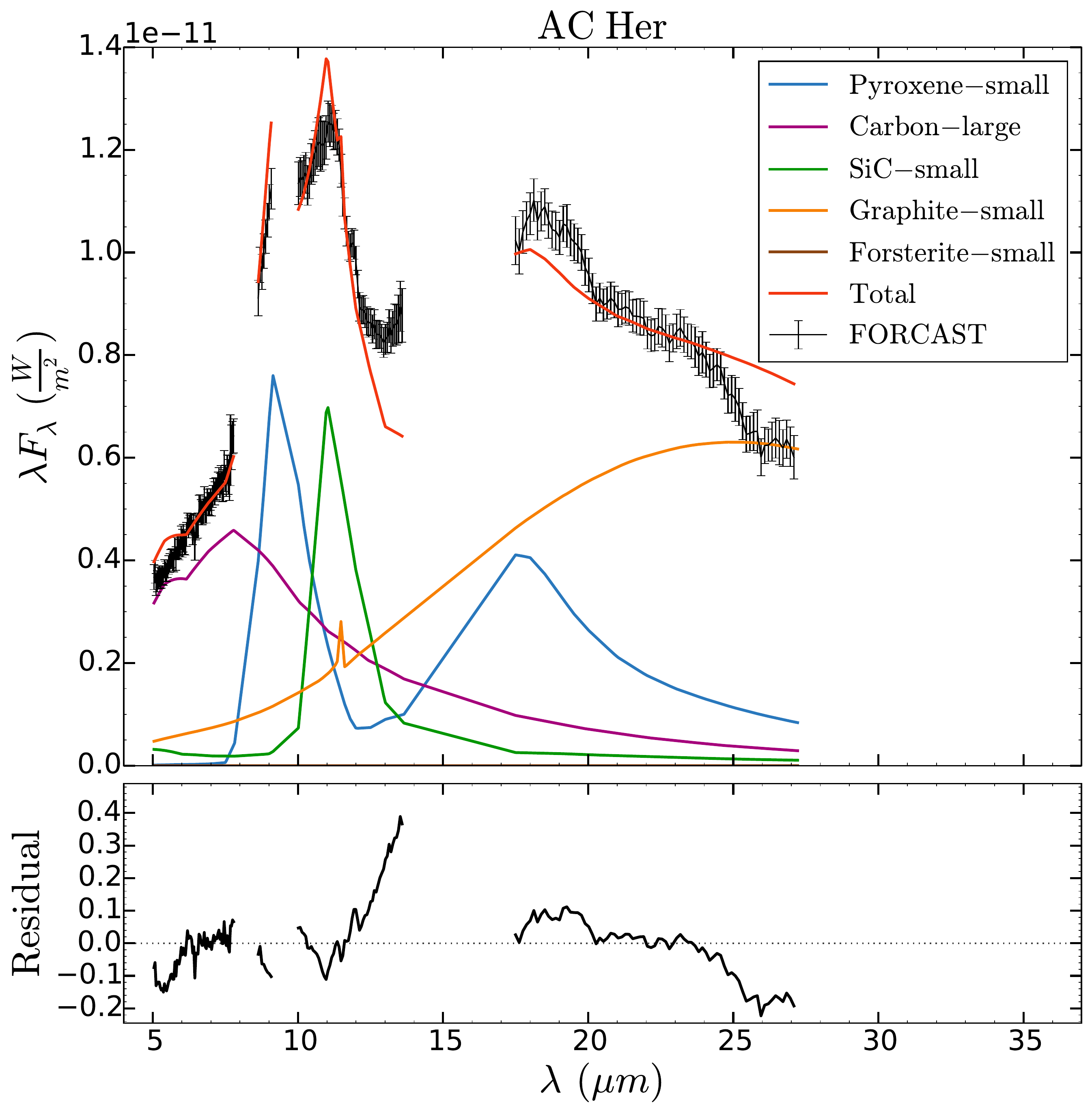}{0.5\textwidth}{(e) Graphite-large removed; $\chi^2_{\rm{red}} = 5.46$}
	\rightfig{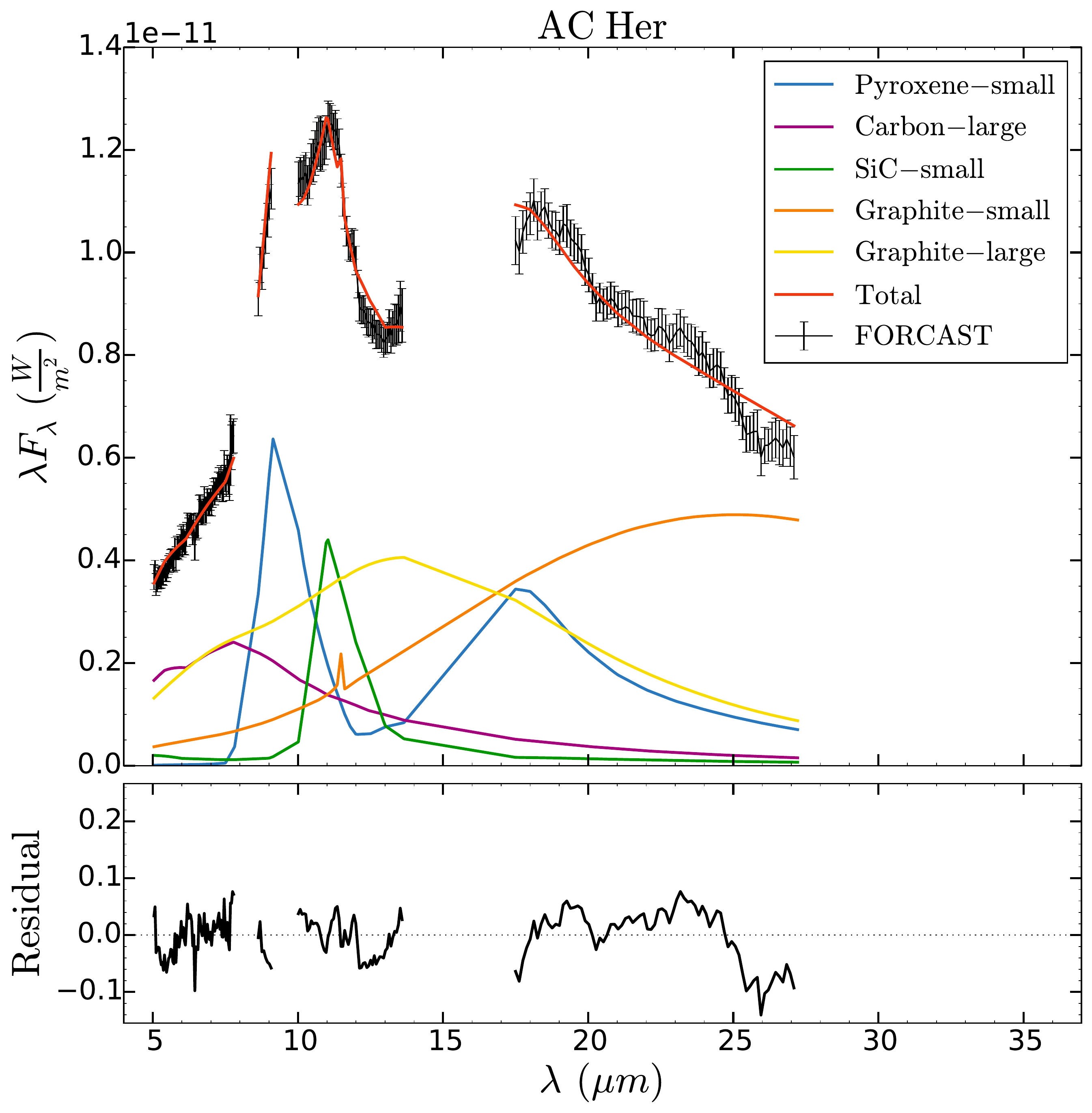}{0.5\textwidth}{(f) Forsterite removed; $\chi^2_{\rm{red}} = 1.01$}}
	\caption{The effect on the fit when \emph{removing} dust species from the best model of AC Her.}
  \label{fig:limit_high_sn_removed}
\end{figure*}

\begin{figure*}
\gridline{\leftfig{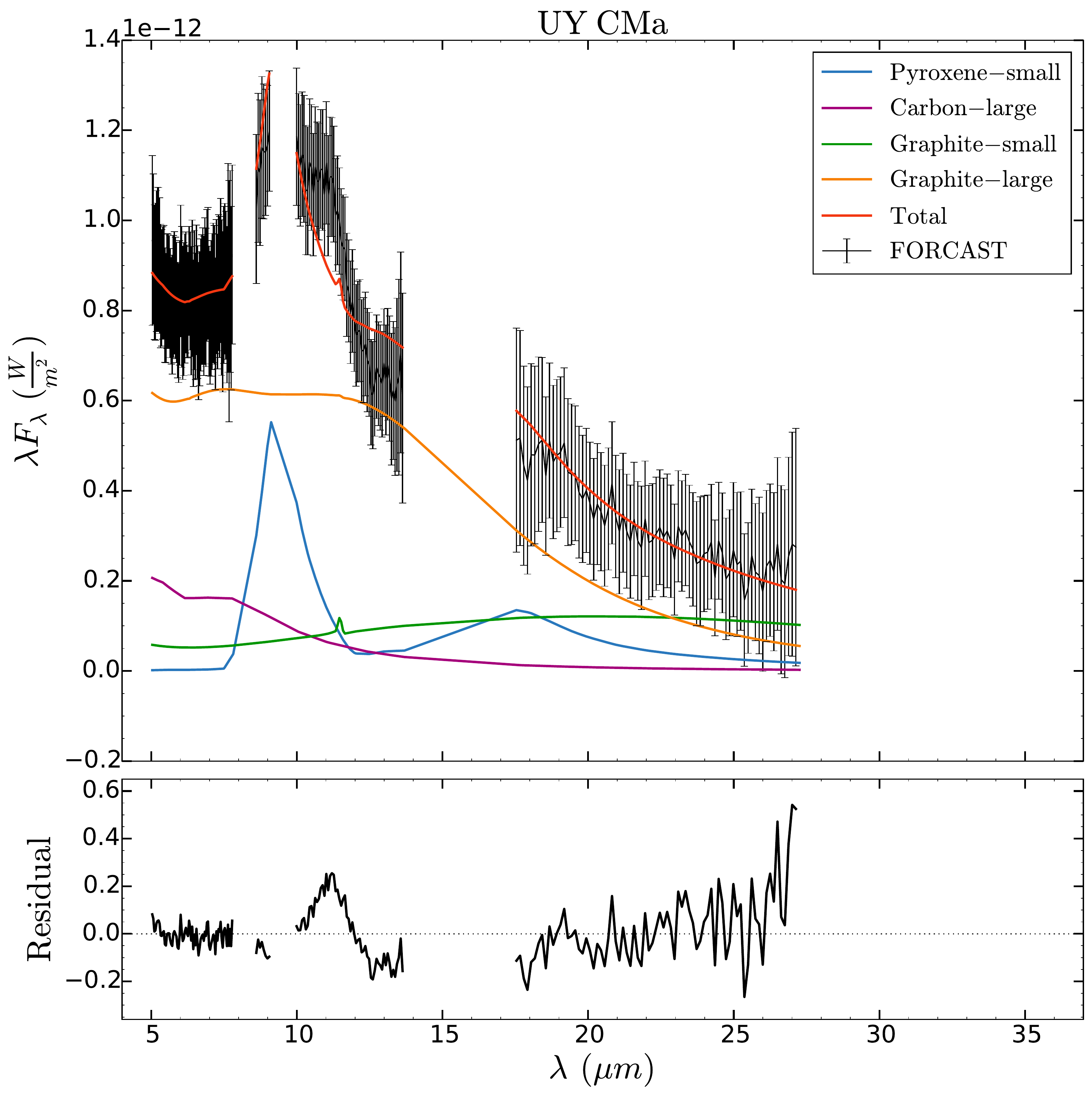}{0.5\textwidth}{(e) Forsterite removed; $\chi^2_{\rm{red}} = 0.26$}}
 \caption{The effect on the fit when \emph{removing} dust species from the best model of UY CMa.}
  \label{fig:limit_low_sn_removed}
\end{figure*}

\begin{figure*}[h!]
\gridline{\leftfig{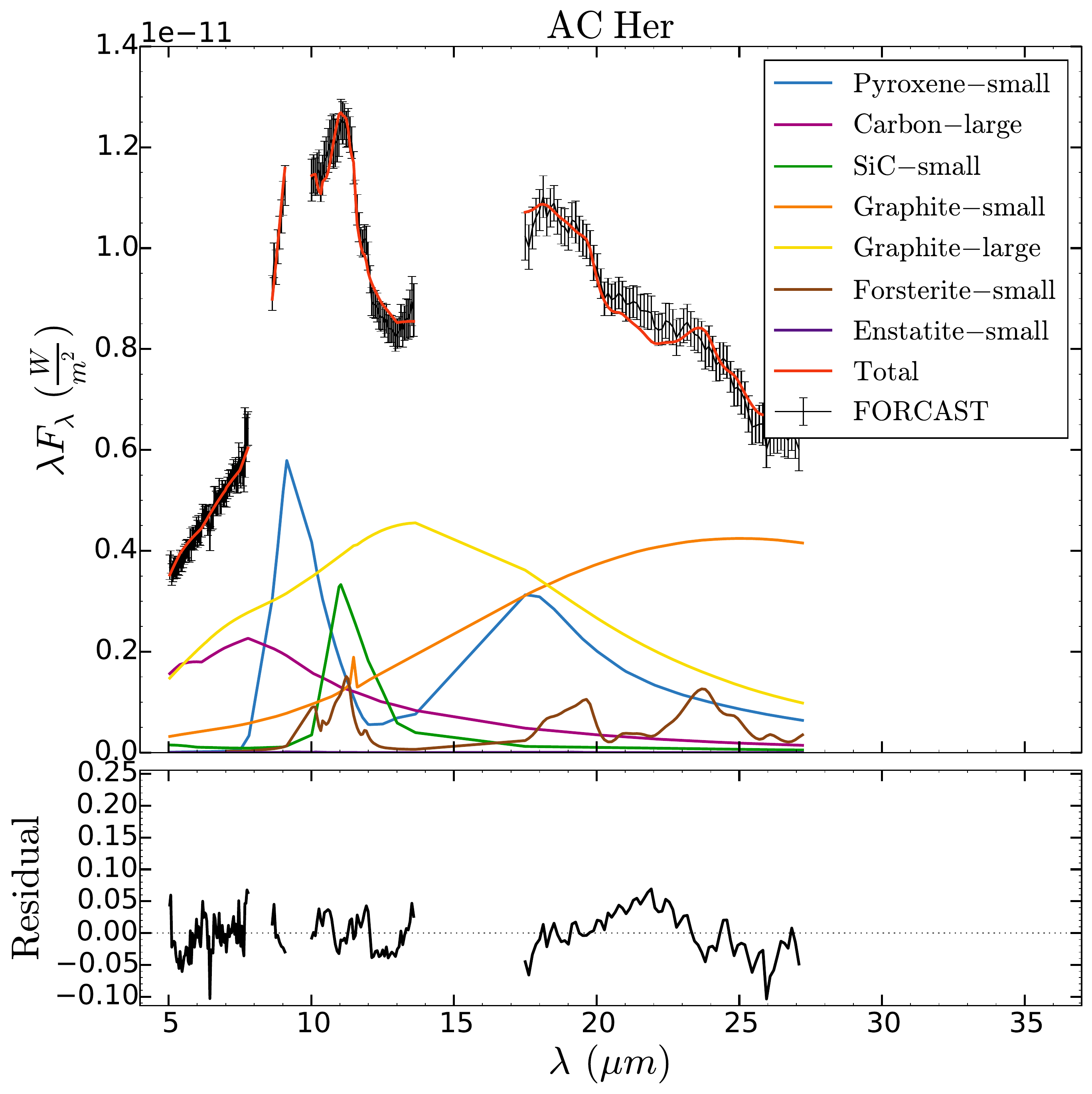}{0.5\textwidth}{(a) Enstatite-small added; $\chi^2_{\rm{red}} = 0.61$}
          \rightfig{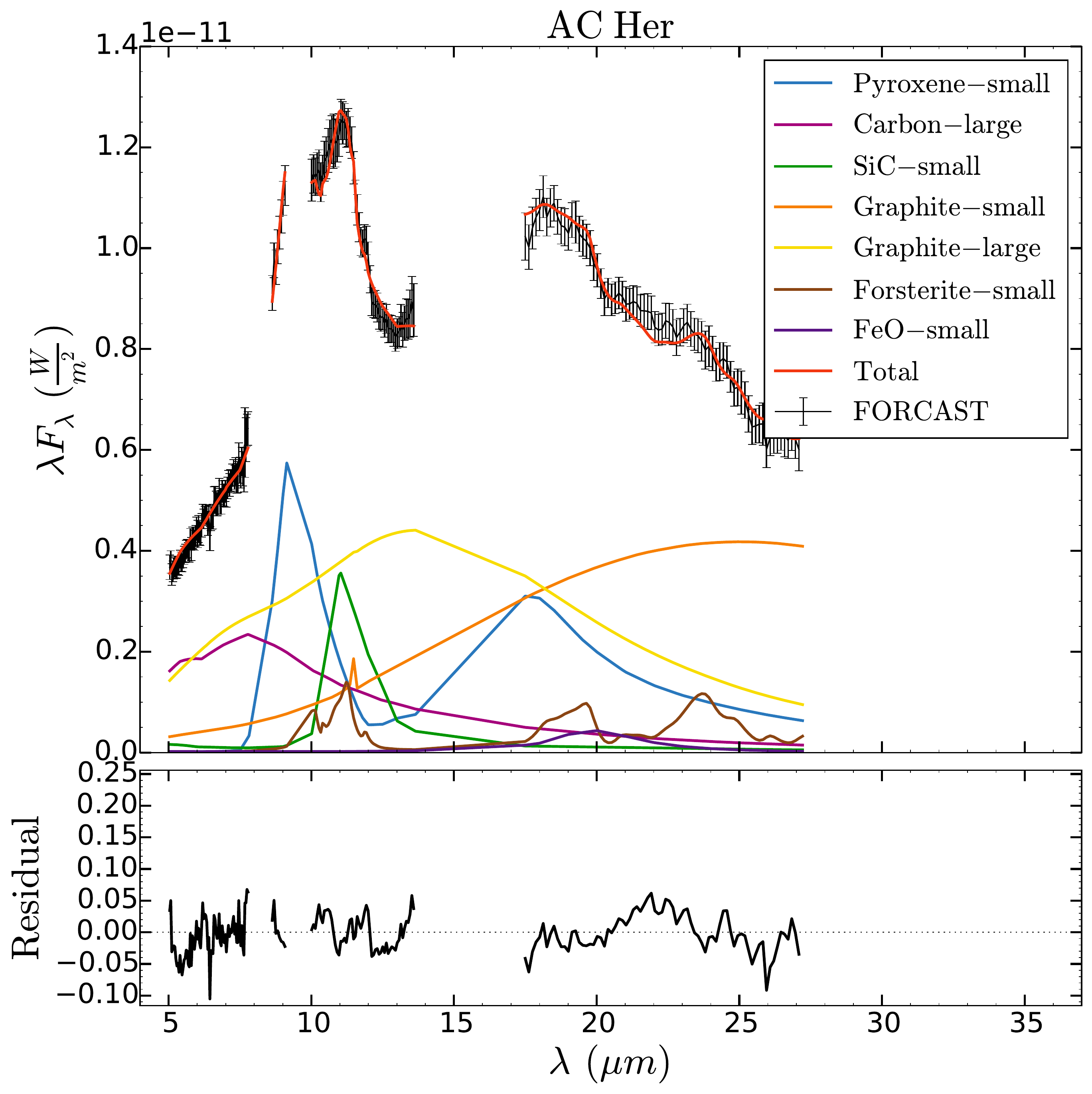}{0.5\textwidth}{(b) FeO-small added; $\chi^2_{\rm{red}} = 0.56$}}
\gridline{\leftfig{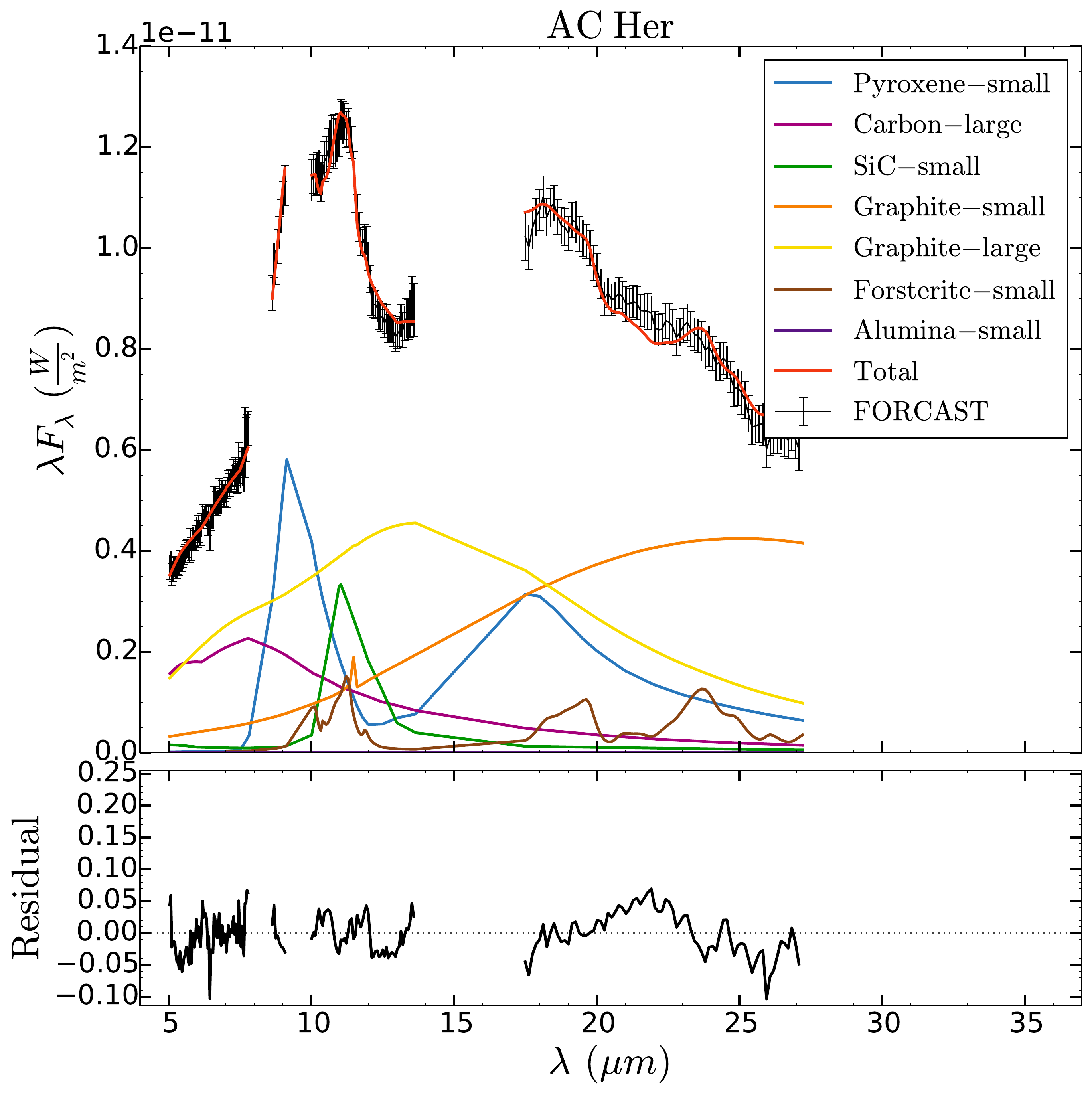}{0.5\textwidth}{(c) Alumina-small added; $\chi^2_{\rm{red}} = 0.61$}
	\rightfig{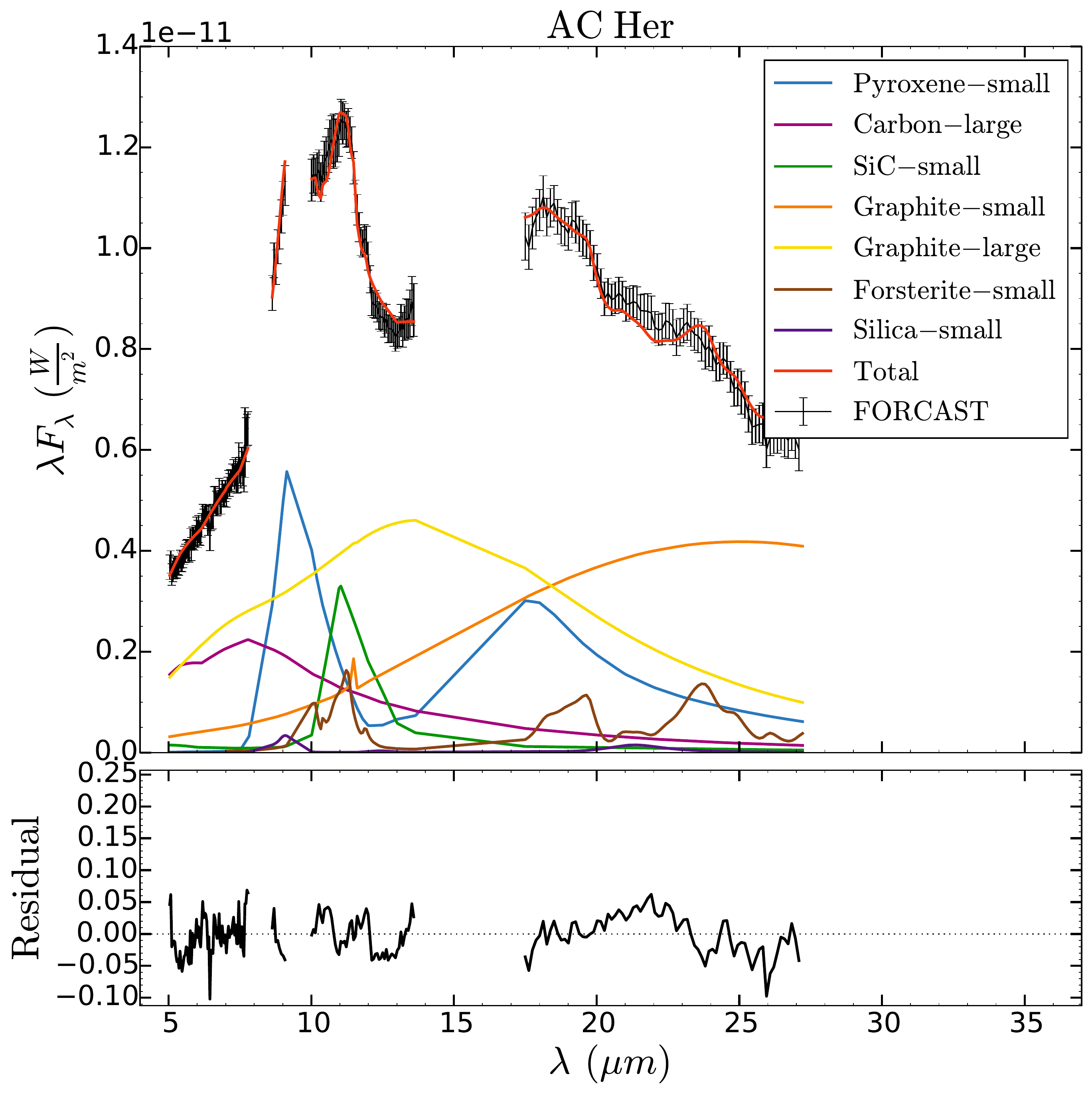}{0.5\textwidth}{(d) Silica-small added; $\chi^2_{\rm{red}} = 0.59$}}
         \caption{The effect on the fit when \emph{adding} dust species to the best model of AC Her.  The best fitting coefficient, $c_i$, for Alumina-small in panel (c) was 0.00.}
  \label{fig:limit_high_sn_added}
\end{figure*}

\begin{figure*}[h!]
\gridline{\leftfig{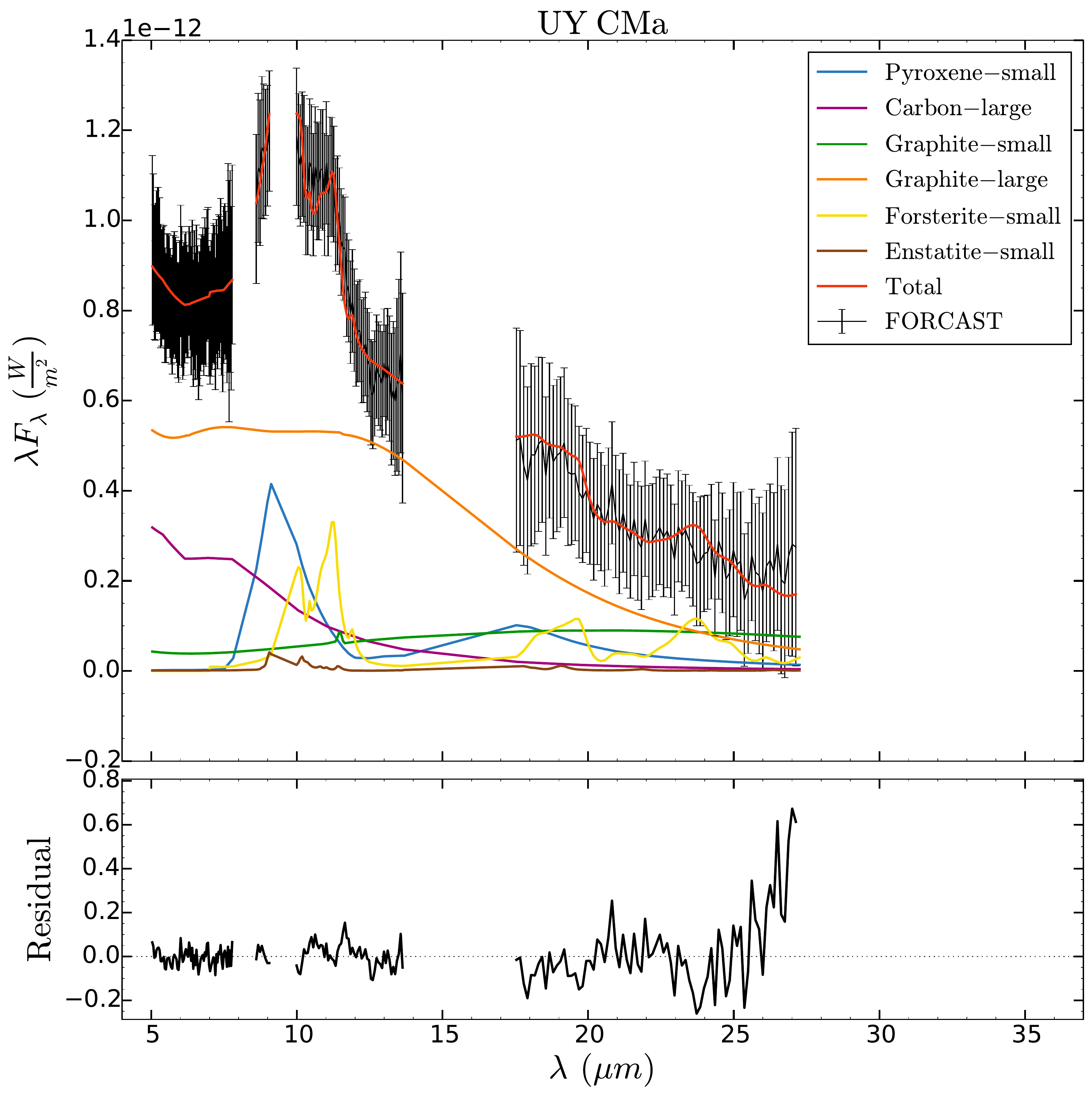}{0.5\textwidth}{(a) Enstatite-small added; $\chi^2_{\rm{red}} = 0.085$}
          \rightfig{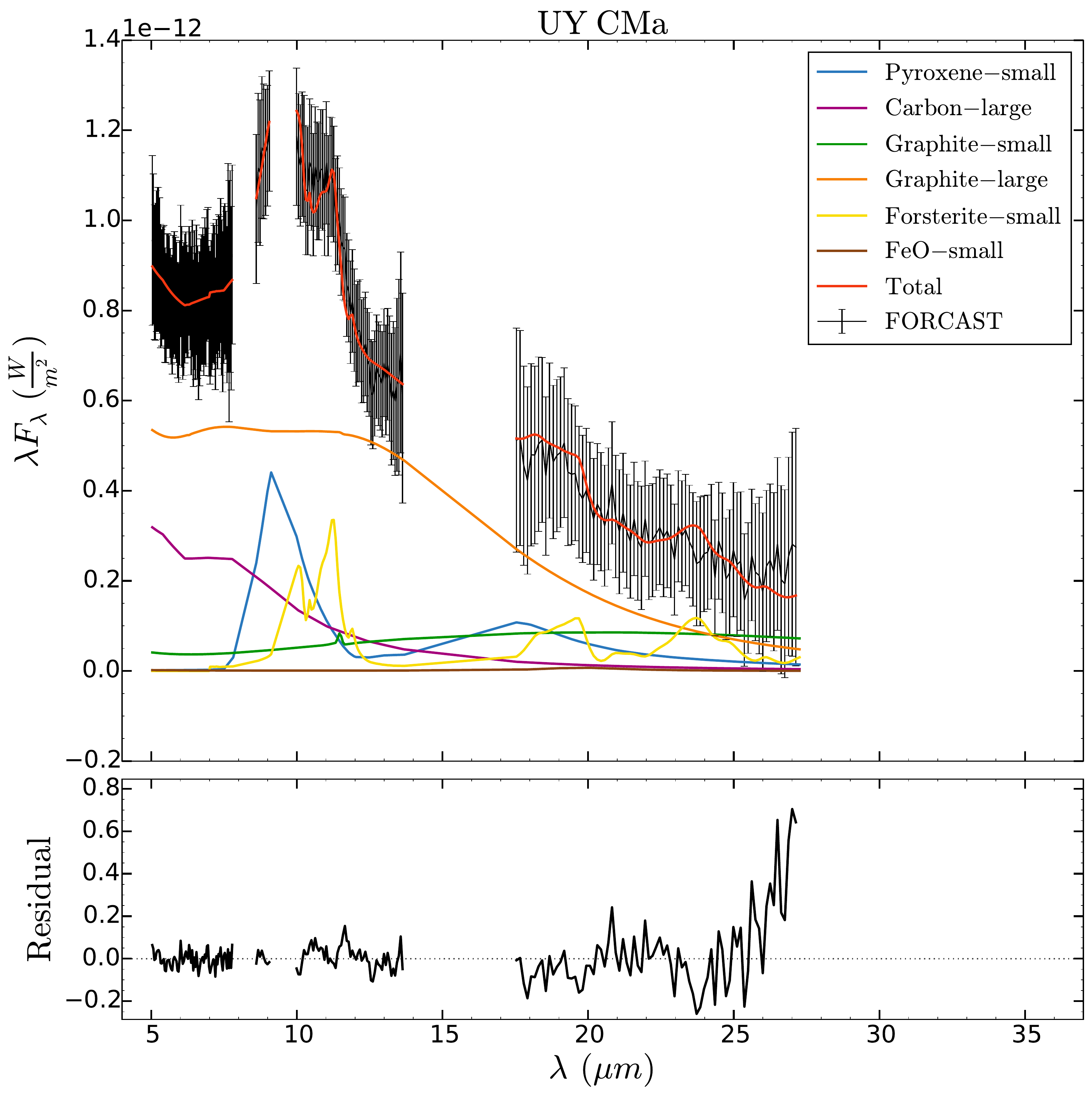}{0.5\textwidth}{(b) FeO-small added; $\chi^2_{\rm{red}} = 0.085$}}
\gridline{\leftfig{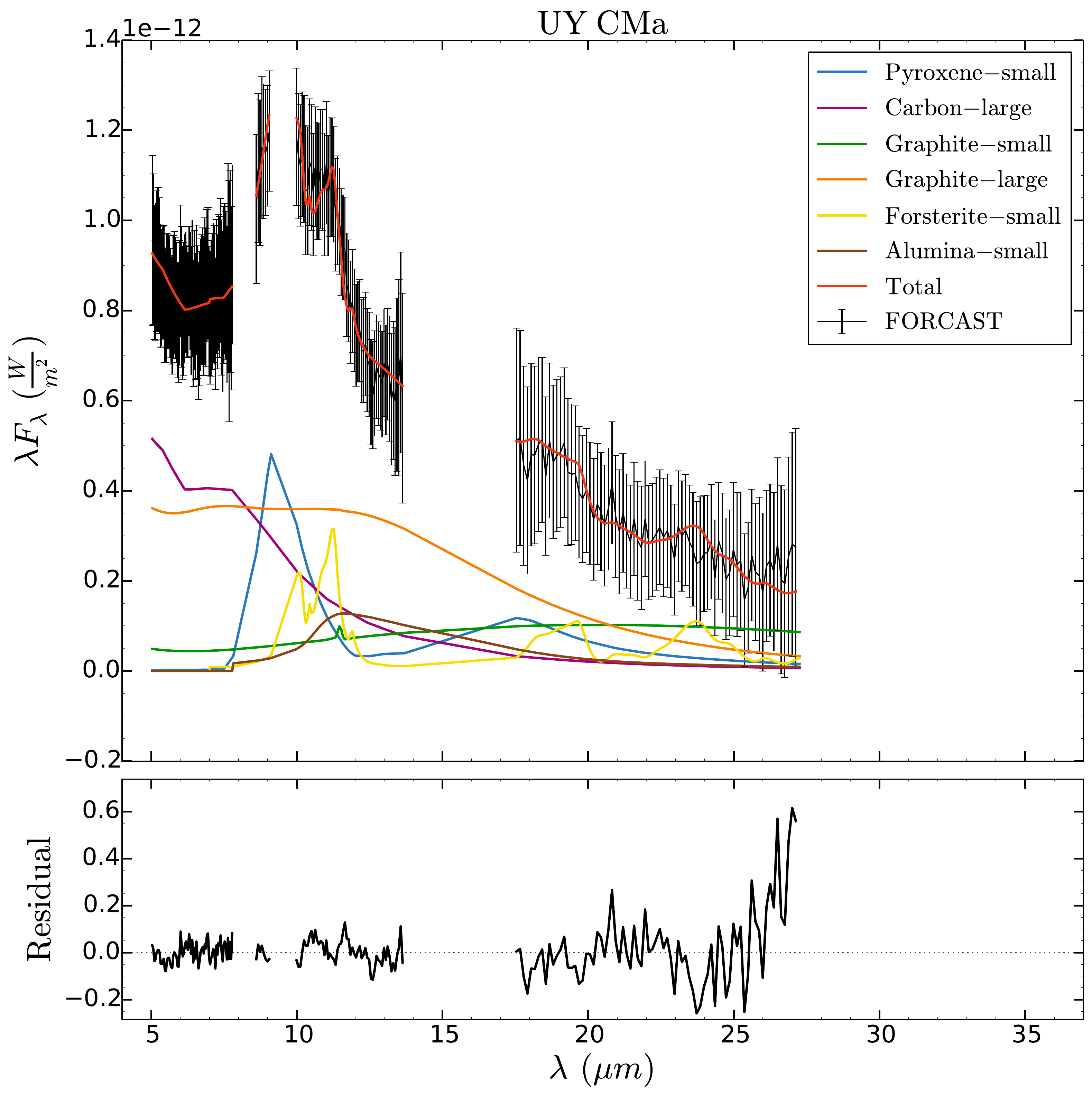}{0.5\textwidth}{(c) Alumina-small added; $\chi^2_{\rm{red}} = 0.080$}
\rightfig{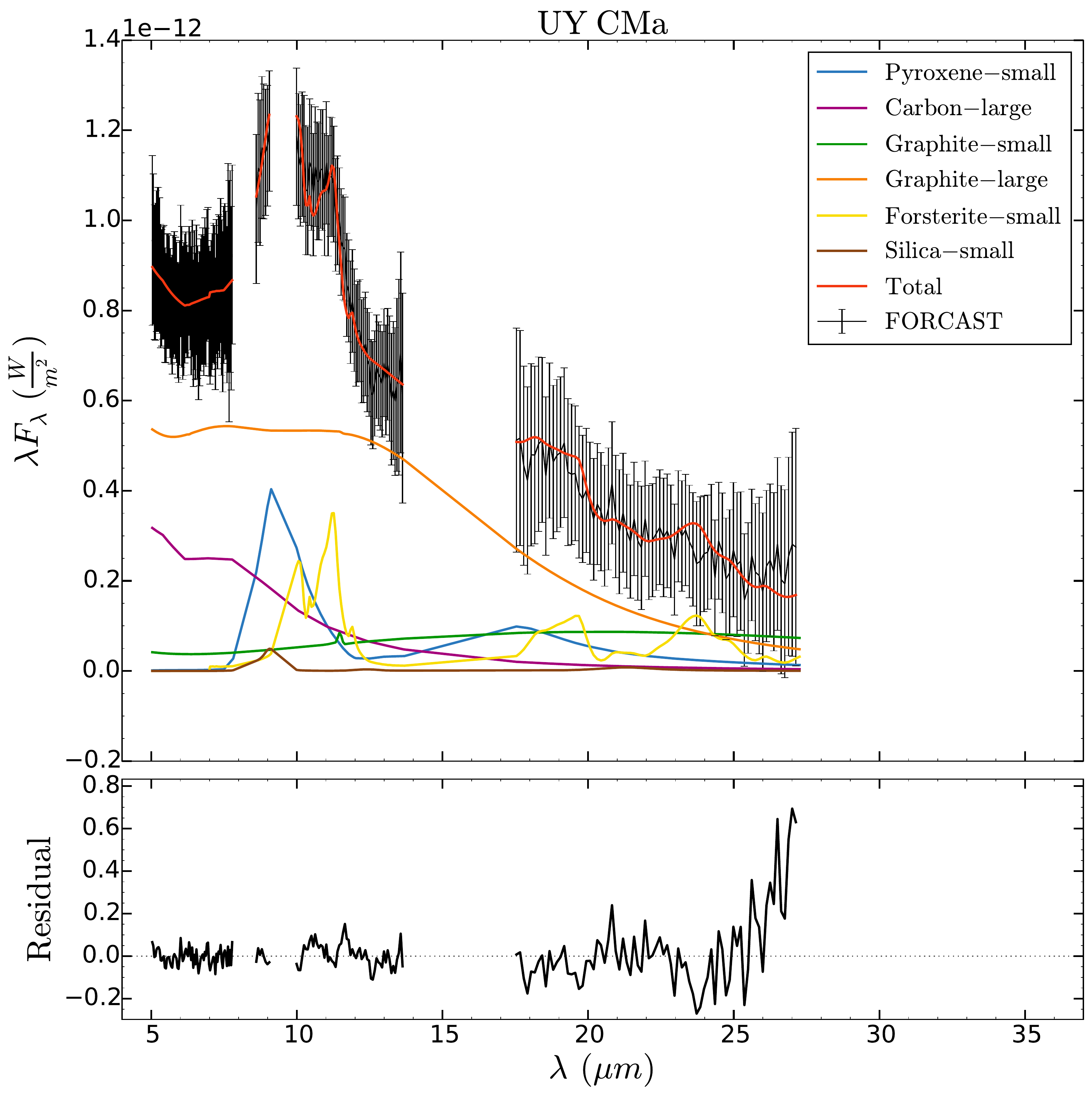}{0.5\textwidth}{(d) Silica-small added; $\chi^2_{\rm{red}} = 0.085$}}
\caption{The effect on the fit when \emph{adding} dust species to the best model of UY CMa.}
  \label{fig:limit_low_sn_added}
\end{figure*}

Because our model included 13 dust species and some of the program stars had relatively low signal-to-noise FORCAST spectra, it was difficult to confirm the uniqueness of our models.  For spectra with high signal-to-noise ratios, excluding dust species from the model had a noticeably negative impact on the goodness of fit (see Figure \ref{fig:limit_high_sn_removed}).  Whereas, for low signal-to-noise spectra the exclusion of dust species changed the overall shape of the fit but the $\chi^2_{\rm{red}}$ values changed very little (see Figure \ref{fig:limit_low_sn_removed}).  For both high and low signal-to-noise spectra, adding additional dust species improves the goodness of fit very little, if at all.  We checked this by including Mg-rich crystalline enstatite \citep{Jaeger98a}, iron oxide (FeO; \citealp{Henning95}), amorphous alumina (porous; \citealp{Begemann97}), and amorphous silica \citep{Henning97} to our model one at a time.  The $\chi^2_{\rm{red}}$ values either remained the same or marginally improved for all of the program stars.  The volume fraction of the added dust species depended on the signal-to-noise ratio of the spectra, with lower signal-to-noise spectra generally including 1-4\% by volume and high signal-to-noise spectra including $\leq$ 1\% by volume.  Figure \ref{fig:limit_high_sn_added} illustrates the effect of adding additional dust species to AC Her (high signal-to-noise spectra) and Figure \ref{fig:limit_low_sn_added} illustrates the effect of adding additional dust species to UY CMa (low signal-to-noise spectra).

\begin{figure*}[h!]
\gridline{\leftfig{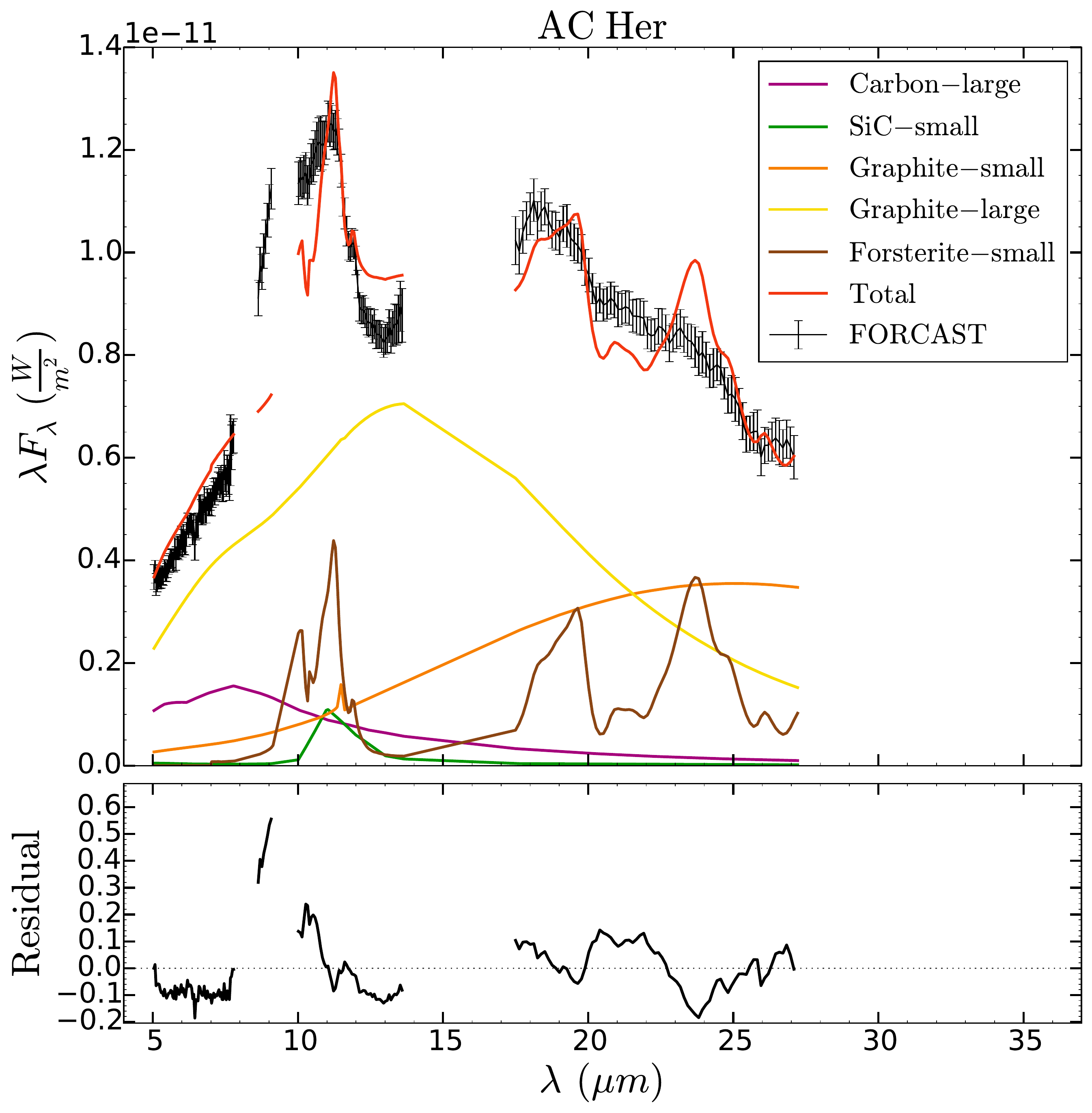}{0.5\textwidth}{(a) Pyroxene-small removed; $\chi^2_{\rm{red}} = 9.13$}
          \rightfig{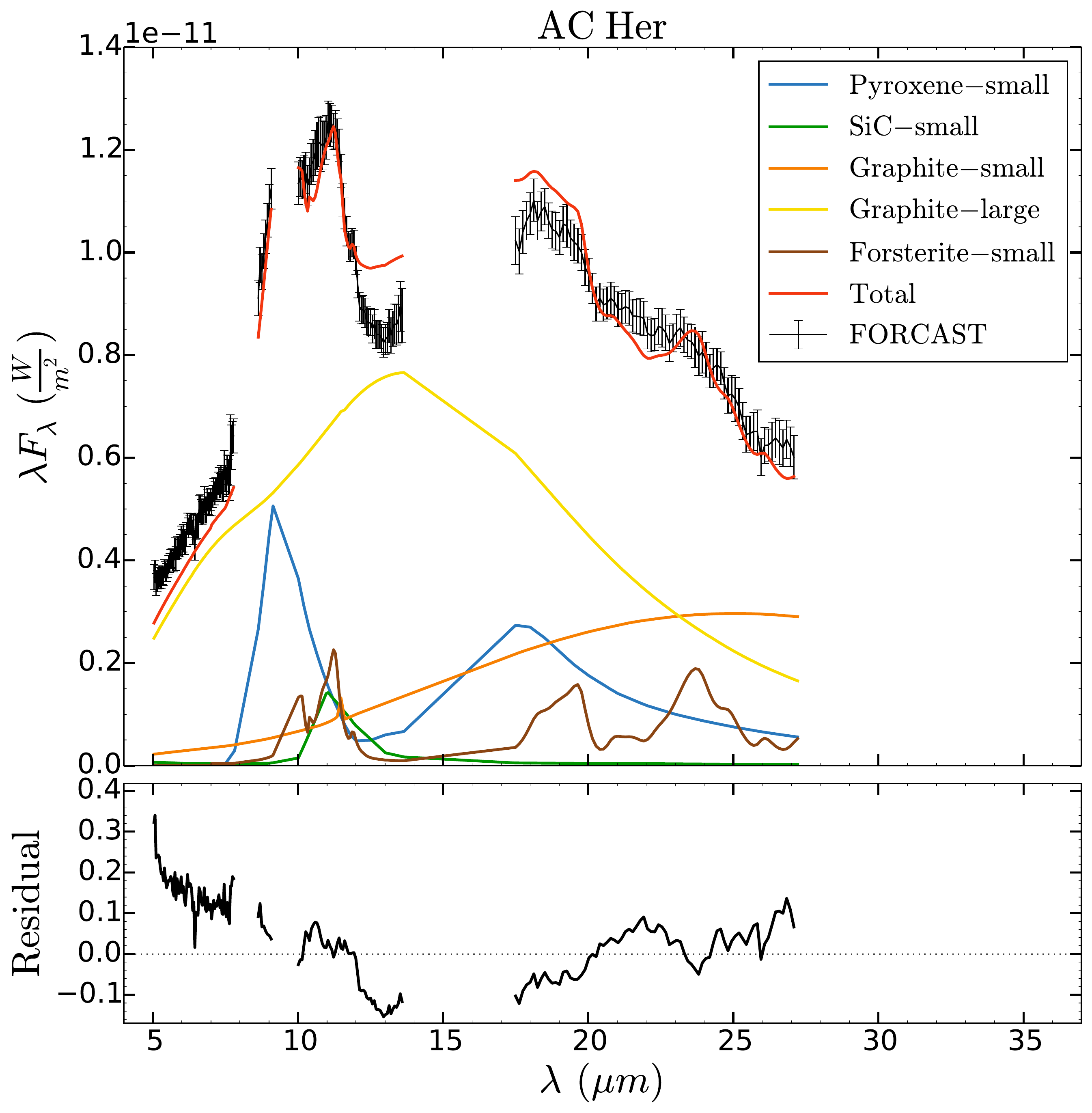}{0.5\textwidth}{(b) Carbon-large removed; $\chi^2_{\rm{red}} = 5.45$}}
\gridline{\leftfig{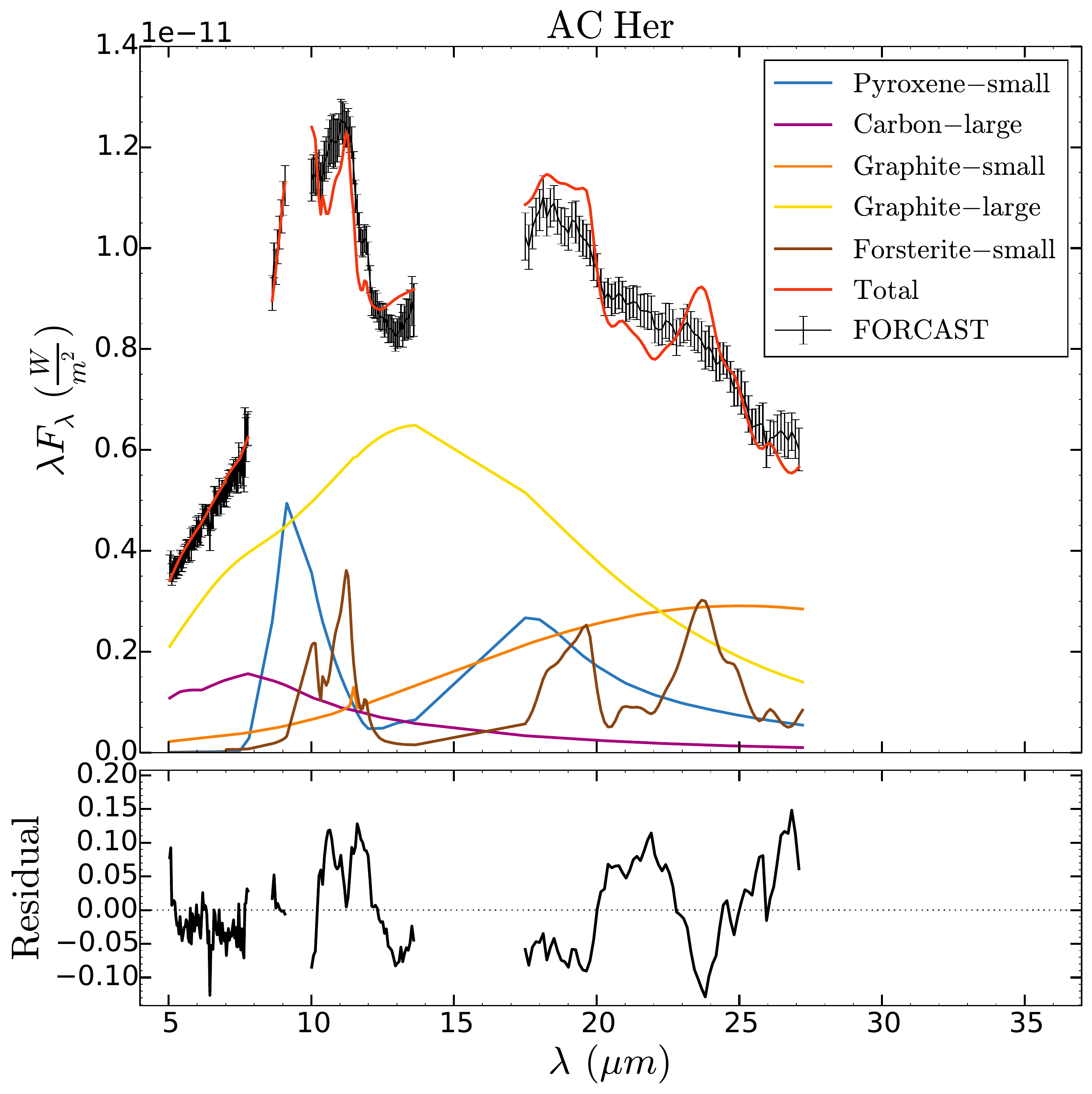}{0.5\textwidth}{(c) SiC-small removed; $\chi^2_{\rm{red}} = 2.37$}
	\rightfig{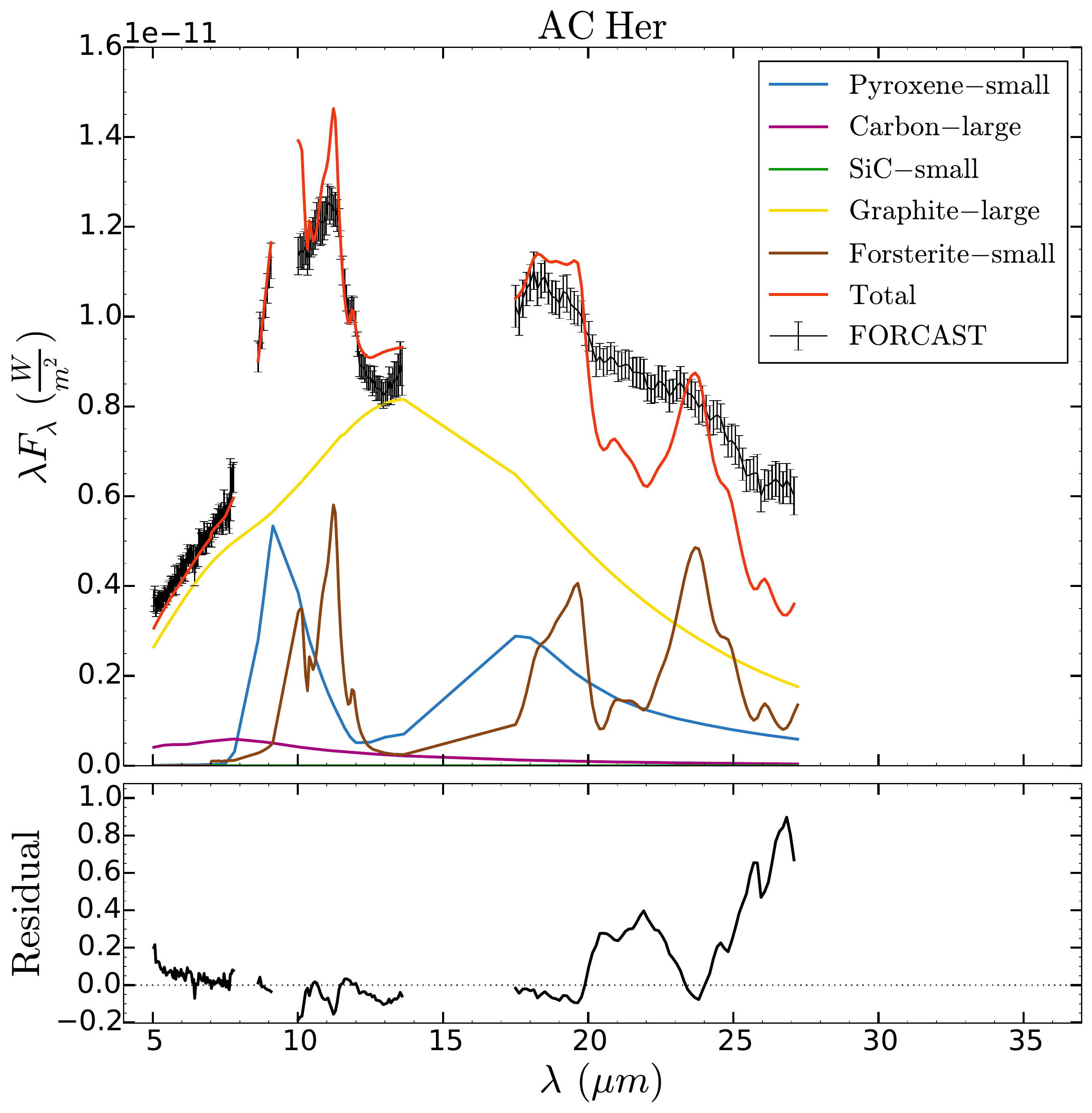}{0.5\textwidth}{(d) Graphite-small removed; $\chi^2_{\rm{red}} = 10.2$}}
\end{figure*}

\begin{figure*}[h!]
\gridline{\leftfig{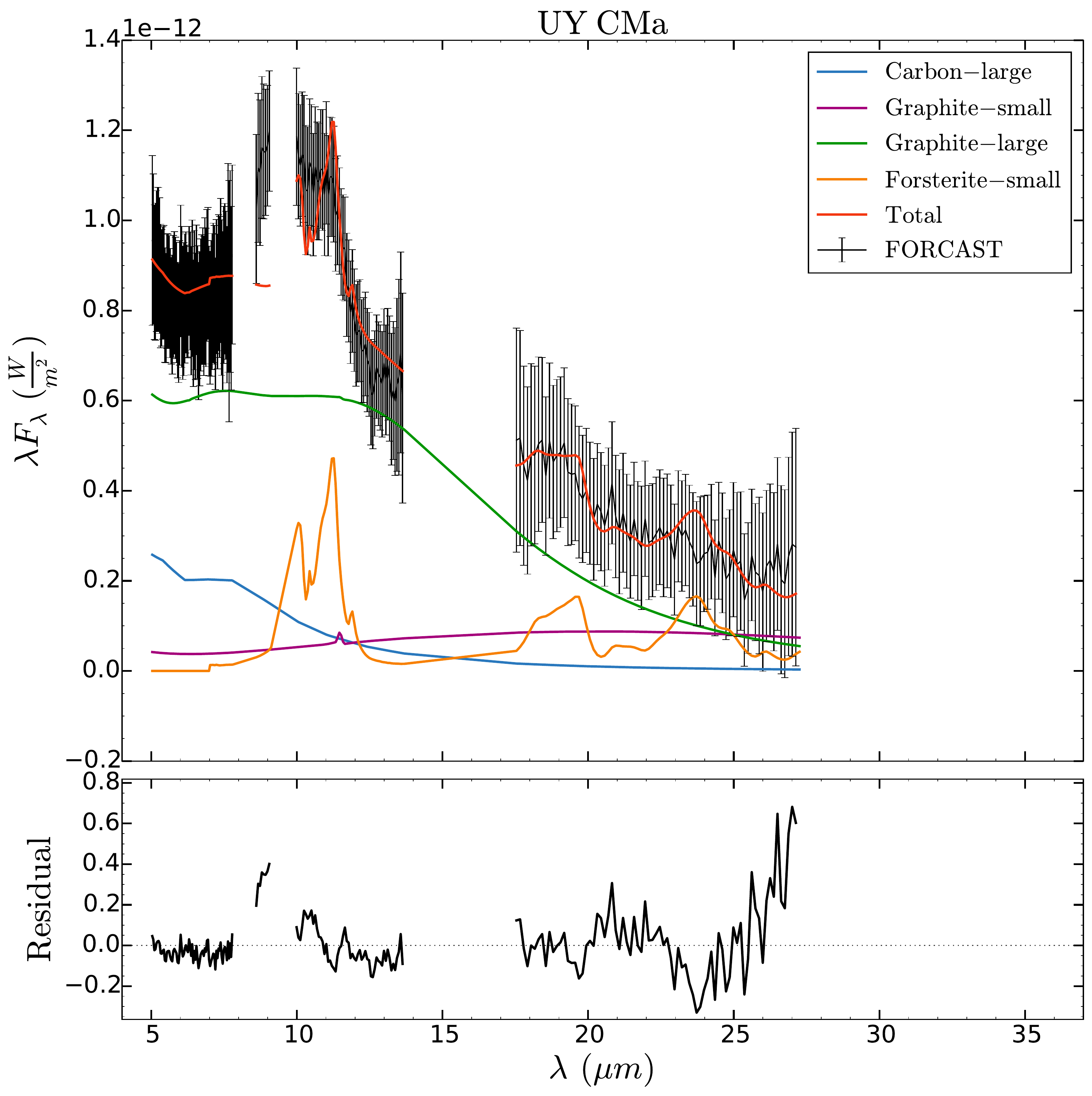}{0.5\textwidth}{(a) Pyroxene-small removed; $\chi^2_{\rm{red}} = 0.34$}
          \rightfig{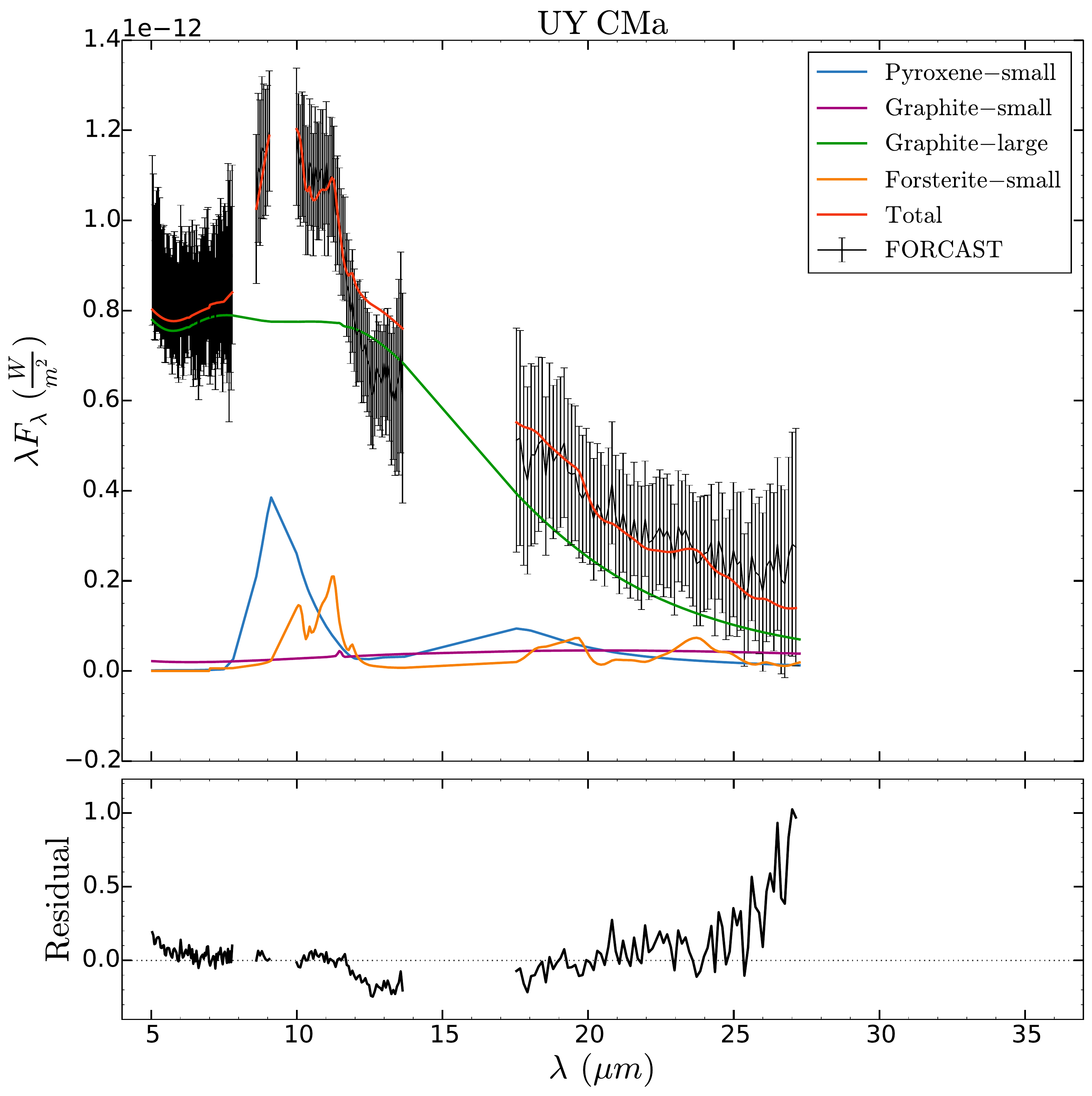}{0.5\textwidth}{(b) Carbon-large removed; $\chi^2_{\rm{red}} = 0.25$}}
\gridline{\leftfig{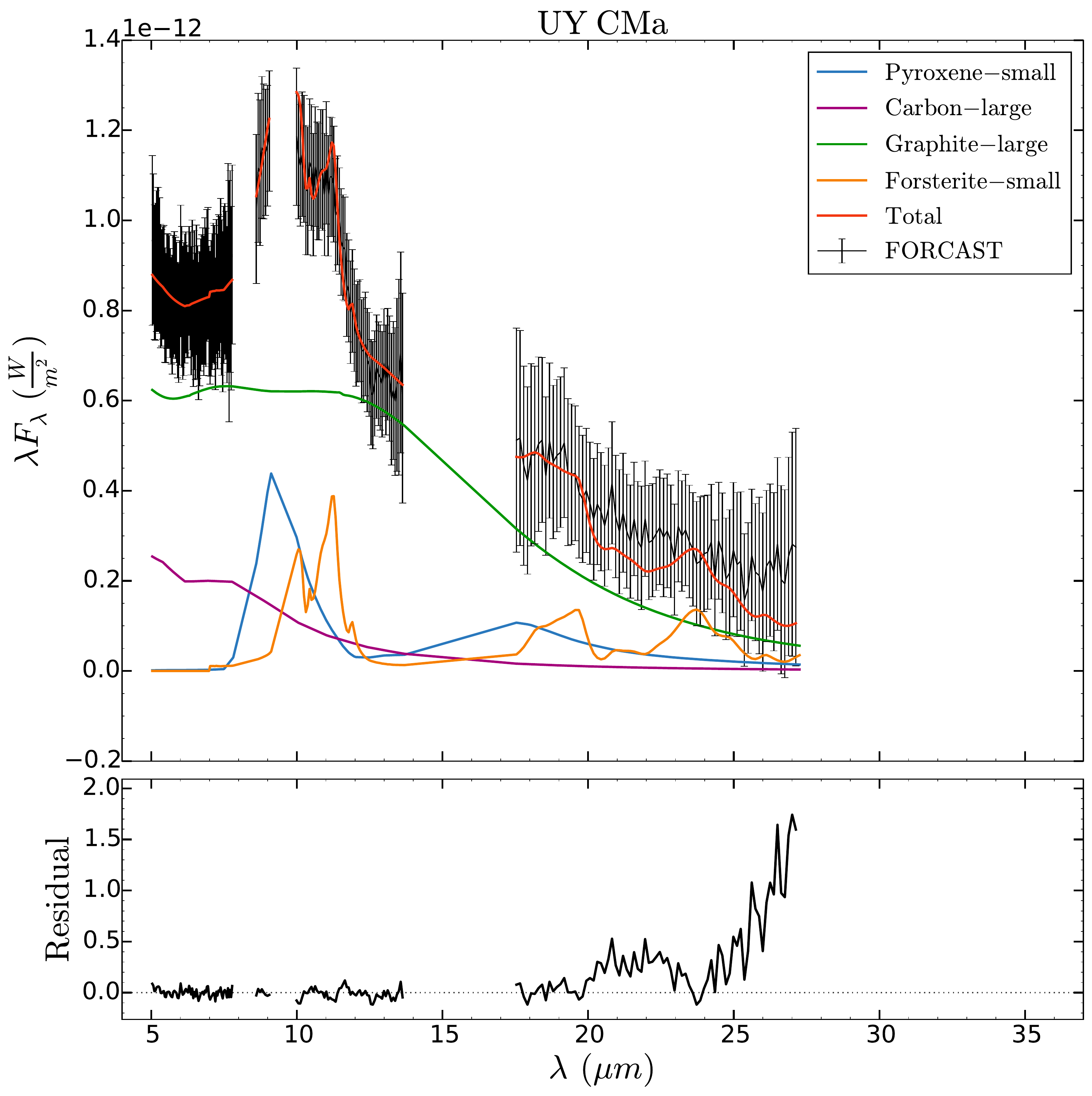}{0.5\textwidth}{(c) Graphite-small removed; $\chi^2_{\rm{red}} = 0.14$}
	\rightfig{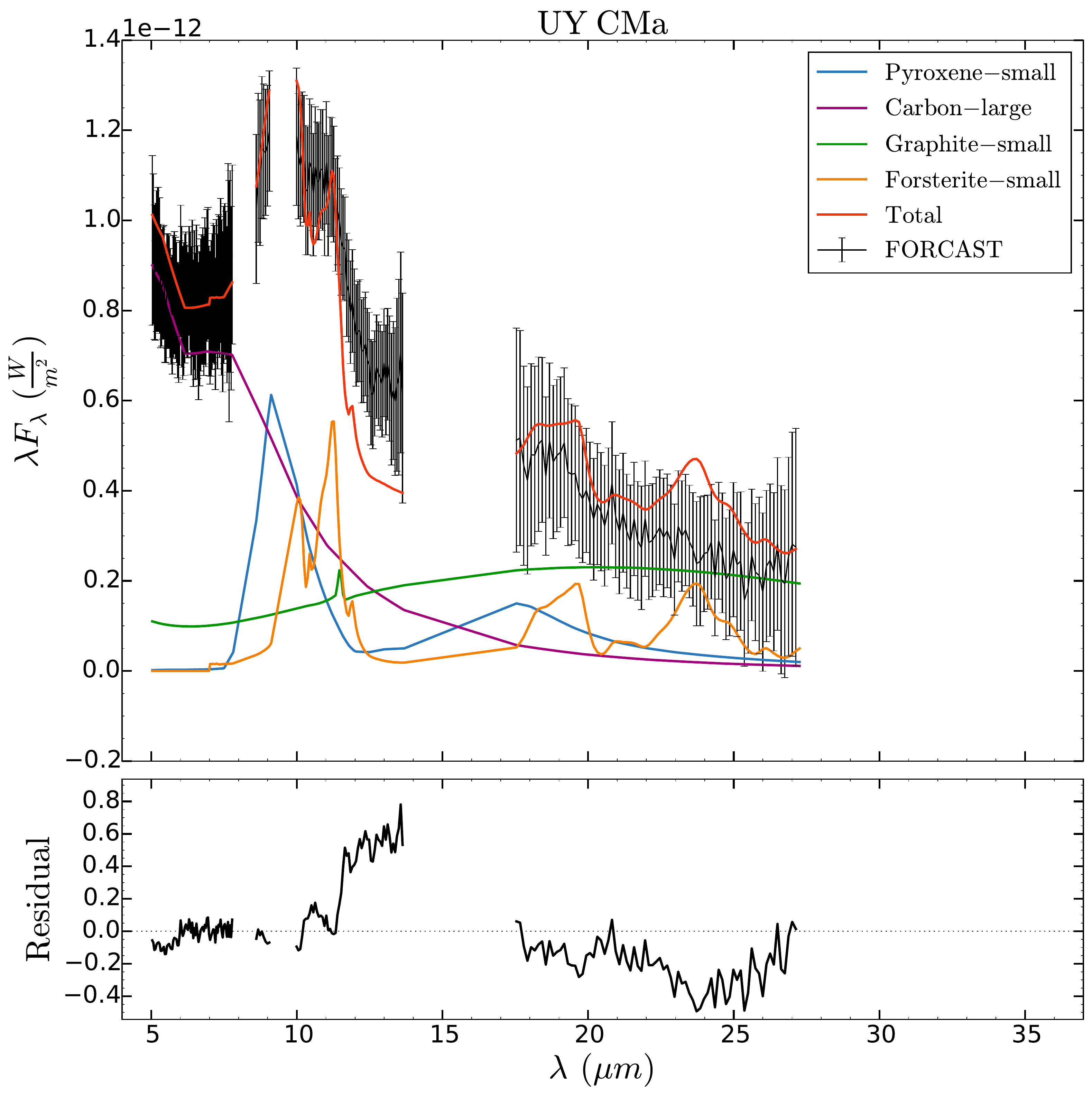}{0.5\textwidth}{(d) Graphite-large removed; $\chi^2_{\rm{red}} = 0.90$}}
\end{figure*}

\vspace{1mm}
\subsection{Crystallinity of the ISM}
An upper limit on the degree of crystallinity of silicates in the diffuse ISM has been estimated by \citet{Kemper04} to be $0.2\% \pm 0.2\%$ by mass.  This estimate is similar to the average crystalline silicate fraction we find for our program stars of $0.4\% \pm 0.05\%$ by volume.  We concur with \citet{Kemper04} that this suggests crystalline material is either diluted in the ISM by other amorphous grain producing processes such as supernovae, or there is an amorphization process that occurs in the ISM on a shorter timescale than the destruction timescale, possibly heavy ion bombardment.  Our mineralogy model of each star predicts that the majority of the dust is in the form of graphite and amorphous carbon.  On average our model predicts $80\% \pm 1\%$ graphite and amorphous carbon, and $57\% \pm 1\%$ graphite around each star by volume.  This large volume fraction of graphite and carbon around post-AGB stars may help explain the 2175 \AA\ bump observed in the interstellar extinction curve which is possibly due to these two species \citep{Rouleau97, Duley98, Bradley05}.

\section{Conclusion}
\label{conclusion}
We have presented a first look at data obtained with SOFIA FORCAST of 15 RV Tauri and 3 SRd variable stars. These data have demonstrated the diversity of dust features present in these systems, possibly tracing the evolutionary track from post-AGB star to PN. These observations of IR excess support the hypothesis that the systems in question are at an advanced stage in their transition to PNe.  Our main conclusions can be summarized as follows:
\begin{itemize}

\item[--] Almost all of the stars observed display a 10 \micron\ and/or 20 \micron\ emission feature.  For most of the stars observed, the FORCAST continua are well described by two Planck functions one at $\sim$ 1000 K and one at $\sim$ 250 K with a majority of the dust ($\sim$ 97\%) in the cooler form.  A single Planck function fit the underlying contiuum of $\rm{o^1\ Cen}$ and V Vul, indicating that these systems may be in the final stages of disk dissipation.

\item[--] Our mineralogy model indicates the presence of both carbon rich and oxygen rich dust species with a majority of the dust, $80\% \pm 1\%$ by volume on average, in the form of amorphous carbon and graphite.  All of the stars display this dual chemistry circumstellar dust.  This requires that either these stars result from a narrow range of masses that terminate AGB evolution just as the carbon exceeds the oxygen abundance or the the formation process is not single star evolution.  We speculate the formation process is common envelope evolution.

\item[--] The spectra do not exhibit any obvious crystalline emission features and our model only predicts UY CMa, RU Cen, and AC Her to have crystalline forsterite at volume fractions of $4\% \pm 0.9\%$, $1\% \pm 0.3\%$, and $1\% \pm 0.1\%$, respectively. 

\item[--] Most of the spectra show a low peak to continuum value (i.e. $< 2.0$) and a low 10/20 \micron\ ratio (i.e. $< 50$) indicating that the grains are relatively large and have undergone significant processing, supporting the hypothesis that the dust is constrained to a Keplerian disk and that we are viewing the heavily processed, central regions of the disk from a nearly face-on orientation.

\item[--] The average composition of the SRd variables contains 8\% more small carbon dust and less graphite (14\% less of the small species and 5\% less of the large) than the average composition of the RV Tauri stars.  Of the three SRd variables modeled in this work, none of them contained the large carbon species--on average the RV Tauri stars contained 13\% by volume.  Overall the average volume fraction of large grains for the SRd variables was 16\% compared to 30\% for the RV Tauri stars.  The paucity of large grains around any of the SRd variables supports the hypothesis that these stars are single star systems. 

\item[--] Between the featureless IR dust species, amorphous carbon is included in more of our models (16 out 17) than metallic iron (4 out of 17).

\end{itemize}
\acknowledgments
The observations were made with the NASA/DLR Stratospheric Observatory for Infrared Astronomy (SOFIA) which is jointly operated by the Universities Space Research Association, Inc. (USRA), under NASA contract NAS2-97001, and the Deutsches SOFIA Institut (DSI) under DLR contract 50 OK 0901 to the University of Stuttgart.

\vspace{5mm}
\facilities{SOFIA (FORCAST)}

\software{pymiecoated \citep{pymiecoated}}
\bibliography{paperNotes}

\begin{thebibliography}{}
\expandafter\ifx\csname natexlab\endcsname\relax\def\natexlab#1{#1}\fi
\providecommand{\url}[1]{\href{#1}{#1}}

\bibitem[{{Acke} {et~al.}(2013){Acke}, {Degroote}, {Lombaert}, {de Vries},
  {Smolders}, {Verhoelst}, {Lagadec}, {Gielen}, {Van Winckel}, \&
  {Waelkens}}]{Acke13}
{Acke}, B., {Degroote}, P., {Lombaert}, R., {et~al.} 2013, \aap, 551, A76

\bibitem[{{Alcolea} \& {Bujarrabal}(1991)}]{Alcolea91}
{Alcolea}, J., \& {Bujarrabal}, V. 1991, \aap, 245, 499

\bibitem[{{Becklin} {et~al.}(2007){Becklin}, {Tielens}, {Gehrz}, \&
  {Callis}}]{Becklin07}
{Becklin}, E.~E., {Tielens}, A.~G.~G.~M., {Gehrz}, R.~D., \& {Callis}, H.~H.~S.
  2007, in \procspie, Vol. 6678, Infrared Spaceborne Remote Sensing and
  Instrumentation XV, 66780A

\bibitem[{{Begemann} {et~al.}(1997){Begemann}, {Dorschner}, {Henning},
  {Mutschke}, {G{\"u}rtler}, {K{\"o}mpe}, \& {Nass}}]{Begemann97}
{Begemann}, B., {Dorschner}, J., {Henning}, T., {et~al.} 1997, \apj, 476, 199

\bibitem[{{Berriman}(2008)}]{Berriman08}
{Berriman}, G.~B. 2008, in \procspie, Vol. 7016, Observatory Operations:
  Strategies, Processes, and Systems II, 701618

\bibitem[{{Blommaert} {et~al.}(2014){Blommaert}, {de Vries}, {Waters},
  {Waelkens}, {Min}, {Van Winckel}, {Molster}, {Decin}, {Groenewegen},
  {Barlow}, {Garc{\'{\i}}a-Lario}, {Kerschbaum}, {Posch}, {Royer}, {Ueta},
  {Vandenbussche}, {Van de Steene}, \& {van Hoof}}]{Blommaert14}
{Blommaert}, J.~A.~D.~L., {de Vries}, B.~L., {Waters}, L.~B.~F.~M., {et~al.}
  2014, \aap, 565, A109

\bibitem[{{Bohren} \& {Huffman}(1983)}]{Bohren83}
{Bohren}, C.~F., \& {Huffman}, D.~R. 1983, {Absorption and Scattering of Light
  by Small Particles} (New York: Wiley)

\bibitem[{{Bouwman} {et~al.}(2001){Bouwman}, {Meeus}, {de Koter}, {Hony},
  {Dominik}, \& {Waters}}]{Bouwman01}
{Bouwman}, J., {Meeus}, G., {de Koter}, A., {et~al.} 2001, \aap, 375, 950

\bibitem[{{Bradley} {et~al.}(2005){Bradley}, {Dai}, {Erni}, {Browning},
  {Graham}, {Weber}, {Smith}, {Hutcheon}, {Ishii}, {Bajt}, {Floss},
  {Stadermann}, \& {Sandford}}]{Bradley05}
{Bradley}, J., {Dai}, Z.~R., {Erni}, R., {et~al.} 2005, Science, 307, 244

\bibitem[{{Bujarrabal} {et~al.}(2013){Bujarrabal}, {Alcolea}, {Van Winckel},
  {Santander-Garc{\'{\i}}a}, \& {Castro-Carrizo}}]{Bujarrabal13}
{Bujarrabal}, V., {Alcolea}, J., {Van Winckel}, H., {Santander-Garc{\'{\i}}a},
  M., \& {Castro-Carrizo}, A. 2013, \aap, 557, A104

\bibitem[{{Bujarrabal} {et~al.}(1988){Bujarrabal}, {Bachiller}, {Alcolea}, \&
  {Martin-Pintado}}]{Bujarrabal88}
{Bujarrabal}, V., {Bachiller}, R., {Alcolea}, J., \& {Martin-Pintado}, J. 1988,
  \aap, 206, L17

\bibitem[{{Bujarrabal} {et~al.}(2015){Bujarrabal}, {Castro-Carrizo}, {Alcolea},
  \& {Van Winckel}}]{Bujarrabal15}
{Bujarrabal}, V., {Castro-Carrizo}, A., {Alcolea}, J., \& {Van Winckel}, H.
  2015, \aap, 575, L7

\bibitem[{{Bujarrabal} {et~al.}(1989){Bujarrabal}, {Gomez-Gonzalez}, \&
  {Planesas}}]{Bujarrabal89}
{Bujarrabal}, V., {Gomez-Gonzalez}, J., \& {Planesas}, P. 1989, \aap, 219, 256

\bibitem[{{Bujarrabal} {et~al.}(2007){Bujarrabal}, {van Winckel}, {Neri},
  {Alcolea}, {Castro-Carrizo}, \& {Deroo}}]{Bujarrabal07}
{Bujarrabal}, V., {van Winckel}, H., {Neri}, R., {et~al.} 2007, \aap, 468, L45

\bibitem[{{Cami} {et~al.}(1998){Cami}, {de Jong}, {Justtannont}, {Yamamura}, \&
  {Waters}}]{Cami98}
{Cami}, J., {de Jong}, T., {Justtannont}, K., {Yamamura}, I., \& {Waters},
  L.~B.~F.~M. 1998, \apss, 255, 339

\bibitem[{{Chan} \& {Kwok}(1990)}]{Chan90}
{Chan}, S.~J., \& {Kwok}, S. 1990, \aap, 237, 354

\bibitem[{{Clarke} {et~al.}(2015){Clarke}, {Vacca}, \& {Shuping}}]{Clarke15}
{Clarke}, M., {Vacca}, W.~D., \& {Shuping}, R.~Y. 2015, in Astronomical Society
  of the Pacific Conference Series, Vol. 495, Astronomical Data Analysis
  Software an Systems XXIV (ADASS XXIV), ed. A.~R. {Taylor} \& E.~{Rosolowsky},
  355

\bibitem[{{Crapsi} {et~al.}(2008){Crapsi}, {van Dishoeck}, {Hogerheijde},
  {Pontoppidan}, \& {Dullemond}}]{Crapsi08}
{Crapsi}, A., {van Dishoeck}, E.~F., {Hogerheijde}, M.~R., {Pontoppidan},
  K.~M., \& {Dullemond}, C.~P. 2008, \aap, 486, 245

\bibitem[{{de Ruyter} {et~al.}(2005){de Ruyter}, {van Winckel}, {Dominik},
  {Waters}, \& {Dejonghe}}]{deRuyter05}
{de Ruyter}, S., {van Winckel}, H., {Dominik}, C., {Waters}, L.~B.~F.~M., \&
  {Dejonghe}, H. 2005, \aap, 435, 161

\bibitem[{{de Ruyter} {et~al.}(2006){de Ruyter}, {van Winckel}, {Maas}, {Lloyd
  Evans}, {Waters}, \& {Dejonghe}}]{deRuyter06}
{de Ruyter}, S., {van Winckel}, H., {Maas}, T., {et~al.} 2006, \aap, 448, 641

\bibitem[{{Deroo} {et~al.}(2005){Deroo}, {Reyniers}, {van Winckel}, {Goriely},
  \& {Siess}}]{Deroo05}
{Deroo}, P., {Reyniers}, M., {van Winckel}, H., {Goriely}, S., \& {Siess}, L.
  2005, \aap, 438, 987

\bibitem[{{Deroo} {et~al.}(2006){Deroo}, {van Winckel}, {Min}, {Waters},
  {Verhoelst}, {Jaffe}, {Morel}, {Paresce}, {Richichi}, {Stee}, \&
  {Wittkowski}}]{Deroo06}
{Deroo}, P., {van Winckel}, H., {Min}, M., {et~al.} 2006, \aap, 450, 181

\bibitem[{{Dorschner} {et~al.}(1995){Dorschner}, {Begemann}, {Henning},
  {Jaeger}, \& {Mutschke}}]{Dorschner95}
{Dorschner}, J., {Begemann}, B., {Henning}, T., {Jaeger}, C., \& {Mutschke}, H.
  1995, \aap, 300, 503

\bibitem[{{Draine} \& {Lee}(1984)}]{Draine84}
{Draine}, B.~T., \& {Lee}, H.~M. 1984, \apj, 285, 89

\bibitem[{{Duley} \& {Seahra}(1998)}]{Duley98}
{Duley}, W.~W., \& {Seahra}, S. 1998, \apj, 507, 874

\bibitem[{{Evans} {et~al.}(2005){Evans}, {Tyne}, {Smith}, {Geballe},
  {Rawlings}, \& {Eyres}}]{Evans05}
{Evans}, A., {Tyne}, V.~H., {Smith}, O., {et~al.} 2005, \mnras, 360, 1483

\bibitem[{{Fokin}(1994)}]{Fokin94}
{Fokin}, A.~B. 1994, \aap, 292, 133

\bibitem[{Foreman-Mackey(2016)}]{corner}
Foreman-Mackey, D. 2016, The Journal of Open Source Software, 24,
  doi:10.21105/joss.00024.
\newblock \url{http://dx.doi.org/10.5281/zenodo.45906}

\bibitem[{{Gehrz}(1972)}]{Gehrz72a}
{Gehrz}, R.~D. 1972, \apj, 178, 715

\bibitem[{{Gehrz} {et~al.}(2009){Gehrz}, {Becklin}, {de Pater}, {Lester},
  {Roellig}, \& {Woodward}}]{Gehrz09}
{Gehrz}, R.~D., {Becklin}, E.~E., {de Pater}, I., {et~al.} 2009, Advances in
  Space Research, 44, 413

\bibitem[{{Gehrz} {et~al.}(1992){Gehrz}, {Jones}, {Woodward}, {Greenhouse},
  {Wagner}, {Harrison}, {Hayward}, \& {Benson}}]{Gehrz92}
{Gehrz}, R.~D., {Jones}, T.~J., {Woodward}, C.~E., {et~al.} 1992, \apj, 400,
  671

\bibitem[{{Gehrz} \& {Ney}(1972)}]{Gehrz72b}
{Gehrz}, R.~D., \& {Ney}, E.~P. 1972, \pasp, 84, 768

\bibitem[{{Gehrz} \& {Woolf}(1970)}]{Gehrz70}
{Gehrz}, R.~D., \& {Woolf}, N.~J. 1970, \apjl, 161, L213

\bibitem[{{Gezer} {et~al.}(2015){Gezer}, {Van Winckel}, {Bozkurt}, {De Smedt},
  {Kamath}, {Hillen}, \& {Manick}}]{Gezer15}
{Gezer}, I., {Van Winckel}, H., {Bozkurt}, Z., {et~al.} 2015, \mnras, 453, 133

\bibitem[{{Gielen} {et~al.}(2008){Gielen}, {van Winckel}, {Min}, {Waters}, \&
  {Lloyd Evans}}]{Gielen08}
{Gielen}, C., {van Winckel}, H., {Min}, M., {Waters}, L.~B.~F.~M., \& {Lloyd
  Evans}, T. 2008, \aap, 490, 725

\bibitem[{{Gielen} {et~al.}(2007){Gielen}, {van Winckel}, {Waters}, {Min}, \&
  {Dominik}}]{Gielen07}
{Gielen}, C., {van Winckel}, H., {Waters}, L.~B.~F.~M., {Min}, M., \&
  {Dominik}, C. 2007, \aap, 475, 629

\bibitem[{{Gielen} {et~al.}(2009){Gielen}, {van Winckel}, {Reyniers},
  {Zijlstra}, {Lloyd Evans}, {Gordon}, {Kemper}, {Indebetouw}, {Marengo},
  {Matsuura}, {Meixner}, {Sloan}, {Tielens}, \& {Woods}}]{Gielen09}
{Gielen}, C., {van Winckel}, H., {Reyniers}, M., {et~al.} 2009, \aap, 508, 1391

\bibitem[{{Gielen} {et~al.}(2011){Gielen}, {Bouwman}, {van Winckel}, {Lloyd
  Evans}, {Woods}, {Kemper}, {Marengo}, {Meixner}, {Sloan}, \&
  {Tielens}}]{Gielen11}
{Gielen}, C., {Bouwman}, J., {van Winckel}, H., {et~al.} 2011, \aap, 533, A99

\bibitem[{{Giridhar} {et~al.}(1998){Giridhar}, {Lambert}, \&
  {Gonzalez}}]{Giridhar98}
{Giridhar}, S., {Lambert}, D.~L., \& {Gonzalez}, G. 1998, \apj, 509, 366

\bibitem[{{Giridhar} {et~al.}(2000){Giridhar}, {Lambert}, \&
  {Gonzalez}}]{Giridhar00}
---. 2000, \apj, 531, 521

\bibitem[{{Giridhar} {et~al.}(2005){Giridhar}, {Lambert}, {Reddy}, {Gonzalez},
  \& {Yong}}]{Giridhar05}
{Giridhar}, S., {Lambert}, D.~L., {Reddy}, B.~E., {Gonzalez}, G., \& {Yong}, D.
  2005, \apj, 627, 432

\bibitem[{{Gonzalez} {et~al.}(1997{\natexlab{a}}){Gonzalez}, {Lambert}, \&
  {Giridhar}}]{Gonzalez97a}
{Gonzalez}, G., {Lambert}, D.~L., \& {Giridhar}, S. 1997{\natexlab{a}}, \apj,
  479, 427

\bibitem[{{Gonzalez} {et~al.}(1997{\natexlab{b}}){Gonzalez}, {Lambert}, \&
  {Giridhar}}]{Gonzalez97b}
---. 1997{\natexlab{b}}, \apj, 481, 452

\bibitem[{{Griffin} {et~al.}(2010){Griffin}, {Abergel}, {Abreu}, {Ade},
  {Andr{\'e}}, {Augueres}, {Babbedge}, {Bae}, {Baillie}, {Baluteau}, {Barlow},
  {Bendo}, {Benielli}, {Bock}, {Bonhomme}, {Brisbin}, {Brockley-Blatt},
  {Caldwell}, {Cara}, {Castro-Rodriguez}, {Cerulli}, {Chanial}, {Chen},
  {Clark}, {Clements}, {Clerc}, {Coker}, {Communal}, {Conversi}, {Cox},
  {Crumb}, {Cunningham}, {Daly}, {Davis}, {de Antoni}, {Delderfield}, {Devin},
  {di Giorgio}, {Didschuns}, {Dohlen}, {Donati}, {Dowell}, {Dowell}, {Duband},
  {Dumaye}, {Emery}, {Ferlet}, {Ferrand}, {Fontignie}, {Fox}, {Franceschini},
  {Frerking}, {Fulton}, {Garcia}, {Gastaud}, {Gear}, {Glenn}, {Goizel},
  {Griffin}, {Grundy}, {Guest}, {Guillemet}, {Hargrave}, {Harwit}, {Hastings},
  {Hatziminaoglou}, {Herman}, {Hinde}, {Hristov}, {Huang}, {Imhof}, {Isaak},
  {Israelsson}, {Ivison}, {Jennings}, {Kiernan}, {King}, {Lange}, {Latter},
  {Laurent}, {Laurent}, {Leeks}, {Lellouch}, {Levenson}, {Li}, {Li},
  {Lilienthal}, {Lim}, {Liu}, {Lu}, {Madden}, {Mainetti}, {Marliani}, {McKay},
  {Mercier}, {Molinari}, {Morris}, {Moseley}, {Mulder}, {Mur}, {Naylor},
  {Nguyen}, {O'Halloran}, {Oliver}, {Olofsson}, {Olofsson}, {Orfei}, {Page},
  {Pain}, {Panuzzo}, {Papageorgiou}, {Parks}, {Parr-Burman}, {Pearce},
  {Pearson}, {P{\'e}rez-Fournon}, {Pinsard}, {Pisano}, {Podosek}, {Pohlen},
  {Polehampton}, {Pouliquen}, {Rigopoulou}, {Rizzo}, {Roseboom}, {Roussel},
  {Rowan-Robinson}, {Rownd}, {Saraceno}, {Sauvage}, {Savage}, {Savini},
  {Sawyer}, {Scharmberg}, {Schmitt}, {Schneider}, {Schulz}, {Schwartz},
  {Shafer}, {Shupe}, {Sibthorpe}, {Sidher}, {Smith}, {Smith}, {Smith},
  {Spencer}, {Stobie}, {Sudiwala}, {Sukhatme}, {Surace}, {Stevens}, {Swinyard},
  {Trichas}, {Tourette}, {Triou}, {Tseng}, {Tucker}, {Turner}, {Vaccari},
  {Valtchanov}, {Vigroux}, {Virique}, {Voellmer}, {Walker}, {Ward}, {Waskett},
  {Weilert}, {Wesson}, {White}, {Whitehouse}, {Wilson}, {Winter}, {Woodcraft},
  {Wright}, {Xu}, {Zavagno}, {Zemcov}, {Zhang}, \& {Zonca}}]{Griffin10}
{Griffin}, M.~J., {Abergel}, A., {Abreu}, A., {et~al.} 2010, \aap, 518, L3

\bibitem[{{Guzman-Ramirez} {et~al.}(2015){Guzman-Ramirez}, {Lagadec}, {Wesson},
  {Zijlstra}, {M{\"u}ller}, {Jones}, {Boffin}, {Sloan}, {Redman}, {Smette},
  {Karakas}, \& {Nyman}}]{Guzman-Ramirez15}
{Guzman-Ramirez}, L., {Lagadec}, E., {Wesson}, R., {et~al.} 2015, \mnras, 451,
  L1

\bibitem[{{Hardy} {et~al.}(2016){Hardy}, {Schreiber}, {Parsons}, {Caceres},
  {Brinkworth}, {Veras}, {G{\"a}nsicke}, {Marsh}, \& {Cieza}}]{Hardy16}
{Hardy}, A., {Schreiber}, M.~R., {Parsons}, S.~G., {et~al.} 2016, \mnras, 459,
  4518

\bibitem[{{He} {et~al.}(2014){He}, {Szczerba}, {Hasegawa}, \& {Schmidt}}]{He14}
{He}, J.~H., {Szczerba}, R., {Hasegawa}, T.~I., \& {Schmidt}, M.~R. 2014,
  \apjs, 210, 26

\bibitem[{{Henning} {et~al.}(1995){Henning}, {Begemann}, {Mutschke}, \&
  {Dorschner}}]{Henning95}
{Henning}, T., {Begemann}, B., {Mutschke}, H., \& {Dorschner}, J. 1995, \aaps,
  112, 143

\bibitem[{{Henning} \& {Mutschke}(1997)}]{Henning97}
{Henning}, T., \& {Mutschke}, H. 1997, \aap, 327, 743

\bibitem[{{Herter} {et~al.}(2012){Herter}, {Adams}, {De Buizer}, {Gull},
  {Schoenwald}, {Henderson}, {Keller}, {Nikola}, {Stacey}, \&
  {Vacca}}]{Herter12}
{Herter}, T.~L., {Adams}, J.~D., {De Buizer}, J.~M., {et~al.} 2012, \apjl, 749,
  L18

\bibitem[{{Hillen} {et~al.}(2015){Hillen}, {de Vries}, {Menu}, {Van Winckel},
  {Min}, \& {Mulders}}]{Hillen15}
{Hillen}, M., {de Vries}, B.~L., {Menu}, J., {et~al.} 2015, \aap, 578, A40

\bibitem[{{Hillen} {et~al.}(2017){Hillen}, {Van Winckel}, {Menu}, {Manick},
  {Debosscher}, {Min}, {de Wit}, {Verhoelst}, {Kamath}, \& {Waters}}]{Hillen17}
{Hillen}, M., {Van Winckel}, H., {Menu}, J., {et~al.} 2017, \aap, 599, A41

\bibitem[{{Hinkle} {et~al.}(2007){Hinkle}, {Brittain}, \& {Lambert}}]{Hinkle07}
{Hinkle}, K.~H., {Brittain}, S.~D., \& {Lambert}, D.~L. 2007, \apj, 664, 501

\bibitem[{{Honda} {et~al.}(2004){Honda}, {Kataza}, {Okamoto}, {Miyata},
  {Yamashita}, {Sako}, {Fujiyoshi}, {Ito}, {Okada}, {Sakon}, \&
  {Onaka}}]{Honda04}
{Honda}, M., {Kataza}, H., {Okamoto}, Y.~K., {et~al.} 2004, \apjl, 610, L49

\bibitem[{{Iben}(1981)}]{Iben81}
{Iben}, Jr., I. 1981, \apj, 246, 278

\bibitem[{{Jaeger} {et~al.}(1998{\natexlab{a}}){Jaeger}, {Molster},
  {Dorschner}, {Henning}, {Mutschke}, \& {Waters}}]{Jaeger98a}
{Jaeger}, C., {Molster}, F.~J., {Dorschner}, J., {et~al.} 1998{\natexlab{a}},
  \aap, 339, 904

\bibitem[{{Jaeger} {et~al.}(1998{\natexlab{b}}){Jaeger}, {Mutschke}, \&
  {Henning}}]{Jaeger98b}
{Jaeger}, C., {Mutschke}, H., \& {Henning}, T. 1998{\natexlab{b}}, \aap, 332,
  291

\bibitem[{{Juh{\'a}sz} {et~al.}(2010){Juh{\'a}sz}, {Bouwman}, {Henning},
  {Acke}, {van den Ancker}, {Meeus}, {Dominik}, {Min}, {Tielens}, \&
  {Waters}}]{Juhasz10}
{Juh{\'a}sz}, A., {Bouwman}, J., {Henning}, T., {et~al.} 2010, \apj, 721, 431

\bibitem[{{Jura}(1986)}]{Jura86}
{Jura}, M. 1986, \apj, 309, 732

\bibitem[{{Karakas} \& {Lattanzio}(2014)}]{Karakas14}
{Karakas}, A.~I., \& {Lattanzio}, J.~C. 2014, \pasa, 31, e030

\bibitem[{{Kashi} \& {Soker}(2011)}]{Kashi11}
{Kashi}, A., \& {Soker}, N. 2011, \mnras, 417, 1466

\bibitem[{{Kastner} {et~al.}(2004){Kastner}, {Huenemoerder}, {Schulz},
  {Canizares}, {Li}, \& {Weintraub}}]{Kastner04}
{Kastner}, J.~H., {Huenemoerder}, D.~P., {Schulz}, N.~S., {et~al.} 2004, \apjl,
  605, L49

\bibitem[{{Kastner} {et~al.}(2016){Kastner}, {Principe}, {Punzi}, {Stelzer},
  {Gorti}, {Pascucci}, \& {Argiroffi}}]{Kastner16}
{Kastner}, J.~H., {Principe}, D.~A., {Punzi}, K., {et~al.} 2016, \aj, 152, 3

\bibitem[{{Kemper} {et~al.}(2002){Kemper}, {de Koter}, {Waters}, {Bouwman}, \&
  {Tielens}}]{Kemper02}
{Kemper}, F., {de Koter}, A., {Waters}, L.~B.~F.~M., {Bouwman}, J., \&
  {Tielens}, A.~G.~G.~M. 2002, \aap, 384, 585

\bibitem[{{Kemper} {et~al.}(2004){Kemper}, {Vriend}, \& {Tielens}}]{Kemper04}
{Kemper}, F., {Vriend}, W.~J., \& {Tielens}, A.~G.~G.~M. 2004, \apj, 609, 826

\bibitem[{{Kemper} {et~al.}(2001){Kemper}, {Waters}, {de Koter}, \&
  {Tielens}}]{Kemper01}
{Kemper}, F., {Waters}, L.~B.~F.~M., {de Koter}, A., \& {Tielens}, A.~G.~G.~M.
  2001, \aap, 369, 132

\bibitem[{{Kiss} {et~al.}(2007){Kiss}, {Derekas}, {Szab{\'o}}, {Bedding}, \&
  {Szabados}}]{Kiss07}
{Kiss}, L.~L., {Derekas}, A., {Szab{\'o}}, G.~M., {Bedding}, T.~R., \&
  {Szabados}, L. 2007, \mnras, 375, 1338

\bibitem[{{Koike} {et~al.}(2003){Koike}, {Chihara}, {Tsuchiyama}, {Suto},
  {Sogawa}, \& {Okuda}}]{Koike03}
{Koike}, C., {Chihara}, H., {Tsuchiyama}, A., {et~al.} 2003, \aap, 399, 1101

\bibitem[{{Kukarkin}(1958)}]{Kukarkin58}
{Kukarkin}, B.~V. 1958, in IAU Symposium, Vol.~5, Comparison of the Large-Scale
  Structure of the Galactic System with that of Other Stellar Systems, ed.
  N.~G. {Roman}, 49

\bibitem[{Leinonen(2012)}]{pymiecoated}
Leinonen, J. 2012, {Python code for calculating Mie scattering from single and
  dual-layered spheres}, v0.2.0,  Python.
\newblock \url{https://pypi.python.org/pypi/pymiecoated}

\bibitem[{{Lisse} {et~al.}(2017){Lisse}, {Christian}, {Wolk}, {G{\"u}nther},
  {Chen}, \& {Grady}}]{Lisse17}
{Lisse}, C.~M., {Christian}, D.~J., {Wolk}, S.~J., {et~al.} 2017, \aj, 153, 62

\bibitem[{{L{\"u}} {et~al.}(2013){L{\"u}}, {Zhu}, \& {Podsiadlowski}}]{Lu13}
{L{\"u}}, G., {Zhu}, C., \& {Podsiadlowski}, P. 2013, \apj, 768, 193

\bibitem[{{Maas} {et~al.}(2005){Maas}, {Van Winckel}, \& {Lloyd
  Evans}}]{Maas05}
{Maas}, T., {Van Winckel}, H., \& {Lloyd Evans}, T. 2005, \aap, 429, 297

\bibitem[{{Maas} {et~al.}(2002){Maas}, {Van Winckel}, \& {Waelkens}}]{Maas02}
{Maas}, T., {Van Winckel}, H., \& {Waelkens}, C. 2002, \aap, 386, 504

\bibitem[{{McDonald} {et~al.}(2015){McDonald}, {Zijlstra}, {Lagadec}, {Sloan},
  {Boyer}, {Matsuura}, {Smith}, {Smith}, {Yates}, {van Loon}, {Jones},
  {Ramstedt}, {Avison}, {Justtanont}, {Olofsson}, {Blommaert}, {Goldman}, \&
  {Groenewegen}}]{McDonald15}
{McDonald}, I., {Zijlstra}, A.~A., {Lagadec}, E., {et~al.} 2015, \mnras, 453,
  4324

\bibitem[{{Min} {et~al.}(2003){Min}, {Hovenier}, \& {de Koter}}]{Min03}
{Min}, M., {Hovenier}, J.~W., \& {de Koter}, A. 2003, \aap, 404, 35

\bibitem[{{Min} {et~al.}(2005){Min}, {Hovenier}, \& {de Koter}}]{Min05}
---. 2005, \aap, 432, 909

\bibitem[{{Mitchell} \& {Robinson}(1981)}]{Mitchell81}
{Mitchell}, R.~M., \& {Robinson}, G. 1981, \mnras, 196, 801

\bibitem[{{Molster} {et~al.}(2002{\natexlab{a}}){Molster}, {Waters}, \&
  {Tielens}}]{Molster02a}
{Molster}, F.~J., {Waters}, L.~B.~F.~M., \& {Tielens}, A.~G.~G.~M.
  2002{\natexlab{a}}, \aap, 382, 222

\bibitem[{{Molster} {et~al.}(2002{\natexlab{b}}){Molster}, {Waters}, {Tielens},
  \& {Barlow}}]{Molster02b}
{Molster}, F.~J., {Waters}, L.~B.~F.~M., {Tielens}, A.~G.~G.~M., \& {Barlow},
  M.~J. 2002{\natexlab{b}}, \aap, 382, 184

\bibitem[{{Molster} {et~al.}(2002{\natexlab{c}}){Molster}, {Waters}, {Tielens},
  {Koike}, \& {Chihara}}]{Molster02c}
{Molster}, F.~J., {Waters}, L.~B.~F.~M., {Tielens}, A.~G.~G.~M., {Koike}, C.,
  \& {Chihara}, H. 2002{\natexlab{c}}, \aap, 382, 241

\bibitem[{{Murakami} {et~al.}(2007){Murakami}, {Baba}, {Barthel}, {Clements},
  {Cohen}, {Doi}, {Enya}, {Figueredo}, {Fujishiro}, {Fujiwara}, {Fujiwara},
  {Garcia-Lario}, {Goto}, {Hasegawa}, {Hibi}, {Hirao}, {Hiromoto}, {Hong},
  {Imai}, {Ishigaki}, {Ishiguro}, {Ishihara}, {Ita}, {Jeong}, {Jeong},
  {Kaneda}, {Kataza}, {Kawada}, {Kawai}, {Kawamura}, {Kessler}, {Kester},
  {Kii}, {Kim}, {Kim}, {Kobayashi}, {Koo}, {Kwon}, {Lee}, {Lorente}, {Makiuti},
  {Matsuhara}, {Matsumoto}, {Matsuo}, {Matsuura}, {M{\"u}ller}, {Murakami},
  {Nagata}, {Nakagawa}, {Naoi}, {Narita}, {Noda}, {Oh}, {Ohnishi}, {Ohyama},
  {Okada}, {Okuda}, {Oliver}, {Onaka}, {Ootsubo}, {Oyabu}, {Pak}, {Park},
  {Pearson}, {Rowan-Robinson}, {Saito}, {Sakon}, {Salama}, {Sato}, {Savage},
  {Serjeant}, {Shibai}, {Shirahata}, {Sohn}, {Suzuki}, {Takagi}, {Takahashi},
  {Tanab{\'e}}, {Takeuchi}, {Takita}, {Thomson}, {Uemizu}, {Ueno}, {Usui},
  {Verdugo}, {Wada}, {Wang}, {Watabe}, {Watarai}, {White}, {Yamamura},
  {Yamauchi}, \& {Yasuda}}]{Murakami07}
{Murakami}, H., {Baba}, H., {Barthel}, P., {et~al.} 2007, \pasj, 59, S369

\bibitem[{{Neugebauer} {et~al.}(1984){Neugebauer}, {Habing}, {van Duinen},
  {Aumann}, {Baud}, {Beichman}, {Beintema}, {Boggess}, {Clegg}, {de Jong},
  {Emerson}, {Gautier}, {Gillett}, {Harris}, {Hauser}, {Houck}, {Jennings},
  {Low}, {Marsden}, {Miley}, {Olnon}, {Pottasch}, {Raimond}, {Rowan-Robinson},
  {Soifer}, {Walker}, {Wesselius}, \& {Young}}]{Neugebauer84}
{Neugebauer}, G., {Habing}, H.~J., {van Duinen}, R., {et~al.} 1984, \apjl, 278,
  L1

\bibitem[{{Nuth} \& {Hecht}(1990)}]{Nuth90}
{Nuth}, III, J.~A., \& {Hecht}, J.~H. 1990, \apss, 163, 79

\bibitem[{{O'Connell}(1961)}]{O'Connell61}
{O'Connell}, D.~J.~K. 1961, Ricerche Astronomiche, 6

\bibitem[{{Payne-Gaposchkin}(1952)}]{Payne-Gaposchkin52}
{Payne-Gaposchkin}, C. 1952, \nat, 170, 223

\bibitem[{{Pegourie}(1988)}]{Pegourie88}
{Pegourie}, B. 1988, \aap, 194, 335

\bibitem[{{Percy}(1993)}]{Percy93}
{Percy}, J.~R. 1993, in Astronomical Society of the Pacific Conference Series,
  Vol.~45, Luminous High-Latitude Stars, ed. D.~D. {Sasselov}, 295

\bibitem[{{Percy} \& {Coffey}(2005)}]{Percy05}
{Percy}, J.~R., \& {Coffey}, J. 2005, Journal of the American Association of
  Variable Star Observers (JAAVSO), 33, 193

\bibitem[{{Percy} {et~al.}(1991){Percy}, {Sasselov}, {Alfred}, \&
  {Scott}}]{Percy91}
{Percy}, J.~R., {Sasselov}, D.~D., {Alfred}, A., \& {Scott}, G. 1991, \apj,
  375, 691

\bibitem[{{Percy} \& {Ursprung}(2006)}]{Percy06}
{Percy}, J.~R., \& {Ursprung}, C. 2006, Journal of the American Association of
  Variable Star Observers (JAAVSO), 34, 125

\bibitem[{{Pilbratt}(2003)}]{Pilbratt03}
{Pilbratt}, G.~L. 2003, in \procspie, Vol. 4850, IR Space Telescopes and
  Instruments, ed. J.~C. {Mather}, 586--597

\bibitem[{{Poglitsch} {et~al.}(2010){Poglitsch}, {Waelkens}, {Geis},
  {Feuchtgruber}, {Vandenbussche}, {Rodriguez}, {Krause}, {Renotte}, {van
  Hoof}, {Saraceno}, {Cepa}, {Kerschbaum}, {Agn{\`e}se}, {Ali}, {Altieri},
  {Andreani}, {Augueres}, {Balog}, {Barl}, {Bauer}, {Belbachir}, {Benedettini},
  {Billot}, {Boulade}, {Bischof}, {Blommaert}, {Callut}, {Cara}, {Cerulli},
  {Cesarsky}, {Contursi}, {Creten}, {De Meester}, {Doublier}, {Doumayrou},
  {Duband}, {Exter}, {Genzel}, {Gillis}, {Gr{\"o}zinger}, {Henning},
  {Herreros}, {Huygen}, {Inguscio}, {Jakob}, {Jamar}, {Jean}, {de Jong},
  {Katterloher}, {Kiss}, {Klaas}, {Lemke}, {Lutz}, {Madden}, {Marquet},
  {Martignac}, {Mazy}, {Merken}, {Montfort}, {Morbidelli}, {M{\"u}ller},
  {Nielbock}, {Okumura}, {Orfei}, {Ottensamer}, {Pezzuto}, {Popesso},
  {Putzeys}, {Regibo}, {Reveret}, {Royer}, {Sauvage}, {Schreiber}, {Stegmaier},
  {Schmitt}, {Schubert}, {Sturm}, {Thiel}, {Tofani}, {Vavrek}, {Wetzstein},
  {Wieprecht}, \& {Wiezorrek}}]{Poglitsch10}
{Poglitsch}, A., {Waelkens}, C., {Geis}, N., {et~al.} 2010, \aap, 518, L2

\bibitem[{{Pollack} {et~al.}(1994){Pollack}, {Hollenbach}, {Beckwith},
  {Simonelli}, {Roush}, \& {Fong}}]{Pollack94}
{Pollack}, J.~B., {Hollenbach}, D., {Beckwith}, S., {et~al.} 1994, \apj, 421,
  615

\bibitem[{{Preston} {et~al.}(1963){Preston}, {Krzeminski}, {Smak}, \&
  {Williams}}]{Preston63}
{Preston}, G.~W., {Krzeminski}, W., {Smak}, J., \& {Williams}, J.~A. 1963,
  \apj, 137, 401

\bibitem[{{Rao} \& {Giridhar}(2014)}]{Rao14}
{Rao}, S.~S., \& {Giridhar}, S. 2014, \rmxaa, 50, 49

\bibitem[{{Robinson} \& {Hyland}(1977)}]{Robinson77}
{Robinson}, G., \& {Hyland}, A.~R. 1977, \mnras, 180, 495

\bibitem[{{Rouleau} {et~al.}(1997){Rouleau}, {Henning}, \&
  {Stognienko}}]{Rouleau97}
{Rouleau}, F., {Henning}, T., \& {Stognienko}, R. 1997, \aap, 322, 633

\bibitem[{{Sakon} {et~al.}(2016){Sakon}, {Sako}, {Onaka}, {Nozawa}, {Kimura},
  {Fujiyoshi}, {Shimonishi}, {Usui}, {Takahashi}, {Ohsawa}, {Arai}, {Uemura},
  {Nagayama}, {Koo}, \& {Kozasa}}]{Sakon16}
{Sakon}, I., {Sako}, S., {Onaka}, T., {et~al.} 2016, \apj, 817, 145

\bibitem[{{Schoenberner}(1983)}]{Schoenberner83}
{Schoenberner}, D. 1983, \apj, 272, 708

\bibitem[{{Shenton} {et~al.}(1995){Shenton}, {Evans}, \&
  {Williams}}]{Shenton95}
{Shenton}, M., {Evans}, A., \& {Williams}, P.~M. 1995, \mnras, 273, 906

\bibitem[{{Shenton} {et~al.}(1992){Shenton}, {Albinson}, {Barrett}, {Davies},
  {Evans}, {Goldsmith}, {Hutchinson}, {Maddison}, \& {Weight}}]{Shenton92}
{Shenton}, M., {Albinson}, J.~S., {Barrett}, P., {et~al.} 1992, \aap, 262, 138

\bibitem[{{Skrutskie} {et~al.}(2006){Skrutskie}, {Cutri}, {Stiening},
  {Weinberg}, {Schneider}, {Carpenter}, {Beichman}, {Capps}, {Chester},
  {Elias}, {Huchra}, {Liebert}, {Lonsdale}, {Monet}, {Price}, {Seitzer},
  {Jarrett}, {Kirkpatrick}, {Gizis}, {Howard}, {Evans}, {Fowler}, {Fullmer},
  {Hurt}, {Light}, {Kopan}, {Marsh}, {McCallon}, {Tam}, {Van Dyk}, \&
  {Wheelock}}]{Skrutskie06}
{Skrutskie}, M.~F., {Cutri}, R.~M., {Stiening}, R., {et~al.} 2006, \aj, 131,
  1163

\bibitem[{{Smith} {et~al.}(1994){Smith}, {Aitken}, \& {Roche}}]{Smith94}
{Smith}, C.~H., {Aitken}, D.~K., \& {Roche}, P.~F. 1994, \mnras, 267, 225

\bibitem[{{Sogawa} \& {Kozasa}(1999)}]{Sogawa99}
{Sogawa}, H., \& {Kozasa}, T. 1999, \apjl, 516, L33

\bibitem[{{Speck} {et~al.}(2009){Speck}, {Corman}, {Wakeman}, {Wheeler}, \&
  {Thompson}}]{Speck09}
{Speck}, A.~K., {Corman}, A.~B., {Wakeman}, K., {Wheeler}, C.~H., \&
  {Thompson}, G. 2009, \apj, 691, 1202

\bibitem[{{Speck} {et~al.}(2005){Speck}, {Thompson}, \& {Hofmeister}}]{Speck05}
{Speck}, A.~K., {Thompson}, G.~D., \& {Hofmeister}, A.~M. 2005, \apj, 634, 426

\bibitem[{{Straniero} {et~al.}(1997){Straniero}, {Chieffi}, {Limongi}, {Busso},
  {Gallino}, \& {Arlandini}}]{Straniero97}
{Straniero}, O., {Chieffi}, A., {Limongi}, M., {et~al.} 1997, \apj, 478, 332

\bibitem[{{Straniero} {et~al.}(2006){Straniero}, {Gallino}, \&
  {Cristallo}}]{Straniero06}
{Straniero}, O., {Gallino}, R., \& {Cristallo}, S. 2006, Nuclear Physics A,
  777, 311

\bibitem[{{Suh}(2000)}]{Suh00}
{Suh}, K.-W. 2000, \mnras, 315, 740

\bibitem[{{Suh}(2002)}]{Suh02}
---. 2002, \mnras, 332, 513

\bibitem[{{Suh}(2016)}]{Suh16}
---. 2016, \apj, 819, 61

\bibitem[{{Sylvester} {et~al.}(1999){Sylvester}, {Kemper}, {Barlow}, {de Jong},
  {Waters}, {Tielens}, \& {Omont}}]{Sylvester99}
{Sylvester}, R.~J., {Kemper}, F., {Barlow}, M.~J., {et~al.} 1999, \aap, 352,
  587

\bibitem[{{Takeuti} \& {Petersen}(1983)}]{Takeuti83}
{Takeuti}, M., \& {Petersen}, J.~O. 1983, \aap, 117, 352

\bibitem[{{Temi} {et~al.}(2012){Temi}, {Marcum}, {Miller}, {Dunham}, {McLean},
  {Wolf}, {Becklin}, {Bida}, {Brewster}, {Casey}, {Collins}, {Horner}, {Jakob},
  {Jensen}, {Killebrew}, {Lampater}, {Mandushev}, {Meyer}, {Pfueller},
  {Reinacher}, {Rho}, {Roellig}, {Savage}, {Smith}, {Teufel}, \&
  {Wiedemann}}]{Temi12}
{Temi}, P., {Marcum}, P.~M., {Miller}, W.~E., {et~al.} 2012, in \procspie, Vol.
  8444, Ground-based and Airborne Telescopes IV, 844414

\bibitem[{{Tielens} {et~al.}(1998){Tielens}, {Waters}, {Molster}, \&
  {Justtanont}}]{Tielens98}
{Tielens}, A.~G.~G.~M., {Waters}, L.~B.~F.~M., {Molster}, F.~J., \&
  {Justtanont}, K. 1998, \apss, 255, 415

\bibitem[{{Tuchman} {et~al.}(1993){Tuchman}, {Lebre}, {Mennessier}, \&
  {Yarri}}]{Tuchman93}
{Tuchman}, Y., {Lebre}, A., {Mennessier}, M.~O., \& {Yarri}, A. 1993, \aap,
  271, 501

\bibitem[{{van Boekel} {et~al.}(2005){van Boekel}, {Min}, {Waters}, {de Koter},
  {Dominik}, {van den Ancker}, \& {Bouwman}}]{vanBoekel05}
{van Boekel}, R., {Min}, M., {Waters}, L.~B.~F.~M., {et~al.} 2005, \aap, 437,
  189

\bibitem[{{van Boekel} {et~al.}(2003){van Boekel}, {Waters}, {Dominik},
  {Bouwman}, {de Koter}, {Dullemond}, \& {Paresce}}]{vanBoekel03}
{van Boekel}, R., {Waters}, L.~B.~F.~M., {Dominik}, C., {et~al.} 2003, \aap,
  400, L21

\bibitem[{{van Winckel}(2003)}]{vanWinckel03}
{van Winckel}, H. 2003, \araa, 41, 391

\bibitem[{{Van Winckel} {et~al.}(1999){Van Winckel}, {Waelkens}, {Fernie}, \&
  {Waters}}]{vanWinckel99}
{Van Winckel}, H., {Waelkens}, C., {Fernie}, J.~D., \& {Waters}, L.~B.~F.~M.
  1999, \aap, 343, 202

\bibitem[{{Van Winckel} {et~al.}(1995){Van Winckel}, {Waelkens}, \&
  {Waters}}]{vanWinckel95}
{Van Winckel}, H., {Waelkens}, C., \& {Waters}, L.~B.~F.~M. 1995, \aap, 293

\bibitem[{{Van Winckel} {et~al.}(1998){Van Winckel}, {Waelkens}, {Waters},
  {Molster}, {Udry}, \& {Bakker}}]{vanWinckel98}
{Van Winckel}, H., {Waelkens}, C., {Waters}, L.~B.~F.~M., {et~al.} 1998, \aap,
  336, L17

\bibitem[{{Waelkens} \& {Waters}(1993)}]{Waelkens93}
{Waelkens}, C., \& {Waters}, L.~B.~F.~M. 1993, in Astronomical Society of the
  Pacific Conference Series, Vol.~45, Luminous High-Latitude Stars, ed. D.~D.
  {Sasselov}, 219

\bibitem[{{Waters} {et~al.}(1992){Waters}, {Trams}, \& {Waelkens}}]{Waters92}
{Waters}, L.~B.~F.~M., {Trams}, N.~R., \& {Waelkens}, C. 1992, \aap, 262, L37

\bibitem[{{Waters} {et~al.}(1993){Waters}, {Waelkens}, {Mayor}, \&
  {Trams}}]{Waters93}
{Waters}, L.~B.~F.~M., {Waelkens}, C., {Mayor}, M., \& {Trams}, N.~R. 1993,
  \aap, 269, 242

\bibitem[{{Wright} {et~al.}(2010){Wright}, {Eisenhardt}, {Mainzer}, {Ressler},
  {Cutri}, {Jarrett}, {Kirkpatrick}, {Padgett}, {McMillan}, {Skrutskie},
  {Stanford}, {Cohen}, {Walker}, {Mather}, {Leisawitz}, {Gautier}, {McLean},
  {Benford}, {Lonsdale}, {Blain}, {Mendez}, {Irace}, {Duval}, {Liu}, {Royer},
  {Heinrichsen}, {Howard}, {Shannon}, {Kendall}, {Walsh}, {Larsen}, {Cardon},
  {Schick}, {Schwalm}, {Abid}, {Fabinsky}, {Naes}, \& {Tsai}}]{Wright10}
{Wright}, E.~L., {Eisenhardt}, P.~R.~M., {Mainzer}, A.~K., {et~al.} 2010, \aj,
  140, 1868

\bibitem[{{Young} {et~al.}(2012){Young}, {Becklin}, {Marcum}, {Roellig}, {De
  Buizer}, {Herter}, {G{\"u}sten}, {Dunham}, {Temi}, {Andersson}, {Backman},
  {Burgdorf}, {Caroff}, {Casey}, {Davidson}, {Erickson}, {Gehrz}, {Harper},
  {Harvey}, {Helton}, {Horner}, {Howard}, {Klein}, {Krabbe}, {McLean}, {Meyer},
  {Miles}, {Morris}, {Reach}, {Rho}, {Richter}, {Roeser}, {Sandell}, {Sankrit},
  {Savage}, {Smith}, {Shuping}, {Vacca}, {Vaillancourt}, {Wolf}, \&
  {Zinnecker}}]{Young12}
{Young}, E.~T., {Becklin}, E.~E., {Marcum}, P.~M., {et~al.} 2012, \apjl, 749,
  L17

\end{thebibliography}

\end{document}